\documentclass[apj,square,numbers]{emulateapj}
\usepackage{graphicx}
\usepackage{latexsym}
\usepackage{epsfig}
\usepackage{color}
\usepackage[colorlinks, linkcolor=blue, citecolor=blue, urlcolor=blue]{hyperref}
\usepackage{apjfonts}

\definecolor{Black}{named}{Black}
\definecolor{Red}{named}{Red}

\shorttitle{Tomography of the $\gamma$-ray diffuse extragalactic signal}
\shortauthors{Xia et al.}

\begin{document}

\title{Tomography of the {\it Fermi}-LAT $\gamma$-ray diffuse extragalactic signal via cross-correlations with galaxy catalogs}

\author{Jun-Qing Xia$^{1,2}$}
\author{Alessandro Cuoco$^{3,4,5}$}
\author{Enzo Branchini$^{6,7,8}$}
\author{Matteo Viel$^{9,10}$}

\affil{$^1$ Key Laboratory of Particle Astrophysics, Institute of High Energy Physics, Chinese Academy of Science, P. O. Box 918-3, Beijing 100049, P. R. China.}
\affil{$^2$ Collaborative Innovation Center of Modern Astronomy and Space Exploration, P. R. China.}
\affil{$^3$ Department of Physics, University of Torino, via P. Giuria 1, 10125 Torino, Italy.}
\affil{$^4$ Istituto Nazionale di Fisica Nucleare, via P. Giuria 1, 10125 Torino, Italy.}
\affil{$^5$ Stockholm University -  AlbaNova University Center, Fysikum, SE-10691, Stockholm, Sweden.}
\affil{$^6$ Dipartimento di Matematica e  Fisica, Universit\`a degli Studi ``Roma Tre'', via della Vasca Navale 84, I-00146 Roma, Italy.}
\affil{$^7$ INFN, Sezione di Roma Tre, via della Vasca Navale 84, I-00146 Roma, Italy.}
\affil{$^8$ INAF, Osservatorio Astronomico di Roma, Monte Porzio Catone, Italy.}
\affil{$^9$ INAF Osservatorio Astronomico di Trieste, Via G. B. Tiepolo 11, I-34141, Trieste, Italy.}
\affil{$^{10}$ INFN, Sezione di Trieste, via Valerio 2, I-34127, Trieste, Italy.}


\begin{abstract}
Building on our previous cross-correlation analysis (Xia et al. 2011)
between the  isotropic $\gamma$-ray background (IGRB) and
different tracers of the large-scale structure of the universe, we
update our results using 60-months of data from the
Large Area Telescope (LAT) on board the Fermi Gamma-ray Space Telescope ({\it Fermi}).  We perform a
cross-correlation analysis both in configuration and spherical
harmonics space between the IGRB  and objects that may
trace the astrophysical sources of the IGRB: QSOs in the Sloan Digital
Sky Survey-DR6, the SDSS-DR8 Main Galaxy Sample,
 Luminous Red Galaxies (LRGs) in the SDSS catalog,
infrared selected galaxies in the Two Micron All Sky Survey (2MASS),
and radio galaxies in the NRAO VLA Sky Survey (NVSS).
The benefit of correlating the {\it Fermi}-LAT signal with catalogs of objects
at various redshifts is to provide   tomographic information on the IGRB
which is crucial to separate the various contributions and to
clarify its origin.

The main result is that, unlike in our previous analysis, we now
observe a \mbox{significant ($>$3.5\, $\sigma$)} cross-correlation signal on angular scales
smaller than $1^{\circ}$
in the NVSS,  2MASS and QSO cases and, at
lower statistical significance ($\sim$3.0\, $\sigma$), with SDSS galaxies.
The  signal is stronger in two energy bands, $E>0.5$~GeV and
$E>1$~GeV, but also seen at $E>10$~GeV.
No cross-correlation signal is detected between {\it Fermi} data and the LRGs.
These results are robust against the choice of the statistical estimator,
estimate of errors, map cleaning procedure and instrumental effects.

Finally, we test the hypothesis that the IGRB observed by {\it Fermi}-LAT
originates from the summed contributions of three types
of  unresolved extragalactic sources:
BL Lacertae objects (BL Lacs), Flat Spectrum Radio Quasars (FSRQs)
and Star-Forming Galaxies (SFGs). We find that a model
in which the IGRB  is mainly   produced by SFGs ($72^{+23}_{-37}$ \%    with $2\sigma$ errors),
with BL Lacs and FSRQs giving a minor contribution,
provides a good fit to the data.
We also consider a possible contribution from Misaligned Active Galactic Nuclei (MAGNs),
and we find that, depending on the details of the model and its uncertainty,
they can also provide a substantial contribution, partly degenerate with the SFG one.

\end{abstract}


\maketitle


\section{Introduction}

The origin of the extragalactic $\gamma$-ray background (EGB) is still
unknown. After its detection and early attempts to unveil its origin
\citep{1972ApJ...177..341K,1973ApJ...186L..99F,
  1982A&A...105..164M,padovani,1996ApJ...464..600S,1998ApJ...494..523S,2004JCAP...04..006K,2004ApJ...613..956S}
major advances have recently been possible thanks to
the {\it Fermi} \mbox{$\gamma$-ray} Space Telescope.
Observations with the
Large Area Telescope ({\it Fermi}-LAT) \citep{2009ApJ...697.1071A} are resolving an ever growing number of sources, making  possible
to characterize their properties,  e.g., \cite{AjelloFSRQs, AjelloBLLs}, and to constrain their contribution to the EGB.
These constraints are further complemented by comparing the unresolved EGB with
semi-analytical models of different types of sources,
e.g., \cite{stecker11,makiya,prodanovic,tamborra014,Achermann12,2013ASSP...34..193R}.
Thanks to {\it Fermi}-LAT a sizable fraction of the EGB is starting to be resolved \citep{IGRBII}.
Therefore, to avoid confusion, it is convenient to use a specific term for the unresolved part
which is the quantity we want to study, to distinguish it from  the resolved
point sources  either masked or subtracted. In the following, we indicate the
unresolved component as
the Isotropic Gamma-Ray Background (IGRB).

The study of the IGRB is hampered by
the presence of spurious contributions, like the Galactic foregrounds, or bright point sources,
that, if not properly subtracted,
contaminate the mean signal and generate systematic errors in the
analysis of the true signal.
These systematic uncertainties can, in principle, be mitigated by
considering the angular correlation properties of the  IGRB \citep{FermiAPS}.
In practice, however, the auto-correlation signal is also quite prone
to the aforementioned systematics, which, being not perfectly isotropic,
affect the measurements on various angular scales.
Instead, an effective way to enhance the signal and filter out systematic effects
is to cross-correlate the IGRB with
several different distributions of extragalactic  objects that may or may not trace
the actual source of the IGRB but certainly do
not correlate with the sources of systematic errors.
This
approach has been proposed and adopted by, e.g., \cite{Cuoco:2007sh}, \cite{2009MNRAS.400.2122A} and \cite{xia11}.
Besides the cross-correlation with catalogs of  extragalactic objects,
a further possibility recently proposed is to cross-correlate the IGRB
with weak gravitational lensing maps \citep{Shirasaki:2014noa, Fornengo:2013rga, Camera:2012cj,Fornengo:2014cya,Camera:2014rja}, which presents the
advantage of tracing the gravitational potential without any bias.  Furthermore, it has been recently shown that cross-correlation
of the IGRB with galaxy catalogs can provide tight constraints on the dark matter annihilation \citep{ando14}.

Results so far have been negative since no clear correlation signal
has been detected with a large statistical significance. For example,
in \cite{xia11} the significance of the correlation between {\it Fermi} data
and SDSS Luminous Red Galaxies (LRGs) was reported to be only at the
2$\sigma$ confidence level.

Nevertheless, these results have been used to set upper limits on the contribution
of different types of potential $\gamma$-ray sources such as blazars and
star-forming galaxies as well as on the mass and cross
section of WIMP dark matter candidates that may also contribute to the
$\gamma$-ray background through their self-annihilation (e.g., \cite{andokomatsu,cholis}).

The goal of this work is to extend and improve on the original study
of \cite{xia11} using the most recent $\gamma$-ray maps obtained with the
{\it Fermi}-LAT.  We estimate the two-point angular cross-correlation function (CCF) and
the cross-angular power spectrum (CAPS) of the {\it Fermi}-LAT IGRB with a
variety of catalogs of objects: SDSS-DR6 quasars
\citep{2009ApJS..180...67R}, SDSS-DR8 Luminous Red Galaxies
\citep{2008arXiv0812.3831A}, NVSS radiogalaxies
\citep{2002MNRAS.337..993B} 2MASS galaxies
\citep{2000AJ....120..298J} and DR8 SDSS
main sample galaxies \citep{dr8}.  These catalogs have in common:  {\it i)} a
large sky coverage that could allow maximizing a potential
cross-correlation signal;  {\it ii)} the fact that they have been already used to
perform quantitative cosmological studies of the Large Scale
Structures (LSS). The fact that they contain different types of sources that span
different ranges of redshifts is important since it increases the sensitivity of
the cross-correlation analysis to  {\it i)} the type of sources that contribute
to the IGRB and {\it ii)} the cosmic epoch in which this contribution has been
provided.

In \cite{xia11} we predicted that after ten years of data taking by the {\it Fermi}-LAT we
would be able to detect a possible contribution to the $\gamma$-ray
background from relatively nearby  ($z \leq 2$) sources  with a
confidence level of $\sim 97$ \%. This prediction was quite conservative
since it was based on the expected Poisson noise level.
Improvements  {\it i)} in the model for the Galactic diffuse signal,
 {\it ii)} in the
characterization of the instrumental point-spread function (PSF)
that allows pushing the analysis to energies lower than 1 GeV
and to scales smaller than $1^{\circ}$, and finally, {\it iii)}
the increase in the number of resolved sources,
allows us to improve our conservative estimate and justifies
our decision to repeat
the cross-correlation analysis using the  5-years
{\it Fermi} maps with energies as small as 0.5 GeV.

Following  \cite{xia11} we compare the results of the cross-correlation analysis with
theoretical predictions obtained under the hypothesis that the diffuse
$\gamma$-ray background has contributions from known extragalactic sources and
set constraints on popular candidates
like galaxies with strong star formation activity and two types of
blazars: the flat spectrum radio quasars (FSRQ) and the BL Lacertae (BL Lac) objects.

In this work we assume a flat Cold Dark Matter model with a
cosmological constant ($\Lambda$CDM) with  cosmological parameters
$\Omega_{\rm b} h^2 = 0.0222$,
$\Omega_{\rm c} h^2 = 0.1189$, $\tau= 0.095$, $h = 0.678$, $\ln{10^{10}A_{\rm s}} = 3.097$ at $k_0=0.05$ Mpc$^{-1}$,
and $n_{\rm s} =0.961$ that are in agreement with recent {\it Planck} results \citep{2013arXiv1303.5079P}.

The layout of the paper is as follows: in Section~\ref{sec:theory} we
briefly review the theoretical background of the cross-correlation
analysis.  In Section~\ref{sec:fermimaps}
we present the {\it Fermi} maps, the various masks
and discuss the procedure adopted to eliminate the potential spurious contributions
to the  extragalactic signal.
The maps of the angular distribution of the extragalactic objects
that we cross-correlate with the {\it Fermi} maps are presented
in Section~\ref{sec:maps}. In Section~\ref{sec:corranalysis}
we describe the statistical estimators used in our cross-correlation analysis,
while in
 Section~\ref{sec:validation} we test the robustness of the results to the cleaning procedure
and to the instrument response modeling and to the data selection. The results are presented in
 Section~\ref{sec:results}, compared to model predictions in
  Section~\ref{sec:chi2}, and discussed in Section~\ref{sec:discussion}
  in which we also summarize our main conclusions.


\section{Theoretical background}
\label{sec:theory}

\begin{table*}
\begin{center}
\caption{
Parameters of the LDDE LFs taken from Ajello {\it et al.} 2012  for FSRQs
and  Ajello {\it et al.} 2014 for BL Lacs.
}
\label{tab:ldde}
\begin{tabular}{lccccccccccccc}

\hline
\hline

Model   & A${}^{a}$         & {$\gamma_1$} &
{L$_*^{b}$}      & {$\gamma_2$} &
{z$_c^*$}                     &
{$p_1^*$}                  & {$\tau$}     &
{$p_2$}              & {$\alpha$} &
{$\mu^*$}              & {$\beta$}    &
{$\sigma$}        \\

\hline

BLLacs1 LDDE & $3.39\times10^{4}$ & $0.27$ &$0.28$ & $1.86$ & $1.34$ & $2.24$ & $4.92$ & $-7.37$ & $4.53\times10^{-2}$ & $2.10$ & $6.46\times10^{-2}$ & $0.26$ \\

BLLacs2 LDDE & $9.20\times10^{2}$ & $1.12$ & $2.43$ & $3.71$ & $1.67$ & $4.50$ & $0.0$ & $-12.88$ & $4.46\times10^{-2}$ & $2.12$ & $6.04\times10^{-2}$ & $0.26$ \\

\hline
FSRQ LDDE$$ & $3.06\times10^{4}$ & $0.21$ &$0.84$ & $1.58$ & $1.47$ & $7.35$ & $0.0$ & $-6.51$ & $0.21$ & $2.44$ & $0.0$ & $0.18$ \\
\hline
\end{tabular} \\
{a}{ In units of $10^{-13}$\,Mpc$^{-3}$  erg$^{-1}$ s.} \, \,
{b}{ In units of $10^{48}$\,erg s$^{-1}$.}
\end{center}
\end{table*}

Here we briefly summarize the theoretical framework adopted in our analysis.
In this work we use the same formalism as in  \cite{xia11} to which we refer the reader for
a more thorough discussion.

Let us consider a  population of $\gamma$-ray sources, $j$, with
power-law energy spectra $I(E) \propto E^{1-\Gamma_j}$
characterized by a luminosity function (LF) $\Phi_j(L_{\gamma},\Gamma,E,z)$
in which we highlight the explicit dependence on the observed $\gamma$-ray energy $E$,
the rest-frame luminosity of the sources $L_{\gamma}$ (generally expressed in erg s$^{-1}$), cosmological redshift $z$,
and photon index $\Gamma_j>1$.
The contribution of  this population to the differential \emph{energy} flux is:
\begin{equation}
\frac{dI_j}{dE}=\frac{c}{4 \pi}
\int \left[ \int_{L_{\rm MIN}}^{L_{\rm MAX}(z)} \!\!\!\!\! \int_{\Gamma_{\rm m}}^{\Gamma_{\rm M}}
\Phi_j(L_{\gamma},\Gamma,(1+z)E,z)L_{\gamma} dL_{\gamma} d\Gamma \right]
\frac{dz}{(1+z)H(z)},
\label{eq:dide}
\end{equation}
where $H(z)=H_0[(1+z)^3\Omega_M+\Omega_{\Lambda}]$ represents the expansion
history in the assumed cosmological model and $(1+z)$ accounts for the
cosmological redshift. All sources along the line of sight contribute to
the integral over $z$.

The integration over $ L_{\gamma}$ is performed within a finite luminosity range. We set the upper value equal to
\begin{equation}
L_{\rm MAX}(z)=4 \pi d_L^2(z) S_{\rm lim} (1+z)^{-2+\Gamma_j},
\end{equation}
where  $d_L$ is the luminosity
distance in the adopted cosmology and $S_{\rm lim}$ is the (energy) flux detection limit.
In general, the flux detection threshold depends on the power-law index. This dependence is strong for
the photon flux and much weaker for the energy flux. For this reason, in this work
we shall ignore the correlation between $S_{\rm lim}$ and $\Gamma_j$.
This implies that resolved sources are excluded and that the integral  (Eq.~\ref{eq:dide}) has  contributions only from
unresolved sources. The lower integration limit, $L_{\rm MIN}$, is taken from recent literature for specific source classes as we shall discuss in the next section.
We note that  the integral converges and
setting $L_{\rm MIN}$ to much smaller values has very little effect on the final results, i.e.,
our results are robust against the value of $L_{\rm MIN}$.
The choice of  $S_{\rm lim}$ depends on the $\gamma$-ray source catalog
used to mask resolved point sources.
In the following we will use the 2FGL \citep{2012ApJS..199...31N}
source catalog and a preliminary version
of the  3FGL \citep{3FGL}
catalog. Typical values of the source detection thresholds
(in units of integrated energy flux above 100 MeV)
are  $5 \times 10^{-12}$  erg cm$^{-2}$s$^{-1}$ and $2.5 \times 10^{-12}$  erg cm$^{-2}$s$^{-1}$
respectively for the 2FGL and 3FGL catalog \citep{3FGL},
with the lower threshold of the 3FGL catalog which is in part due to the
larger dataset used (4 years vs. 2 years) and in part to the improved characterization
of the response of the LAT.
In practice, however,  the luminosity density $\rho_{\gamma}(z)$
that we shall use to characterize the contribution of a given type of sources to the energy flux
is weakly dependent on the detection threshold, i.e., similar results are obtained with
the 3FGL and 2FGL thresholds, even if the former has a deeper reach in flux than the latter.
This reflects the fact that below 100 GeV the bulk of the EGB  is still unresolved in both the 2FGL and 3FGL catalogs.
For this same reason, the energy density is insensitive to
the precise modeling of the detection efficiency, which is not exactly a step function but
represents rather a smooth transition in flux between zero and full efficiency.

In the integration over $\Gamma$ we assume that the
intrinsic distribution of photon indices is a Gaussian, which
implies that for a given redshift $z$ and luminosity $L_{\gamma}$ the LF has the form
\begin{equation}
\label{eq:index}
\Phi(L_{\gamma},z,\Gamma) \propto  e^{-\frac{ (\Gamma-\mu(L_{\gamma}))^2}{2\sigma^2}} \; ,
\end{equation}
where $\mu$ and $\sigma$ are, respectively, the mean and dispersion of the distribution.
The mean  is allowed to be a function of the source luminosity (expressed in units of $10^{48}$ erg s$^{-1}$):
\begin{equation}
\label{eq:blazseq}
\mu(L_{\gamma}) = \mu^* + \beta \times ( \log_{10} (L_{\gamma}) - 46).
\end{equation}
Since in the luminosity range of interest  $\sigma\ll\mu$, as we will see, then
we can approximate the $\Gamma$ distribution with a Dirac delta centred on $\Gamma=\mu$. With this approximation
the integrated flux  $I_j(>E) $ can be
 expressed as
\begin{equation}
I_j(>E) \equiv \int^{\infty}_{E} \frac{dI_j}{dE} dE  =
\frac{cE^{2-\Gamma_j}}{4\pi}  \int   \rho_{\gamma}(z) dz   \; ,
\label{eq:integratedflux}
\end{equation}
where
\begin{equation}
\rho_{\gamma}(z) \equiv \int_{L_{\rm MIN}}^{L_{\rm MAX}(z)} \Phi_j(L_{\gamma},z)L_{\gamma}  \frac{(1+z)^{-\mu_j(L_{\gamma})}}{H(z)} dL_{\gamma}  \;
\label{eq:density}
\end{equation}
is  the  the mean luminosity density at $z$ and
$\Phi_j(L_{\gamma},z)\equiv \Phi_j(L_{\gamma},z,\Gamma=\mu(L_{\gamma}))$.
In this paper we deal with maps of photon counts rather than energy
flux; the photon flux (above energy $E$) being simply $(2-\Gamma_j)/(1-\Gamma_j) \times I_j(>E)/E$.
We will consider maps of integrated flux above three energy
thresholds: $I(>E={\rm 0.5\ GeV})$, $I(>E={\rm 1\ GeV})$ and $I(>E={\rm 10\ GeV})$.

Variations in the number density of unresolved sources, $n_{\gamma}(z,{\bf x})$,
are responsible for the local fluctuation in the $\gamma$-ray luminosity density, $\rho_{\gamma}(z,{\bf x})$
and, therefore, in the integrated $\gamma$-ray flux.
If the luminosity is proportional to the number of sources then the two fluctuations in $n_{\gamma}$ and $\rho_{\gamma}$
are related through
 \begin{equation}
\delta_{\gamma}(z,{\bf x}) \equiv \frac{\rho_{\gamma}(z,{\bf x})-\rho_{\gamma}(z)}
{\rho_{\gamma}(z)}=
\frac{n_{\gamma}(z,{\bf x})-n_{\gamma}(z)}{n_{\gamma}(z)}
\equiv \delta_{n_{\gamma}}(z, {\bf x})  \;,
\end{equation}
where the mean number density of sources is $n_{\gamma} \equiv \int  \Phi(L) dL$.
We further assume a linear mapping, called  biasing, between mass density  $\rho_m$ and the number density of objects:
\begin{equation}
\delta_{n_{\gamma}}(z,{\bf x})\equiv b_{\gamma}(z)\delta_m(z,{\bf x})= b_{\gamma}(z)
\frac{\rho_m(z,{\bf x})-\rho_m(z)} {\rho_m(z)},
\label{eq:bias}
\end{equation}
where we allow for a redshift-dependent bias parameter, $b_{\gamma}(z)$.

Fluctuations in the integrated flux along the generic
direction ${\bf n}$
can be obtained from (\ref{eq:integratedflux})
and the linear biasing prescription (\ref{eq:bias}):
 \begin{equation}
\delta I( {\bf n} ) \equiv \frac{I(>E,{\bf n}) -   I(>E)}{I(>E)}=
\frac{\int  \rho_{\gamma}(z)b_{\gamma}(z) \delta_m(z,{\bf x})dz}
        {I(>E)} \;.
\label{eq:deltaflux}
\end{equation}
where $I \equiv I(>E)$ indicates the $\gamma$-ray mean flux and
$I( {\bf n} )\equiv I( >E,{\bf n} )$ is the energy flux along the generic
direction ${\bf n}$.

Our goal is to investigate the cross-correlation between the diffuse IGRB maps
and the sky-projected spatial
distribution of different types of extragalactic objects.
The angular cross-power spectrum between the extragalactic background
$I( >E,{\bf n} )$ and
the fluctuation  of discrete sources, $j$, can be expressed as
\begin{equation}
C_l^{I,j} =\frac{2}{\pi}
\int k^2 P(k) [G_l^I(k)]
[G_l^j(k)]  dk \;,
\label{eq:angularspectrum}
\end{equation}
where $P(k)$ is the  power spectrum of density fluctuations, $l$ is the multipole of the spherical harmonics expansion
and the functions $G(k)$ specify the contribution of each field to the cross-correlation signal.
More specifically, the contribution of the IGRB is given by
\begin{equation}
G_l^I(k)=\int  \rho_{\gamma}(z)b_{\gamma}(z) D(z) j_l[k \chi(z)]dz
\label{eq:autocorr}
\end{equation}
where $j_l[k \chi(z)]$ are spherical Bessel functions, $D(z)$ is the
linear growth factor of density fluctuations and $\chi(z)$ is the
comoving distance to redshift $z$.
The analogous quantity for the number density fluctuations in a population of discrete objects is
\begin{equation}
G_l^j(k)=\int \frac{dN(z)}{dz} b_j(z) D(z) j_l[k \chi(z)]dz,
\label{eq:crossocorr}
\end{equation}
where $dN(z)/dz$ is the redshift distribution of the objects.
Here we make the hypothesis that these objects,
which do not necessarily coincide  with the sources of the IGRB,
 also trace the underlying mass density field modulo
some $z$-dependent linear bias parameter
$b_j(z) $.

Note that in the cross-correlation we consider the integrated flux $I( >E,{\bf n} )$ rather than its dimensionless
analogous $\delta I({\bf n})$ given in
Eq.~\ref{eq:deltaflux}. With this choice  the cross-correlation signal is robust to
any spurious monopole term arising from an incorrect subtraction of the
model Galactic diffuse signal or charged particle contamination.
One implication of this choice is that our model cross-correlation signal (\ref{eq:angularspectrum})
is dependent on the mean integrated flux,  $I( >E)$. For this reason, and to account for
uncertainties in the estimate of the mean IGRB signal, we allow for some freedom in the
normalization of the luminosity function of the putative $\gamma$-ray sources and, accordingly,
add an additional free parameter in the model.

In this work we also estimate the angular two-point cross-correlation function of the flux maps and discrete object
catalogs which is simply
the Legendre transform of the angular power spectrum
\begin{equation}
\left \langle  I( >E,{\bf n_1} )  \delta_j( {\bf n_2} ) \right \rangle = \sum_l\frac{2l+1}{4\pi}C_l^{I,j} P_l[\cos(\theta)] \;,
\label{eq:2point}
\end{equation}
where  $P_l[x]$ are the Legendre polynomials and $\theta$ is the separation angle between directions
$ {\bf n_1}$ and  ${\bf n_2}$.
The angular  two-point correlation function and power spectrum are two ways of expressing the same information. However, in practice,
the two statistics are somewhat complementary as they  probe different scales with different efficiency and
their respective estimators are prone to different types of biases. For this reason we shall compute both quantities.


\begin{figure}
\centering
\epsfig{file=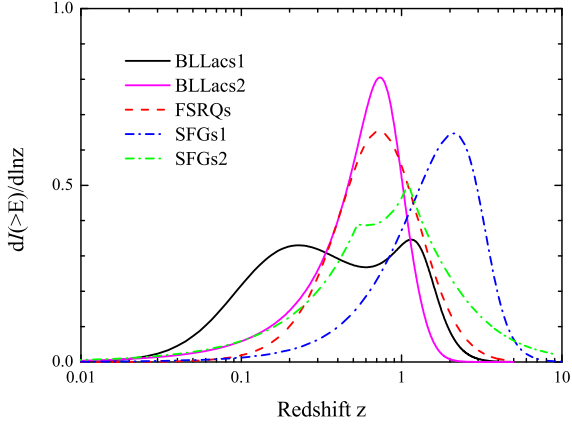, angle=0, width=0.52 \textwidth}
\caption{Integrated $\gamma$-ray flux per logarithmic redshift bin $dI(>E)/ d\ln z$
as a function of $z$ for  three different source classes:  FSRQs
(red, dashed), BL Lacs (black or magenta, continuous) and star-forming galaxies
(blue or green, dot-dashed).}
\label{fig:zdist1}
\end{figure}

\subsection{Modeling the mean flux and the cross-correlation signal}
\label{sec:model}

One of the aims of this work is to compare the measured cross-correlation signal with
model predictions  obtained under the assumption that some specific type of unresolved sources
contributes to the IGRB. We note that even auto-correlation studies can provide constraints
on the nature and spatial clustering of the sources contributing to the signal (e.g., \citealp{persic89}).
In our case, we are required to model: $i)$ the
correlation properties of the underlying mass density field; $ii)$ its relation with
discrete tracers, i.e., the biasing prescription; $iii)$  the mean IGRB flux.
To model  the cross-correlation signal we consider  the cosmologically evolving
 mass density power spectrum, $P(k,z)$,
obtained from the public code {\tt CAMB} \citep{2000ApJ...538..473L} for the linear part and the
{\tt Halofit} built-in routine for non-linear evolution \citep{2003MNRAS.341.1311S}.
In addition, we use the linear growth factor $D(z)$ and the comoving distance $\chi(z)$
appropriate for the cosmological model adopted.

To model the bias and the mean IGRB signal we need to select a class of objects which are likely to contribute to the IGRB
and specify the energy spectrum, luminosity function and fraction of IGRB contributed by the source, $f_j$.
Here we consider three different candidates: {\it FSRQs}, {\it BL Lacs} and star-forming galaxies ({\it SFGs})

\begin{enumerate}

\item {\it FSRQs} are a type of AGN (blazars) with a
relativistic jet pointing close to the line of sight. \cite{AjelloFSRQs}
have recently determined the $\gamma$-ray luminosity function
of these objects which they have parametrized in the framework of a
Luminosity Dependent Density Evolution (LDDE) model:
\begin{equation}
\Phi(L_{\gamma},z=0, \Gamma) =\frac{A}{\ln(10)L_{\gamma}}
\left[\left(\frac{L_{\gamma}}{L_{*}}\right)^{\gamma_1}+
\left(\frac{L_{\gamma}}{L_{*}}\right)^{\gamma2}
\right]^{-1}\times e^{-\frac{ (\Gamma-\mu(L_{\gamma}))^2}{2\sigma^2}} \; .
\label{eq:lf0}
\end{equation}
The term in parentheses, a smoothly-joined  double power-law function, represents the
luminosity function of the local FSRQs
and the exponential term is the same photon index distribution as Eq.~\ref{eq:index}.
In the LDDE model the luminosity function at the redshift $z$ can be expressed as
\begin{equation}
\Phi(L_{\gamma},z,\Gamma) = \Phi(L_{\gamma},z=0,\Gamma) \times e(z,L_{\gamma}) \; ,
\end{equation}
where
\begin{equation}
e(z,L_{\gamma})= \left[
\left( \frac{1+z}{1+z_c(L_{\gamma})}\right)^{-p_1(L_{\gamma})} +
\left( \frac{1+z}{1+z_c(L_{\gamma})}\right)^{-p_2} 	
 \right]^{-1}
\label{eq:evol}
\end{equation}
with
\begin{equation}
p_1(L_{\gamma}) = p_1^* + \tau \times (\log_{10}(L_{\gamma})-46)
\label{eq:p1}
\end{equation}
and
\begin{equation}
z_c(L_{\gamma})= z_c^*\cdot (L_{\gamma}/10^{48})^{\alpha} \,.
\label{eq:zpeak}
\end{equation}
$z_c$ represents  the  luminosity-dependent redshift
at which the evolution changes from positive to negative
and $z_c^*$ is the evolutionary peak for an object
with a luminosity  of 10$^{48}$\,erg s$^{-1}$.
This LDDE luminosity function model  is specified by the
12  parameters listed in Table~\ref{tab:ldde} with the particular values determined by
 \cite{AjelloFSRQs} by fitting $\gamma$--ray data.
In the fit, the authors  have set $\beta=\tau=0$, i.e., they have assumed that
neither the overall shape of the luminosity  function
nor the spectral index depend on the luminosity of the sources, $L_{\gamma}$.
Note that the evolutionary term $e(z,L_{\gamma})$ in Eq.~\ref{eq:evol} is not equal to unity at $z=0$.
To derive the density $\rho_{\gamma}(z)$ in Eq.~\ref{eq:density} required to calculate the correlations,
we set $L_{\rm MIN}=10^{44}$ erg s$^{-1}$ as recommended in  \cite{AjelloFSRQs}, although,
as already explained, choosing a lower value or even zero does not significantly affect
the results.

The parameters of the luminosity function uniquely determine the
contribution of FSRQs to the mean diffuse IGRB signal. However, as anticipated,  in this work we prefer to
keep the normalisation of the luminosity function free.
This additional degree of freedom  is meant to absorb experimental errors in the measurement
of the mean diffuse IGRB signal and uncertainties in modeling the clustering of the sources.
The resulting redshift distribution of the \mbox{$\gamma$-ray} flux contributed by FSRQs
shown in Fig.~\ref{fig:zdist1}  is rather broad
and peaks at $z\sim 0.5$.

The last ingredient of the model is the bias of the sources. Here we adopt
the redshift-dependent AGN bias proposed by \cite{2009MNRAS.396..423B}
in the  framework of the semi-analytic models of AGN-black holes
co-evolution: $b_{FSRQ}(z)=0.42+0.04(1+z)+0.25(1+z)^2$.
To test the robustness of our results to the biasing scheme we have considered
two alternative bias models:
{\it i)} the rather unphysical case of a constant bias $b_{FSRQ}=1.04$
obtained by considering it equal to the bias model of \cite{2009MNRAS.396..423B} estimated at $z=0.5$.
{\it ii)} a linear, $z$-dependent model in which the bias is set equal to that
of a $10^{13} \, {\rm M}_{\odot}$ halo. This latter choice reproduces the bias of X-ray selected
AGN estimated by \cite{K013} and represents an upper limit, since the bias of optically selected
AGN is matched by the bias of $10^{13} \, {\rm M}_{\odot}$ halos.

\item {\it BL Lacs}. These sources  represent
a different  sub-class of blazars with, on average, lower luminosities than FSRQs.
To model their luminosity function we adopt the LDDE functional form  as in Eq.~\ref{eq:lf0}
and set the free parameters according to the best fit to $\gamma$--ray data
obtained by  \cite{AjelloBLLs}.
We consider two possibilities corresponding to
the two sets of parameters in  Table~\ref{tab:ldde}.
In the first case, that we dub BLLacs1, the authors let all parameters free to vary.
In the model BLLacs2, they instead set $\tau=0$, i.e., basically switching off the
dependence  on the luminosity of the $p_1$ index in the evolutionary term.
It should be stressed that the BLLacs1 model represents a better fit
to the data in  \cite{AjelloBLLs}. Nonetheless, we consider also
model BLLacs2 to study the  robustness of the
interpretation of the cross-correlation in terms of BL Lacs.  In accordance with  \cite{AjelloBLLs}  we set  $L_{\rm MIN}=7\times10^{43}$ erg s$^{-1}$
for the calculation of the $\rho_{\gamma}(z)$ integral.

The redshift distribution of the $\gamma$-ray flux of BL Lacs
for the two models considered is shown in Fig.~\ref{fig:zdist1}.
In both cases the distribution is rather broad.
However, in BLLacs1  the  distribution is more local and peaks at
$z\sim0.1$, whereas in  BLLacs2  the peak is at
much higher redshift ($z\sim0.7$) although  with a significant tail at low $z$.
We assume that the biasing of BL Lacs is the same in both models and equal to that of  FSRQs, i.e.,
$b_{BLLac}=b_{FSRQ}$. The robustness of the results on this choice is also tested using the same
 alternative bias models considered for FSRQs.

\begin{figure}
\centering
\epsfig{file=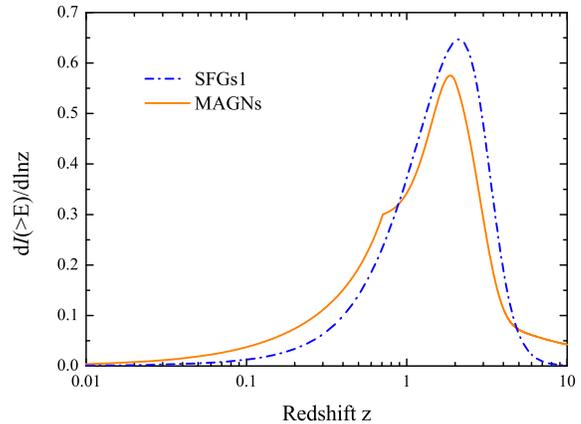, angle=0, width=0.52 \textwidth}
\caption{Comparison between the $\gamma$-ray flux per logarithmic  redshift bin $d I(>E)/d \ln z$.}
\label{fig:zdistMAGNs}
\end{figure}

\item {\it  SFGs}.
As reference  and comparison with our  previous work  \citep{xia11}
we adopt the phenomenological model  of  \cite{2010arXiv1003.3647F} in which the
 $\gamma$-ray emission in SFGs over cosmic time
 is proportional to their star-formation rate  and the gas mass-fraction, both
 normalised  to the present values in the Milky Way (MW).
The energy spectrum is assumed to be similar to the one observed in the MW
which can be modeled approximately as a broken power law with photon indexes
$\Gamma \sim 1.5$ and $\Gamma \sim 2.475$ above and below $\sim 500$ MeV, respectively.
The contribution to the IGRB, shown in Fig.~\ref{fig:zdist1}, is spread over a
wide redshift range and peaks
at $z\sim1$, with a high-redshift tail more extended than that of  BL Lacs and FSRQs.
We dub this model SFGs1.

For the sake of completeness
we consider also a second model (SFGs2), originally proposed by \cite{Achermann12} and
recently revised by \cite{tamborra014}.
The ingredients of this model are the SFG LF obtained from infrared observations
 \citep{rodighero10, gruppioni013} and the empirical relation
between luminosity in the infrared and  $\gamma$-ray bands
calibrated using a  samples of local galaxies observed in both bands
and assumed to be valid at all redshifts \citep{Achermann12}.
{The most recent IR observations of  \cite{gruppioni013} have enabled the
measurement of the LFs  of different sub-populations of galaxies.
Specifically,   \cite{gruppioni013} subdivide  the infrared galaxies 
into normal spiral galaxies (SP), starburst galaxies (SB), and galaxies hosting an AGN 
but whose infrared emission is still dominated by star-forming activity (SF-AGN),
and provide the LFs separately for each sub-population.
We model the $\gamma$-ray emission separately for the three  populations
by assuming  a power-law energy spectrum with
$\Gamma =2.475$ for SP and SF-AGN
and of $\Gamma =2.2$ for SB \citep{Achermann12}.
To test if the redshift distribution of the  $\gamma$-ray emission
is robust with respect to the assumed spectral shape of galaxies
we use the LF of the whole population of infrared objects
from \cite{gruppioni013}. 
Assuming a single global index $\Gamma =2.475$ for this
population does not significantly modify the results.
It has to be noted that besides the contribution from
SP, SB and SF-AGN this global LF contains also
a contribution from pure AGNs, which is, however, very subdominant.
The IGRB contribution of SP + SB + SF-AGN as a function of redshift is plotted in Fig.~\ref{fig:zdist1}.}
The distribution is significantly different than in model SFGs1 since in this case
the IGRB is mostly contributed by low-redshift galaxies.
The discrepancy between the two distributions reflects a fundamental
difference between the two models. Both models use
the luminosity density in the infrared band,
in \cite{2010arXiv1003.3647F} assuming  that is a good tracer of the star formation history
(which is a common assumption, see, for example, the discussion in \cite{rodighero10}),
and in \cite{Achermann12} assuming it is a tracer of the $\gamma$-ray LF itself.
However, model SFGs1 further contemplates the possibility of a time-dependent
 gas quenching that reduces the  $\gamma$-ray emission at low redshift.
As a result we expect that the two models predict  very different cross-correlation
signals,   despite having similar  $\gamma$-ray  luminosities integrated over redshift  \citep{tamborra014}.

Finally, based on the observations in  \citet{2004PhRvD..69h3524A} and theoretical arguments \citep{wilman08}
we
assume for both models that SFGs trace the
underlying mass density field with no bias (i.e., $b_{SFG}=1$).
We also consider an alternative case in which the bias of the SFGs is set equal to that
of a Milky Way-sized halo of $10^{12}\, {\rm M}_{\odot}$.

\end{enumerate}

\subsection{Misaligned AGNs}

Another type of source that  potentially contributes to the IGRB is misaligned AGNs (MAGN) \citep{DiMauro:2013xta,Inoue:2011bm}.
Similarly to the case of SFGs, too few MAGNs have been detected in $\gamma$ rays to determine their LF directly in this band.
Instead, the $\gamma$-ray LF is inferred from that measured in some other band by exploiting the observed relation between the
luminosities in the two bands. For MAGN this is done by considering the radio band, i.e., by using the observed MAGN radio LF and
the radio  to  $\gamma$ luminosity relation. The latter relation is calibrated on a local sample of objects for which observations in
both bands are available and then extrapolated at all redshifts.

We did follow this procedure and used both the  radio luminosity function of \cite{2001MNRAS.322..536W} and the radio $\gamma$-ray luminosity
relation derived in \cite{DiMauro:2013xta} to obtain the MAGN contribution to the IGRB at all redshifts. The result is shown in
Fig.~\ref{fig:zdistMAGNs}, which is the analogous to Fig.~\ref{fig:zdist1} but featuring only the MAGN and the SFGs1 models.
The main point here is the similarity between the two distributions that peak at $z\sim 2$ and extend to much higher redshift.
We also note that while the SFG distribution is smooth, the MAGN exhibits a feature at $z\sim 1$. This reflects the composite nature of
the  radio-band LF that receives contributions from different types of objects with sharp breaks in their redshift distributions.

For the scope of our analysis, the fact that the two redshift distributions are similar implies a potential degeneracy in the
model predictions. This degeneracy can be broken only if the linear bias parameters of the two populations, upon which the amplitude of the
clustering depends, are different. In fact, we do expect that the bias of the MAGN, which are typically associated with massive dark matter halos,
is higher than that of the SFGs1 and,  as a consequence, the MAGN bias is higher than that of SFGs at all redshifts
(see e.g., \cite{wilman08}, \cite{lindsay14} and references therein).
This together with the fact that the bias is an increasing function of redshift and that both populations contribute to the IGRB out to
high redshifts make it possible, in principle, to discriminate between the two populations through a cross-correlation analysis.
In practice, however, the bias of both populations is ill-constrained by present observations.

For this reason, in the present analysis, we focus on those objects for which the contribution to the IGRB and its anisotropy are more robust
to the uncertainties in the bias. In fact, we have considered blazars, since their $dI(>E)/d \ln z$ is suppressed at high redshift
where the bias of the objects is expected to increase, and SFG, since their bias is expected to be close to unity even at high redshifts.
For the very same reason we have decided not to include MAGN in our model: they can be found at very high redshift and their bias
is large and rapidly increases with redshift.
Of course this does not mean that MAGNs do not contribute to the correlation signal. Only that our analysis will not be able to
discriminate between SFG and MAGN contributions. Therefore it should be kept in mind that the SFG contribution, which we 
will study in the following, may actually include a MAGN signal as well and that our model cross-correlation signal, entirely based
on blazars and SFGs, could be underestimated at high redshifts.

\subsection{Further theoretical contributions}
\label{sec:1halotheory}

{ All of our models assume that the discrete sources sample, according to a deterministic bias relation, the
underlying mass density field whose two-point clustering properties can be quantified
 by a nonlinear power spectrum that can be obtained using
 {\tt CAMB} + {\tt Halofit}.
 This is an approximation that ignores
the presence of substructures within virialized halos and, consequently, underestimates the power on small scales.
In the language of the halo model \citep{Cooray:2002dia}, we underestimate the {\it 1-halo} term contribution to the correlation signal.
Theoretical modeling of this term is challenging since it depends on several characteristics both of the
catalog and of the $\gamma$-ray emitting population  which are quite uncertain.
For example, it is necessary to specify how the galaxies of a catalog are distributed on average within
a DM halo of given mass, and, analogously, how $\gamma$-ray sources of different luminosities
populate the same DM halo. Both quantities can be modeled, but within large uncertainties.
Further details are discussed, for example, in \cite{ando14}, \cite{Camera:2014rja} and \cite{Ando:2006mt}.
On the contrary, the shape of the {\it 1-halo} term is  easier to model. 
Assuming point-like DM halos, this term would simply be a constant in multipole space,
or a delta-like term at $\theta=0^{\circ}$ in the CCF.
Small distortions from a constant arise considering the finite extent of DM halos,
and are typically important at very high multipoles ($\gtrsim$ 1000) for
halos at low redshift  ($\lesssim$ 0.1).
In the following, we will thus adopt a phenomenological approach and model
this term as constant in $\ell$ with a free normalization, and check against
the data if there is a preference for the inclusion of a {\it 1-halo}  correlation.

A second assumption of our model is that the sources responsible for the $\gamma$-ray signal and
the various astrophysical sources that we cross-correlate trace the same mass density with different bias relations, i.e., they
are not the same population. There is, however, the possibility that they are the same population seen at two different
wavelengths  (or two populations that largely  overlap each other). In this case one would expect a strong cross-correlation peak in the CCF
 at $\theta=0^{\circ}$ corresponding, again, to a constant in multipole space.
With enough angular resolution, this possibility could be distinguished, in principle, from a pure  {\it 1-halo} term, 
due to the distortion induced by the finite extent of the DM halo in the latter
 (with the possible caveat of a positive correlation signal at $\theta \gtrsim 0^{\circ}$
 that may arise even in this case when the emission in two different bands originates from two separate regions of the same object, like,
 for example, possibly
 in the case of the nucleus and the jet of an AGN).
Typically, however, the LAT PSF  is  too  large to allow discriminating between the two cases.
We will thus model both terms as  constant in multipole space, and  consider their
sum as a single contribution whose presence will be tested in the data.
We will indicate these contributions collectively as   {\it 1-halo}-like terms.

}


\section{{\it Fermi}-LAT maps}
\label{sec:fermimaps}

In this section we describe the IGRB maps obtained from 5  years of
{\it Fermi}-LAT data taking and the masks and templates used to subtract  contributions from  {\it i)} $\gamma$--ray resolved sources,
{\it ii)} Galactic diffuse emission due to interactions of cosmic rays with the interstellar medium and
{\it iii)} additional Galactic emission located at high  Galactic latitude in prominent structures
such as the {\it Fermi} Bubbles  \citep{2010arXiv1005.5480S,Fermi-LAT:2014sfa} and Loop~I  \citep{2009arXiv0912.3478C}.
The validity of the masking procedure, its effectiveness in removing foreground and resolved source contributions to the IGRB signal, and its
impact on  cross-correlation analysis are assessed in Section~\ref{sec:validation}.

{\it Fermi}-LAT is the primary instrument onboard the {\it Fermi} Gamma-ray Space
Telescope launched in June 2008 \citep{2009ApJ...697.1071A}.
It is a pair-conversion telescope covering the energy range between
20 MeV and $> 300$ GeV. Due to its excellent angular resolution ($\sim 0.1^{\circ}$
above 10 GeV), very large field of view ($\sim 2.4$ sr), and
very efficient rejection of background from charged particles, it is
the best experiment to investigate the nature of the IGRB in the GeV energy range.

For our analysis we have used  60 months of data
from August 4th 2008 to August 4th 2013.
More specifically, we have used the {\verb"P7REP_CLEAN"}
 event selection\footnote{See http://www.slac.stanford.edu/exp/glast/groups/canda/lat\_Performance.htm
   for a definition of the  \texttt{P7REP} and \texttt{P7} event selections and their characteristics.}
  in order to ensure
a low level of cosmic-ray (CR) background contamination.
Further, to greatly reduce the contamination from the bright  Earth limb emission
we exclude photons detected {\it i)}
with measured zenith angle
larger than 100$^{\circ}$ or {\it ii)}
when the rocking angle of the LAT with respect to the zenith
was larger than 52$^{\circ}$.
In order to generate the final flux maps we have
produced  the corresponding exposure maps using the standard routines from
the LAT \emph{Science Tools}\footnote{http://fermi.gsfc.nasa.gov/ssc/data/analysis/documentation/Cicerone/} version 09-32-05,
using the latest recommended {\verb"P7REP_CLEAN_V15"} instrument response functions (IRFs).
We use both back-converting and front-converting events and we checked the robustness
of the results using either data subsample (see Section~\ref{sec:frontback}).
The GaRDiAn package \citep{FermiLAT:2012aa,Ackermann:2009zz} was then used to pixelize both
photon count and exposure maps in HEALPix\footnote{http://healpix.jpl.nasa.gov/}
format \citep{2005ApJ...622..759G}. The maps contain
$N_{\rm pix} = 3\, 145\, 728 $ pixels with an angular size  of $\sim 0.11^{\circ}\times0.11^{\circ}$
corresponding to the HEALPix resolution parameter  $N_{\rm side}=512$.
Finally,  the flux maps are obtained by dividing the count maps by exposure maps in three energy
ranges: $E>500$ MeV, $E>1$ GeV and  $E>10$ GeV

\begin{figure}
\centering
\epsfig{file=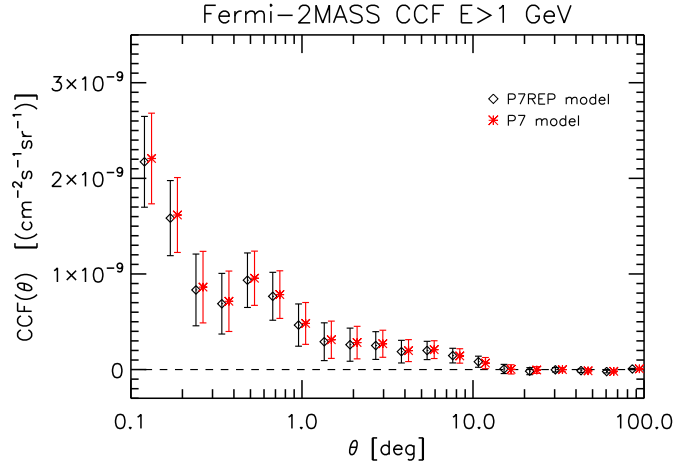, angle=0, width=0.50\textwidth}
\caption{Comparison of Fermi-2MASS $E>$ 1 GeV CCF  for two different Galactic foreground models.}
\label{fig:diff_comp_P7_P7REP}
\end{figure}

\begin{figure}
\centering
\epsfig{file=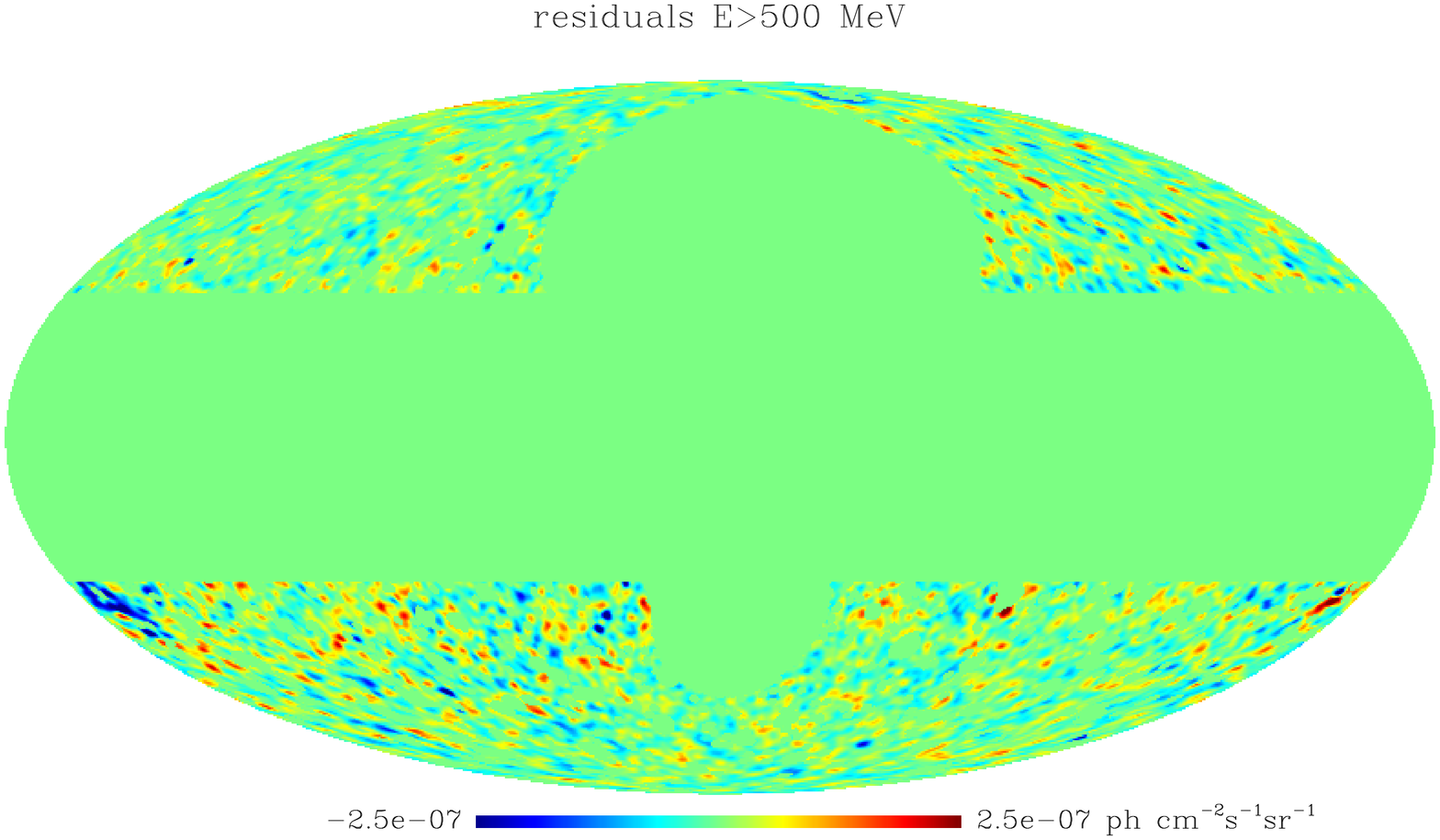, angle=90, width=0.45 \textwidth}
\epsfig{file=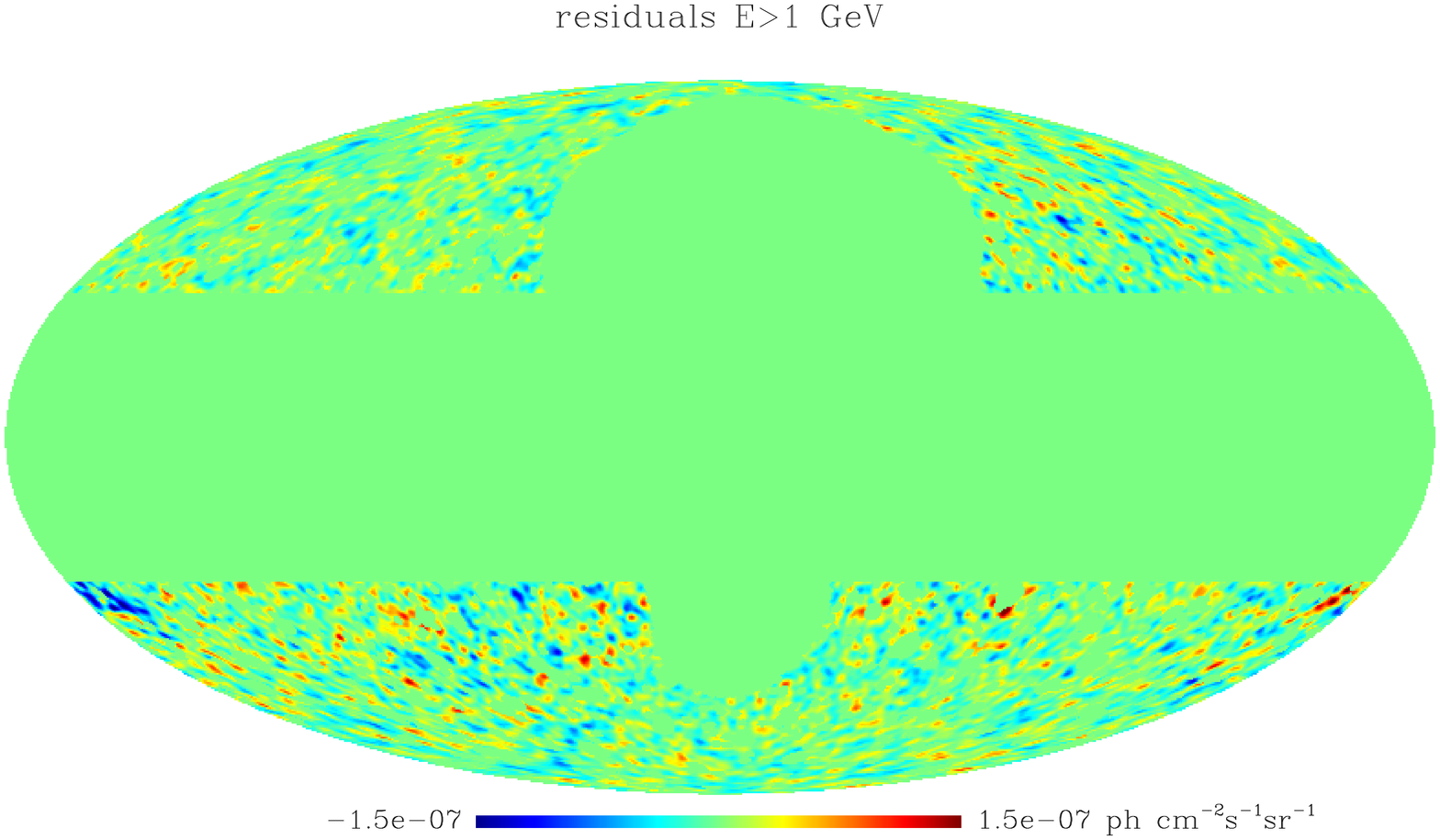, angle=90, width=0.45 \textwidth}
\epsfig{file=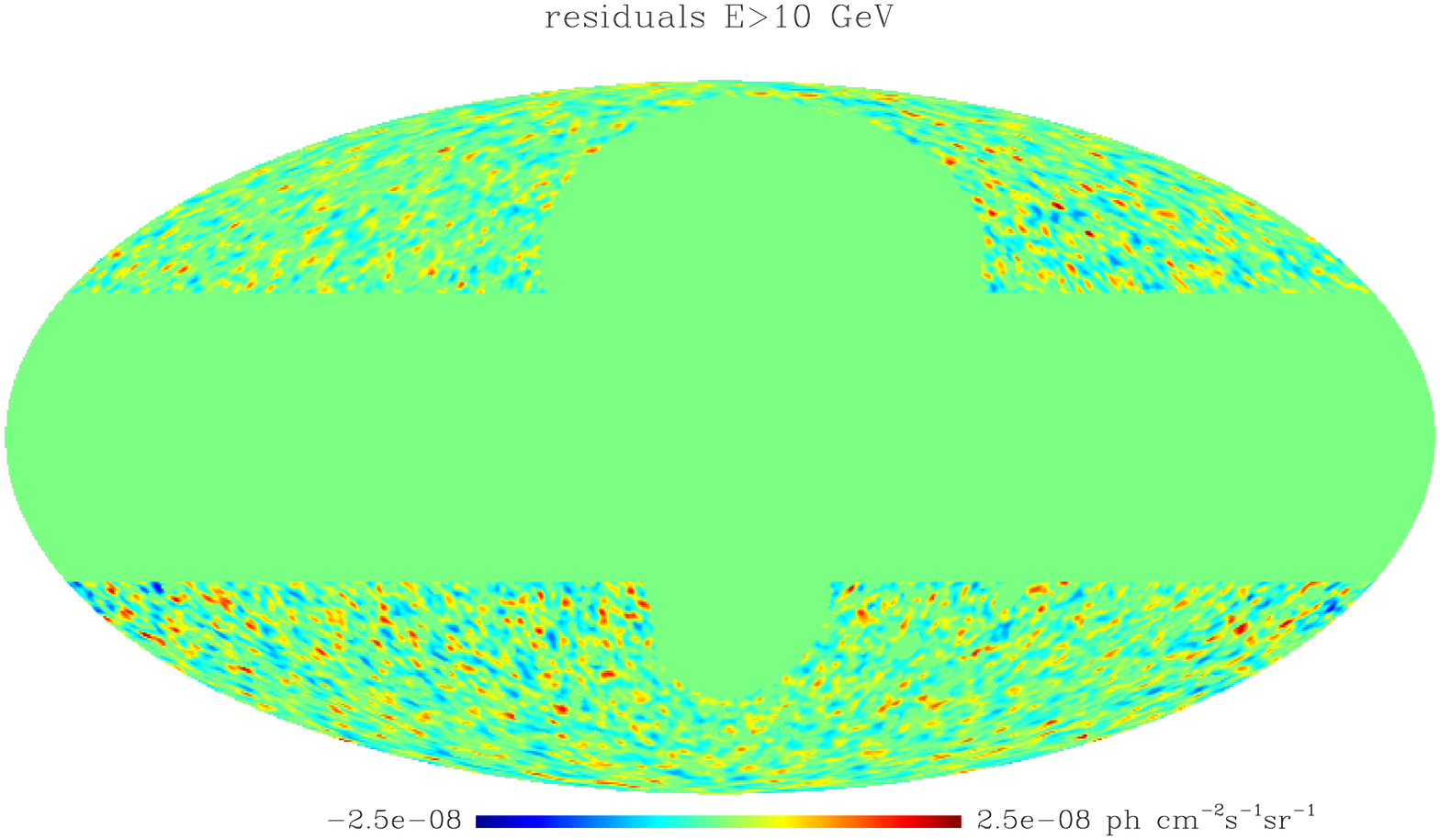, angle=90, width=0.45 \textwidth}
\epsfig{file=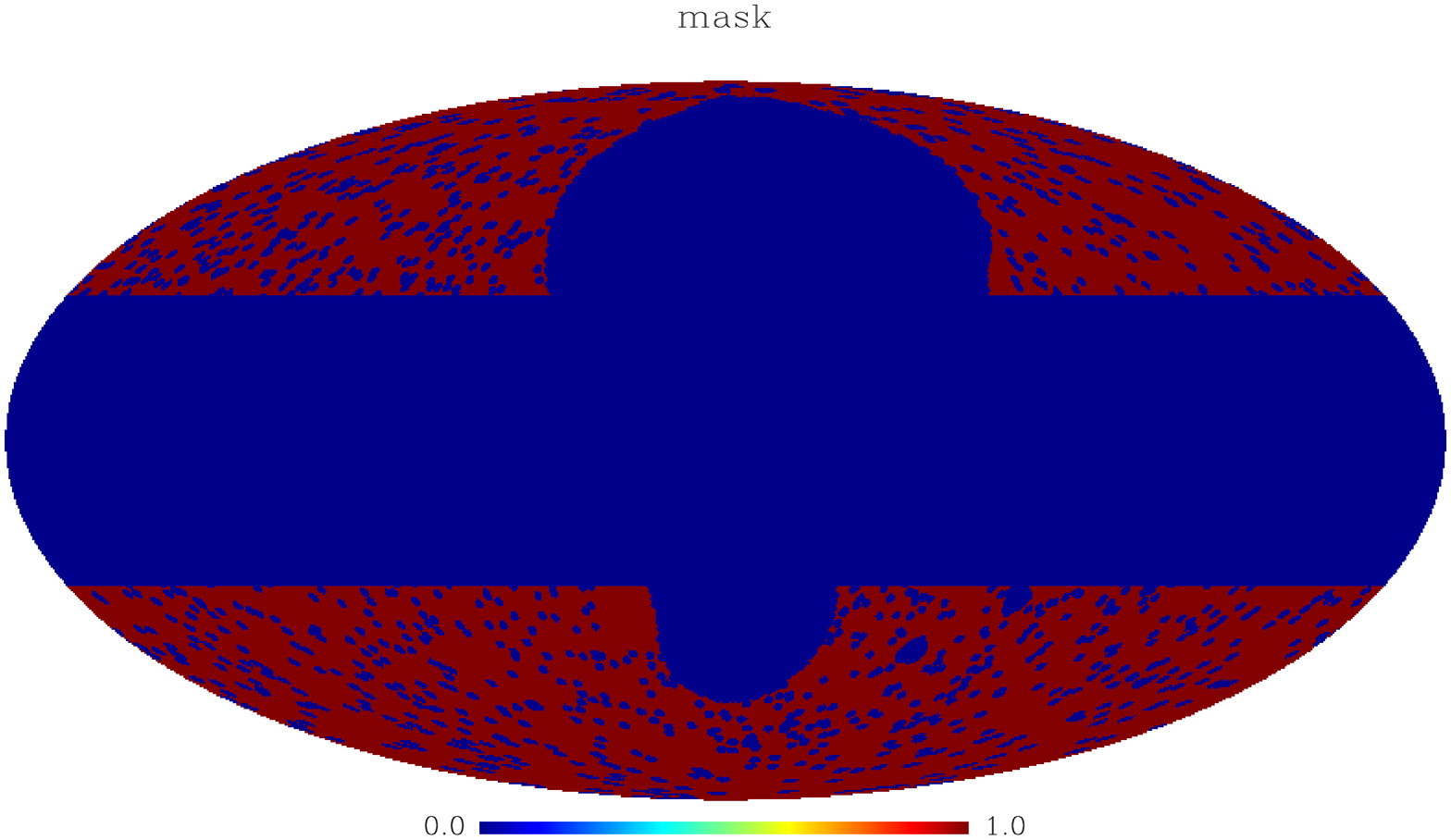, angle=90, width=0.45 \textwidth}
\vspace{-0.3cm}
\caption{Upper three plots: Fluctuations of the residual $\gamma$-ray photon flux maps
obtained from 60 months of {\it Fermi}-LAT data for energies $E>500$ MeV
(top), $E>1$ GeV (second from the top),  $E>10$ GeV (second from the
bottom). Foreground emission from Galactic diffuse, Sun and Moon have been subtracted from
the data as well as multipoles as large as $l=10$.
Different colors indicate fluctuations of different amplitude according to the color code
scheme in the plots. The flat-color areas across the Galactic plane and around the {\it Fermi} Bubbles and Loop~I
correspond to the mask and have been ignored in the correlation analysis. The mask, which also accounts for resolved
sources,  is further shown in the bottom plot.
The maps are in Galactic coordinates and have a resolution $N_{\rm side}=512$.
For visualization, but not during the analysis,
the maps  have been further
smoothed with  $1^\circ$ Gaussian filter to remove small scale Poisson noise.
}
\label{fig:Fermi_maps}
\end{figure}

To reduce the impact of the Galactic emission on our analysis focused on the IGRB,
we apply a Galactic latitude cut \mbox{$|b|> 30^{\circ}$} in order to mask the bright emission along
the Galactic plane.  Moreover, we also exclude the region associated
to the {\it Fermi} Bubbles and the Loop~I structure.
In \cite{xia11} we have experimented with different latitude cuts and found that \mbox{$|b|>30^{\circ}$}
represents the best compromise between pixels statistics and Galactic contamination.
The corresponding mask, obtained from the tabulated contours of the  {\it Fermi} Bubbles given
in  \cite{2010arXiv1005.5480S} is shown in the bottom panel of Fig.~\ref{fig:Fermi_maps}
as the bulge-like central region together with the  horizontal strip mask corresponding to the latitude cut.
The mask also features a number of smaller holes placed at the position of all resolved
sources  in a preliminary version of  the 3FGL catalog. 
In the $E>1$ GeV maps all sources are masked out with
a disk of $1^{\circ}$  radius. For $E>0.5$ GeV we used larger disks of $2^{\circ}$ but only for the
500 sources with  the highest integrated flux above 100 MeV in the catalog,
while the remaining ones are still masked with disks of $1^{\circ}$.
Finally, to exclude the contribution from the
Small and Large Magellanic Clouds, which are more extended, we have
used two larger circles with  $3^{\circ}$ radius.
To test the robustness of our results on the subtraction of resolved sources we
have also built a similar mask using  the previous 2FGL catalog  (\verb"gll_psc_v08.fit"\footnote{http://fermi.gsfc.nasa.gov/ssc/data/access/lat/2yr\_catalog/} ).
Further details  are given in Section \ref{sec:montecarlo}.
When cross-correlating with a given galaxy catalog, the mask
specific to that catalog is further employed.
The masks for the catalogs we use can be seen in \cite{xia11}.

Although we select a part of the sky at high Galactic latitude,  the Galactic diffuse emission
in this region is still significant and needs to be removed.
For this purpose, and to check the robustness to this correction,
we use two models of Galactic diffuse emission:
\verb"ring_2year_P76_v0.fits"\footnote{http://fermi.gsfc.nasa.gov/ssc/data/access/lat/BackgroundModels.html \label{foot4}}
 and  \verb"gll_iem_v05_rev1.fit"${}^{\ref{foot4}}$,
which we subtract from the observed emission to obtain the \emph{cleaned} maps.
Both models are based on a fit of the LAT data to templates of the H\,{\sc i} and CO gas distribution in the Galaxy
as well as  on an inverse Compton model obtained with the GALPROP code\footnote{A more detailed
description can be found at
http://fermi.gsfc.nasa.gov/ssc/data/access/lat/BackgroundModels.html}.
The first model \verb"ring_2year_P76_v0.fits" is tuned to 2 years of \verb"P7"
data and further uses uniform flat patches
to model some features of the diffuse sky such as the {\it Fermi} Bubbles and Loop~I.
The model  \verb"gll_iem_v05_rev1.fit" is based on 4 years of \verb"P7REP"
data and adopts an alternative procedure to account for residual diffuse emission
involving templates of residual emission obtained in early stages of model fitting
to construct the final model.
Since part of the data has been introduced in some form as an \textit{ a posteriori}
model component\footnote{http://fermi.gsfc.nasa.gov/ssc/data/access/lat/ {}~Model\_details/FSSC\_model\_diffus\_reprocessed\_v12.pdf} \citep{JMtemplate14},
the  \verb"gll_iem_v05_rev1.fit"  model
is not recommended for diffuse emission studies.
However, since these additional templates  only affects regions that are within
our masked areas, we can safely use also this model  to test the robustness of
the results with respect to the modeling of the Galactic emission.
Indeed, the comparison between the two models shows actually that they are very similar in our
region of interest  and they give almost identical residuals.
This was expected since in our region of interest  the diffuse emission is basically
accounted for by a single
template based on the local H\,{\sc i} emission.
Not surprisingly, the correlation functions that we obtain when using this model agree at the percent level with the results obtained with the other model, as shown in Fig.~\ref{fig:diff_comp_P7_P7REP}.
For definiteness, we set  \verb"ring_2year_P76_v0.fits" as our baseline model.
Note that, in general, it is not recommended to use the model \verb"ring_2year_P76_v0.fits",
tuned on  \verb"P7" data, with  \verb"P7REP" data, even though
in this particular case, as shown in Fig.~\ref{fig:diff_comp_P7_P7REP},
the differences  between the results derived from the two models  are small.

Finally, we have also subtracted the contributions from solar and lunar emission
along the ecliptic. For this purpose we used  the appropriate
routines of the LAT {\it Science Tools} and selected options to obtain
templates consistent
with the data selection and IRFs
 choices described above.
The model of the energy and spatial emission from the Sun and Moon have been
taken from the related papers \citep{Abdo:2011xn,2012ApJ...758..140A}, tabulated into
the files \verb"solar_profile_v2r0.fits"\footnote{http://fermi.gsfc.nasa.gov/ssc/data/analysis/scitools/solar\_template.html \label{footsunmoon}}
and   \verb"lunar_profile_v2r0.fits"${}^{\ref{footsunmoon}}$.

The practical procedure we use to obtain the maps of residual  photon counts is to use
 \verb"GaRDiAn"
to convolve the  Galactic emission model and the Sun and Moon templates with the
exposure maps and the PSF which are then subtracted from the observed counts.
Residual flux maps are then obtained by further dividing the residual photon counts by the exposure maps.

As the Galactic diffuse emission models are not exact, cleaning is not perfect and the
residual flux maps  are still contaminated by
 spurious signal, especially on large angular scales.
The impact on cross-correlation analyses is expected to be small since Galactic foreground emission
is not expected to correlate with the extragalactic signal that we want to investigate. Possible
counter examples, like extinction effects, are small and will be discussed in Section~\ref{sec:maps}
in the context of optical extragalactic surveys.
Nonetheless
we  follow
\cite{xia11} and apply a  cleaning procedure that, using   HEALPix tools, removes all
contributions from multipoles  up to $\ell=10$ including the monopole and dipole.

The resulting residual photon flux maps, which we dub $\ell10$-maps,
for the three energy ranges considered in this work
are shown in the three upper panels of Fig.~\ref{fig:Fermi_maps}.
The masked area, also shown in the bottom plot, is represented by the uniform strip
across the Galactic plane and further extending around the {\it Fermi} Bubbles and Loop~I.
Fluctuations have amplitude in the range $\pm 1.5 \cdot 10^{-7}$
ph cm$^{-1}$ s$^{-1} $ sr$^{-1}$ for the case $E>1$ GeV according to the color code
shown in the plots. Note that for visualization,
but not during the analysis, the maps have been smoothed with a $1^\circ$ Gaussian filter
to remove small-scale Poisson noise.
The model seems to slightly over-subtract the $\gamma$-ray emission around  $(l,b)\sim(175^{\circ},-35^{\circ})$,
corresponding to the gas- and dust-rich (and thus difficult to model) Taurus Molecular region.
Note, however, that when performing the cross-correlation, this region is
masked by the further mask specific to the catalog, except in the NVSS case (see below),
so that no bias in the results is expected from this feature.


\section{Maps of Discrete Sources}
\label{sec:maps}

In this section we describe the different catalogs of extragalactic objects that
we cross-correlate with the $\ell10$-maps of the diffuse IGRB
obtained after the cleaning procedure described in Section~\ref{sec:fermimaps}.

In this work we have considered five different catalogs:  {\it i)}
the SDSS DR6 quasar
catalog released by ~\citet{2009ApJS..180...67R} that should
trace the FSRQ population,
 {\it ii)} the
IR-selected 2 Micron All-Sky Survey (2MASS)
extended source catalog \citep{2000AJ....120..298J}
to trace SFGs,
  {\it iii)} the NRAO VLA Sky Survey
(NVSS) catalog of radio galaxies
\citep{1998AJ....115.1693C} that we regard as alternative tracers to the
FSRQs,
{\it iv)} the DR8 SDSS catalog of Luminous Red Galaxies  \citep{2008arXiv0812.3831A},
which should trace an intrinsically fainter, more local AGN population
like the BL Lacs, and
{\it v)} the DR8 SDSS
main galaxy sample \citep{dr8} as a potential additional tracer of
SFGs.

\begin{figure}
\centering
\epsfig{file=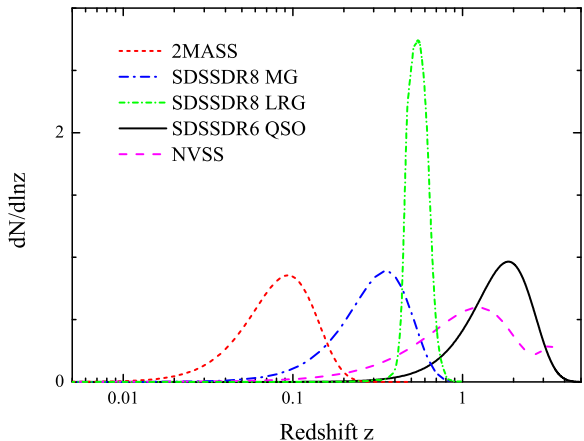, angle=0, width=0.55\textwidth}
\caption{Redshift distributions, $dN/d\ln z$, of the different
  types of objects considered for our cross-correlation analysis. SDSS
  DR6 QSOs (black, continuous line), 2MASS galaxies (red,
  short-dashed), NVSS galaxies (magenta, long-dashed), SDSS DR8 LRGs
  (green, short, dot-dashed) and SDSS DR8 Main Galaxy Sample (blue,
  long, dot-dashed) }
\label{fig:lss_dndz}
\end{figure}

The redshift distributions, $dN/d\ln z$, of the various sources are shown
in Fig.~\ref{fig:lss_dndz}, and described in more detail in the next subsections.
The various distributions peak at quite different redshifts, with
2MASS representing the most local population and SDSS DR6 QSOs
the most distant one. These characteristics in principle enable breaking down the
cross-correlation analysis in different redshift ranges, effectively allowing a {\it tomographic}
investigation of the IGRB origin.
A detailed description of these
catalogs, except the SDSS-DR8 main galaxy sample, can be found in \cite{xia11}.  Below we
briefly summarise the main features of each sample.

\subsection{SDSS DR6 QSO}
\label{sec:QSO}

The SDSS DR6 quasar catalog (\citet{2009ApJS..180...67R}, hereafter DR6-QSO) contains about
$N_{\rm q}\approx10^6$ quasars with photometric redshifts between
$0.065$ and $6.075$, covering almost all of the north Galactic hemisphere
plus three narrow stripes in the south, for a total
area of  8417 deg$^2$ (corresponding to $\sim20\%$  of the whole sky). The
DR6-QSO dataset extends previous similar SDSS datasets
 \citep{2004ApJS..155..257R,2006ApJ...638..622M}. The main
improvements are due to the fact that this catalog contains QSOs at higher
redshift and also contains putative QSOs flagged as objects with ultraviolet excess (UVX objects). We refer the reader to
\cite{2009ApJS..180...67R} for a detailed description of the object
selection performed with the non-parametric Bayesian classification kernel
density estimator (NBC-KDE) algorithm.

In this work we used objects listed in the electronically published table
with a ``good object'' flag in the range $[0,6]$.
The higher the value, the more probable for the object to be a real
QSO \citep{2009ApJS..180...67R}. We only consider the quasar candidates
selected via the UV-excess-only criteria ``uvxts=1'', i.e., objects clearly
showing a UV excess which represents a characteristic QSO spectral  signature.
After this selection we are left with $N_{\rm q}\approx6\times10^5$ quasar candidates.

In order to determine the actual sky coverage of the DR6
survey and generate the corresponding geometry mask we Monte Carlo sample
the observed areas with a sufficiently large number
of objects using the DR6 database to ensure roughly uniform
sampling on the SDSS CasJobs\footnote{http://skyserver.sdss3.org/CasJobs/} website.
Following \cite{2009JCAP...09..003X}
we combine the pixelized mask geometry with a foreground mask
obtained by removing all pixels with the {\it g}-band Galactic extinction $A_{\rm
g}\equiv3.793\times E(B-V) > 0.18$ to
minimize the impact of Galactic reddening.

The redshift distribution function $dN/dz$ of the DR6-QSO sample
in Fig.~\ref{fig:lss_dndz} is well
approximated by the analytic function:
\begin{equation}
\frac{dN}{dz}(z)=\frac{\beta}{\Gamma(\frac{m+1}{\beta})}\frac{z^m}{z^{m+1}_0}
\exp\left[-\left(\frac{z}{z_0}\right)^\beta\right]~,\label{reddis}
\end{equation}
where three free parameters values are $m=2.00$, $\beta=2.20$, and
$z_0=1.62$ \citep{2009JCAP...09..003X}.
In addition, to calculate theoretical predictions (Eq.~\ref{eq:crossocorr})
we follow \cite{2008PhRvD..77l3520G,2009JCAP...09..003X}
and assume a constant, linear bias model
$b_S=2.3$.

\subsection{2MASS}
\label{sec:2mass}

The 2MASS extended source catalog is an almost-all-sky survey that
contains $\sim770000$ galaxies with mean redshift $\langle
z\rangle\approx0.072$. In this work we have selected objects with
apparent isophotal magnitude $12.0<K'_{20}<14.0$, where the prime
symbol indicates that magnitudes are corrected for Galactic extinction
using $K'_{20}=K_{20}-A_k$, with $A_{\rm k}=0.367\times E(B-V)$. We
select objects with a uniform detection threshold (${\rm
  use_{-}src}=1$), remove known artefacts (${\rm cc_{-}flag} \neq a$
and ${\rm cc_{-}flag} \neq z$), and exclude regions with severe
reddening, $A_{\rm k}> 0.05$, \cite{1998ApJ...500..525S}.
This procedure leaves approximately 67\% of the sky unmasked. The
redshift distribution of the selected objects can be approximated with
the same functional form used for DR6 QSOs with parameters $m=1.90$, $\beta=1.75$,
and $z_0=0.07$ \citep{2008PhRvD..77l3520G}. The value of the linear
bias of 2MASS galaxies has been set equal to $b_S=1.4$
\citep{2007MNRAS.377.1085R}.

The possible incompleteness of the 2MASS catalog at faint magnitudes might
affect our cross-correlation analysis. For this reason we have
repeated the analysis using smaller 2MASS samples with
more conservative magnitude cuts: $K'_{20} < 13.9$, $K'_{20} < 13.7$
and $K'_{20} < 13.5$.  More specifically, we computed the two-point
cross-correlation function between these 2MASS maps and the {\it Fermi} $E>1$
GeV residual map and find that the CCF results do not change significantly, i.e.,
the possible incompleteness of the larger catalog does not induce any
spurious correlation signal. Therefore, in our analysis we
use the larger 2MASS sample cut at $K'_{20} = 14$.

\subsection{NVSS}
\label{sec:nvss}

The NRAO VLA Sky Survey (NVSS, \citealp{1998AJ....115.1693C}) offers the most
extensive sky coverage among the catalogs considered here
 ($82\%$ of the sky to a completeness limit of about 3
mJy at 1.4 GHz)
and contains $1.8\times10^6$ sources. We
include in our analysis only NVSS sources brighter than 10 mJy
since the surface density distribution of fainter sources suffers
from declination-dependent fluctuations \citep{2002MNRAS.337..993B}. We also
exclude the Galactic Plane strip at $|b| < 5^{\circ}$ where the catalog may be
substantially affected by Galactic emissions.
This additional cut is redundant with the one applied to the LAT maps.
It is applied to compute the
NVSS source surface density at this flux threshold which turns out to be  16.9 deg$^{-2}$.

The redshift distribution at this flux limit has been
determined by \cite{2008MNRAS.385.1297B}.
Their sample, complete to a flux
density of 7.2 mJy, contains 110 sources with $S\geq 10$ mJy, of
which 78 (71\%) have spectroscopic redshifts, 23 have redshift
estimates from the $K-z$ relation for radio sources, and 9 were not
detected in the $K$-band and therefore have a lower limit to $z$. We
adopt the smooth parametrization of this redshift distribution
given by \cite{2010A&ARv..18....1D}:
\begin{equation}
\frac{dN}{dz}(z)=1.29 + 32.37z - 32.89z^2 + 11.13z^3 - 1.25z^4~.
\end{equation}

For the bias of the NVSS sources  we adopt
a linear model $b_S=1.8$  \citep{2012MNRAS.426.2581G}.
Note that a comprehensive analysis of the NVSS autocorrelation function
is provided by \cite{2010ApJ...717L..17X}.

\subsection{SDSS DR8 LRG}
\label{sec:lrg}

To sample the spatial distribution of the LRGs we use
the photometric LRG catalog from the final imaging of
SDSS DR8 instead of the MegaZ LRGs sample since the latter
has an excess of power on large scales
with respect to the $\Lambda$CDM model \citep{2011PhRvL.106x1301T}.
The sample used here was selected using the same criteria as the
SDSS-III BOSS ``CMASS'' sample defined in
\citet{2011MNRAS.417.1350R}. They applied  colour cuts to account for seeing effects, dust
extinction, sky brightness, airmass, and possible stellar
contamination.

\citet{2012ApJ...761...14H} further excluded
regions where $E(B-V) > 0.08$ for the dust extinction, when the seeing
in the $i$-band $>2.0''$ in FWHM, and additionally masked
regions affected by  bright stars. This selection yields a sample with
$\sim8\times 10^5$ LRGs and leaves approximately 22\% of the sky unmasked.

Photometric redshifts of this sample are calibrated using
about 100,000 BOSS spectra as a training sample for the photometric
catalog. The resulting redshift range is $0.45 < z < 0.65$ with a mean
redshift  $\bar{z}\sim0.5$, as shown in Fig.~\ref{fig:lss_dndz}.
Also in this case, and following \citet{2011MNRAS.417.1350R, 2014MNRAS.438.1724H}
we assume a linear bias parameter  $b_S = 2.1$.

\begin{figure}
\centering
\epsfig{file=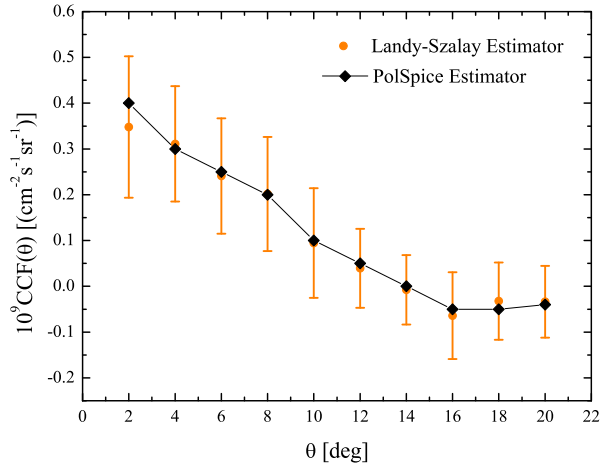, angle=0, width=0.5\textwidth}
\caption{
CCFs between {\it Fermi} maps $E>1$ GeV and
2MASS galaxies computed using PolSpice (black diamonds) and LS (orange dots) estimators. Error bars are computed by using the Jackknife resampling method.}
\label{fig:ccf_estimator}
\end{figure}

\subsection{SDSS DR8 Main Galaxy Sample}
\label{sec:SDSSmain}

We consider the sample of photometrically-selected ``main" galaxies
extracted from from the SDSS-DR8 catalog.
The selection is performed according to \citet{2008PhRvD..77l3520G}: we consider
 objects with  extinction-corrected $r$-band magnitude in the range $18<r<21$
 and with redshifts in the range $0.1<z<0.9$. Further, we only include objects with
 redshift errors smaller than $\sigma_z = 0.5z$, which leaves us with about
 $4.2\times 10^7$ sources with redshifts distributed around a median value $\bar{z}\sim0.35$.
In addition, we adopt a foreground mask to minimise the effect of Galactic extinction
by excluding all galaxies within pixels in which the $r$-band Galactic
extinction $A_r > 0.18$. Finally, we have about 35 million sources for the analyses.

The redshift distribution $dN/dz$  of the SDSS galaxies
 can be approximated with the same  functional form
  used for DR6 QSOs and 2MASS galaxies  with parameters
$m=1.5$, $\beta=2.3$, and $z_0=0.34$.
Following \citep{2012MNRAS.426.2581G} we use
a constant bias parameter $b_S = 1.2$.


\begin{figure}
\centering
\epsfig{file=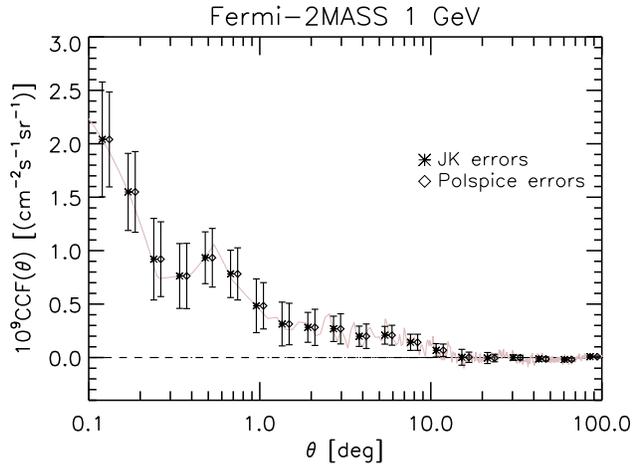, angle=0, width=0.50 \textwidth}
\caption{Comparison of the error bars of the CCF between {\it Fermi} maps $E>1$ GeV and
2MASS galaxies computed using the PolSpice covariance matrix and the Jackknife  resampling method.
CCFs for the Jackknife sub-samples are calculated with the PolSpice estimator.
The thin line shows the unbinned CCF.
\label{fig:errorcompare}}
\end{figure}

\section{Cross-Correlation Analysis}
\label{sec:corranalysis}

In this section we describe the cross-correlation
between the residual IGRB flux maps and the
angular distribution of extragalactic sources in the catalogs
described in  Section~\ref{sec:maps}.
All the maps we use are in HEALPix format with resolution of $N_{\rm  side} = 512$.

Our analysis relies on the latest version  v02-09-00 of 	
  PolSpice,\footnote{See  http://www2.iap.fr/users/hivon/software/PolSpice/}
  a publicly available
statistical tool to estimate both  the angular two-point cross-correlation function
$\hat{C}^{\rm fg}(\theta)$ and
 the cross angular power spectrum
$\hat{C}^{\rm fg}(l)$
of any two diffuse datasets, f and g,  pixelized on a sphere.
The code is  based on the fast Spherical Harmonic Transforms allowed by iso-latitude pixelisations
and automatically corrects for the effects of the masks.
Datasets and masks in the form of  HEALPix sky maps are input to the code.
The output consists of the angular two-point correlation function, the angular power spectrum and
its covariance matrix, which account for the effect of incomplete sky coverage and from the
nominal beam window function and average pixel function. In our calculations, we
 perform the correlation analyses in the multipole and angular ranges
$\ell\in[10,1000]$  and $\theta\in[0.1^{\circ},100^{\circ}]$  for CAPS and CCF, respectively.

The PolSpice estimator has been described in detail and thoroughly tested
as a tool to measure both the spectrum and the covariance matrix of a
sky map \citep{szapudi01,chon04,efstathiou04,challinor05}.
Since in our previous work  \citep{xia11} we based our analysis on the Landy-Szalay (LS)
estimator \citep{ls93} and computed errors using a Jackknife (JK)  resampling technique (see below), we
have checked the consistency of the two approaches and compared the outputs from the
different estimators.

\begin{figure}
\centering
\epsfig{file=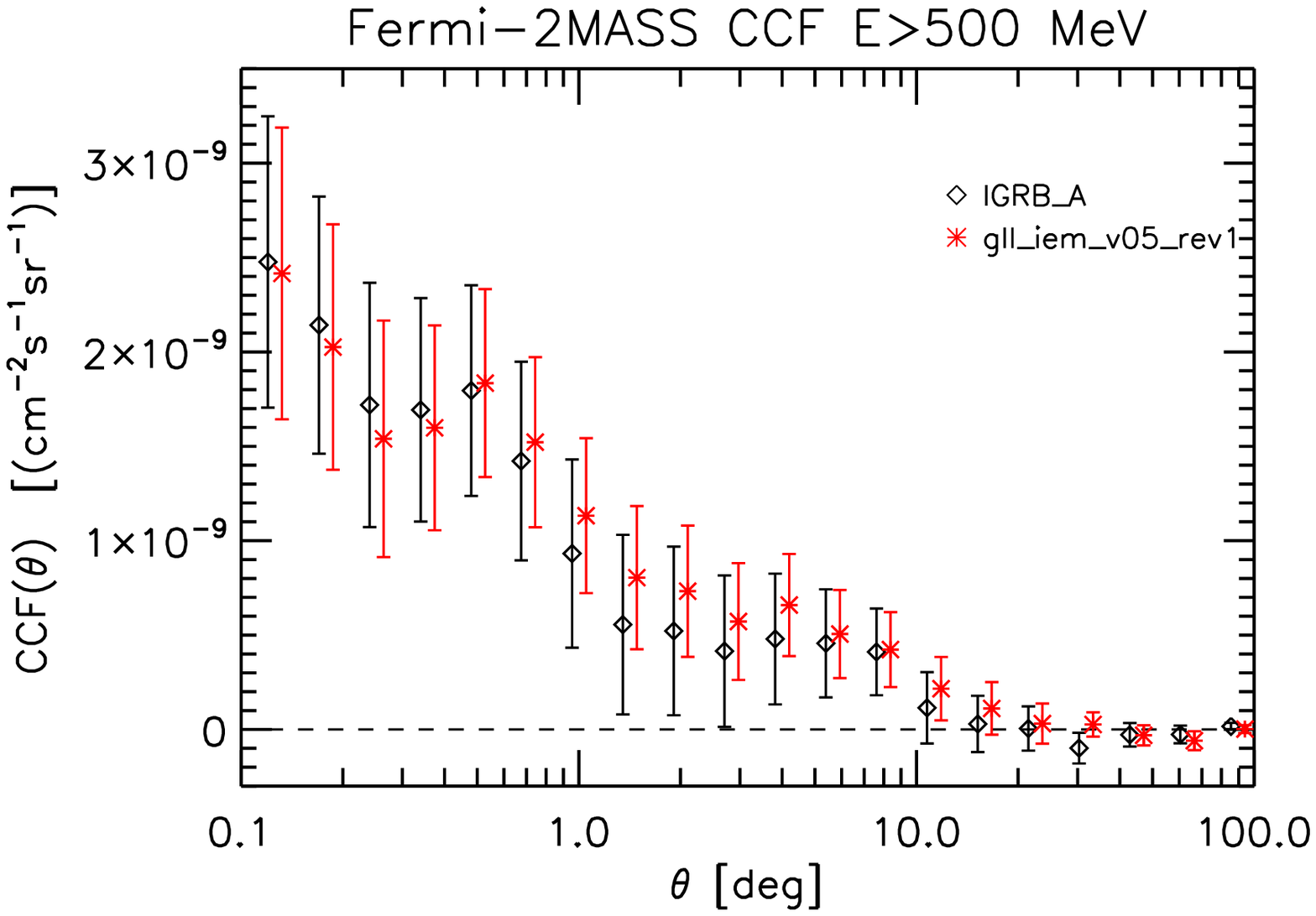, angle=0, width=0.45 \textwidth}
\epsfig{file=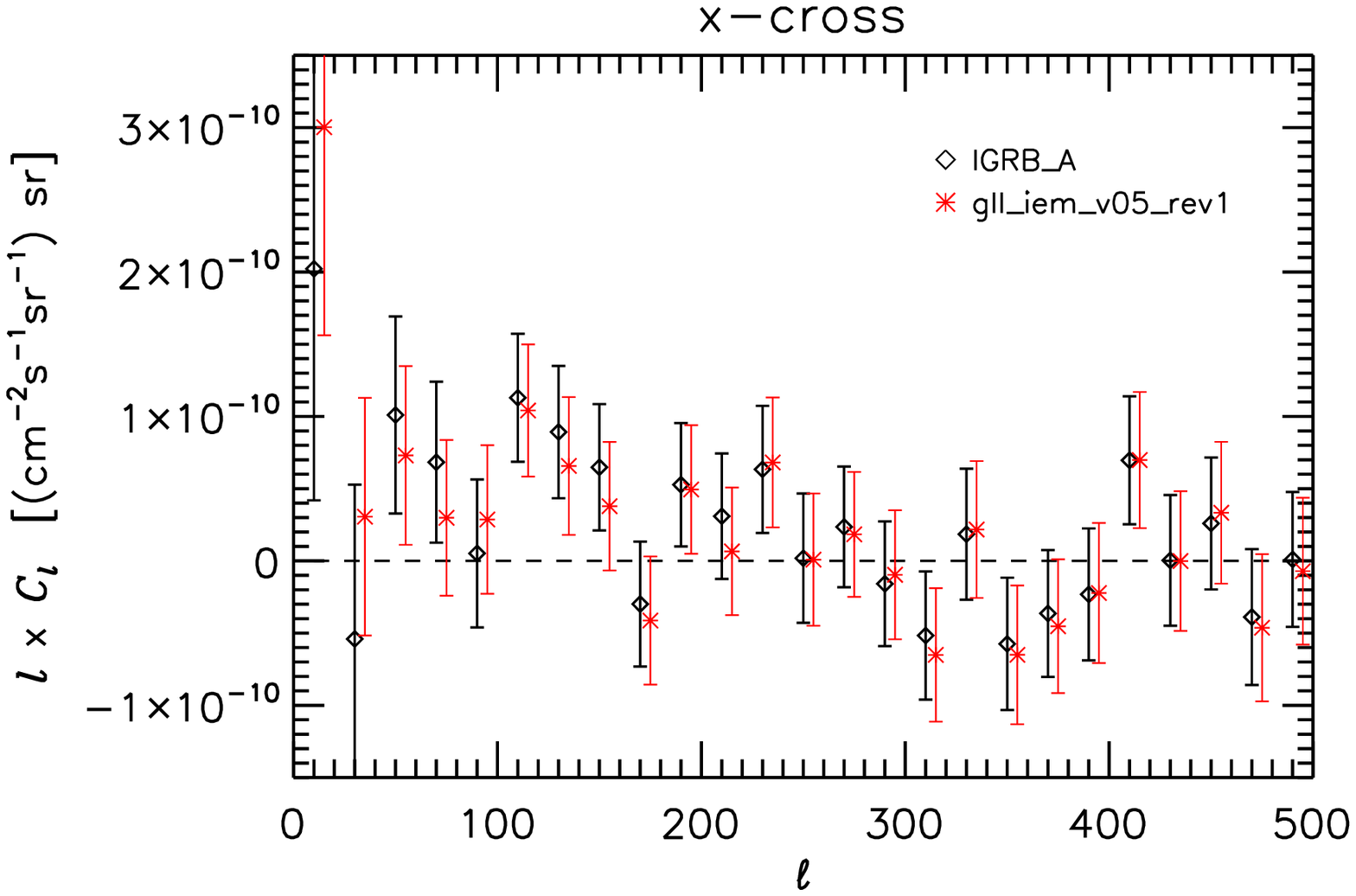, angle=0, width=0.45 \textwidth}
\caption{Comparison of {\it Fermi}-2MASS E$>500$ MeV CCF (lower panel) and CAPS (upper panel) for two different Galactic foreground models.}
\label{fig:galmodels_test}
\end{figure}

In  Fig.~\ref{fig:ccf_estimator}  we  compare the CCF between 2MASS and
{\it Fermi} data above 1 GeV estimated using  PolSpice  with the same CCF estimated using  LS.
The agreement between the two methods is at about 10\% in the first angular bin and
few \% in the other bins, well within the
amplitude of the 1-$\sigma$ random errors, and demonstrates that our results are robust to the choice of the
estimator for two-point statistics. Note that, only in this particular case, the angular binning
 is linear, dictated by the LS estimation method which,
being particularly computationally expensive, is applied using maps
with a coarse $N_{\rm side}=64$ pixelization.

Analogously, in  \cite{xia11}, in order to estimate the covariance matrix we used
the Jackknife  resampling method \citep{2002ApJ...579...48S}, which divides the data into $M$ patches
and estimates the covariance matrix as
\begin{equation}
C_{\theta\theta'}^{\rm JK}=\frac{M-1}{M}\sum^{M}_{k=1}\left[\xi^{\rm obs}_k(\theta)-\xi^{\rm mean}(\theta)\right]
\left[\xi^{\rm obs}_k(\theta')-\xi^{\rm mean}(\theta')\right]~,
\end{equation}
where $\xi^{\rm obs}_k(\theta)$ is the correlation function estimated in the k-th subsample
and $\xi^{\rm mean}(\theta)$ is the mean correlation function over the $M$ subsamples.
PolSpice provides an estimate for the covariance matrix of the CAPS, $\bar{M}_{\ell_1\ell_2}$
\citep{2004MNRAS.349..603E}. From this, and using the procedure described in \cite{2013arXiv1303.5075P}  we obtain the covariance matrix for the CCF:
 \begin{equation}
C_{\theta\theta'}^{\rm PS}=\sum_\ell\sum_{\ell'}\frac{2\ell+1}{4\pi}\frac{2\ell'+1}{4\pi}P_\ell(\cos(\theta))P_{\ell'}(\cos(\theta'))
\bar{M}_{\ell_1\ell_2}~ \; ,
\end{equation}
which is then averaged over the angular separations $\theta$ and $\theta'$ within each bin to obtain a binned covariance matrix.
In Fig.~\ref{fig:errorcompare} we show the same CCF plotted in Fig.~\ref{fig:ccf_estimator} with two sets of error bars corresponding to the
(square root of the) diagonal elements of the Polspice  and Jackknife covariance matrices.
The agreement between the two sets of error bars  is excellent,  as  the agreement between
 the off-diagonal elements (not shown) for which the largest difference does not exceed 10\%.

\begin{figure}
\centering
\epsfig{file=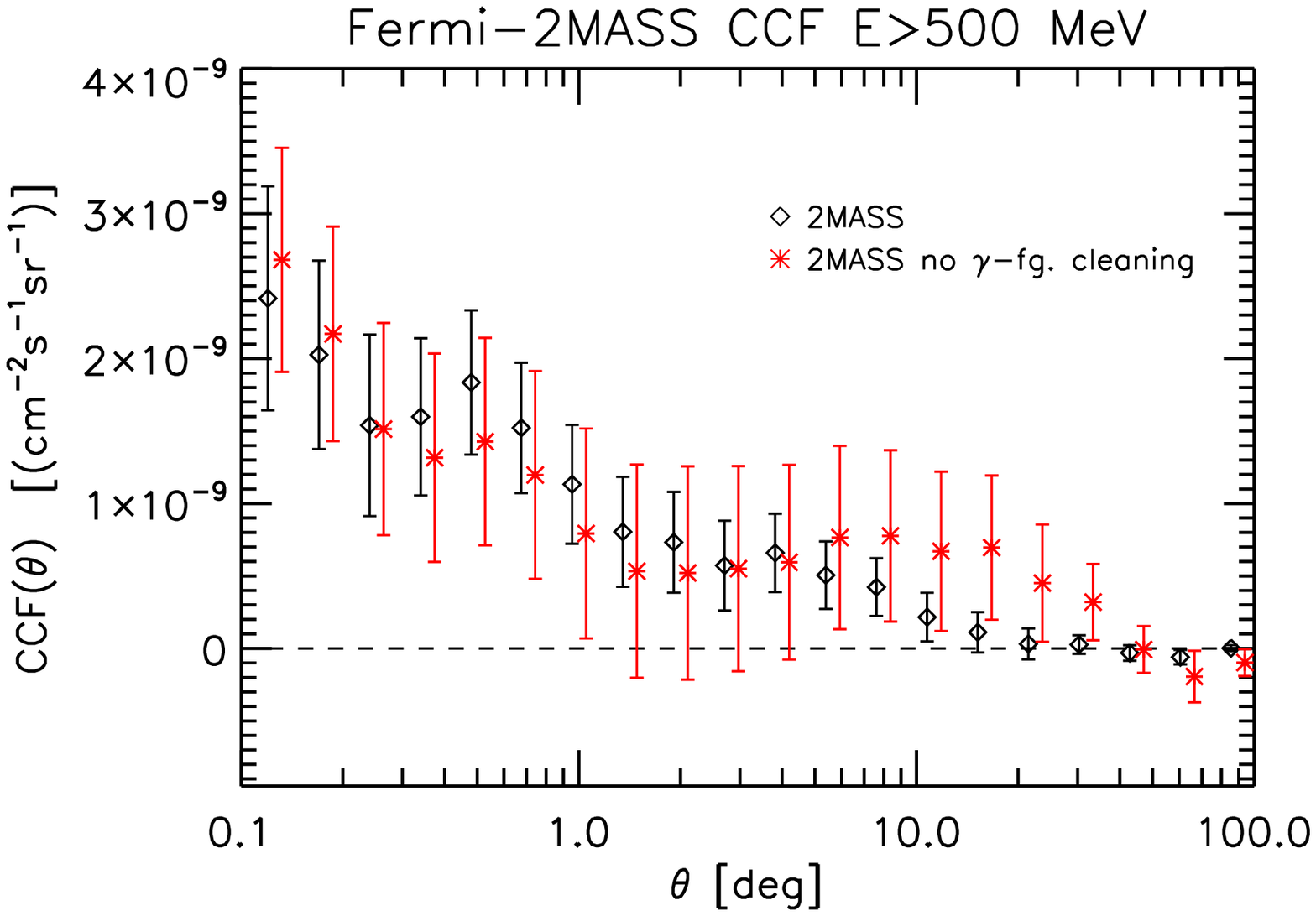, angle=0, width=0.45 \textwidth}
\epsfig{file=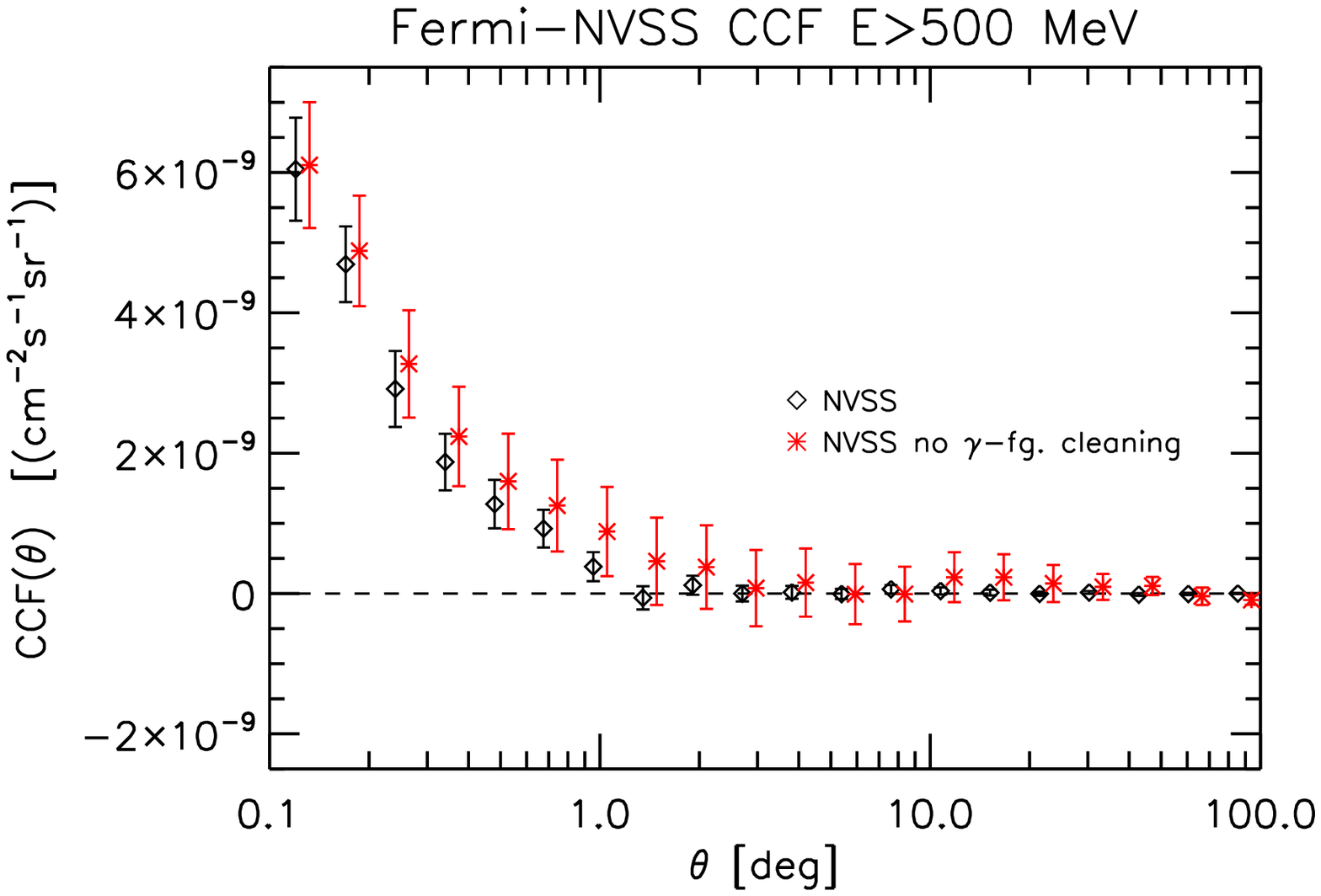, angle=0, width=0.45 \textwidth}
\caption{Comparison between  CCFs with and without Galactic foreground cleaning for the E$>500$ MeV {\it Fermi}-2MASS  case (upper panel) and  {\it Fermi}-NVSS case (lower panel).}
\label{fig:galmodels_test_v2}
\end{figure}

\section{Validation and checks}
\label{sec:validation}

In this section we assess the validity of the different steps of our analysis and assess the robustness of
our results.
For brevity we  only present a subset of cross-correlation analyses involving
the {\it Fermi}-LAT and 2MASS  maps.
However, we have performed the very same
robustness tests for all cross-correlation analyses described in this work.

\subsection{Test with different Galactic diffuse models}
\label{sec:galmodel}

The cleaning procedure described in Section~\ref{sec:fermimaps} is potentially  prone to systematic errors that
may affect the correlation analysis.
We searched for these systematic effects by using two different emission models also described in Section~\ref{sec:fermimaps}
to correct for  Galactic emission. The results show that the correlation signal does not change
appreciably when using either model.

Given the importance of this issue we performed a third test in which we adopted an alternative
Galactic emission model fully based on GALPROP.
In this test we used the GALPROP `model A' described in the {\it Fermi}-LAT collaboration
paper \citep{IGRBII} which is one of the models used to assess the
systematic uncertainties in the derivation of  IGRB energy spectrum
due the Galactic diffuse emission modeling.
The model consists of two components, inverse Compton emission
and gas emission (pion decay plus bremmstrahlung).
Together with an isotropic template we normalize this
3-component model by fitting
 the high Galactic latitude $\gamma$-ray sky (more precisely using the mask of Fig.~\ref{fig:Fermi_maps}),
 leaving the normalization of all the 3 components free to vary in the fit.
The tuned model is adopted as in Section \ref{sec:fermimaps} to clean the Galactic diffuse emission.
The usual $\ell$10 cleaning procedure is then used to derive the final residual
map from which we calculate the CCFs and CAPS.
In Fig.~\ref{fig:galmodels_test} we compare the  CCFs and CAPS between $\gamma$ rays and
2MASS for the  \verb"gll_iem_v05_rev1.fit" model already used in Section \ref{sec:fermimaps}
to  the alternative model results.
From the plots it can be seen that some difference exists  at low multipoles (below $\sim$40).
However the difference is within the 1 $\sigma$ error bars and decreases at small scales (high multipoles), where the signal is higher.
This shows that the results of our cross-correlation analysis are robust with respect to the modeling of  the Galactic diffuse emission.

Finally, for completeness, we show in Fig.~\ref{fig:galmodels_test_v2} the CCF between the $E>$500 $\gamma$-ray
map \emph{without any foreground cleaning} and 2MASS and NVSS.
The plot shows that removing the Galactic emission is important to reduce
the size of the error bars. On the other hand, even without any removal,
the correlation is not biased, as expected given the fact that
no correlation between the Galactic emission and LSS is in principle present.
This in turn also implies that
an imperfect Galactic emission removal would  similarly not introduce any bias,
although the error bars could be not optimal.

\subsection{Use of different Galactic and point sources masks}
\label{sec:montecarlo}
Incorrect masking is another potential source of systematic errors. To tests the robustness of our results
against the choice of the mask and to check the possible presence of Galactic foreground contamination
we have
varied the Galactic latitude cut from $b=20^{\circ}$ to $b=60^{\circ}$ in steps of $\Delta b=10^{\circ}$,
as in \cite{xia11}. In addition, we performed the cross-correlation analysis using only the northern or the southern hemispheres of the
maps.
In all the cases we explored the results turned out to be consistent within the errors that clearly increase
with the size of the masked region.

Inefficient excision of discrete sources is yet another potential concern. We estimate its impact
on our analysis by using different masks corresponding to excluding sources from
two different catalogs: 2FGL and a preliminary version of the 3FGL. Moreover, to excise sources we used two criteria: {\it i)}
we masked out circles of $1^{\circ}$ radius centered on all sources and {\it ii)}
we masked out circles of $2^{\circ}$ radius for the  500  brightest sources and
circles of $1^{\circ}$ radius
for the others.
The results of the correlation analyses turned out to be insensitive to the
choice of the source mask.
Clearly, increasing the size of the masked areas decreases the risk of contamination
but also decreases the significance of the correlation signal.
Therefore, in an attempt to compromise between maximizing the statistical significance and
minimizing contamination in the different cross-correlation analyses we proceed as follows:
 {\it i)} for the cross-correlation with the NVSS and 2MASS catalogs and for $E>500$ MeV,
which represents the case of large surveyed area and large contamination due to the broadening
of the {\it Fermi}-LAT PSF below 1 GeV,
we adopt the most conservative source mask based on the 3FGL catalog and
with larger ($2^{\circ}$ radius)  circles,  {\it ii)} for NVSS and 2MASS and  $E>1$ GeV, with a lower contamination,
we use 3FGL and  $1^{\circ}$, {\it iii)} for all SDSS-based catalogs, which
 have the smallest sky coverage,
we have considered a less aggressive $2^{\circ}$
2FGL mask for $E>500$ MeV, and {\it iv)}
a  $1^{\circ}$ 2FGL mask for higher energy cuts.

While the results are robust against the choice of the source mask, their significance can change appreciably.
For the NVSS and 2MASS catalogs, using different source masks varies the size of the error bars
by 20-30\%, so the choice of the mask is not critical in these cases.
For the various SDSS-based catalogs, instead, using  3FGL rather than 2FGL significantly
reduces the significance of the results and, for this reason, we opt for the least conservative
mask.

\begin{figure}
\centering
\epsfig{file=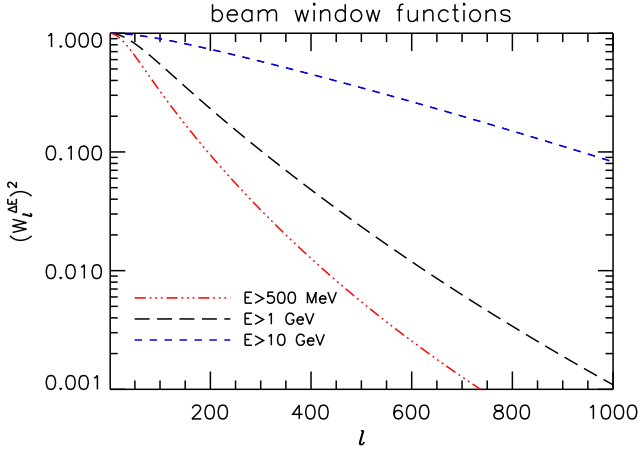, angle=0, width=0.49 \textwidth}
\caption{Effective squared  window functions of the beam, $(W_l^{\Delta E})^2$,
in the  3 energy ranges $E>500$ MeV  (red, dot-dashed), $E>1$ GeV (black, long-dashed), $E>10$ GeV (blue, short-dashed).
\label{fig:beamwindow}}
\end{figure}

\subsection{Robustness to $\gamma$--ray event  conversion}
\label{sec:frontback}

In our analysis, to maximize statistics, we consider both   front-  and back-converting  $\gamma$--ray events.
However, the two types of events have different characteristics, most notably
back-converting events have a larger PSF.
To check whether the nature of the conversion affects our results we divided the
$\gamma$--ray datasets into two subsamples, each one
containing only front-converting or back-converting events.
The resulting maps were cleaned using the same procedure, i.e.,
convolving the various model templates with the appropriate IRFs.

As a result of halving the number of events, the significance of the correlation is somewhat
reduced for each subsample, in particular for the SDSS catalogs.
Within the increased error bars, however, we do not observe
any bias among the three datasets (front only, back only and front plus back).
We conclude that possible systematic differences between the front
and back datasets are below the present statistical uncertainties,
and  we thus decided to perform the analysis using both type of events jointly.

\begin{figure}
\centering
\epsfig{file=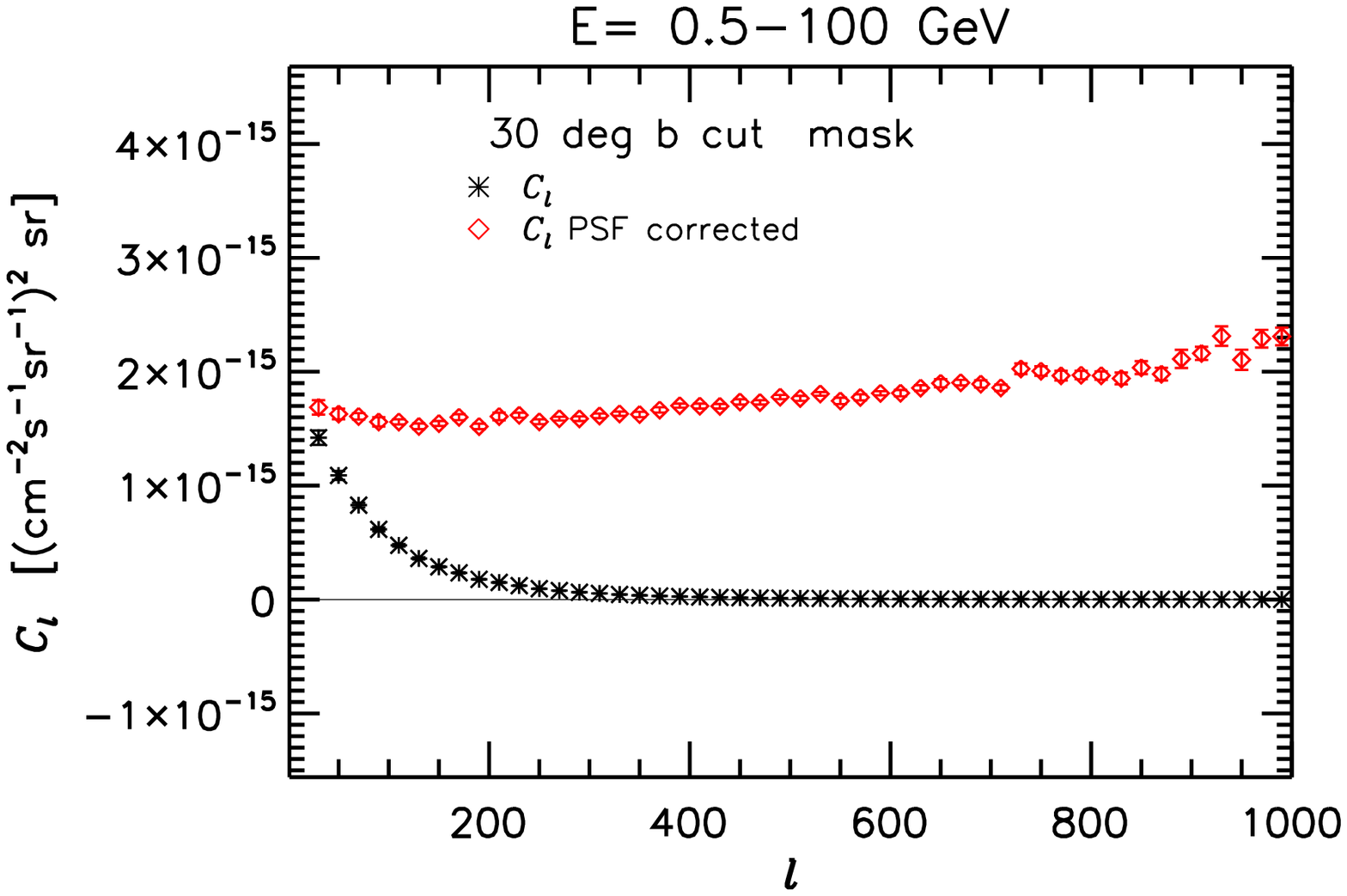, angle=0, width=0.45 \textwidth}
\epsfig{file=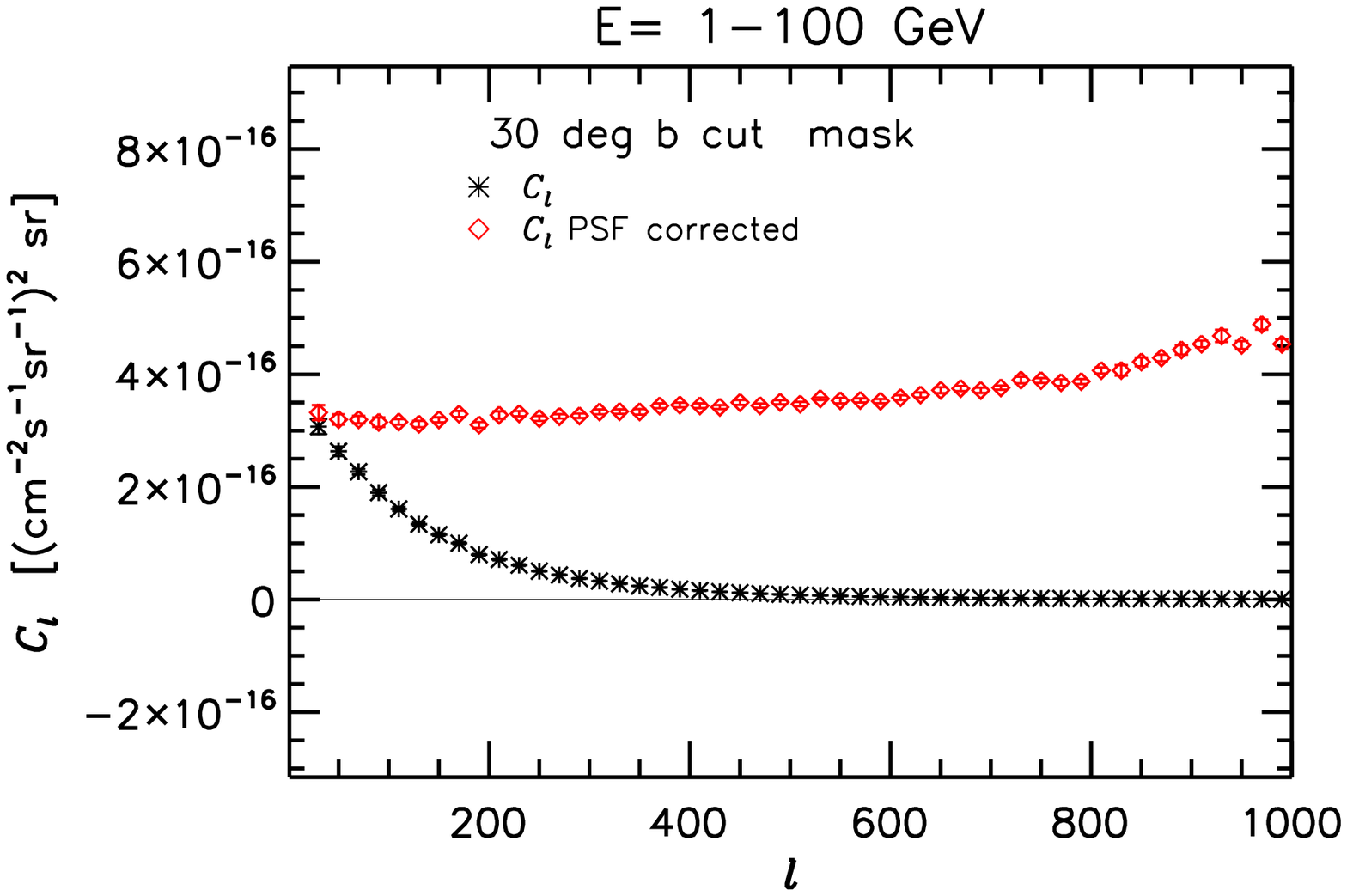, angle=0, width=0.45 \textwidth}
\epsfig{file=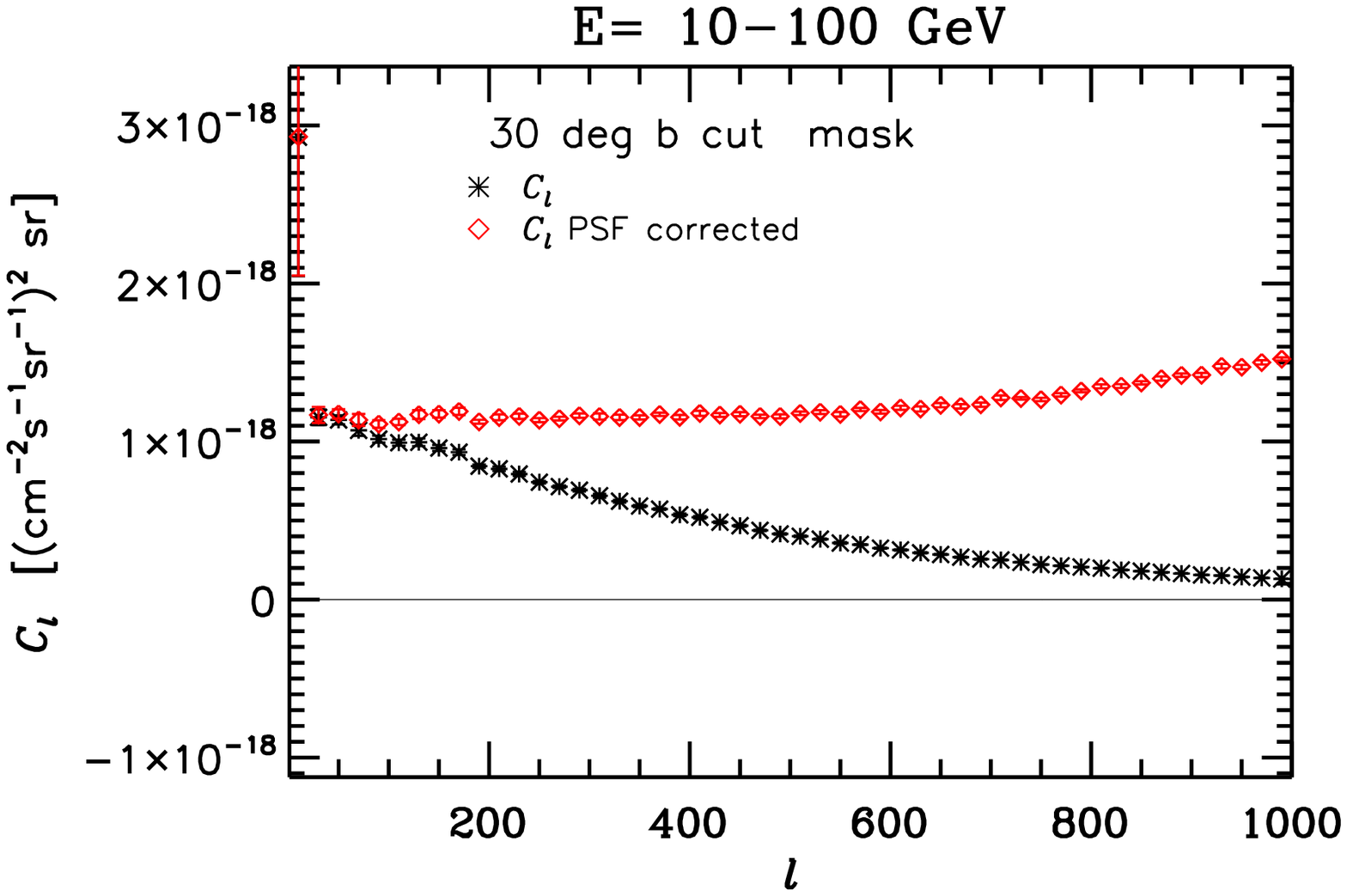, angle=0, width=0.45 \textwidth}
\caption{Measured auto power spectrum of the LAT maps at $|b|>30^\circ$ (black asterisks) and ratio between
the APS and the average beam window squared  $(W_l^{\Delta E})^2$ (red, open dots)  for 3 energy bands.}
\label{fig:beam_test}
\end{figure}

\begin{figure*}
\centering \epsfig{file=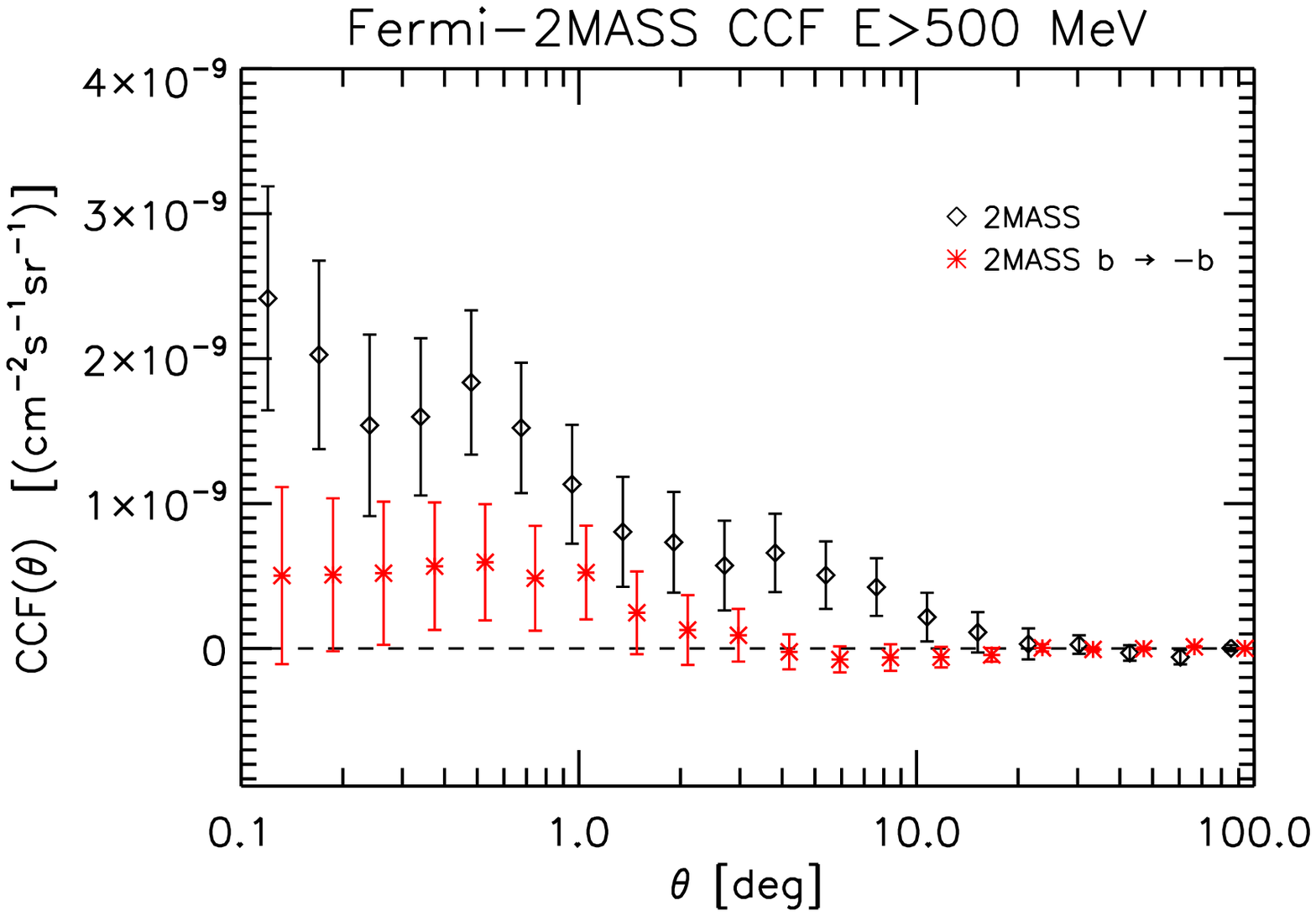, angle=0, width=0.33 \textwidth}
\centering \epsfig{file=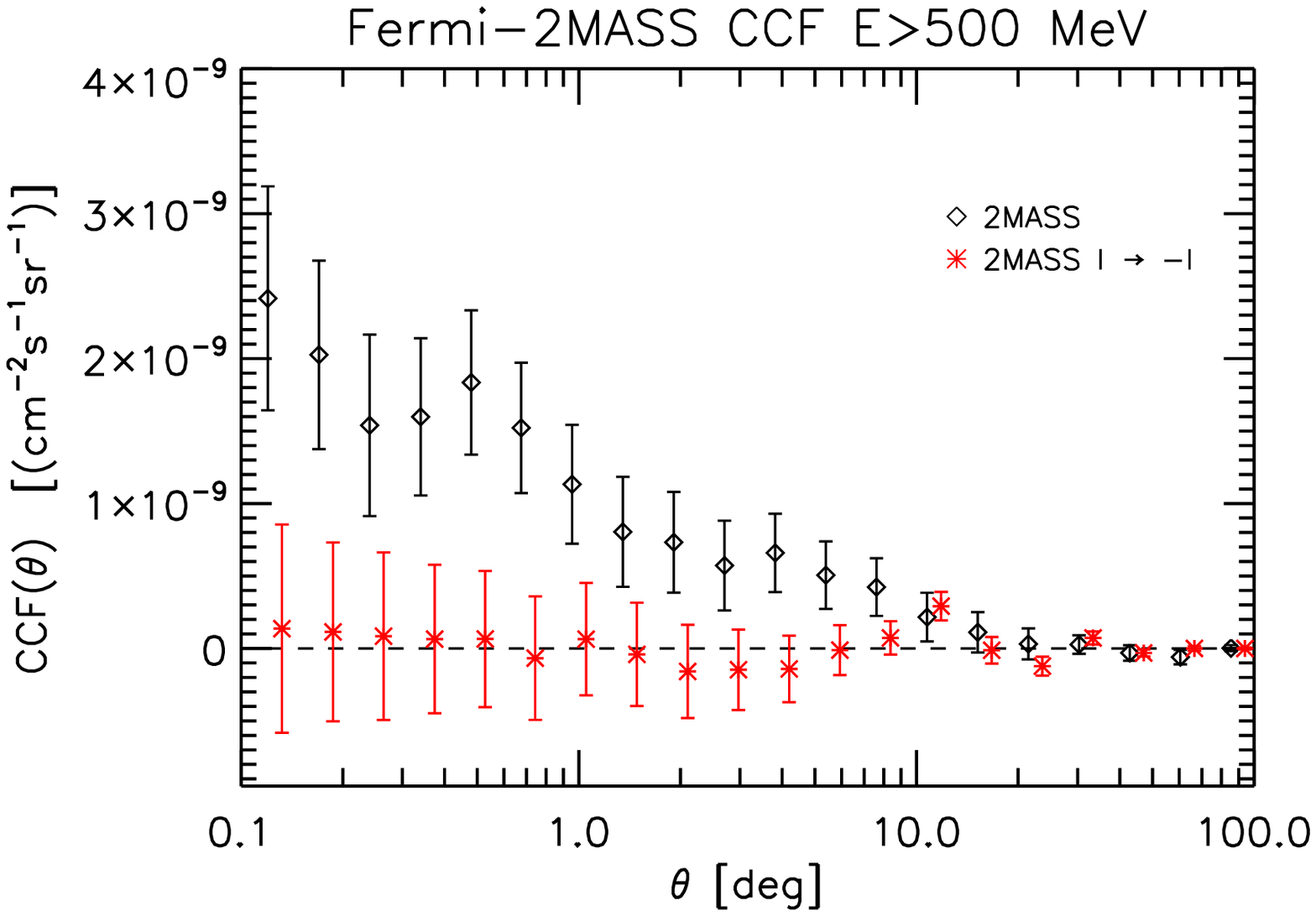, angle=0, width=0.33 \textwidth}
\centering \epsfig{file=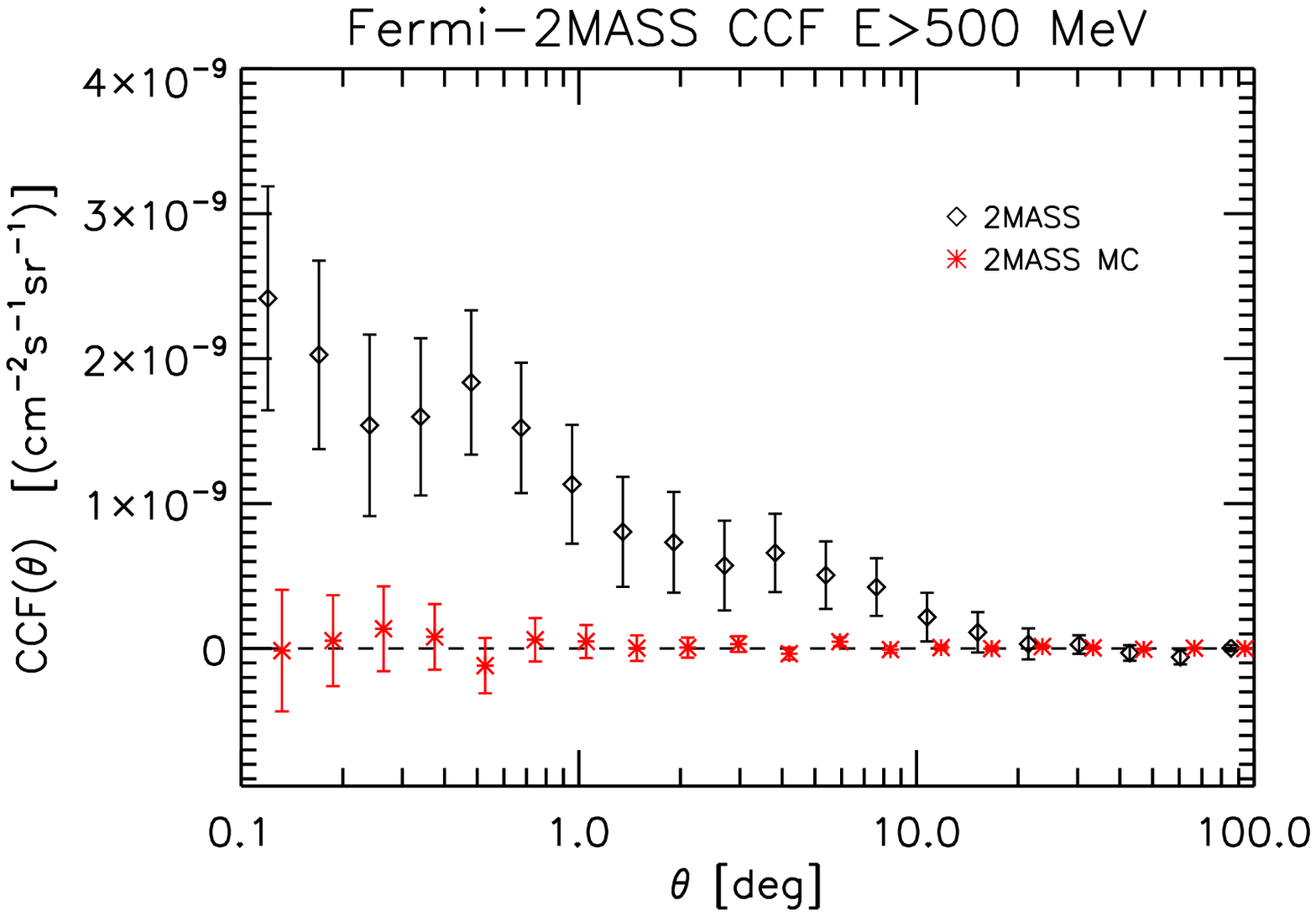, angle=0, width=0.33 \textwidth}
%
%
\centering \epsfig{file=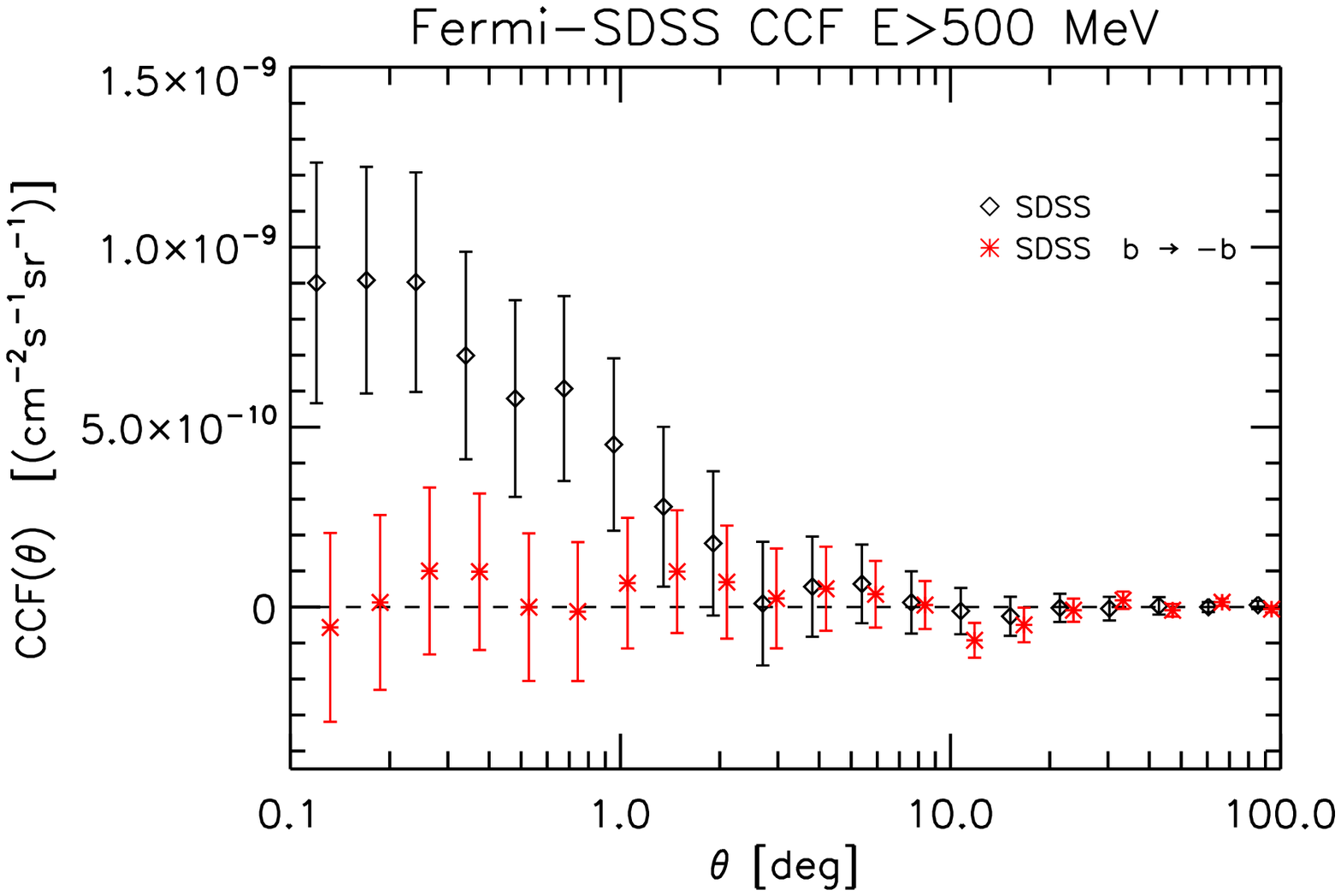, angle=0, width=0.33 \textwidth}
\centering \epsfig{file=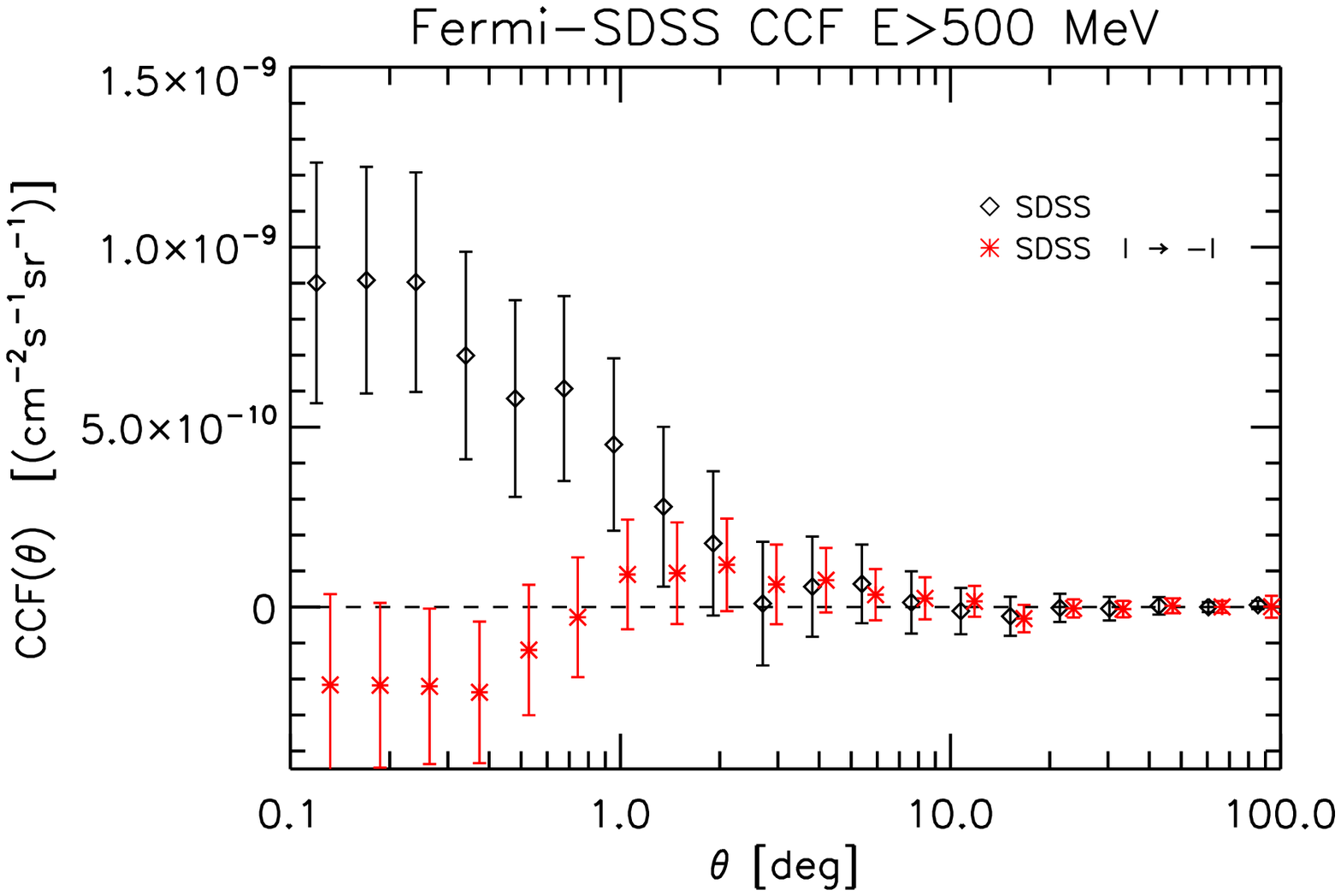, angle=0, width=0.33 \textwidth}
\centering \epsfig{file=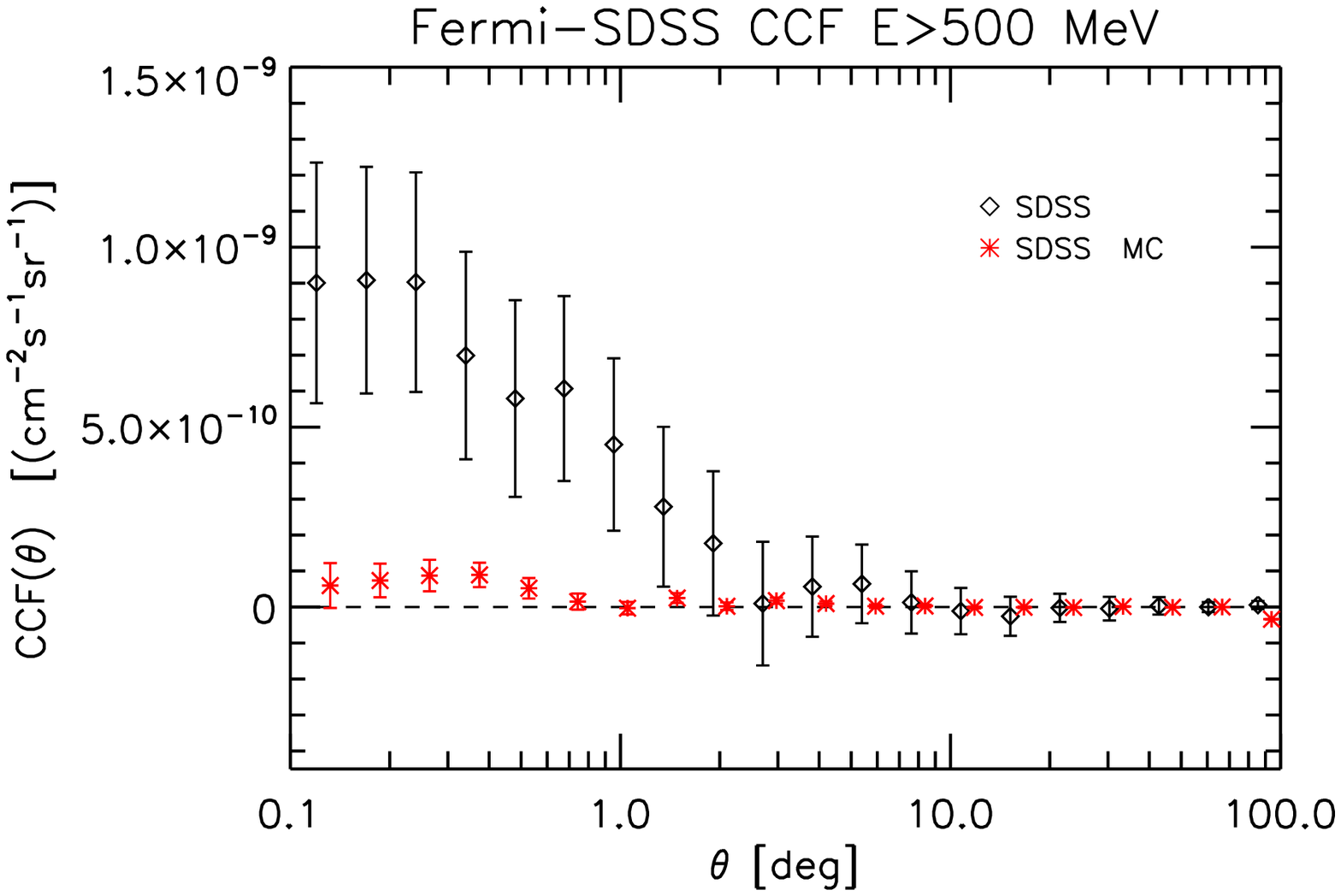, angle=0, width=0.33 \textwidth}
%
%
\centering \epsfig{file=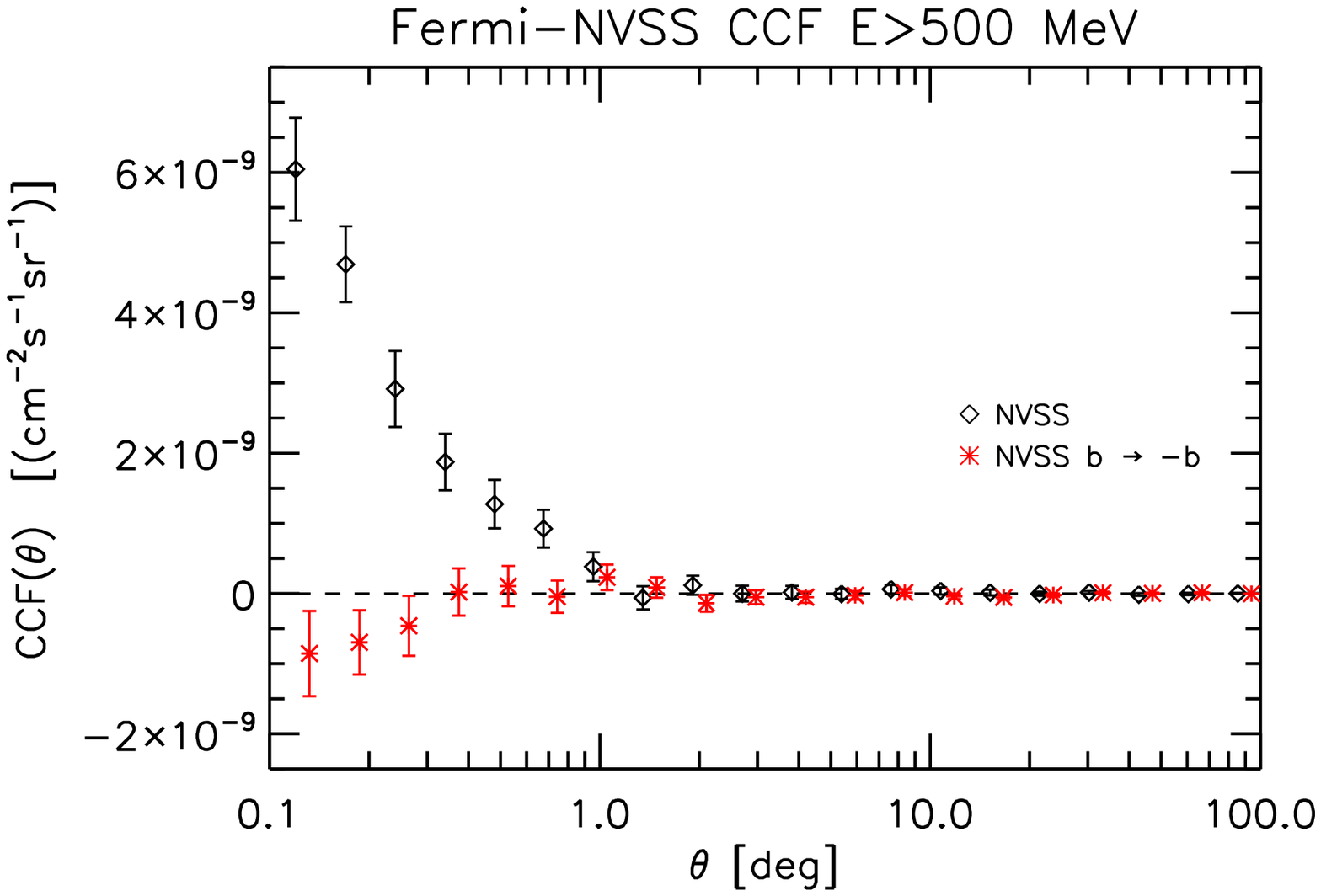, angle=0, width=0.33 \textwidth}
\centering \epsfig{file=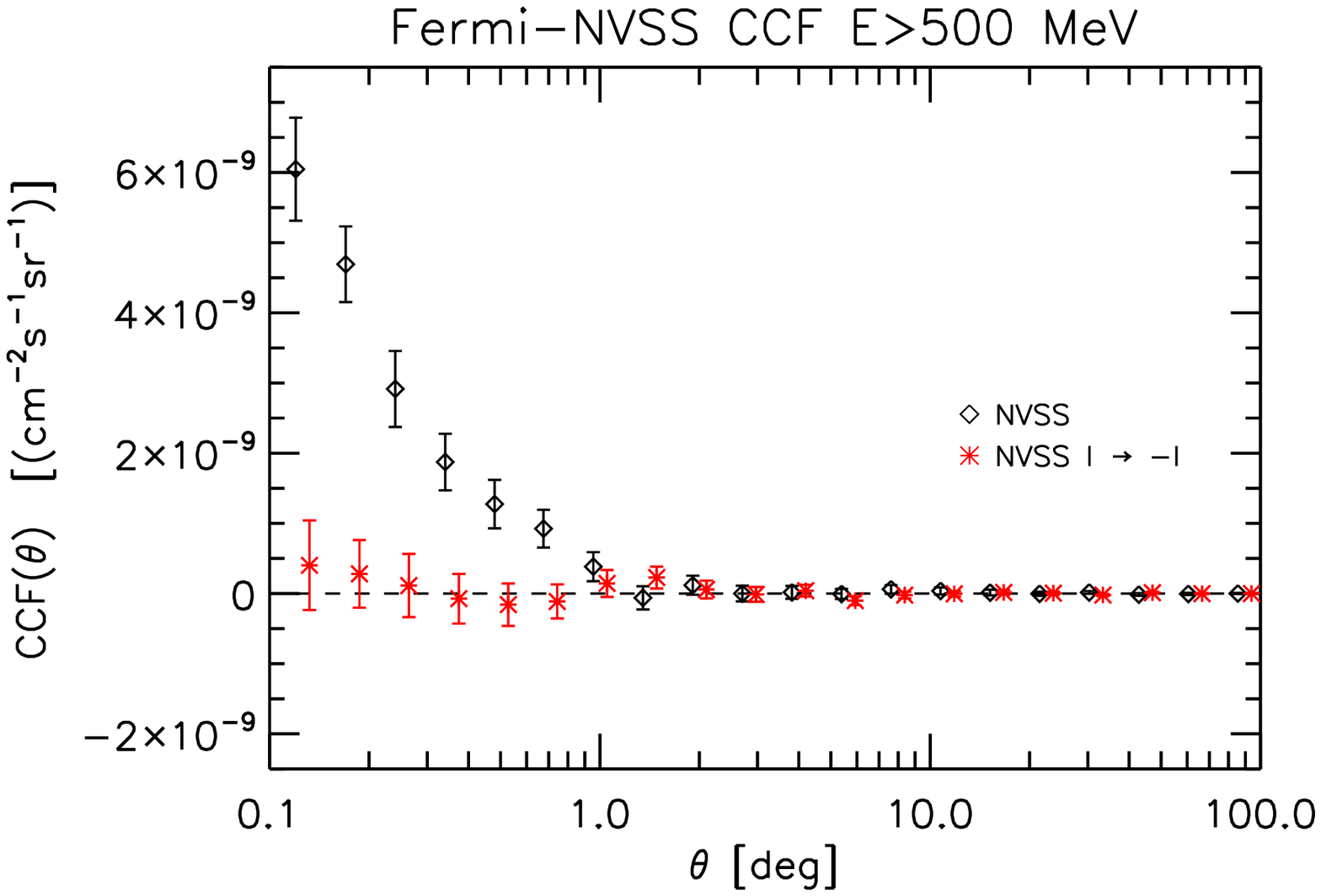, angle=0, width=0.33 \textwidth}
\centering \epsfig{file=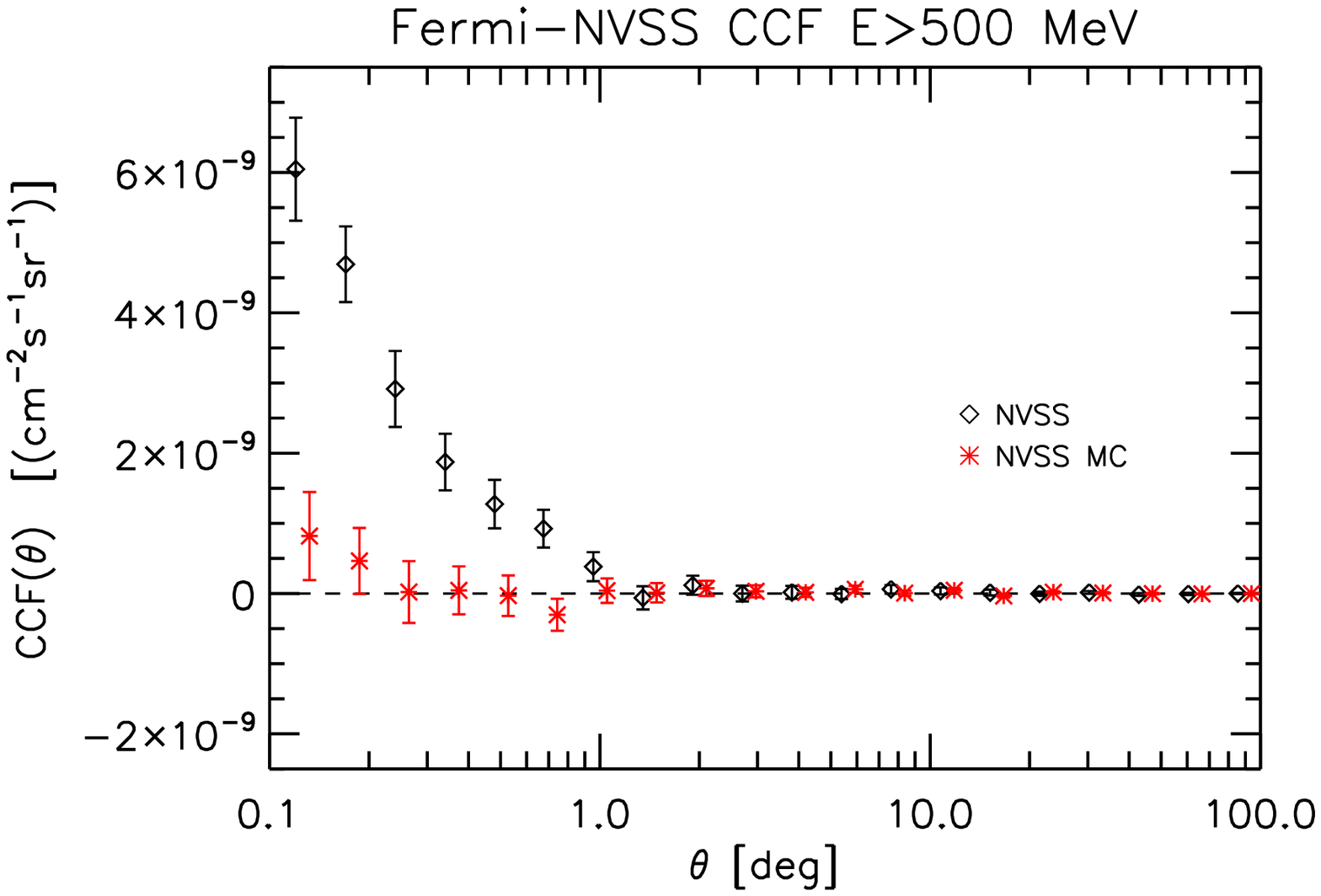, angle=0, width=0.33 \textwidth}
%
\caption{Cross-correlations between the {\it Fermi} $E>$500 MeV map and 3 mock realization of each of the 3 catalogs 2MASS, SDSS main galaxy sample and NVSS, compared with the correlations with the true catalogs.
The three mock realizations refer to the cases of catalog galaxies with scrambled Galactic coordinates
($b \rightarrow-b$ and $l \rightarrow -l$) and catalog galaxies  randomly
distributed (MC (Monte Carlo) label in the plots) over the catalog sky-area
(see text more details).}
\label{fig:null_test}
\end{figure*}

\subsection{Sensitivity to the PSF of the detector}
\label{sec:psf}

As we shall see, a significant fraction of the cross-correlation signal observed
in our analysis originates from small angular scales
comparable with  the angular extent of the LAT PSF.
As a consequence, we need to estimate the effect of the PSF and include it explicitly in our analysis.

The PSF  smears out the signal from small to large angular scales, hence reducing the
amplitude of both the CCF and CAPS at small angular separations and large multipoles, respectively.
However this effect can be  modeled if the PSF of the telescope, or the instrumental beam, is measured
accurately.  In the case of the LAT the beam depends on the energy, and the PSF can be determined
either from observations by stacking the images of bright point sources \citep{2013ApJ...765...54A} or
from a Monte Carlo  simulation of the detector performance \citep{2012ApJS..203....4A}.
The characterization of the PSF has improved with the \verb"P7" and \verb"P7REP" data release and a discrepancy
at high energies
($>$ few GeVs) between the Monte Carlo PSF and the in-flight PSF present for the \verb"P6" data
is now significantly reduced \citep{2012ApJS..203....4A}.
The beam shape is part of the IRFs and can be estimated using the LAT Science Tools. In particular,
 we used the tool \verb"gtpsf" to obtain the PSF as a function of energy and angular separation of the
 photon from its  true arrival direction. As the latter is a function of one angle only we are
 neglecting the ellipticity of the beam which, in any case, turns out to be negligible.
 It is more convenient to consider the effect of the beam in
 harmonic space, where it can be expressed as a multiplication rather than
 in  configuration space, where it would be a convolution.
 Indeed, if $C_l(E)$ represents the true CAPS at a given energy, then
 the measured one is  $\tilde{C}_l(E)=W_l(E) C_l(E)$, where $W_l(E)$ is the
 beam window
function. The latter can be expressed as  a Legendre transform:
\begin{equation}
W_l(E)=2\pi \int_{-1}^{1} \!\!  d\cos(\theta) \ P_l(\cos \theta) {\rm PSF}(\theta, E) \; ,
\end{equation}
where $P_l(x)$ is the Legendre polynomial of order $l$ and
PSF$(\theta, E)$ is the shape of the beam.
Since we are analyzing data integrated over a fairly large energy bin
within which the PSF can vary significantly, the effective
window function for the bin will be  a weighted average over the energy range:
\begin{equation}
W_l(E_1\!\! <\!\! E\!\! <\!\! E_2)=\frac{1}{N} \int_{E_1}^{E_2} \!\!  dE \ W_l(E) \frac{dN}{dE}(E) \; ,
\end{equation}
where $dN/dE$ represents the differential number of photon counts
in the region of the sky we want to analyze,   $N = \int_{E_1}^{E_2}  dE (dN/dE)(E)$,   and    $[E_1,E_2]$ is the
energy bin considered.
Finally, there is a further window function to take into account due to the fact
that we use a pixelized map, which is the pixel window function itself $W_{{\rm pix},l}$.
The pixel window function depends on the size of the pixel (and on the shape of the pixel itself).
It can be easily extracted using the appropriate HEALPix tools.
The final window function is then given by $W_l(E_1\!\! <\!\! E\!\! <\!\! E_2) \cdot W_{{\rm pix},l}$.
The effective window functions for the 3 energy ranges considered in this work
are shown in Fig.~\ref{fig:beamwindow}.

\begin{figure*}
\centering \epsfig{file=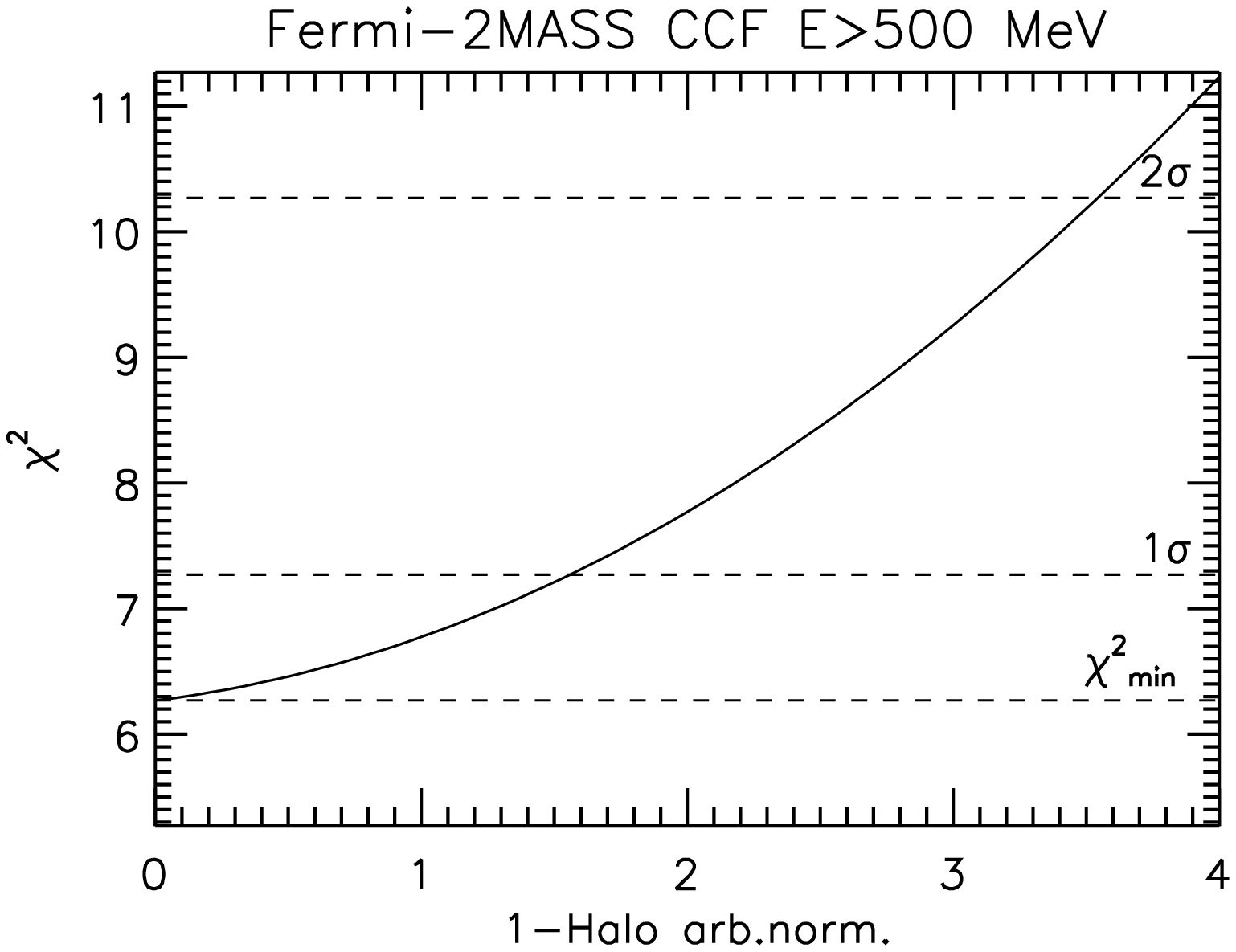, angle=0, width=0.20 \textwidth}
\hspace{-0.2cm}
\centering \epsfig{file=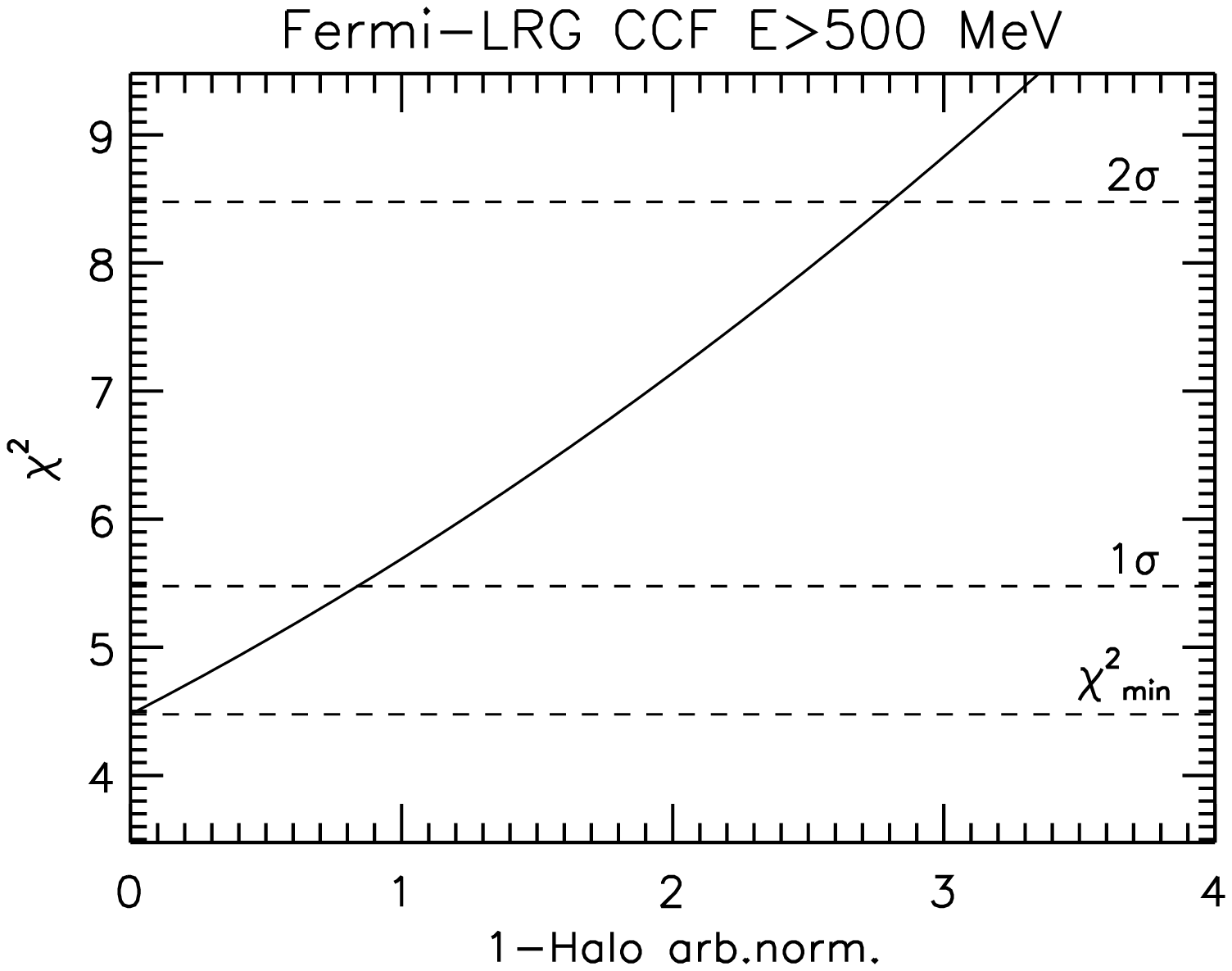, angle=0, width=0.20 \textwidth}
\hspace{-0.2cm}
\centering \epsfig{file=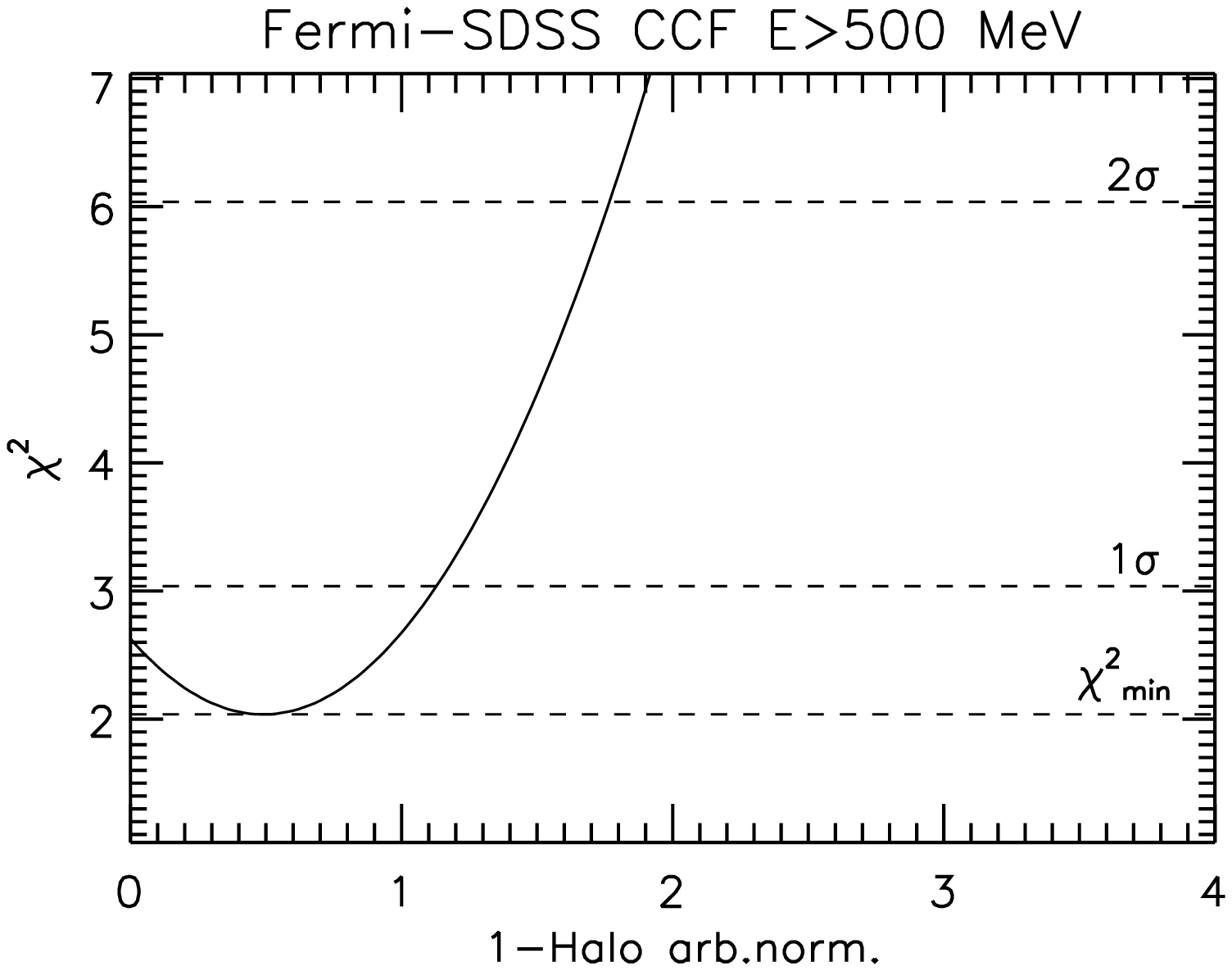, angle=0, width=0.20 \textwidth}
\hspace{-0.2cm}
\centering \epsfig{file=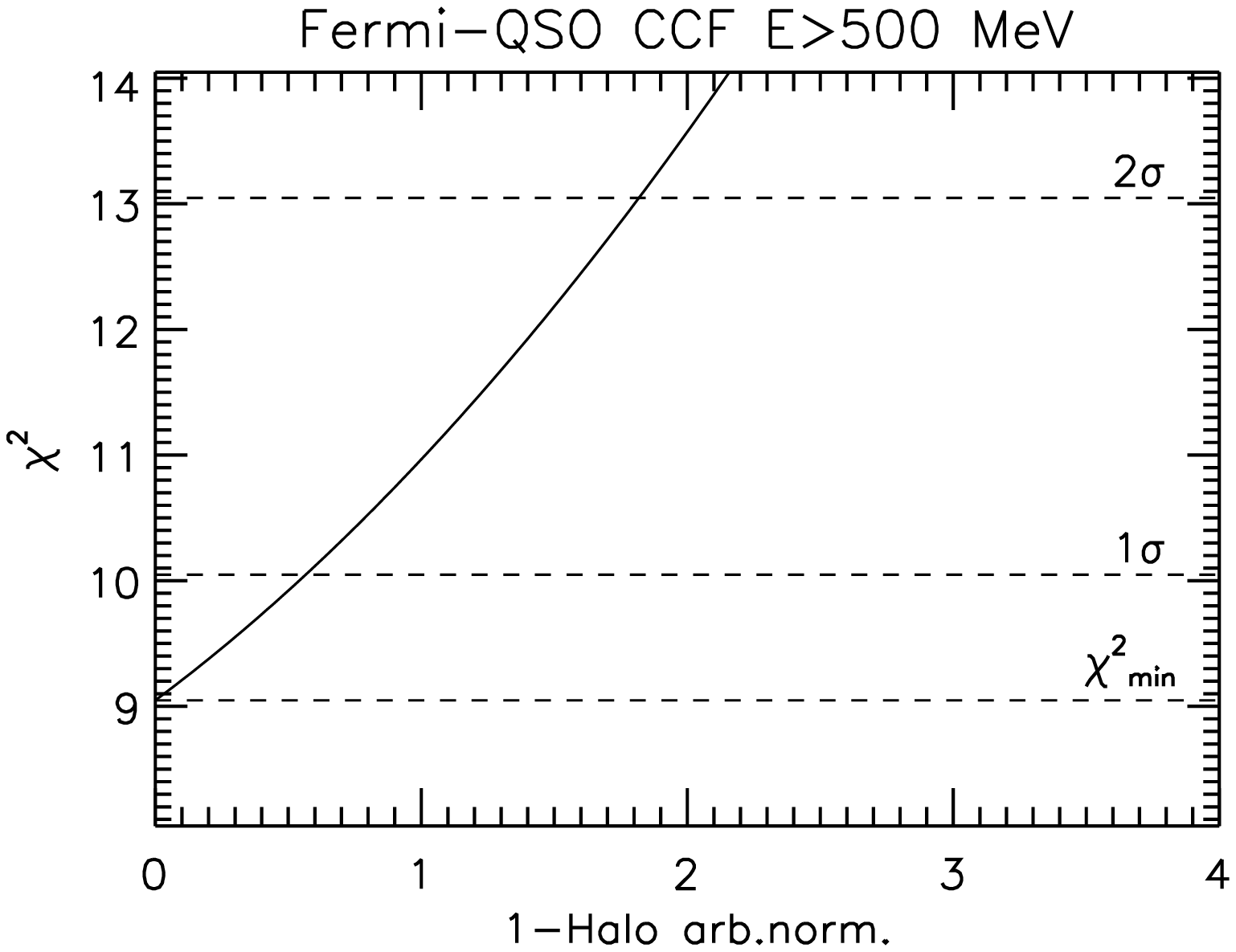, angle=0, width=0.20 \textwidth}
\hspace{-0.2cm}
\centering \epsfig{file=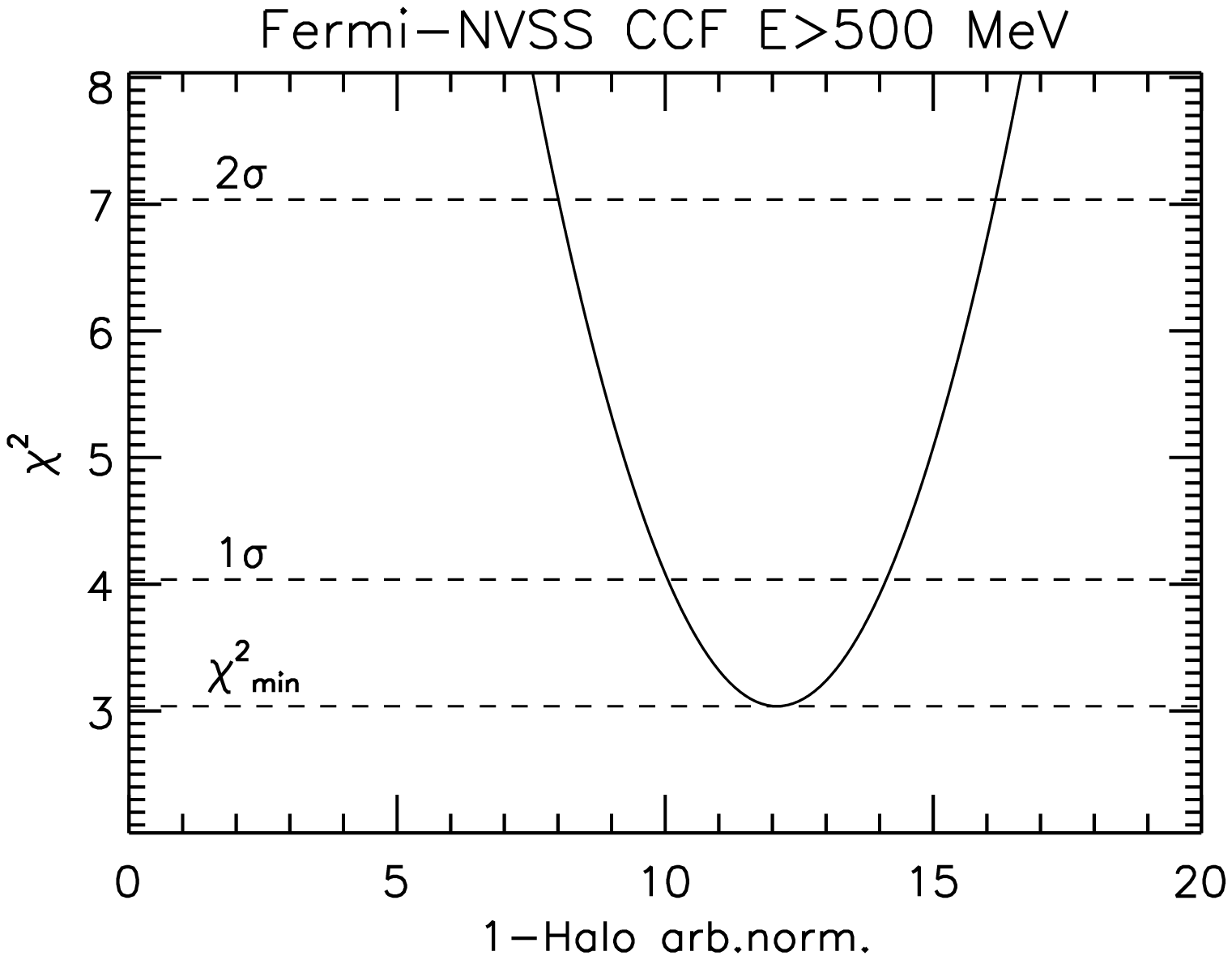, angle=0, width=0.20 \textwidth}
%
%
\centering \epsfig{file=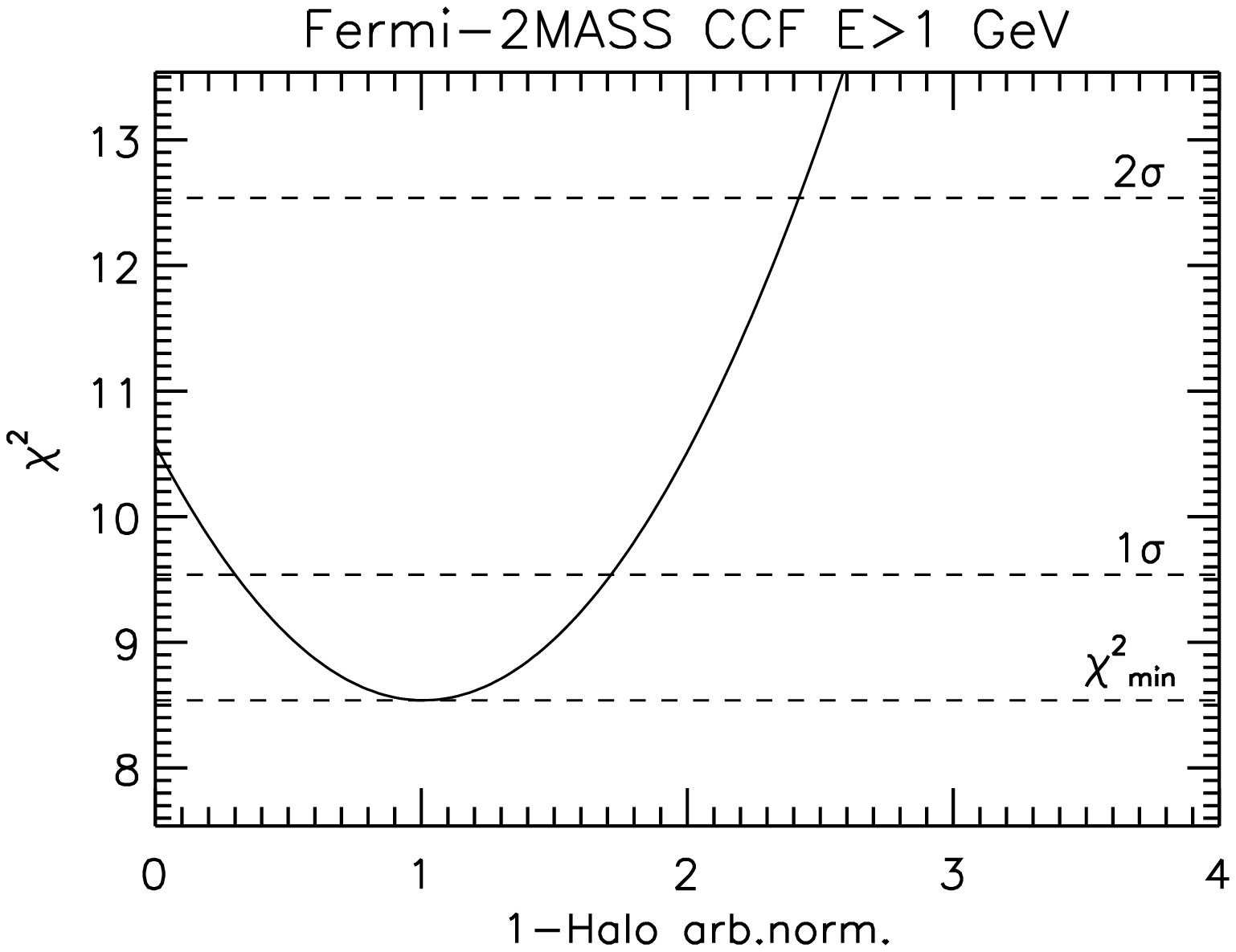, angle=0, width=0.20 \textwidth}
\hspace{-0.2cm}
\centering \epsfig{file=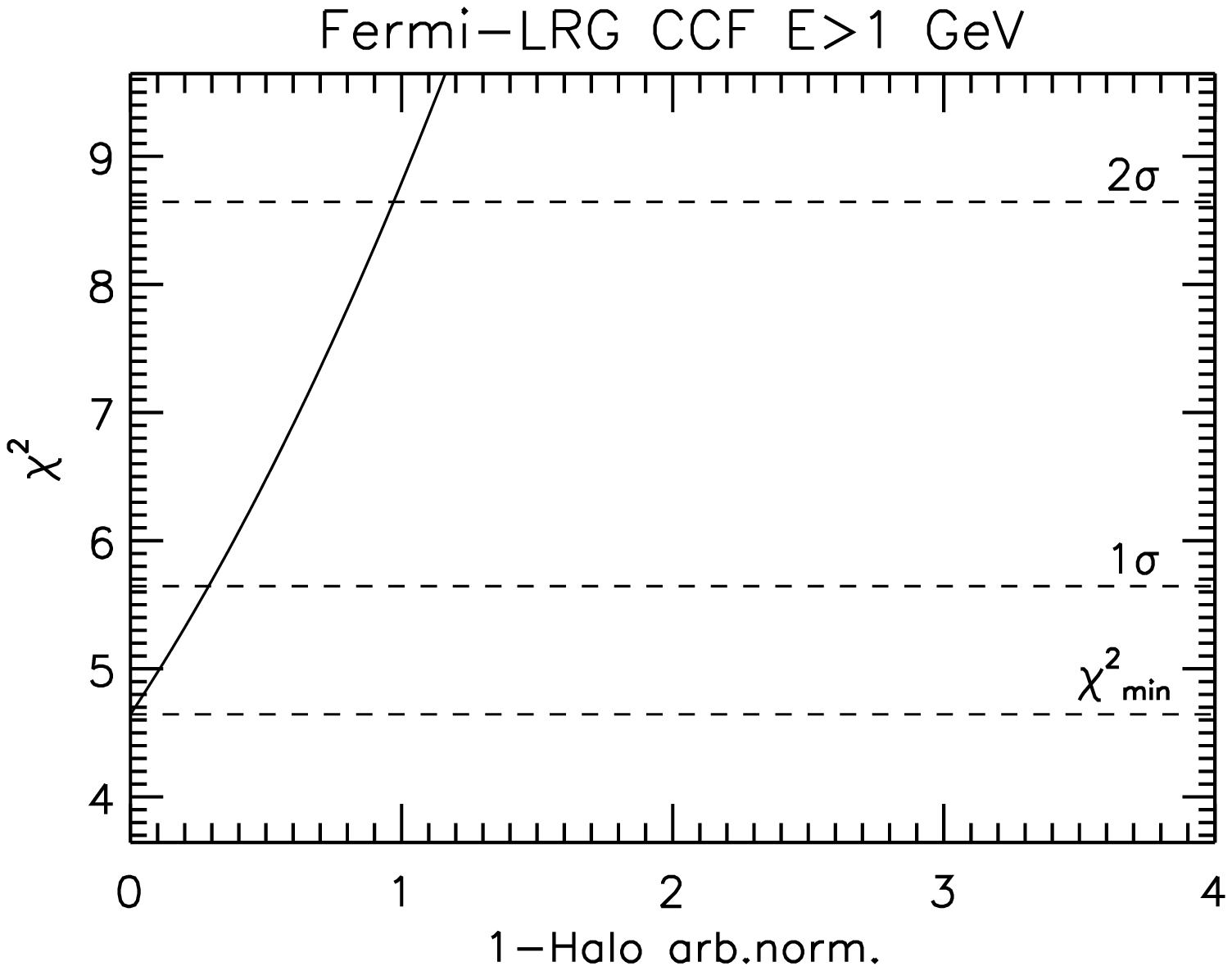, angle=0, width=0.20 \textwidth}
\hspace{-0.2cm}
\centering \epsfig{file=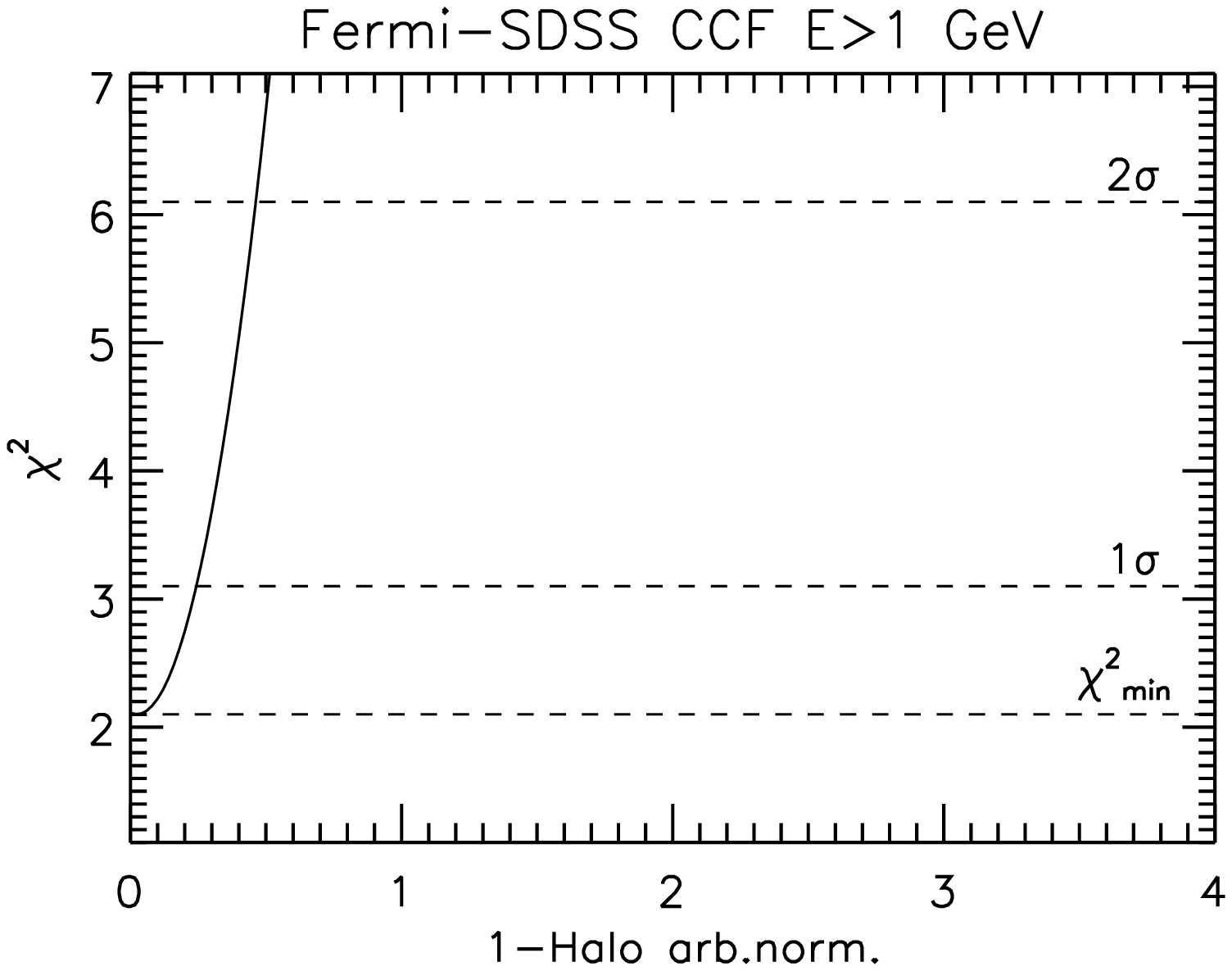, angle=0, width=0.20 \textwidth}
\hspace{-0.2cm}
\centering \epsfig{file=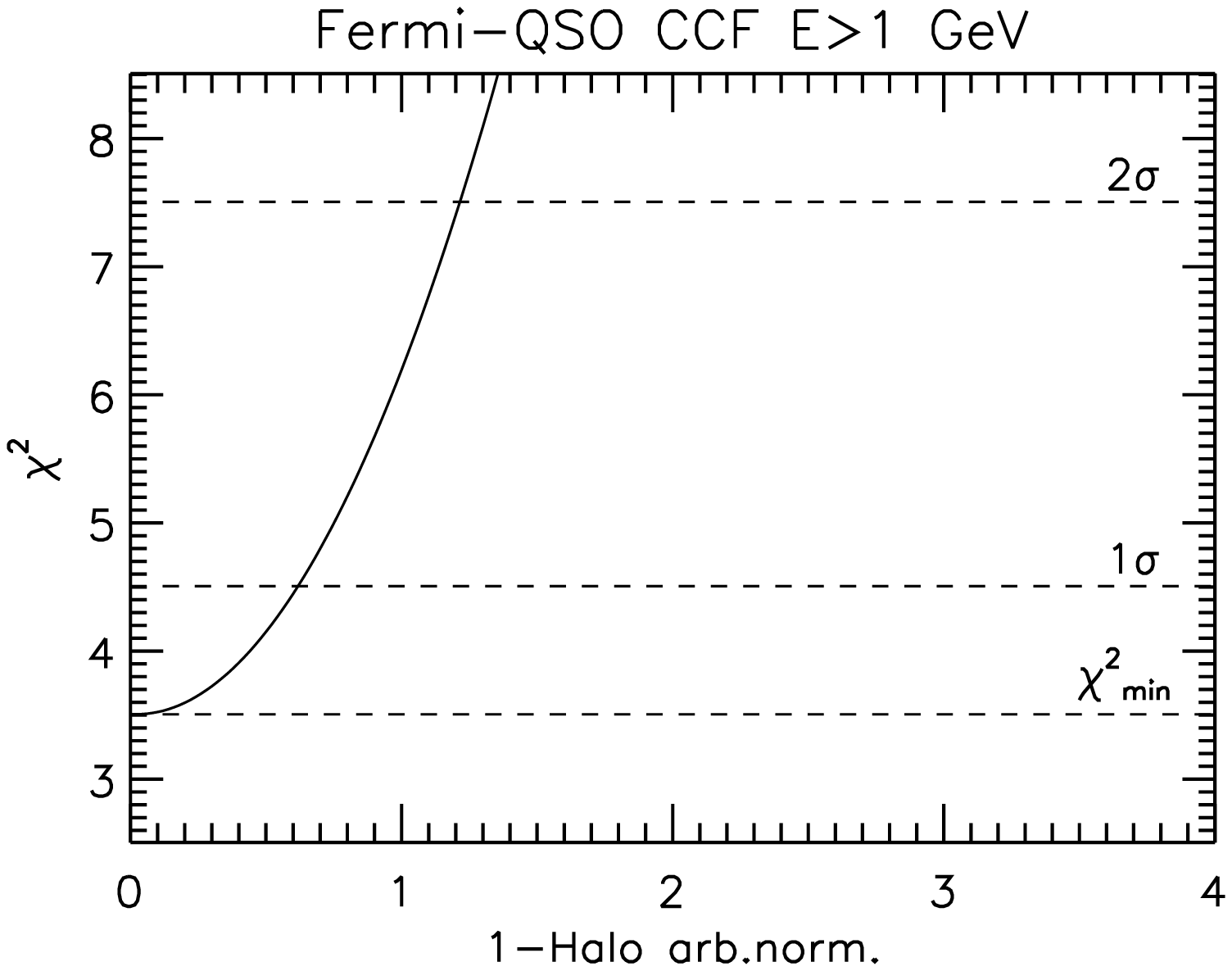, angle=0, width=0.20 \textwidth}
\hspace{-0.2cm}
\centering \epsfig{file=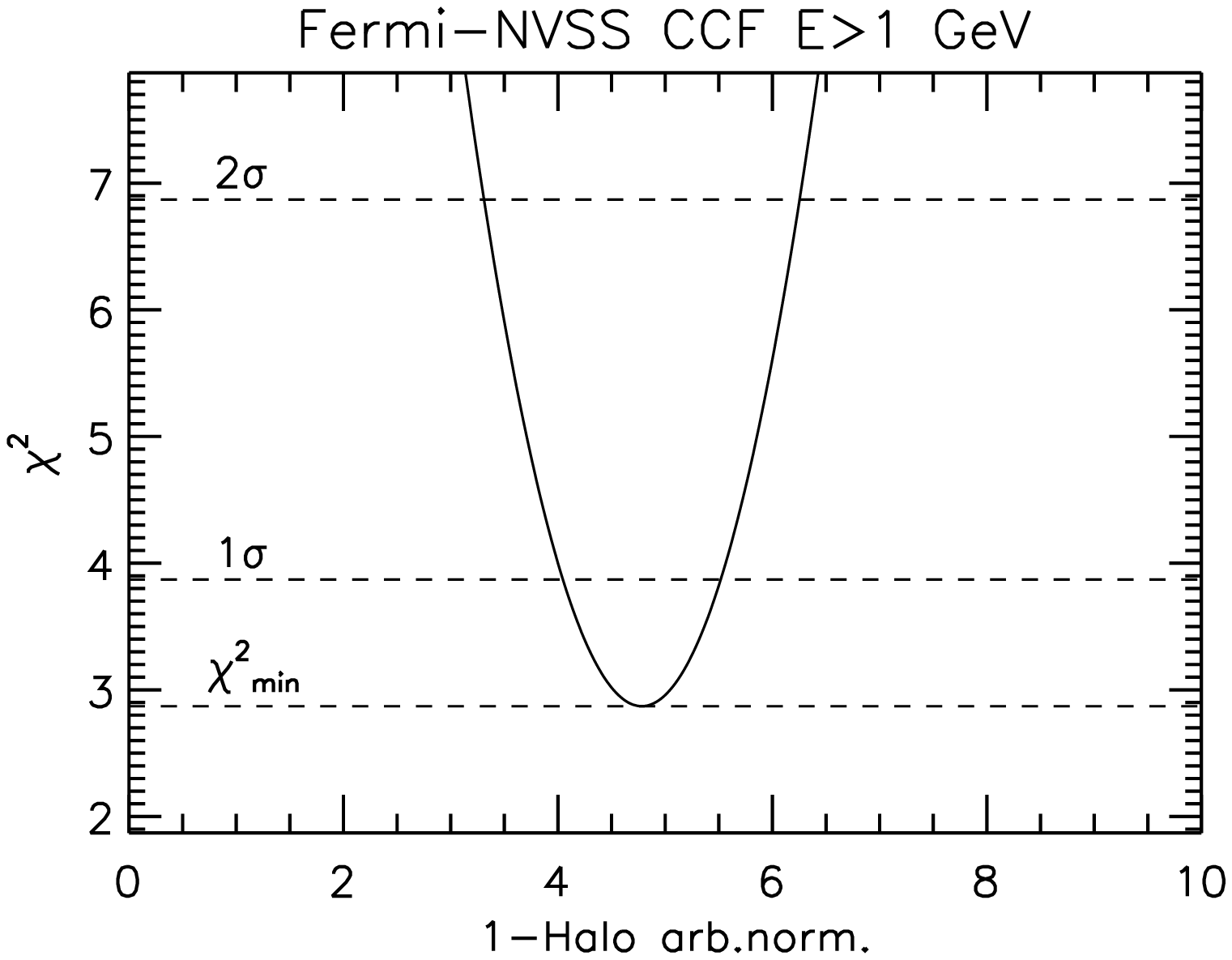, angle=0, width=0.20 \textwidth}
%
%
%
\centering \epsfig{file=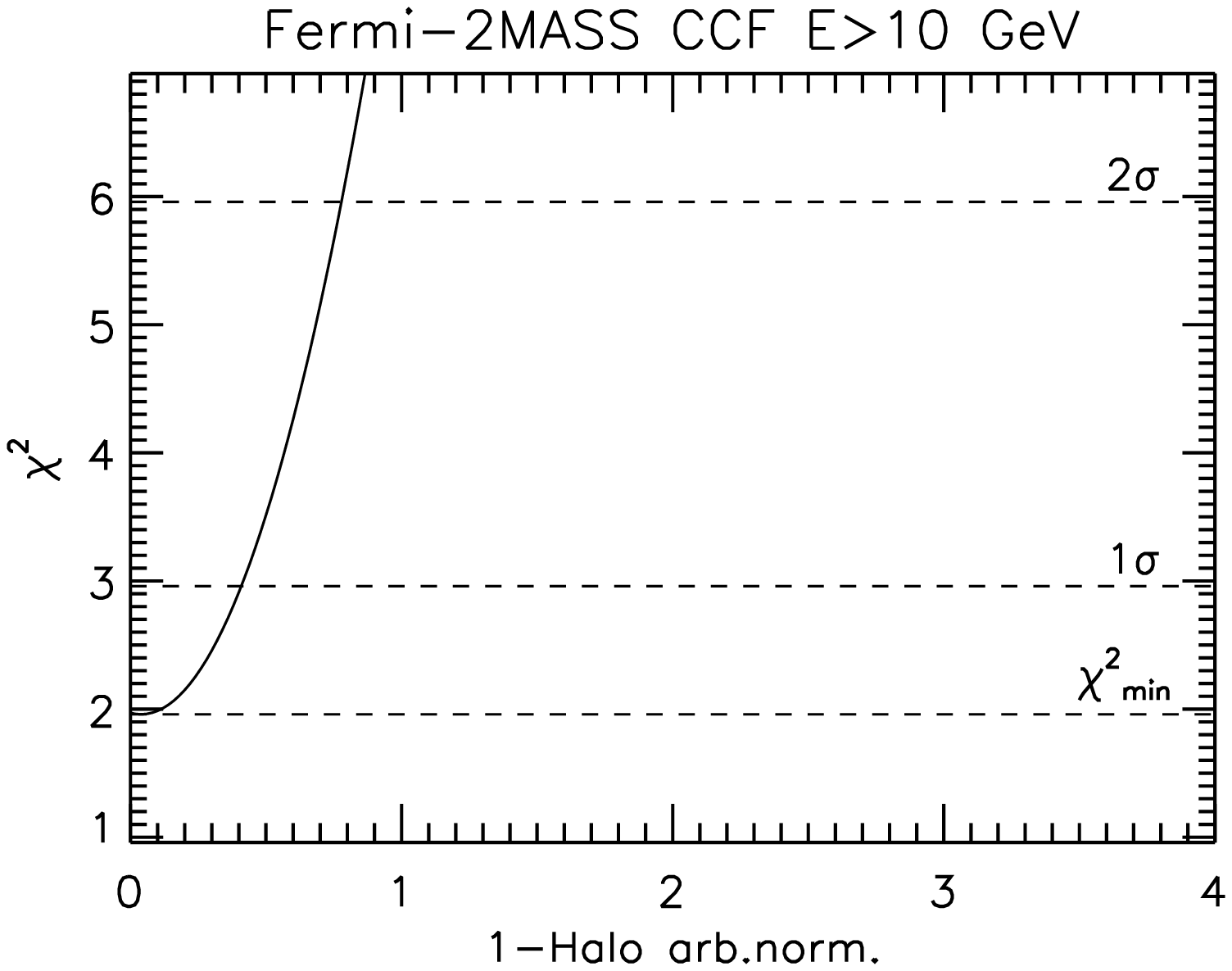, angle=0, width=0.20 \textwidth}
\hspace{-0.2cm}
\centering \epsfig{file=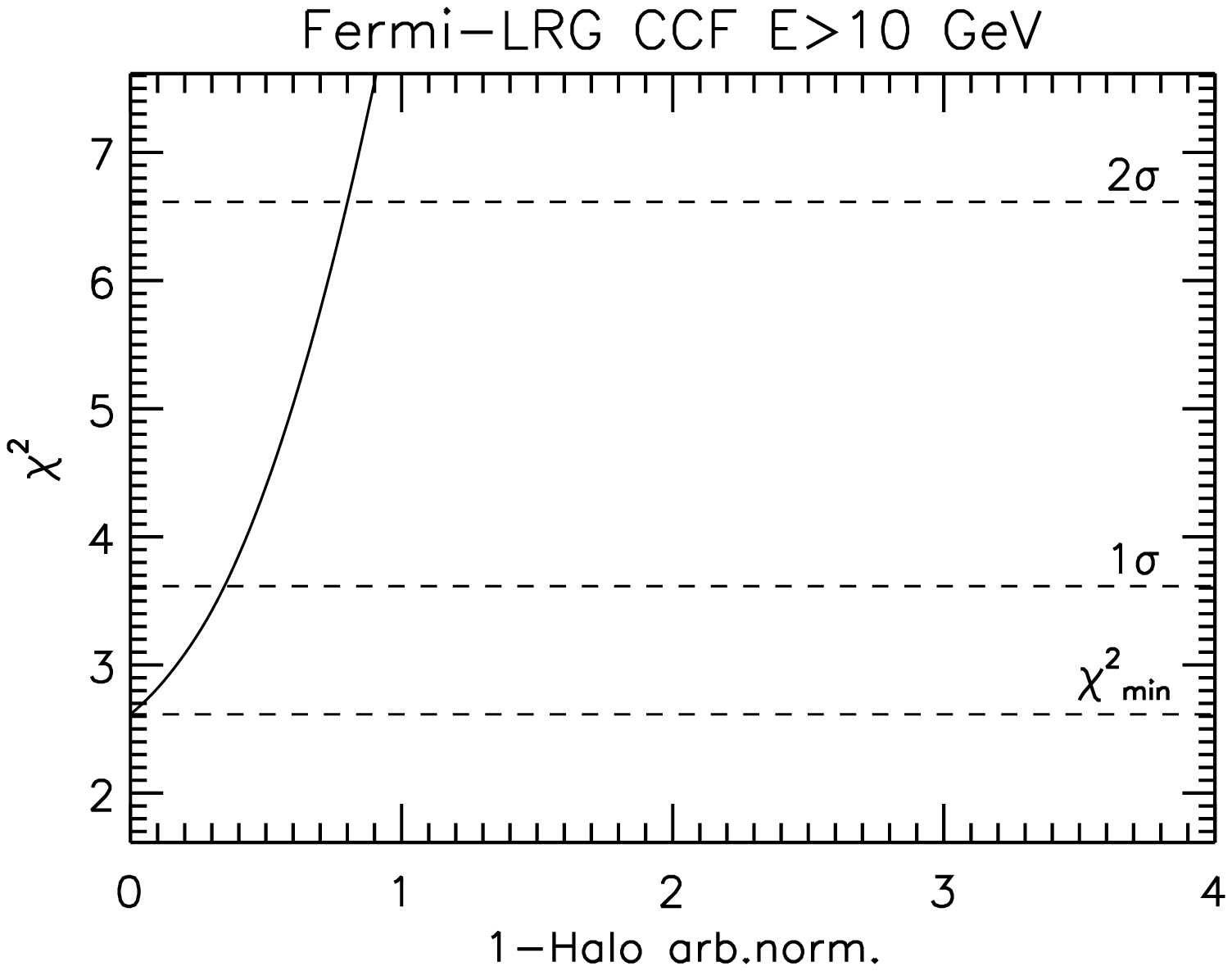, angle=0, width=0.20 \textwidth}
\hspace{-0.2cm}
\centering \epsfig{file=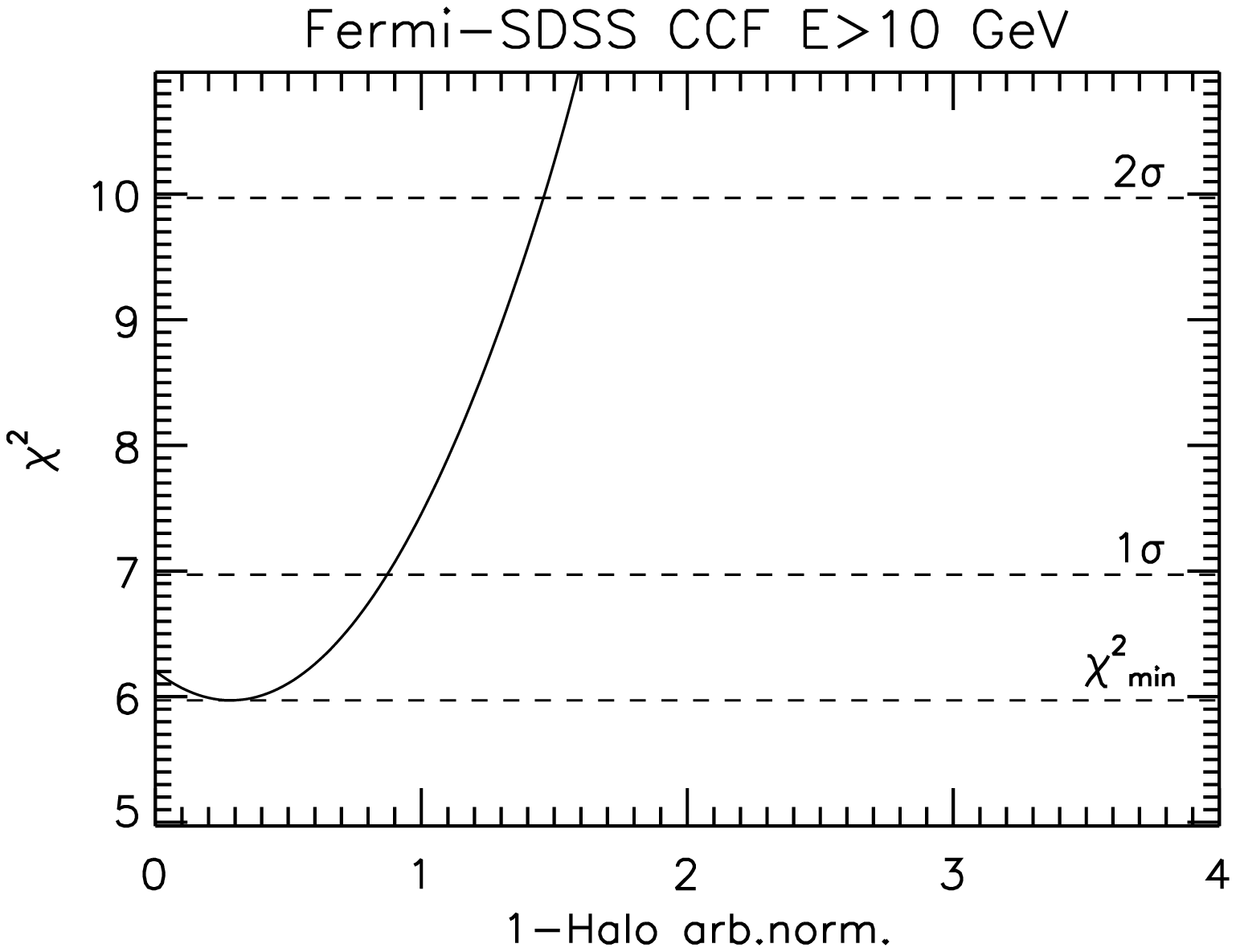, angle=0, width=0.20 \textwidth}
\hspace{-0.2cm}
\centering \epsfig{file=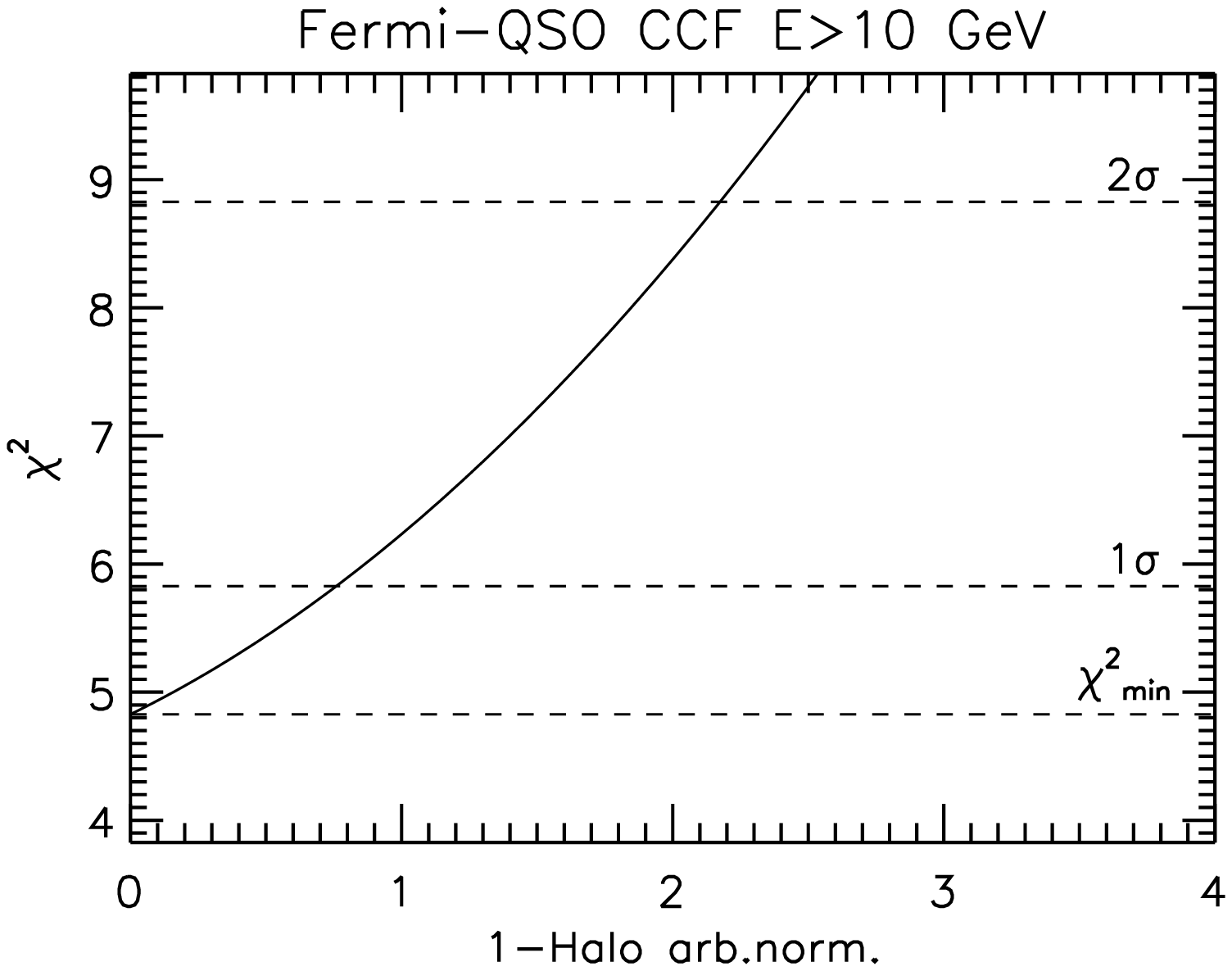, angle=0, width=0.20 \textwidth}
\hspace{-0.2cm}
\centering \epsfig{file=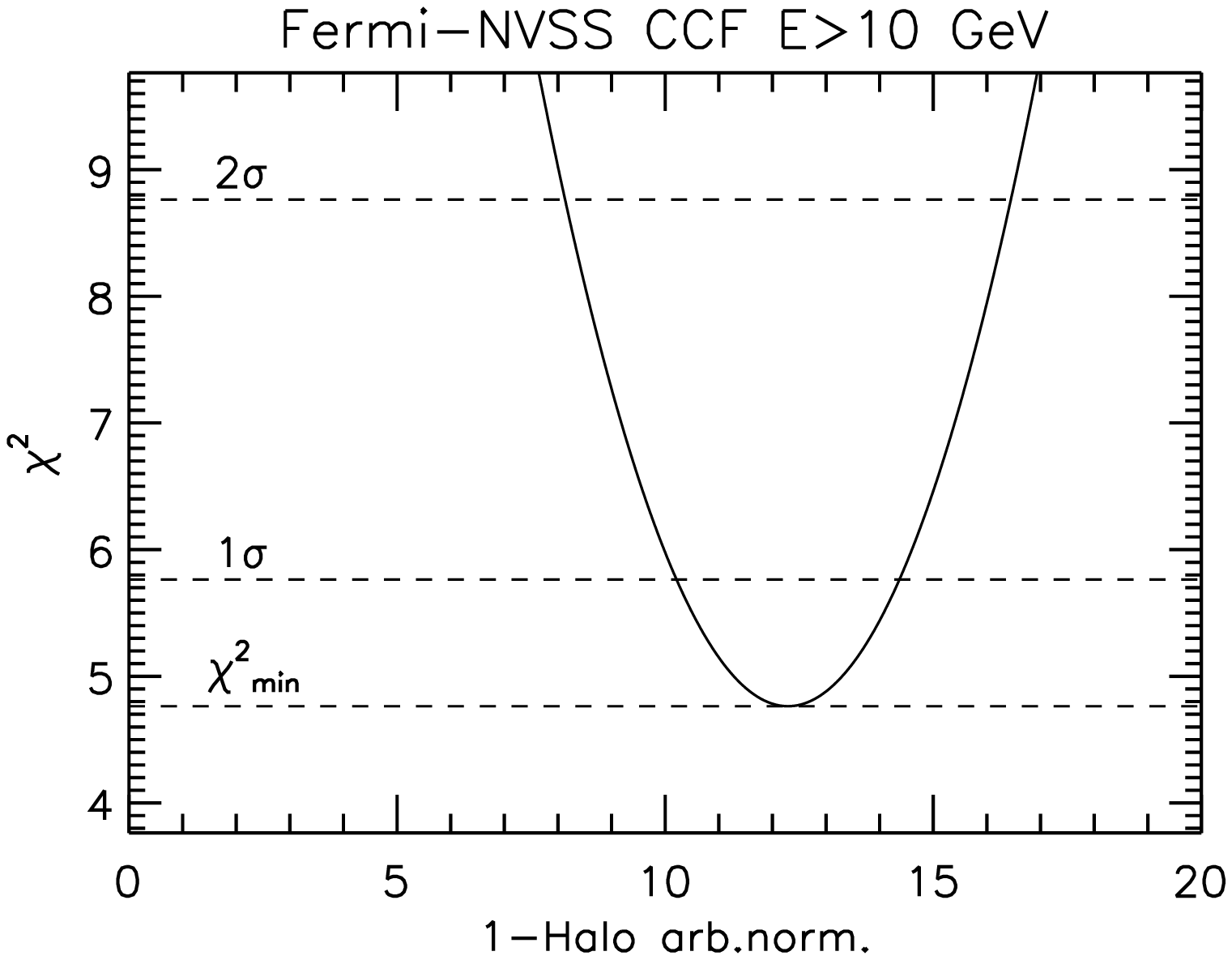, angle=0, width=0.20 \textwidth}
%
%
\caption{1-dimension $\chi^2$ profile  of the {\it 1-halo}-like normalization (in arbitrary units) from the joint {\it 1-halo}-like and SFGs1 2-dimensional fit.
Each panel shows the case of the fit to a single catalog and energy band CCF.}
\label{fig:1Halo}
\end{figure*}

Once determined, the effective window function can be fed into
PolSpice to recover the true CAPS and the CCF from the measured ones.
However, we find that the algorithm used to perform the deconvolution is quite unstable, especially
in the CCF reconstruction.
For this reason we take the opposite approach and instead of deconvolving the signal, we
convolve the model predictions and compare them to the measurement.
More explicitly, if $C_l$ is the model  CAPS
and $W_l^{\Delta E}$ is the estimated effective window function in the bin $\Delta E$,
then the convolved CAPS is
\begin{equation}
C_l^W = C_l  \, W_l^{\Delta E} \;,
\end{equation}
and, analogously, the convolved CCF is
\begin{equation}
{\rm CCF}^W(\theta)=\sum_l  \frac{(2l+1)}{4\pi} C_l^W P_l(\cos \theta) \;.
\end{equation}

To  test the validity of our PSF correction procedure we proceed as follows:
  {\it i)} we consider the high Galactic latitudes  $|b|>30^{\circ}$ region of the sky (to reduce
the impact of the Galactic contamination) but without masking the locations of
known point sources;
{\it ii)} we calculate  the   auto power spectrum for this region (APS, not the CAPS);
{\it iii)} we then apply the pipeline described in the above paragraphs to obtain an empirical estimate of the window function
in each of the three energy bands considered.
The resulting APS
in this case will be dominated by the bright point sources and is expected to match the
Legendre transform of the window function (squared): $(W_l^{\Delta E})^2$.
The results are shown in the three panels of  Fig.~\ref{fig:beam_test}.
Since the APS of the map, represented by black asterisks, is expected to match the window function
then the flatness of the ratio between the APS and  $(W_l^{\Delta E})^2$
indicates that our hypothesis is correct and that our estimated window function is robust
in all three energy ranges and at all multipoles, apart from a small overestimation at very high $l$.
Note that, in contrast to \cite{FermiAPS} where the $\gamma$-ray APS is also considered,
we neglect here the effect of Poisson shot noise which is sub-dominant at all multipoles,
given the very strong APS signal from bright point sources.
Instead, no shot noise term needs to be considered for
CAPS, since the uncorrelated noises from the two maps being cross-correlated
do not produce a net non-null noise CAPS, contrary to the APS case.
The slight over-estimate of the APS PSF correction in the 700-1000 $l$-range at the level of
20-30\% turns into a 10-15\% systematic effect for the CAPS, where the PSF correction
is given by $W_l^{\Delta E}$ rather than $(W_l^{\Delta E})^2$.
On the other hand, as we will see in section~\ref{sec:results} all CAPS are compatible with zero
in this multipole range, except for a weak signal for NVSS, so that the error is dominated
by statistical random errors. We will thus neglect the above systematic effect.

\subsection{Cosmic-ray  Contamination}

The IGRB maps we obtained in Section \ref{sec:fermimaps}
contain, besides true \mbox{$\gamma$-ray} events,  also some contamination
from cosmic rays which have been mis-classified as $\gamma$ rays.
With the \verb"P7REP_CLEAN"  event selection that we used,  the cosmic-ray
 contamination is at the level of 15-20\% of the IGRB flux
above 1 GeV, rising to 40-50\% at 500 MeV (see Fig.~28 in \cite{2012ApJS..203....4A}).
Since the contaminant cosmic rays
are not expected to correlate with  cosmological structures,
they do not induce systematic errors in our analysis.
Instead, they will only increase random error
because the signal-to-noise ratio of the $\gamma$-ray signal will be reduced.
Nonetheless, to verify this hypothesis
we used the IGRB maps  produced with the
\verb"P7REP_ULTRACLEAN"
selection, which has a slightly reduced  cosmic-ray
 contamination with respect to \verb"P7REP_CLEAN",
at the level of 10-15\% of the IGRB flux
above 1 GeV and 30-40\% at 500 MeV \citep{2012ApJS..203....4A}.
We computed CAPS and CCFs for this case and found that the
results are  indistinguishable from those obtained with the
\verb"P7REP_CLEAN"  selection.
A more stringent test could be performed using special
event selection criteria designed to further reduce
the CR contamination for specific studies of the IGRB
spectrum as in \cite{2010PhRvL.104j1101A} and \cite{IGRBII}.
These selections should in principle allow to reduce the error bars.
However, these selection criteria introduce more conservative cuts to reduce
the background. Consequently, the benefit of the better cleaning is counteracted by
the effective area reduction, resulting typically in no effective reduction of the error bars.
Indeed, the optimal selection should balance purity
and photon statistics. However, the search for such compromise is  beyond the scope of our analysis.

\begin{figure*}
\centering \epsfig{file=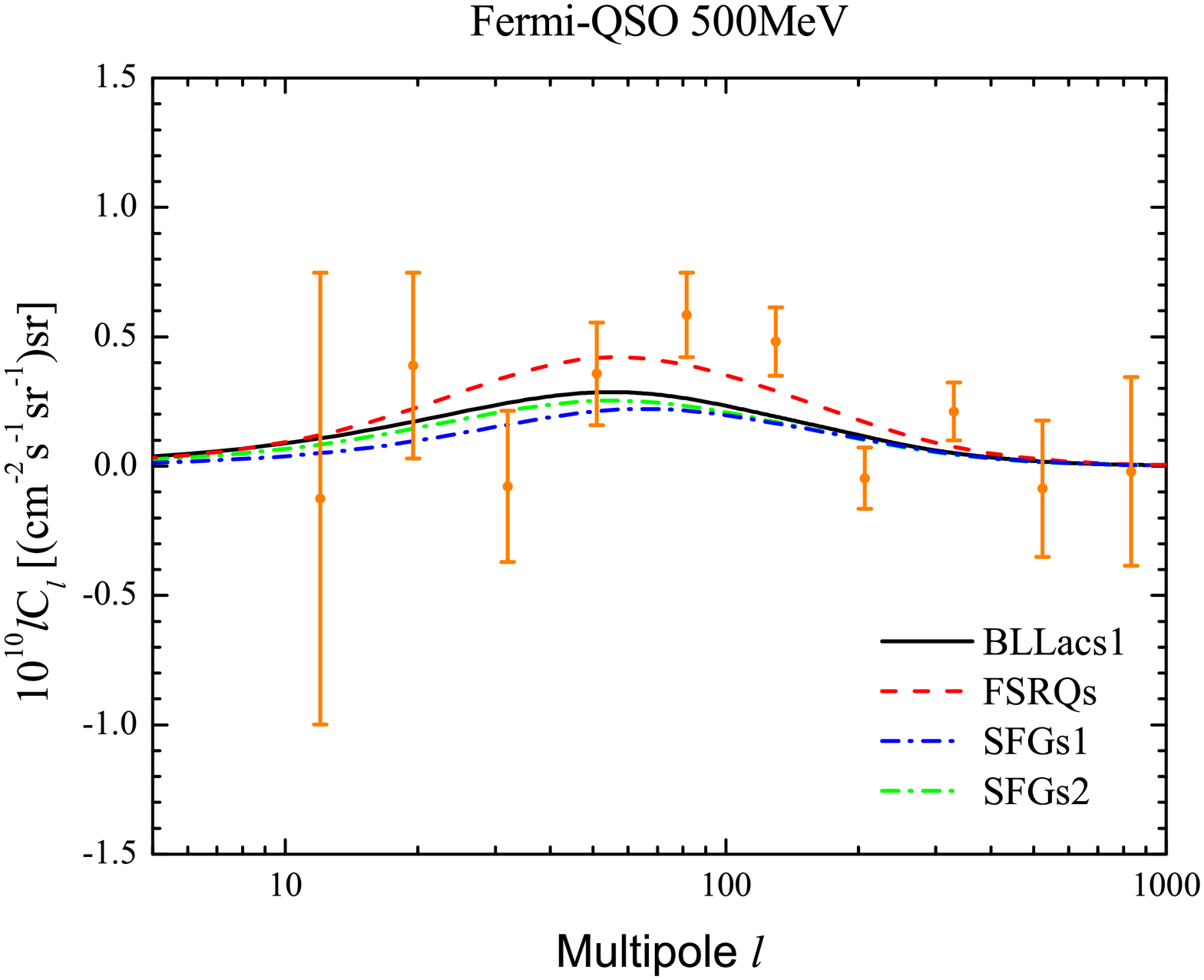, angle=0, width=0.33 \textwidth}
\centering \epsfig{file=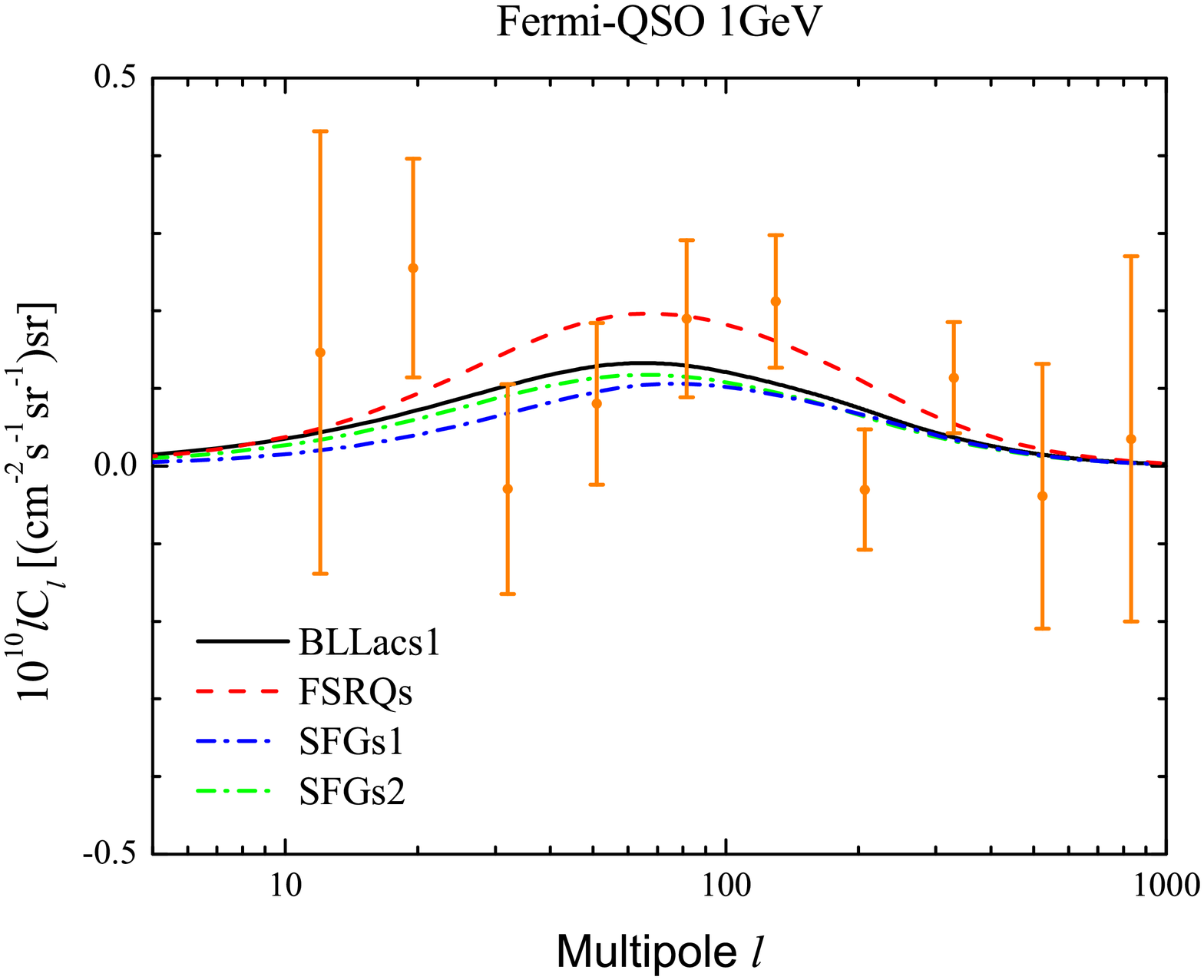, angle=0, width=0.33 \textwidth}
\centering \epsfig{file=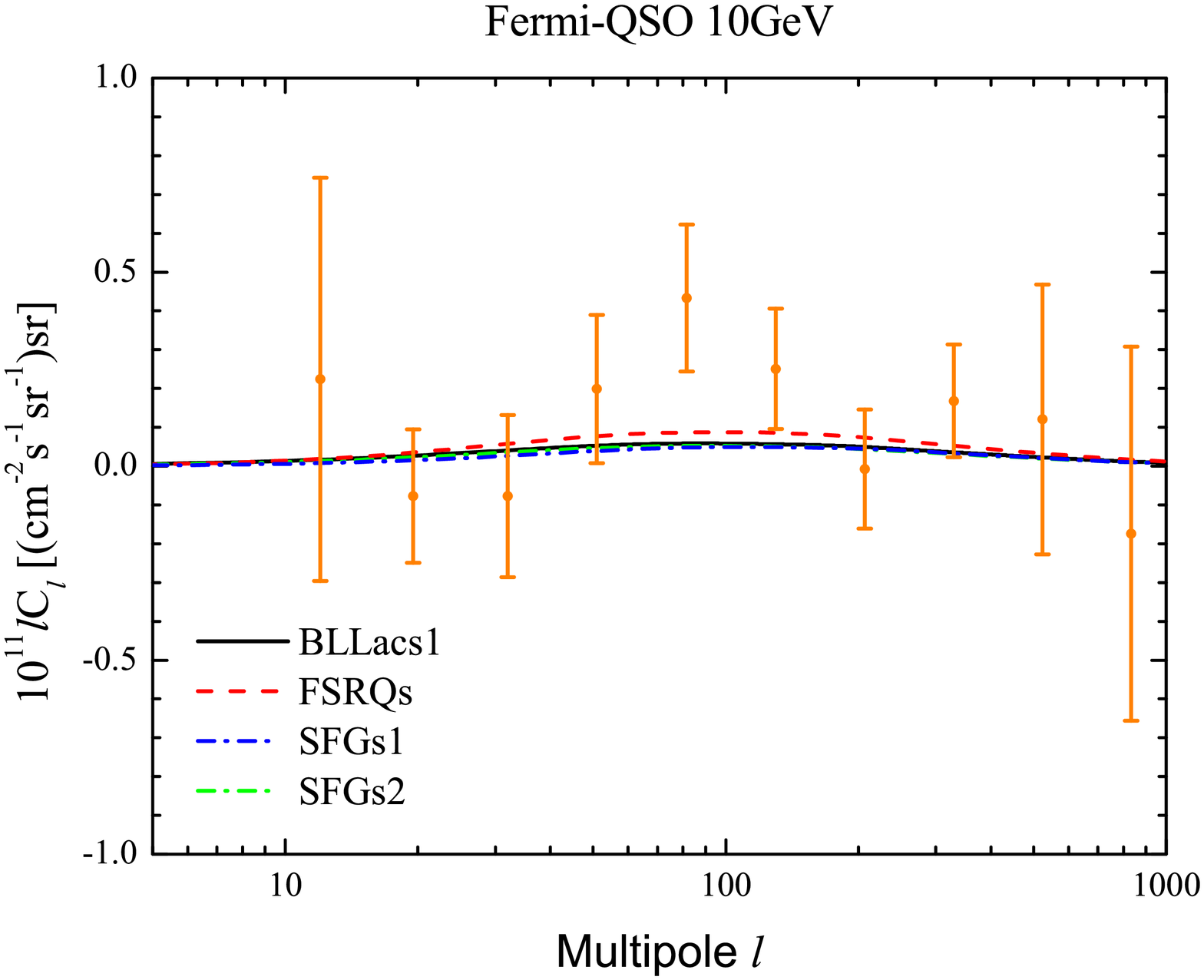, angle=0, width=0.33 \textwidth}
\centering \epsfig{file=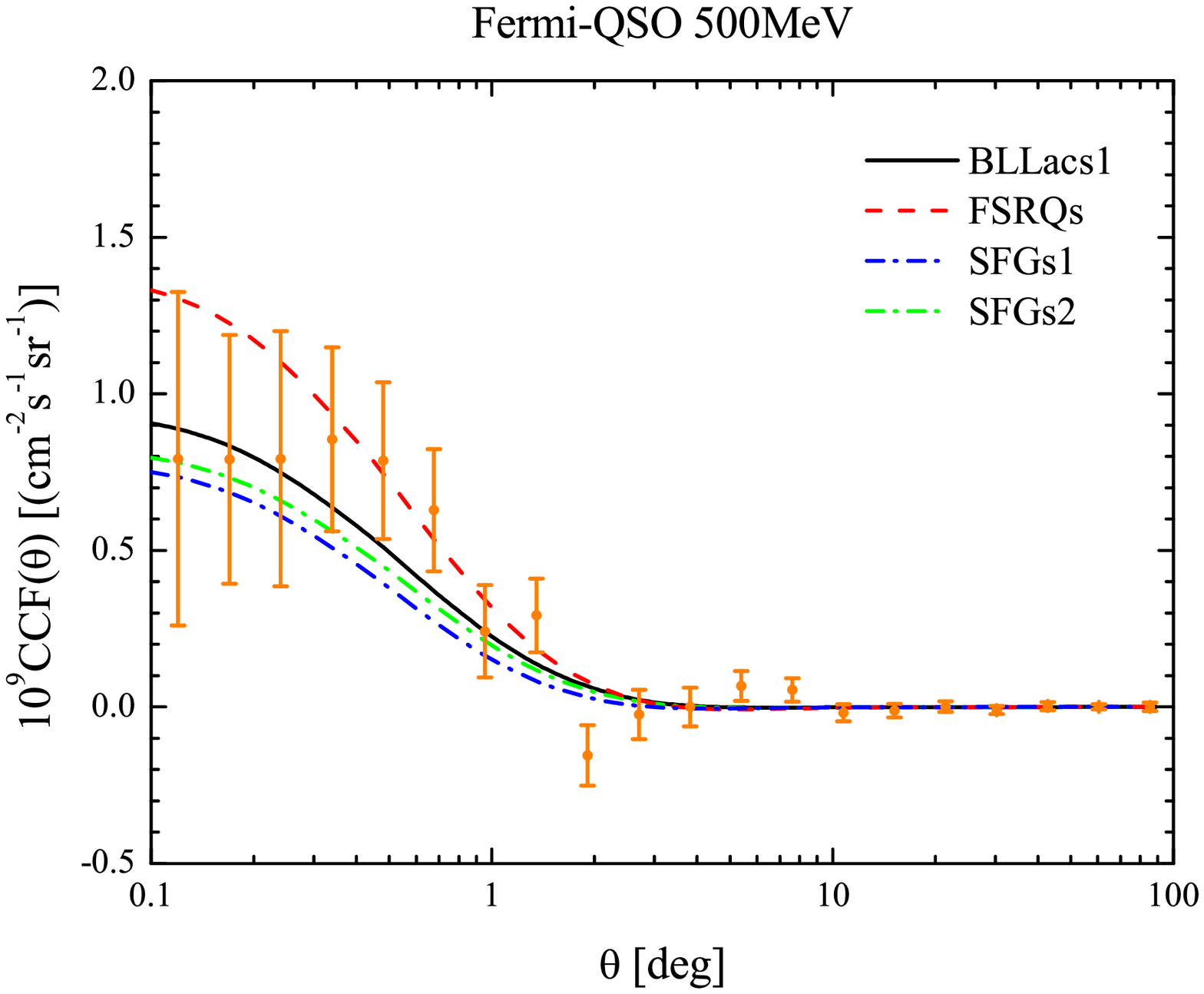, angle=0, width=0.33 \textwidth}
\centering \epsfig{file=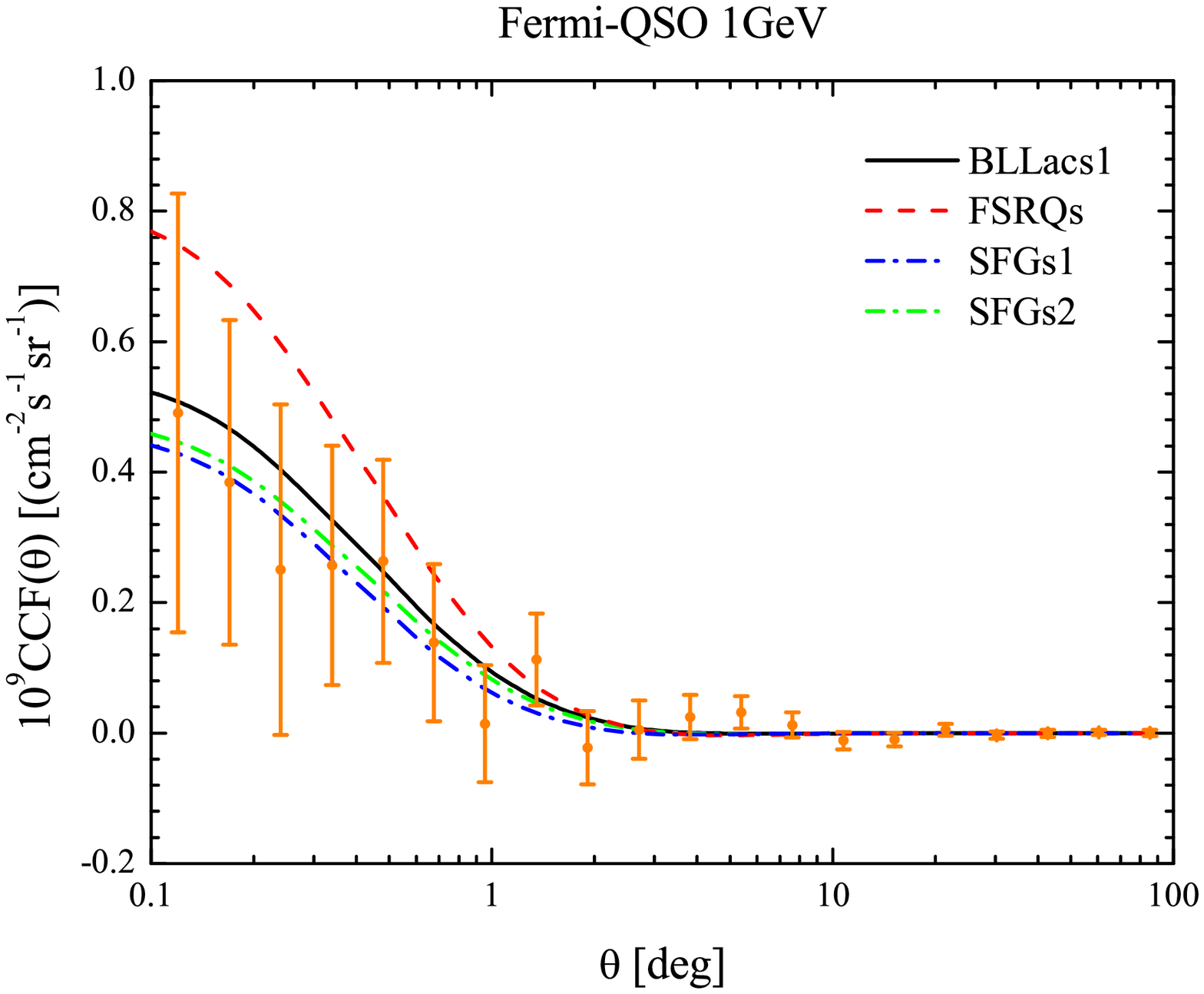, angle=0, width=0.33 \textwidth}
\centering \epsfig{file=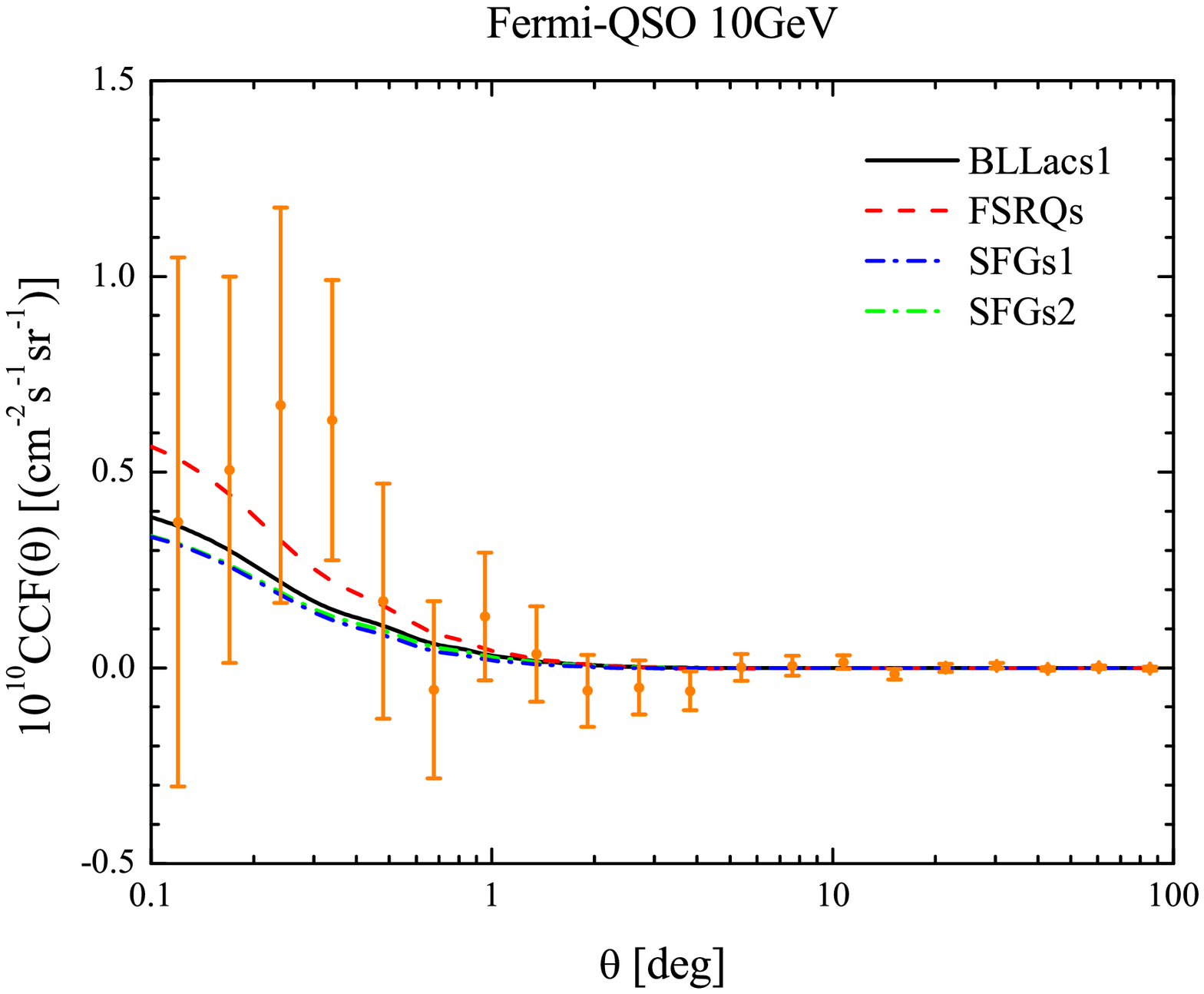, angle=0, width=0.33 \textwidth}
\caption{
CAPS (upper panels) and CCF (lower panels) estimated from the SDSS DR6 QSOs
  map and the {\it Fermi}-LAT IGRB maps in
  three energy bands. The three panels refer to three energy cuts $E>0.5$ GeV (left panels), $E>1$ GeV
  (middle panels) and $E>10$ GeV (right panels).  Error bars on the data points (orange dots) represent the
  diagonal elements of the PolSpice covariance matrix.
  Model predictions for different types of sources are represented by continuous curves: FSRQs (red, dashed), BL Lacs (black, solid)
  star-forming galaxies (blue and green, dot-dashed)  All the models are \emph{a priori} models (i.e., not fitted)
  normalized assuming that the given source class contributes 100\% of the IGRB.
\label{fig:qso_ccf_fermi}}
\end{figure*}

\subsection{No-signal tests}

To check the robustness of the results we performed further tests using
mock catalogs with no cross-correlation with LSS, verifying
that the computed CCFs are compatible with a null signal.

In Fig.~\ref{fig:null_test}
we show the cross-correlation between the {\it Fermi} $E>$500 MeV map and 3 mock realization of each of the 3 catalogs 2MASS, SDSS main galaxy sample and NVSS.
The correlations are compared with the ones with the true catalogs.
For each catalog 2 mock realizations were built scrambling the Galactic coordinates
of each galaxy of the sample, changing $b \rightarrow-b$
 in one case and $l \rightarrow -l$ in the other.
These two realizations preserve the intrinsic clustering of the catalog but remove the cross-correlation with LSS.
To compute the CCF we use the corresponding scrambled coordinate mask of the given catalog.
A third realization was performed creating a Monte Carlo catalog redistributing the galaxies of the
catalog randomly over the sky-area covered by the catalog.
In this case the new catalog contains no intrinsic clustering. To compute the CCF we use in this case the original
catalog mask.

The plots Fig.~\ref{fig:null_test} show that the correlation present for the true catalog disappears
when the mock catalogs are used, as expected.
We note that the size of the error bars are typically smaller than the true catalog CCF error bars
 when the Monte Carlo catalog is used,
but not in the case of the scrambled-coordinates catalogs.
This is likely due to the fact that the Monte Carlo catalogs do not contain intrinsic clustering
as opposed to the true and scrambled-coordinates catalogs, and
emphasizes the importance of the errors cross-checks we performed in section~\ref{sec:corranalysis}.

\begin{figure*}
\centering \epsfig{file=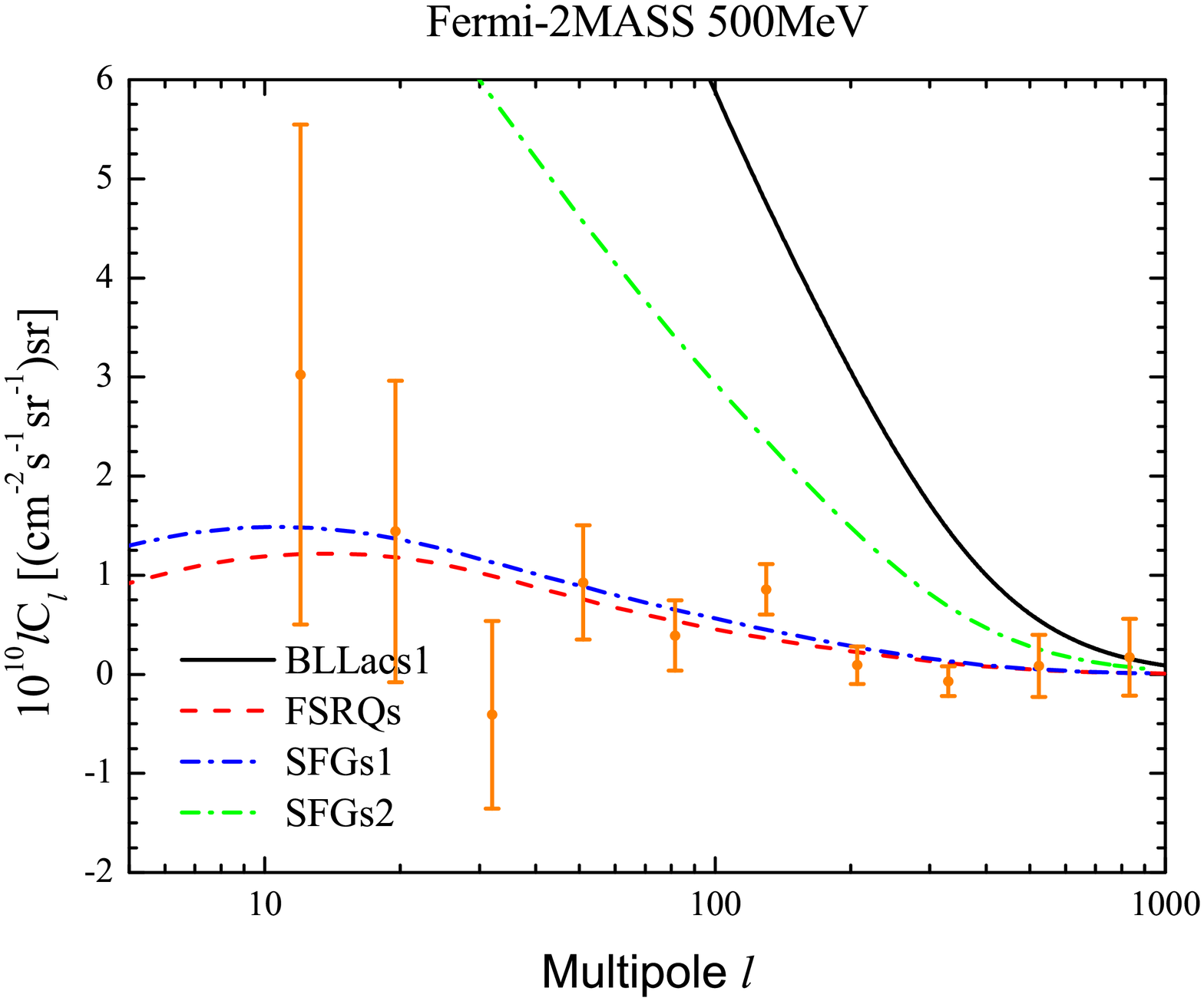, angle=0, width=0.33 \textwidth}
\centering \epsfig{file=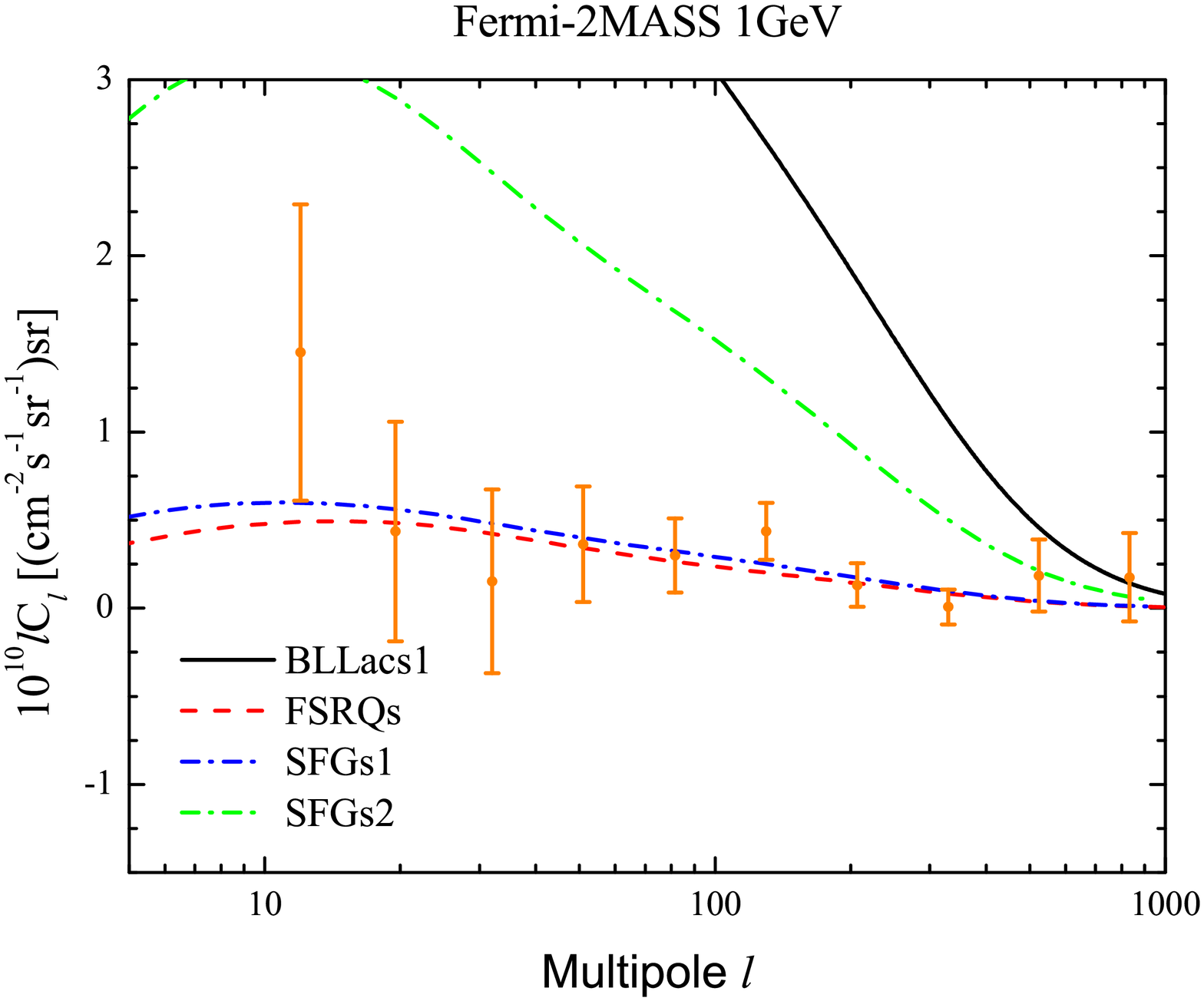, angle=0, width=0.33 \textwidth}
\centering \epsfig{file=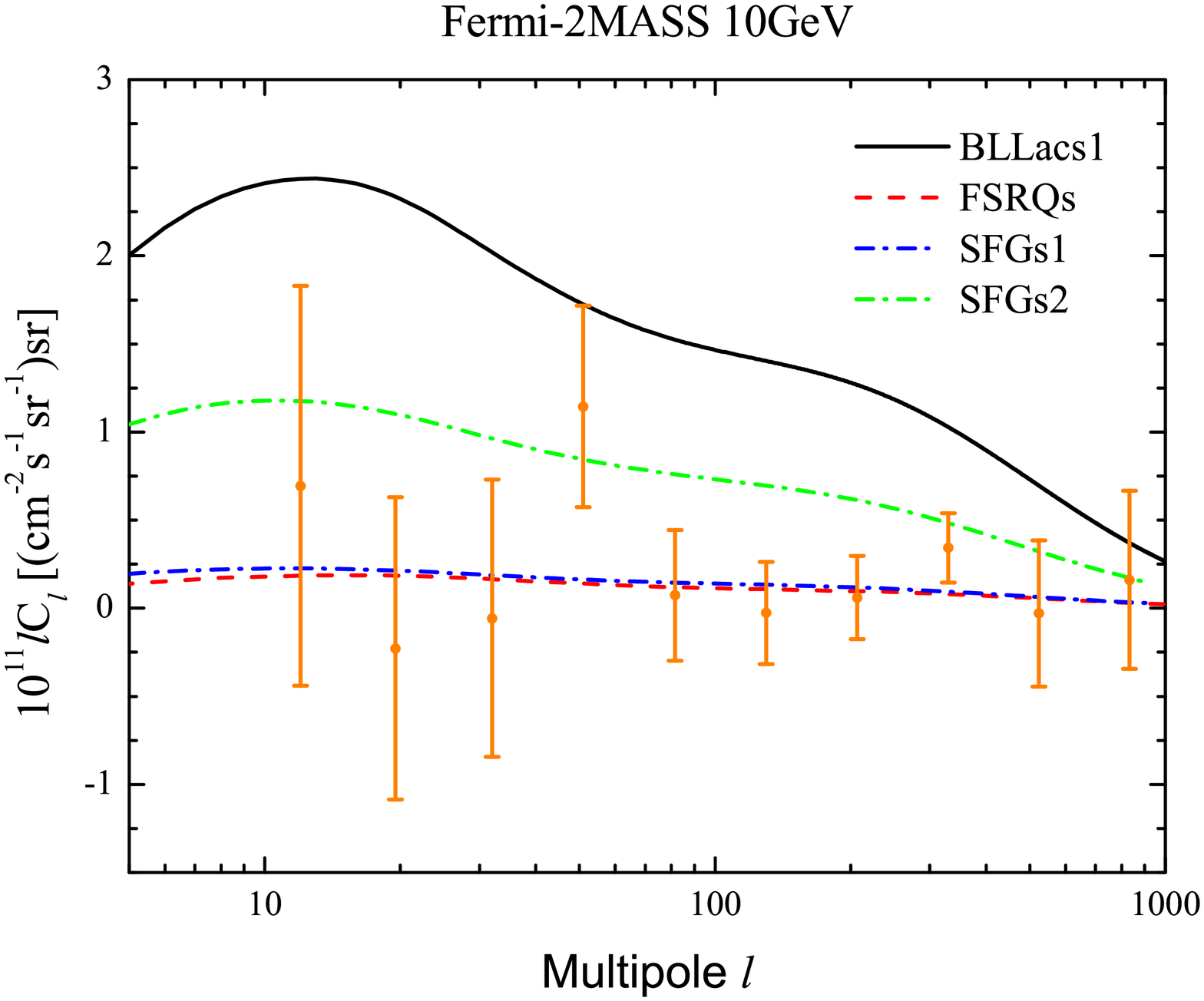, angle=0, width=0.33 \textwidth}
\centering \epsfig{file=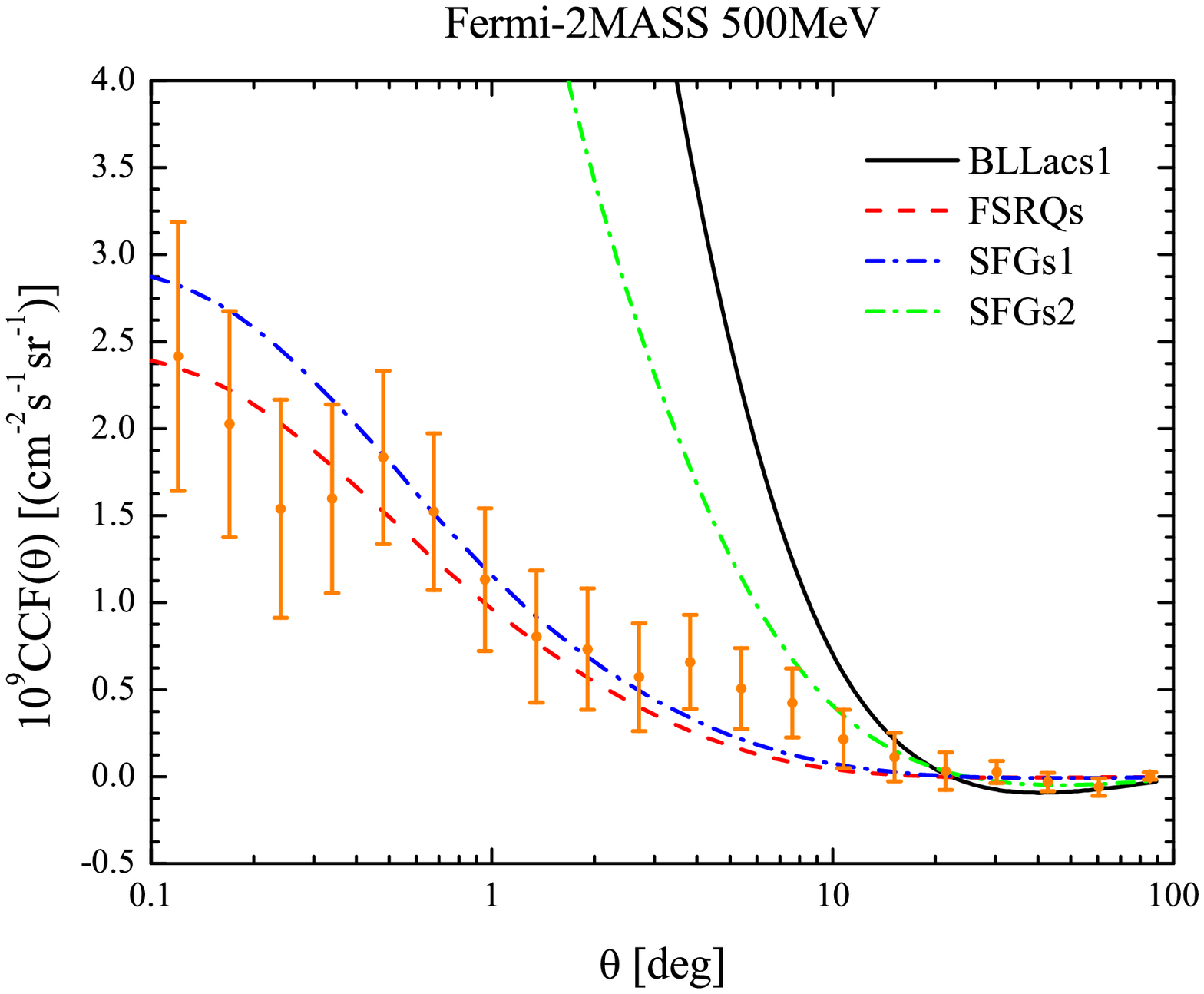, angle=0, width=0.33 \textwidth}
\centering \epsfig{file=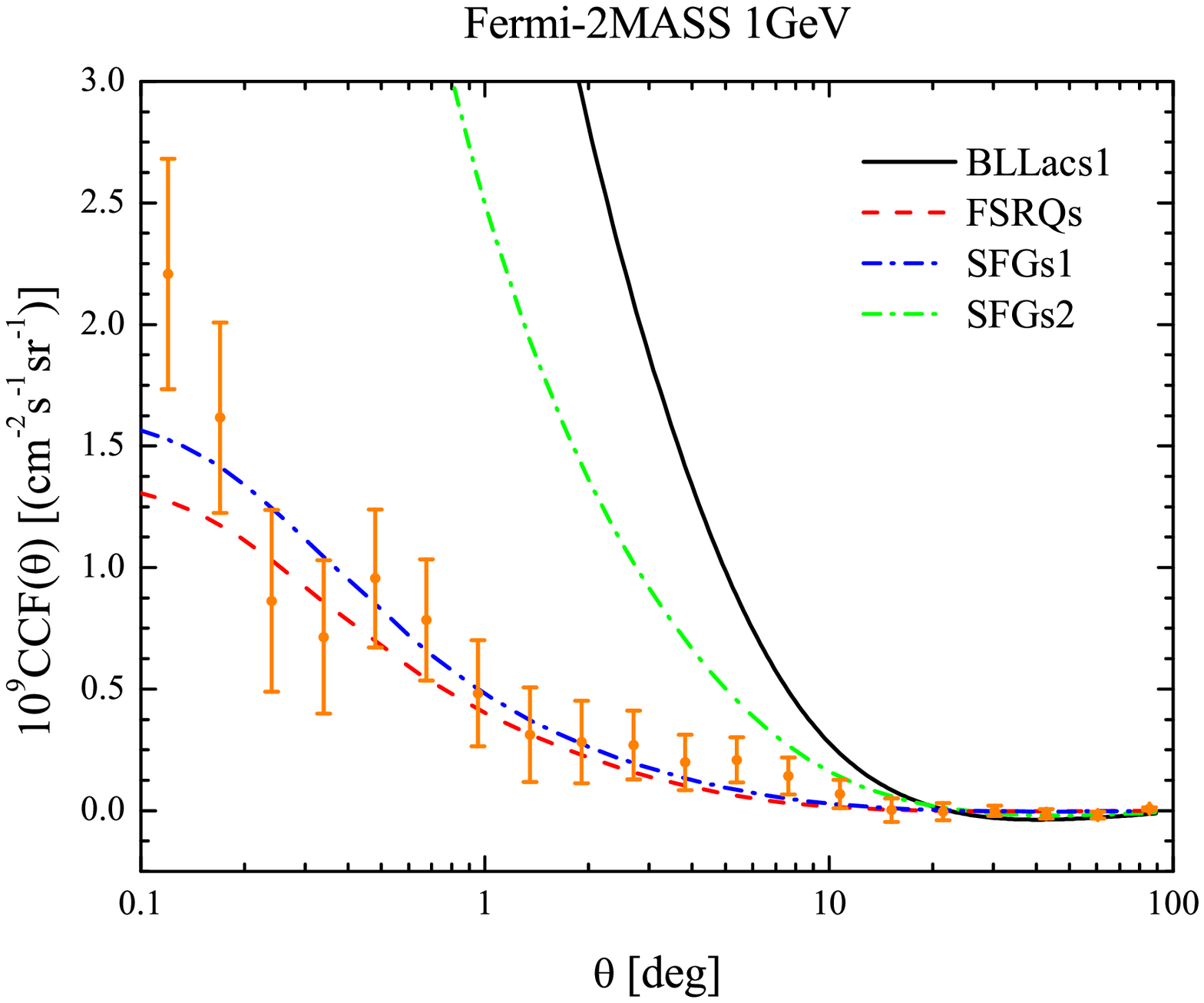, angle=0, width=0.33 \textwidth}
\centering \epsfig{file=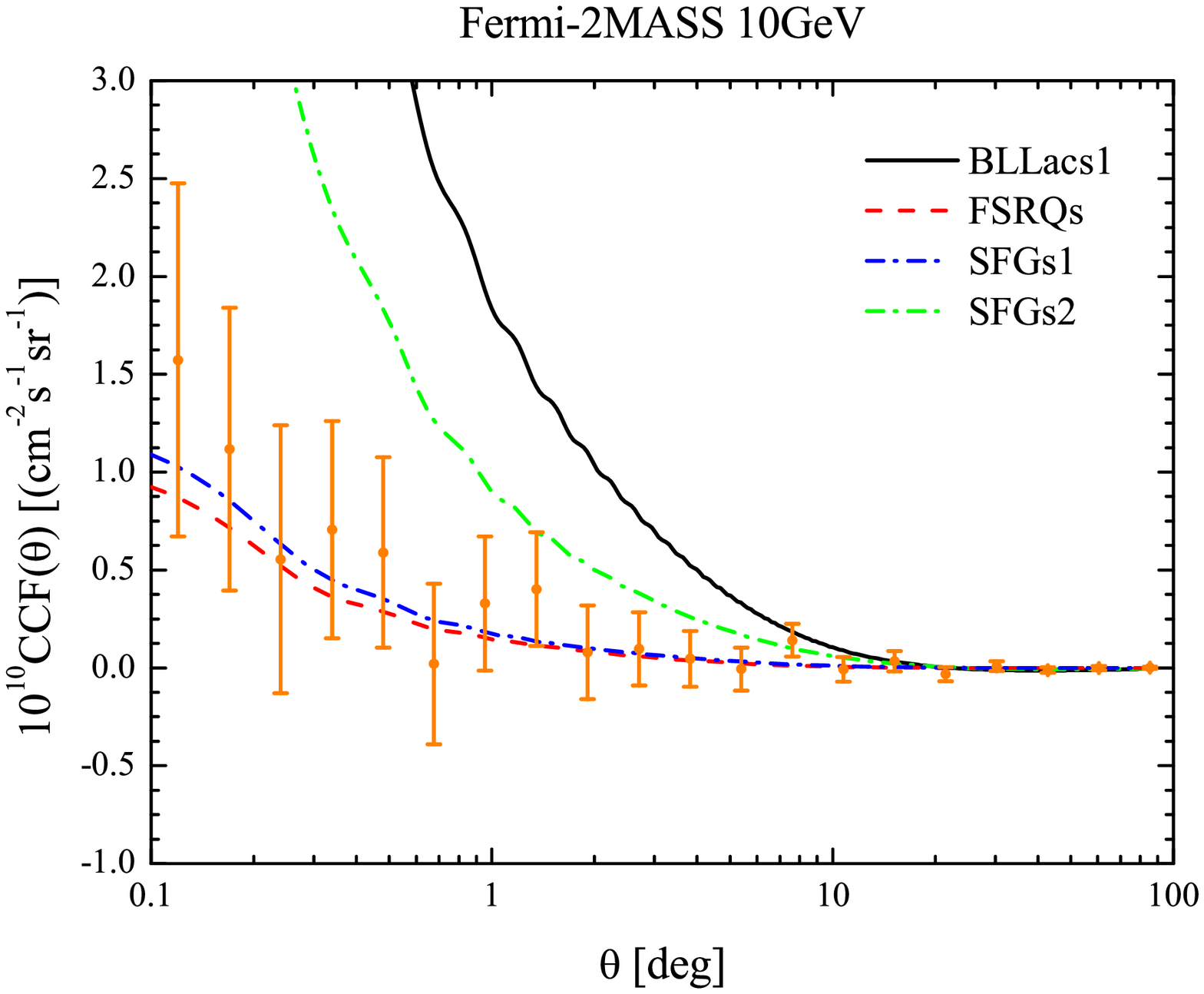, angle=0, width=0.33 \textwidth}
\caption{Analogous to Fig.~\ref{fig:qso_ccf_fermi} using 2MASS galaxies}
\label{fig:2mass_ccf_fermi}
\end{figure*}

\section{Results}
\label{sec:results}

In this section we show the results of our cross-correlation analysis both in configuration space (i.e., the angular CCF)
and in harmonic space (i.e., the CAPS) obtained by combining the cleaned {\it Fermi}-LAT IGRB maps with
the angular distributions of objects in the various catalogs presented in Section~\ref{sec:maps}.
The analysis is carried out in three different energy bands, and the results are compared with theoretical
predictions obtained under the assumption that the extragalactic, diffuse IGRB signal is generated
by a combination of three different types of unresolved sources,  BL Lacs, FSRQs and SFGs, described in
Section~\ref{sec:model}.
As explained in Section \ref{sec:psf} we do not attempt to deconvolve the measured correlation function
from the instrument PSF. The predicted correlations are, instead, convolved
with the PSF itself and then compared with the measurements.
As we anticipated in previous sections, we take
the relative contributions of the different types of sources to the IGRB  as a free parameter of the model.
In the next section we will perform a quantitative analysis to constrain
these parameters. In this section and in the following plots, we use \emph{a priori} models assuming that
each source class contributes 100\% of the observed IGRB spectrum,
with the purpose of an equal-footing and quick comparison between the data and the different models.
In fact, we assume a reference ``IGRB'' to normalize the model predictions since,
as already discussed in the previous sections, the results of our correlation
analysis  do not depend on  the measured
IGRB and its uncertainties.
More precisely, our reference IGRB  is  $I(E)=I_0 (E/E_0)^{-2.4}$
with $E_0=100$ MeV and $I_0=1.44\times10^{-7}$ MeV cm$^{-2}$s$^{-1}$sr$^{-1}$,
which is consistent with the measured one \citep{2010PhRvL.104j1101A}.
As a result, the integrated intensities of the IGRB in the 3 energy ranges that we consider here are
$1.0\times10^{-6}$,  $4.0\times10^{-7}$, $1.5\times10^{-8}$ cm$^{-2}$s$^{-1}$sr$^{-1}$,
for $E>500$ MeV, $E>1$ GeV, $E>10$ GeV, respectively.

In the following section we illustrate the results of cross-correlating the individual catalogs. Unlike in the analysis presented by
\citet{xia11},    we now observe a significant cross-correlation signal that, in Section~\ref{sec:chi2},
we compare with model predictions to infer the nature of the sources that contribute to the IGRB.

To asses the significance of the signals we use the usual likelihood ratio test assuming
a gaussian likelihood  $L\propto \exp (-\chi^2/2)$ with
\begin{equation}
   \chi^2 = \sum_{i\, j}   (d_i -m_i(f_{\rm sfg}) )  C^{-1}_{i\,j}  (d_j -m_j(f_{\rm sfg}) ) \, ,
\label{eq:chi2_v1}
\end{equation}
where $C_{i\,j}$ is the covariance matrix among  the different angular or multipole bins $i$ computed using PolSpice,
 $d_i$ represents the data, i.e., the
CCF or CAPS measured at the bin $i$, and
$m_i(f_{\rm sfg})$ is the model prediction  which depends from the parameter $f_{\rm sfg}$,
i.e., the normalization of the model CCF or CAPS (see also Section~\ref{sec:chi2}).
We use as model the SFGs1 model with free normalization, although we note
that the $\chi^2_{\rm bf}$  and  the significances calculated using the other models are very similar.
In Eq.~\ref{eq:chi2_v1} the sum extends over 10 angular bins logarithmically spaced between $\theta= 0.1^\circ$ and $100^{\circ}$
for the CCF and
over 10 multipole bins logarithmically spaced between $\ell= 10$ and $\ell=1000$
for the CAPS.
The resulting test statistics (TS) in this case simplifies as TS$=\chi^2_0 -\chi^2_{\rm bf}$,
where $\chi^2_0$ is the $\chi^2$ of the data with respect to
the null hypothesis (CCF($\theta$)=0 or CAPS($\ell$)=0)
and $\chi^2_{\rm bf}$ is the best-fit $\chi^2$ of the data
with respect to the model.
The derived TS significances are shown in  Table~\ref{tab:sigmasCCF} for the CCFs and
 Table~\ref{tab:sigmasCAPS}  for the CAPS and are commented in the sub-sections
 below for each catalog. The tables also report  $\chi^2_{\rm bf}$ and
 the significances in $\sigma$s assuming $\sigma=\sqrt{TS}$.

\begin{figure*}
\centering \epsfig{file=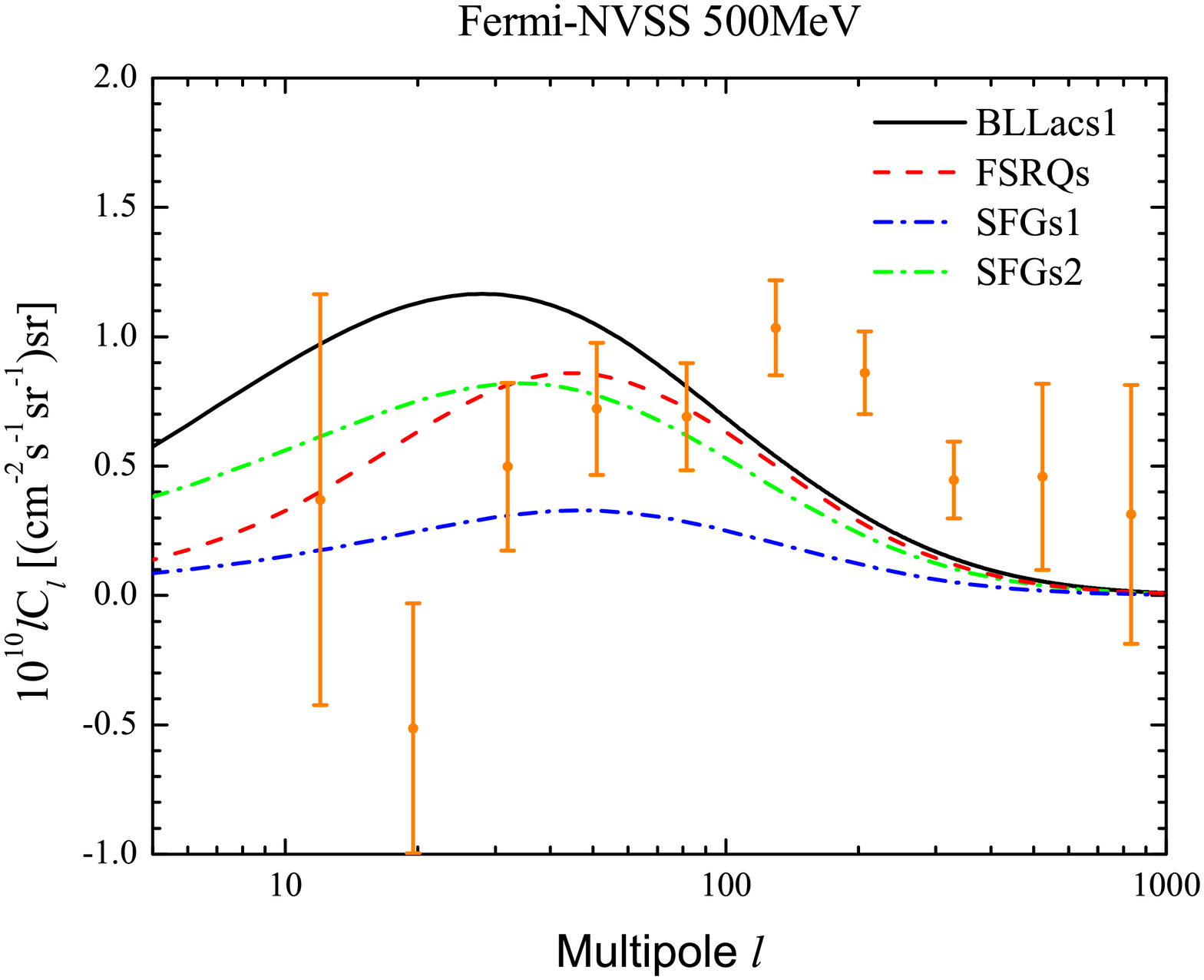, angle=0, width=0.33 \textwidth}
\centering \epsfig{file=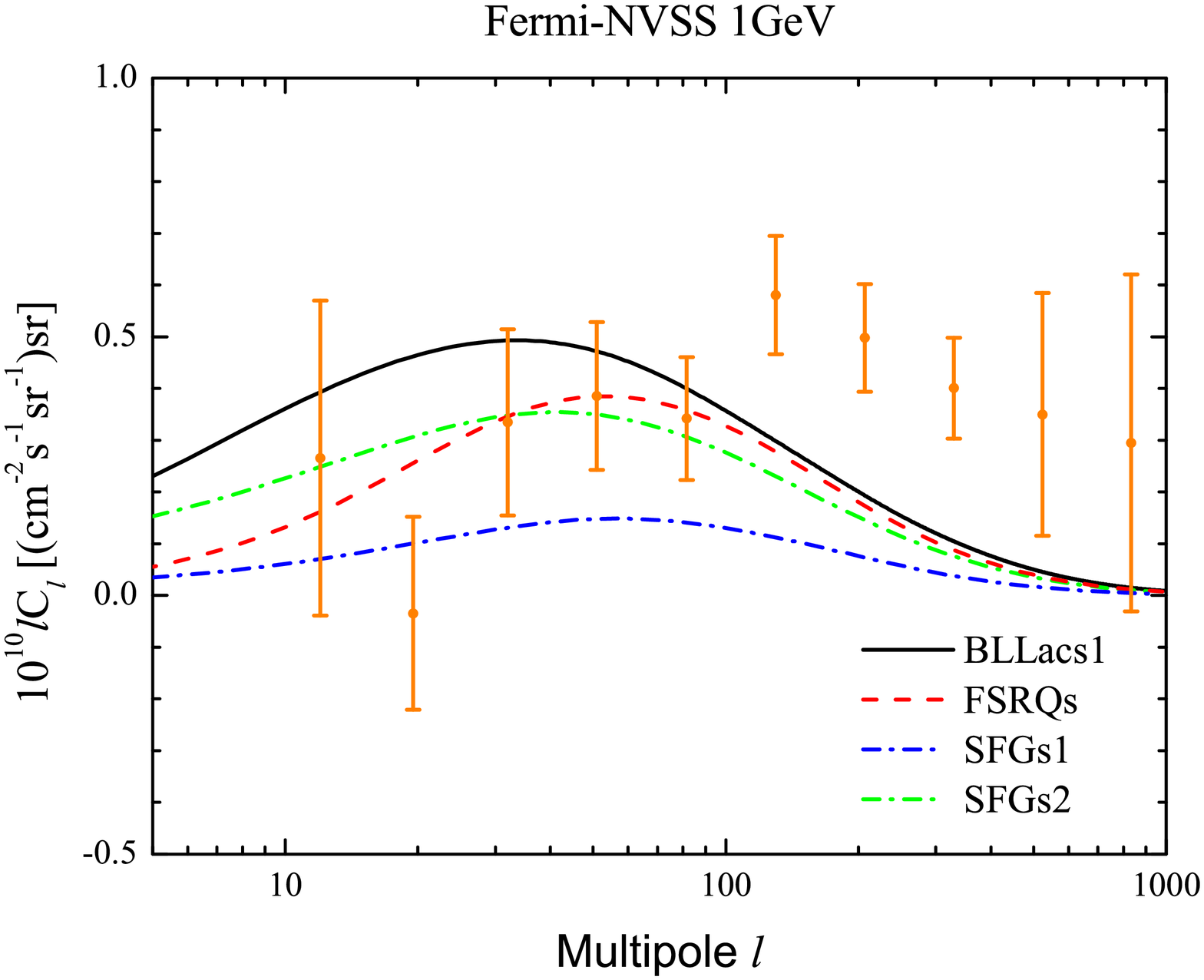, angle=0, width=0.33 \textwidth}
\centering \epsfig{file=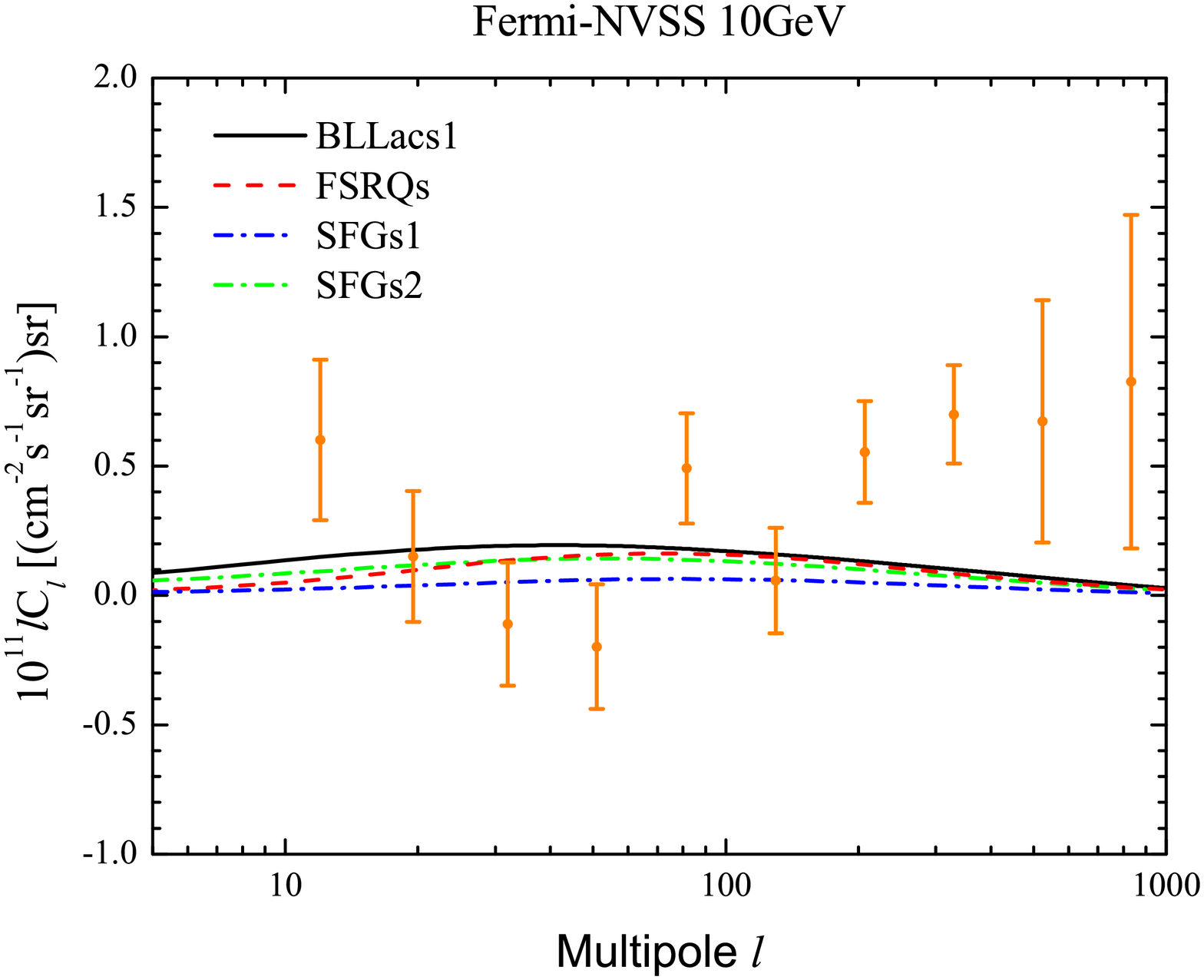, angle=0, width=0.33 \textwidth}
\centering \epsfig{file=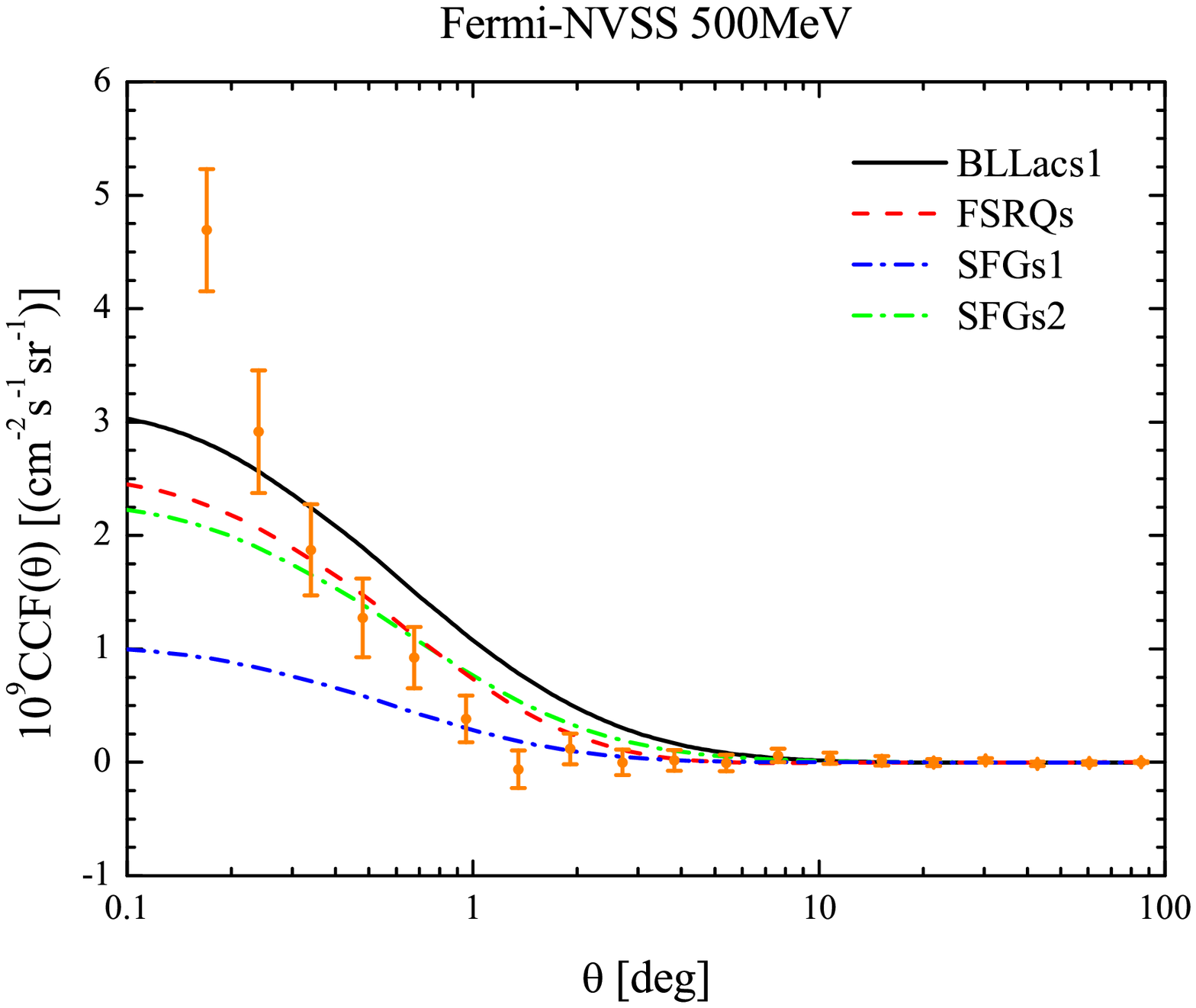, angle=0, width=0.33 \textwidth}
\centering \epsfig{file=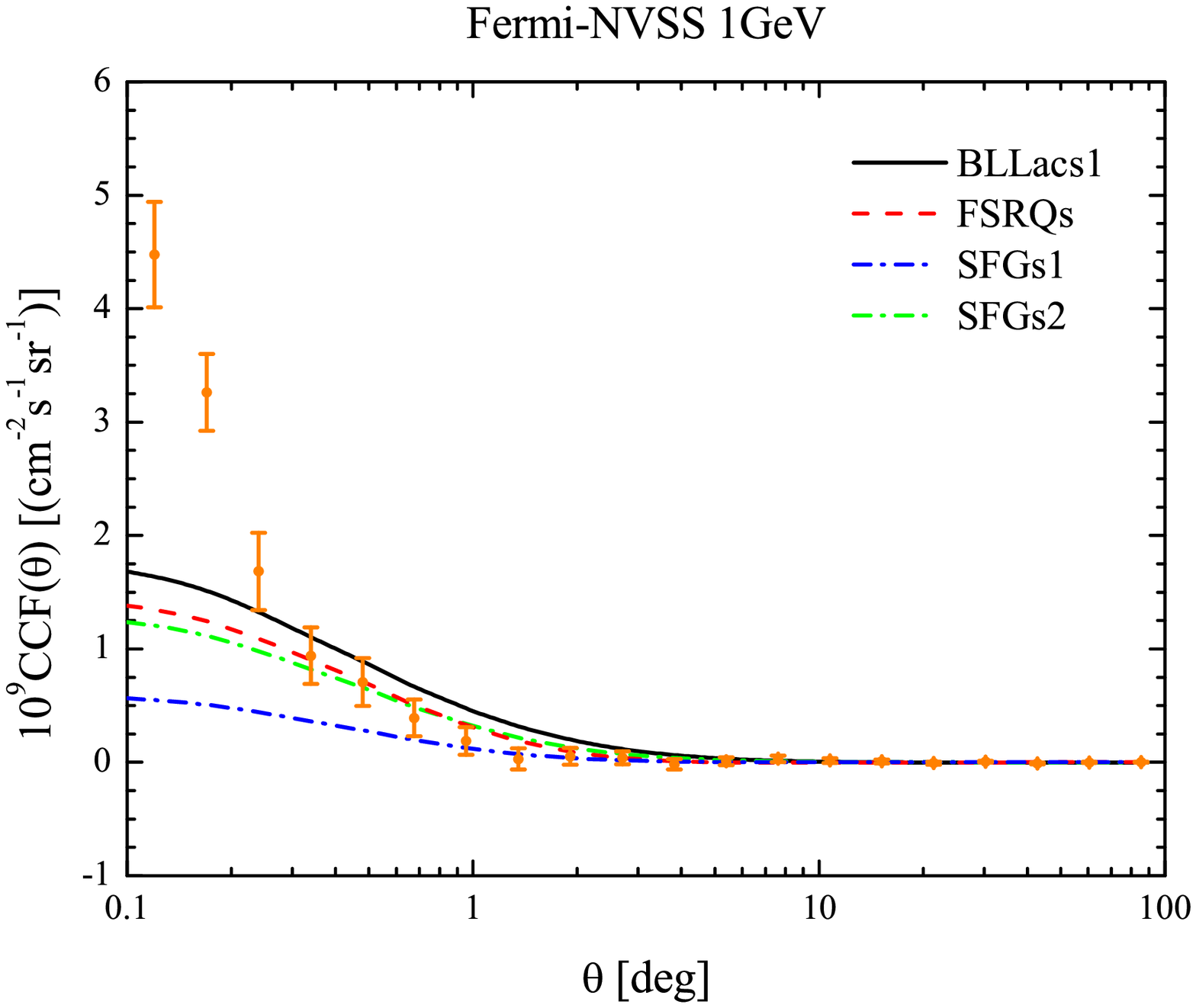, angle=0, width=0.33 \textwidth}
\centering \epsfig{file=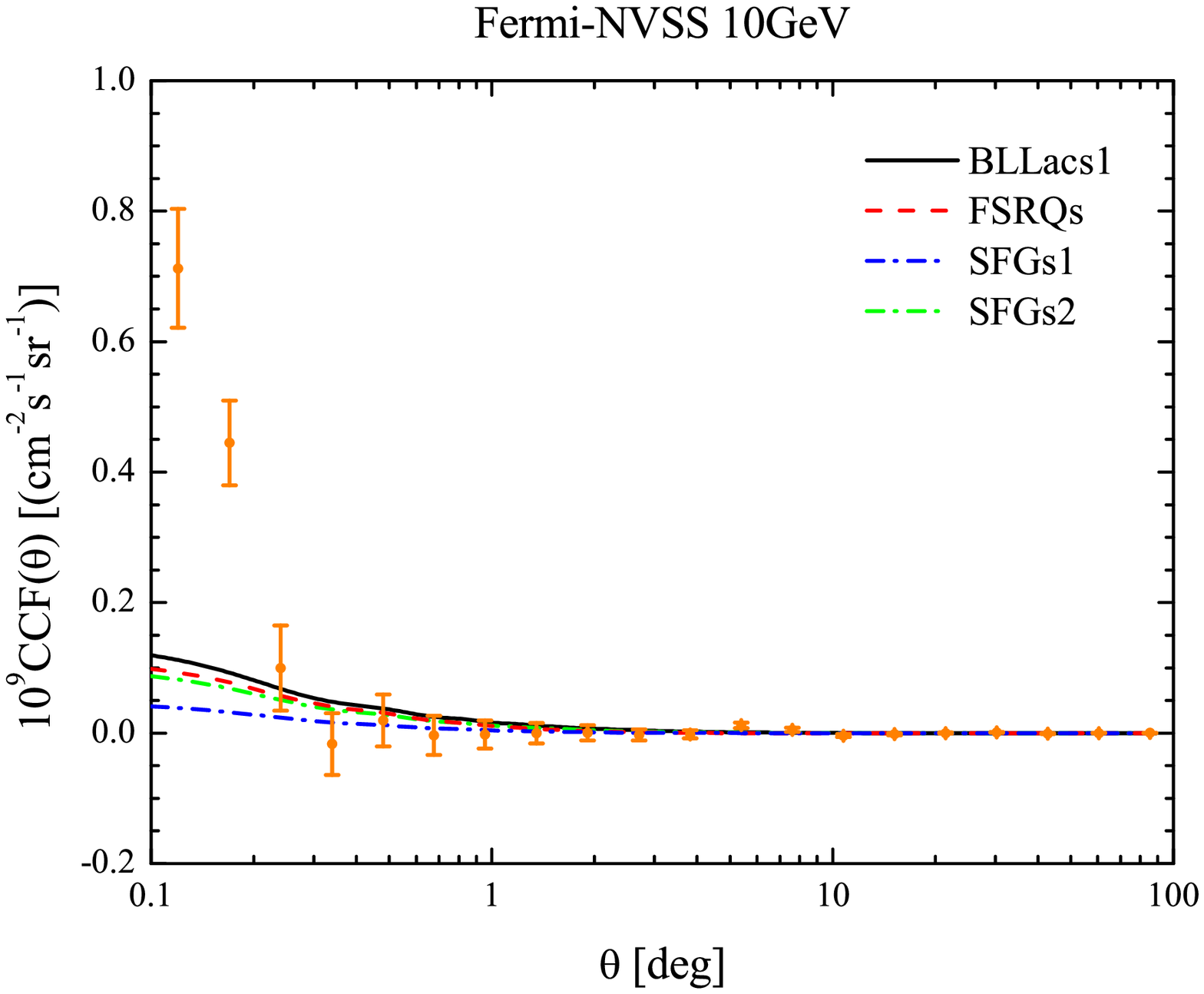, angle=0, width=0.33 \textwidth}
\caption{Analogous to fig.~\ref{fig:qso_ccf_fermi} using NVSS galaxies}
\label{fig:nvss_ccf_fermi}
\end{figure*}

\begin{figure*}
\centering \epsfig{file=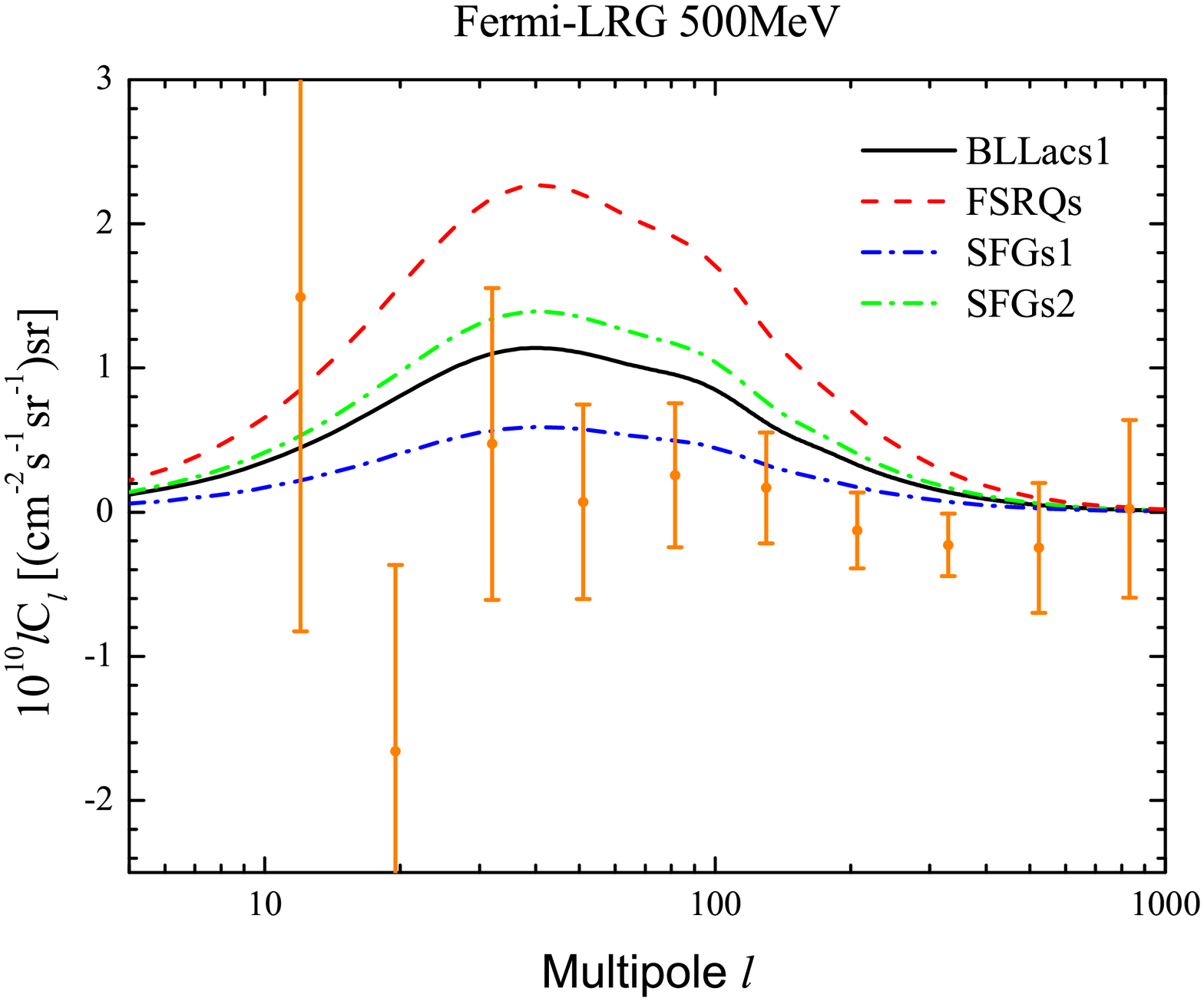, angle=0, width=0.33 \textwidth}
\centering \epsfig{file=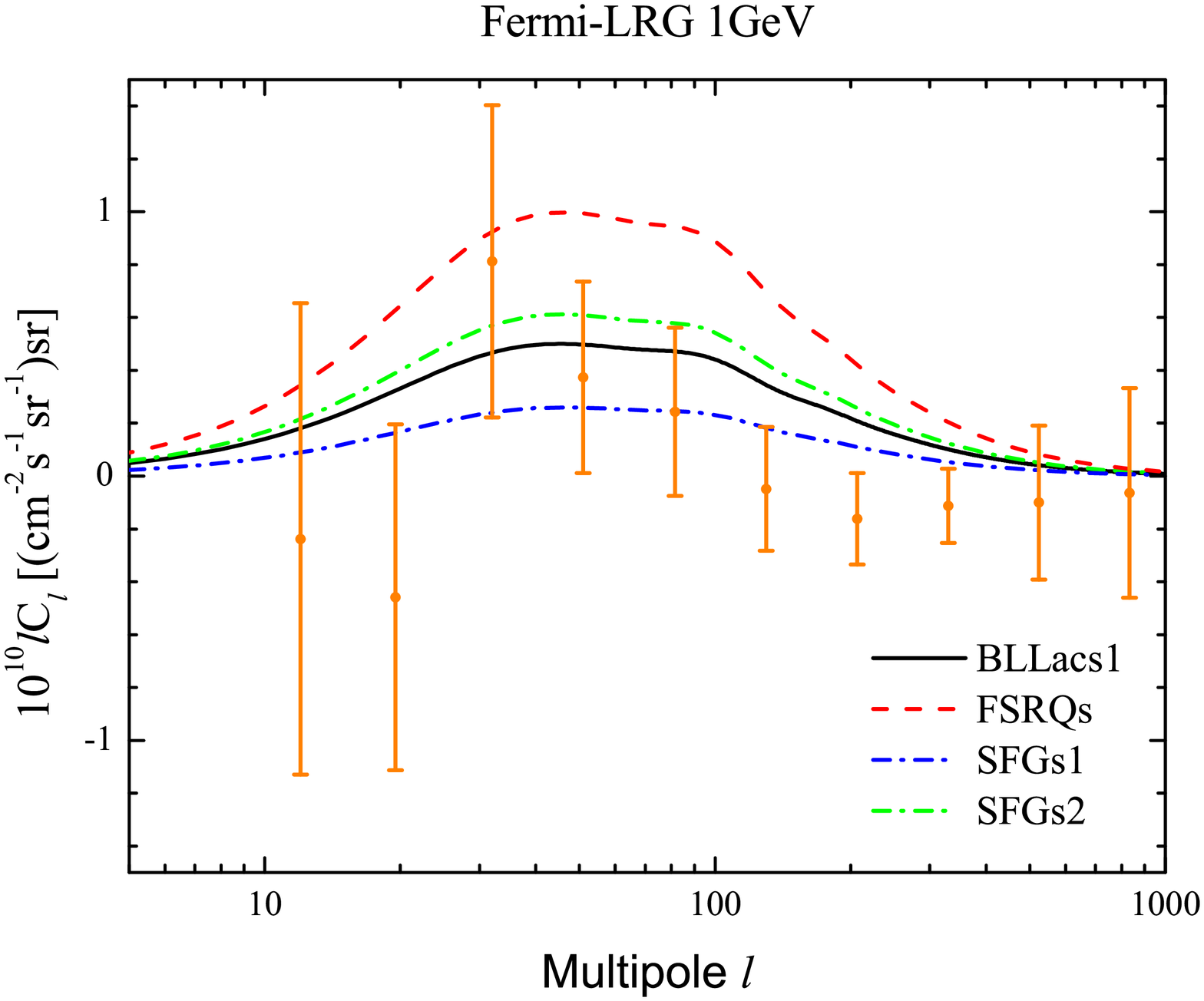, angle=0, width=0.33 \textwidth}
\centering \epsfig{file=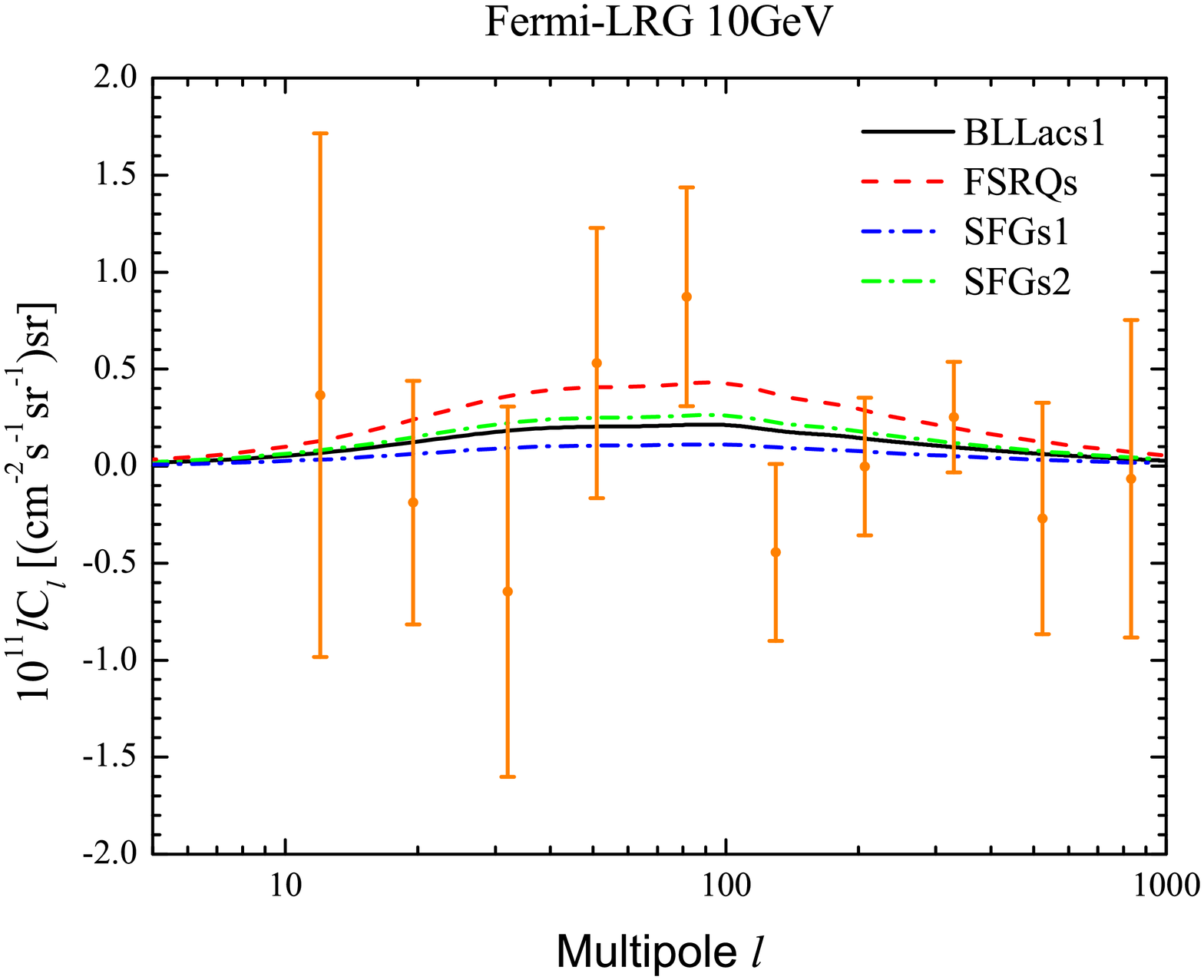, angle=0, width=0.33 \textwidth}
\centering \epsfig{file=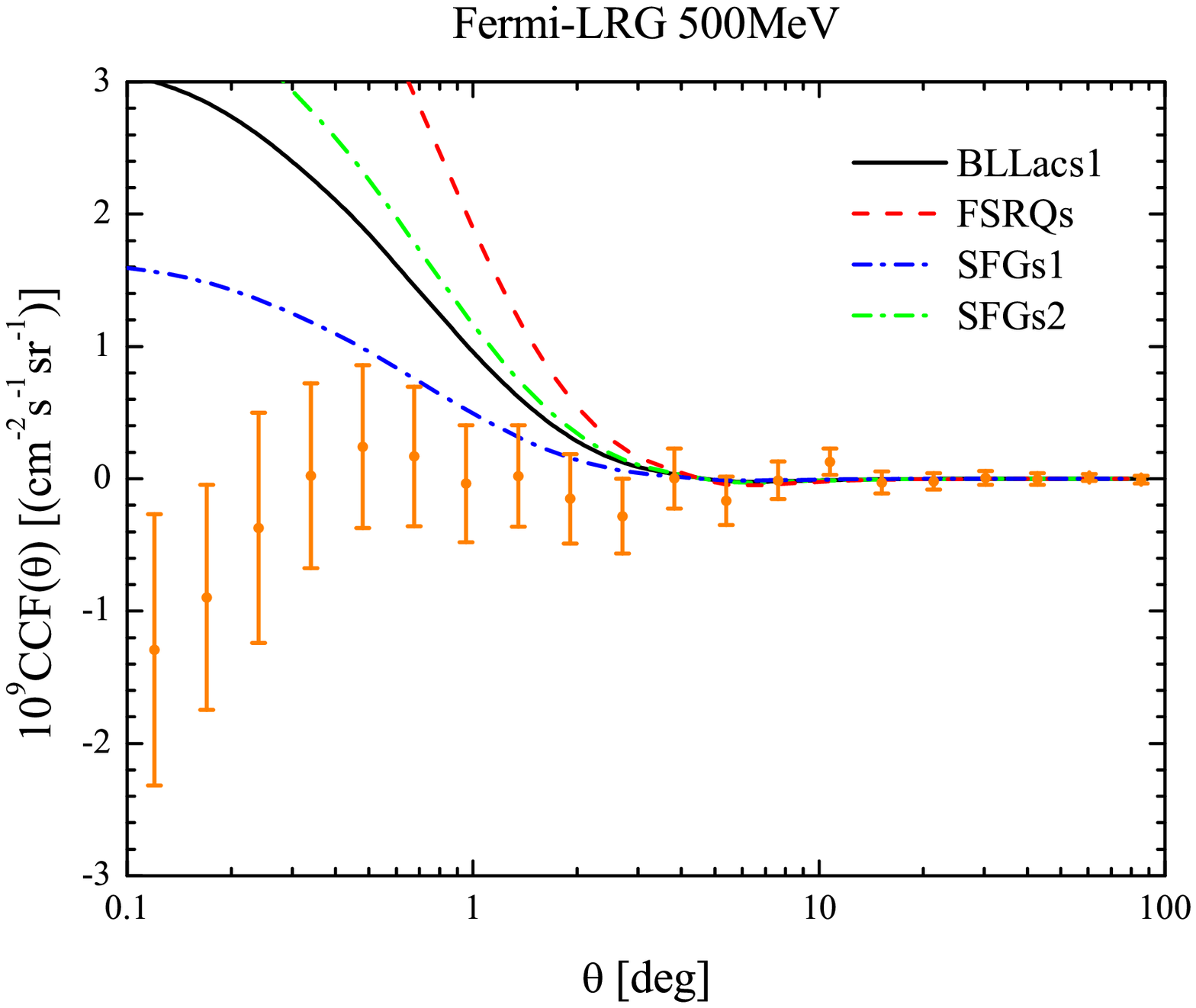, angle=0, width=0.33 \textwidth}
\centering \epsfig{file=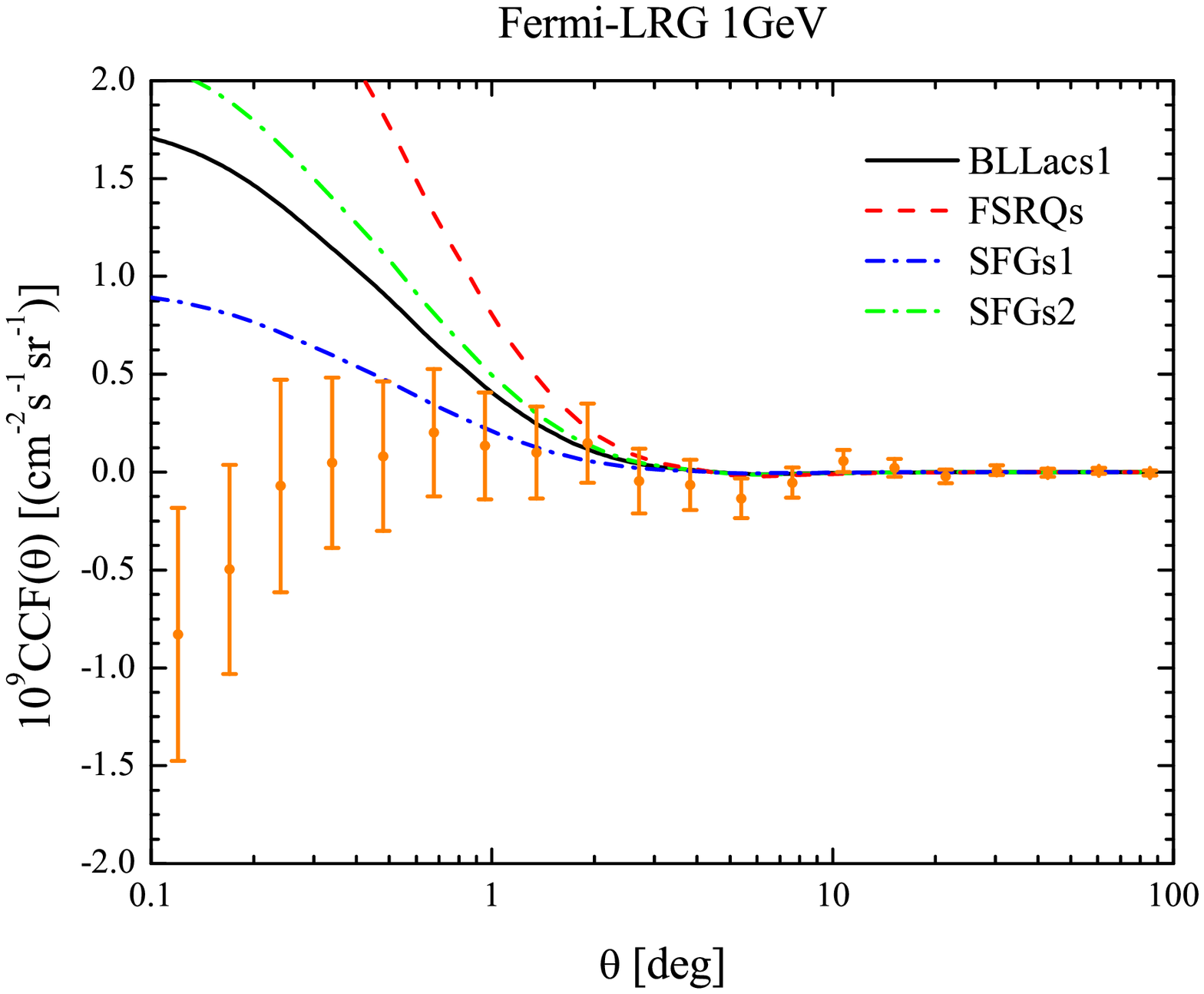, angle=0, width=0.33 \textwidth}
\centering \epsfig{file=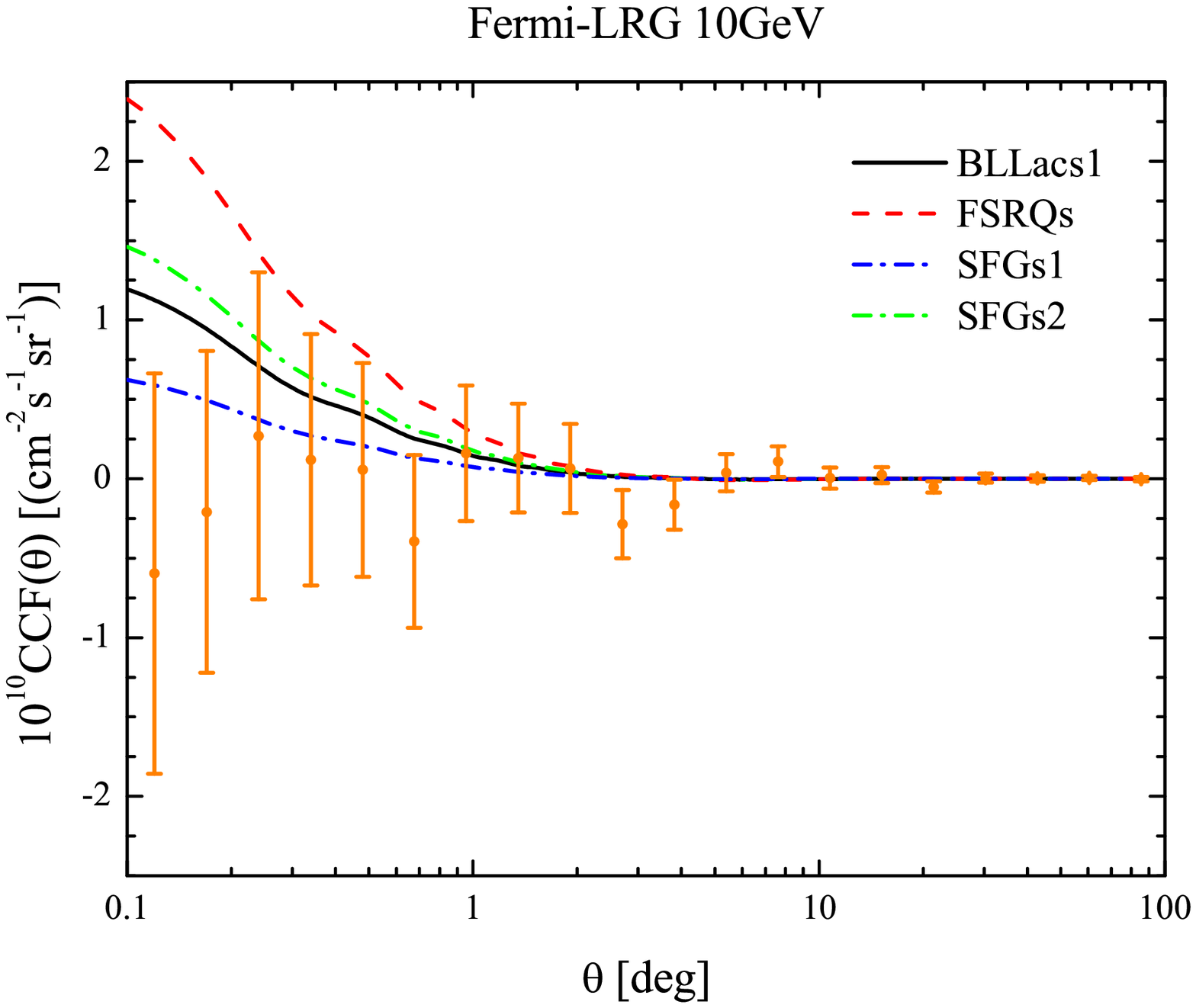, angle=0, width=0.33 \textwidth}
\caption{Analogous to fig.~\ref{fig:qso_ccf_fermi} using SDSS DR8 Luminous Red galaxies}
\label{fig:lrg_ccf_fermi}
\end{figure*}

\subsection{1-halo-like term}
\label{sec:1halodata}

As discussed in section \ref{sec:1halotheory}, a further contribution to the cross-correlation
can arise from a {\it 1-halo} term or if part of the sources of a given catalog
are also themselves $\gamma$-ray emitters.
We denoted these terms collectively as {\it 1-halo}-like.
To test empirically the possibility of the presence of a {\it 1-halo}-like term we adopt the 
following procedure.
For each catalog and for each energy band we perform a two-parameter
fit using a similar $\chi^2$ as in Eq.~\ref{eq:chi2_v1}, but modeling $m_i$
as the sum of the SFGs1 model with free normalization plus
a further term proportional to the PSF profile, i.e.,  $\propto W_\ell^{\Delta E}$ for the CAPS,
and  to the related harmonic transform for the CCF.
The latter is representative of a correlation at zero angular separation which
is spread at larger angular scales by the effect of the PSF, as expected for a {\it 1-halo}-like term.
 Again, the results change only marginally if a model different  from the SFGs1
model is used.

In Fig.~\ref{fig:1Halo} we show the 1-dimensional $\chi^2$ of the normalization
of the {\it 1-halo}-like term profiled over the remaining parameter,
i.e., the function obtained from  $\chi^2(f_{\rm sfg},f_{\rm 1h})$ after
minimizing over $f_{\rm sfg}$. 
The plots refer to the fit to the CCFs. A fit to the CAPS gives similar results.
It can be seen that for all the energy bands and the four catalogs SDSS DR6 QSO,
2MASS, SDSS LRGs and SDSS MG the significance of this extra term is typically
below 1 $\sigma$,  reaching the largest significance of little more than  1 $\sigma$ 
only in the case of the correlation  of 2MASS with $\gamma$ rays above 1 GeV. 
The only exception is the NVSS case, where, clearly, a strong preference for a {\it 1-halo}-like
term is present.
We will discuss further the NVSS case in the dedicated section below,
while,  given the lack of significant indications
of the presence of {\it 1-halo}-like contributions, we will not consider  the other catalogs
further in the following and in the global fitting described in Section \ref{sec:chi2}.

\subsection{Cross-correlation with SDSS DR6 QSOs}
\label{sec:ccfqso}

The DR6 QSO catalog contains  AGN at high redshifts that should preferentially trace
bright FSRQs, whose redshift distribution also extends to high redshifts.
The results of the cross-correlation analysis are shown in Fig.~\ref{fig:qso_ccf_fermi}.
For readability, in all the plots of this and the following sections,
we only show the predictions for the BLLacs1 model, since
the predictions for the BLLacs2 model are rather similar.
The upper panels show the CAPS in three energy bands (increasing from left to right). We plot the corresponding CCFs in the lower panels.
At small angular scales $\theta < 1^{\circ}$ we observe a cross-correlation signal which is more significant
in the low energy band ($\sim4.5\, \sigma$ for the CCF and $\sim5\, \sigma$ for the CAPS), where photon statistics are higher.
The fact that a weaker signal \mbox{(2-3 $\sigma$)} is also present for  $E>1$ GeV suggests that
the cross-correlation is genuine and not an artifact from systematic errors in the cleaning procedure.

The observed CCF  is
perfectly consistent with the theoretical predictions of all the \emph{a priori} models considered: BL Lacs, FSRQs
and SFGs for all three energy ranges. This is not entirely surprising since the $dI(>E)/dz$
of all these models overlap significantly with the
$dN/dz$ of the DR6 QSOs. The similarity of the model predictions
implies that BL Lacs, FSRQs, SFGs and DR6 QSOs all trace the underlying mass density field at high redshifts.

We note that the SDSS DR6 catalog
of QSOs is prone to a systematic error that we did not
investigate in the previous sections: contamination by stars.
To investigate this issue and  assess the magnitude of the effect,
we have extracted a large number ($\sim
8\times10^4$) of SDSS DR6 stars
with apparent magnitudes in the range
$16.9 < g < 17.1$ from the CasJobs website.
We then estimated  the cross-correlation
between this star catalog and the {\it Fermi} maps.
The resulting CCFs turned out to be consistent with zero, showing that
any residual stellar contamination does not correlate with the IGRB and
does not contribute to the cross-correlation signal.

\subsection{Cross-correlation with 2MASS galaxies}
\label{sec:ccf2mass}

The 2MASS survey catalog is the most local sample that we have considered. These near-infrared-selected galaxies
are likely to trace the local SFG  population rather than the  AGN population. The results of the analysis are shown in
Fig.~\ref{fig:2mass_ccf_fermi}.
We observe a signal in the CCF at angular separations smaller than $\sim 10^{\circ}$ with a significance $\sim3.5\, \sigma$,
and a signal in the CAPS  with a similar $3.5\, \sigma$ significance which appears to result mainly from multipoles smaller than $\sim 200$.
The angular extent
and the amplitude of this signature depends on the energy band.
Intriguingly the  significance remain stable at $E>1$ GeV, even slightly increasing,
especially  in the case of the CCF, possibly indicating a signal peaking at around GeV, as expected
for the case of nearby SFGs.

The comparison with the models excludes, with high significance,
that BL Lac could give a dominant contribution to the IGRB diffuse emission
at low redshift. In this respect both the CCF and the CAPS provide strong constraints.
This result is in agreement with similar independent findings, based
on the population studies of resolved BL Lacs \citep{2010ApJ...720..435A} and the anisotropy of the IGRB \citep{cuoco12},
which both indicate a low contribution  BL Lacs ($<$20-30\%) at least up to $\sim$50 GeV.
Above 50 GeV the contribution is more uncertain and can be more significant \citep{AjelloBLLs}.
In this respect, our constraints in the energy range $E>10$ GeV are weak
and do not provide a direct test.
Conversely, both SFGs1 and  FSRQs are equally good candidates for the IGRB in the energy range explored here.
The close match with the data stems from the fact that they have similar $dI(>E)/dz$ at low redshifts.
As expected, the predictions for the SFGs2 model are significantly different from the
SFGs1 one and do not fit the data.
This implies that in model SFGs2
the contribution from star-forming galaxies
should be  $<$20-30\% of the total, as for the BL Lacs.
As discussed in Section~\ref{sec:model} the large differences between the
two SFG models originates from the different distributions $dI(>E)/dz$
with the distribution for SFGs2 peaking at very low redshift as opposed to the SFGs1 one
that extends to high redshift. The implications of these differences are discussed in
 Sections~\ref{sec:chi2} and \ref{sec:discussion}.

\begin{figure*}
\centering \epsfig{file=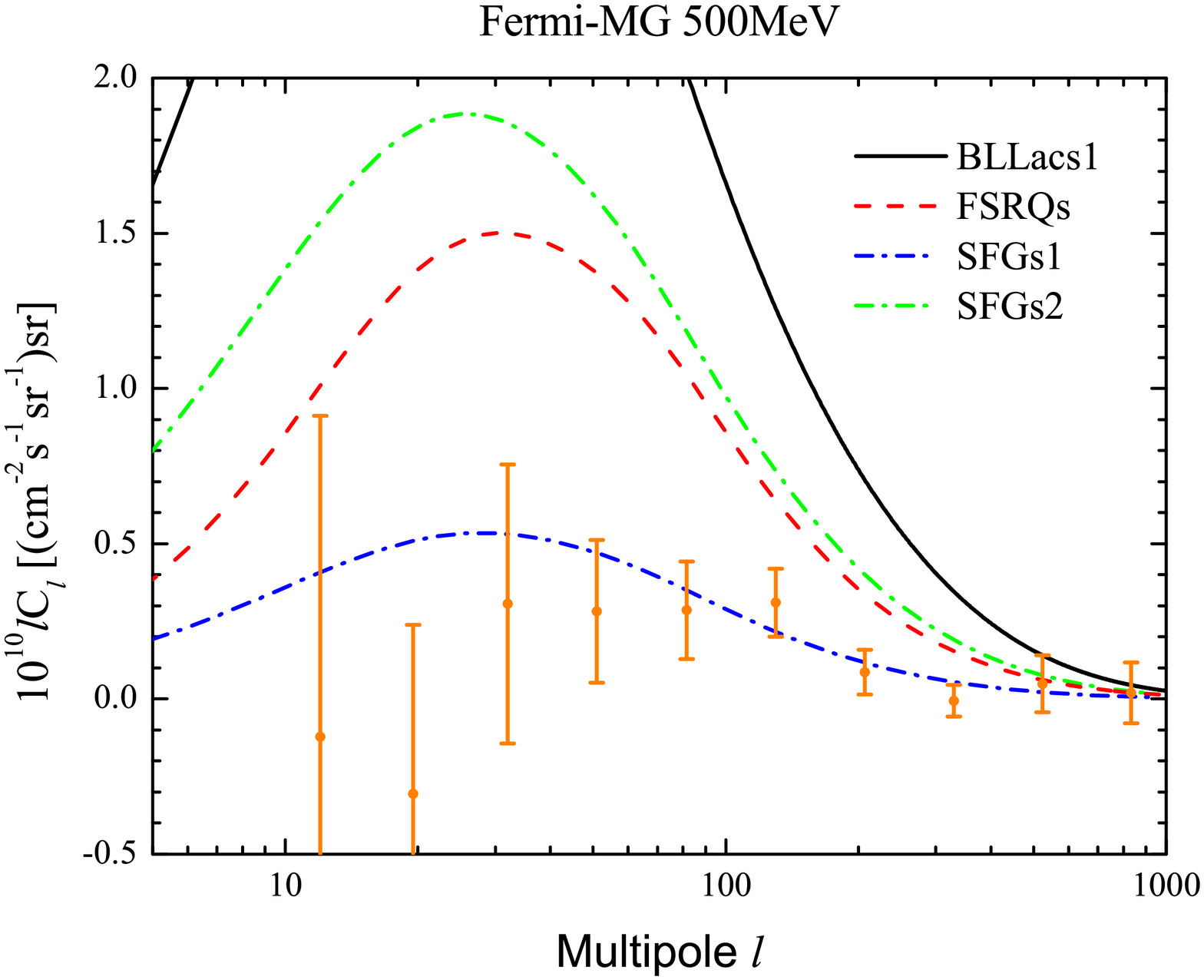, angle=0, width=0.33 \textwidth}
\centering \epsfig{file=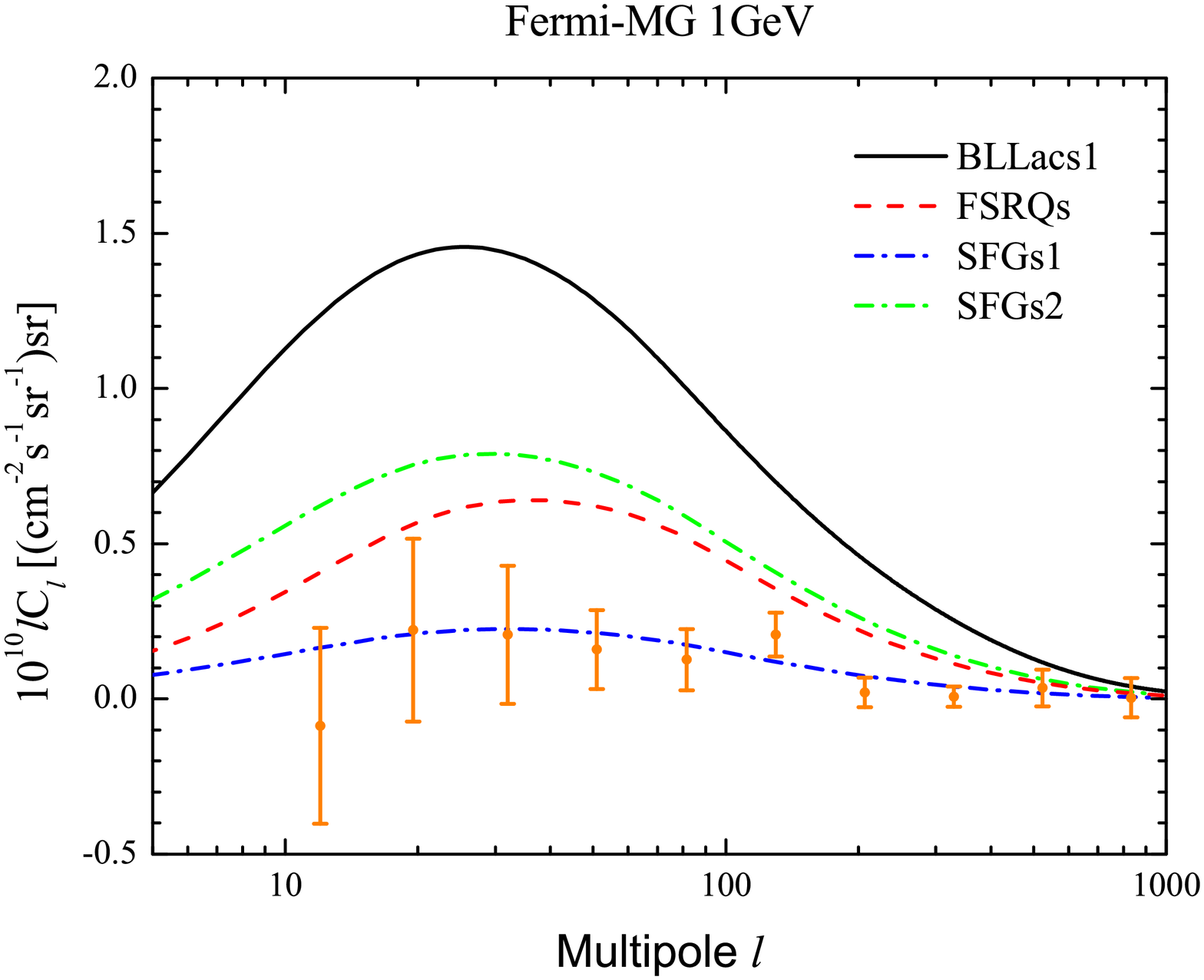, angle=0, width=0.33 \textwidth}
\centering \epsfig{file=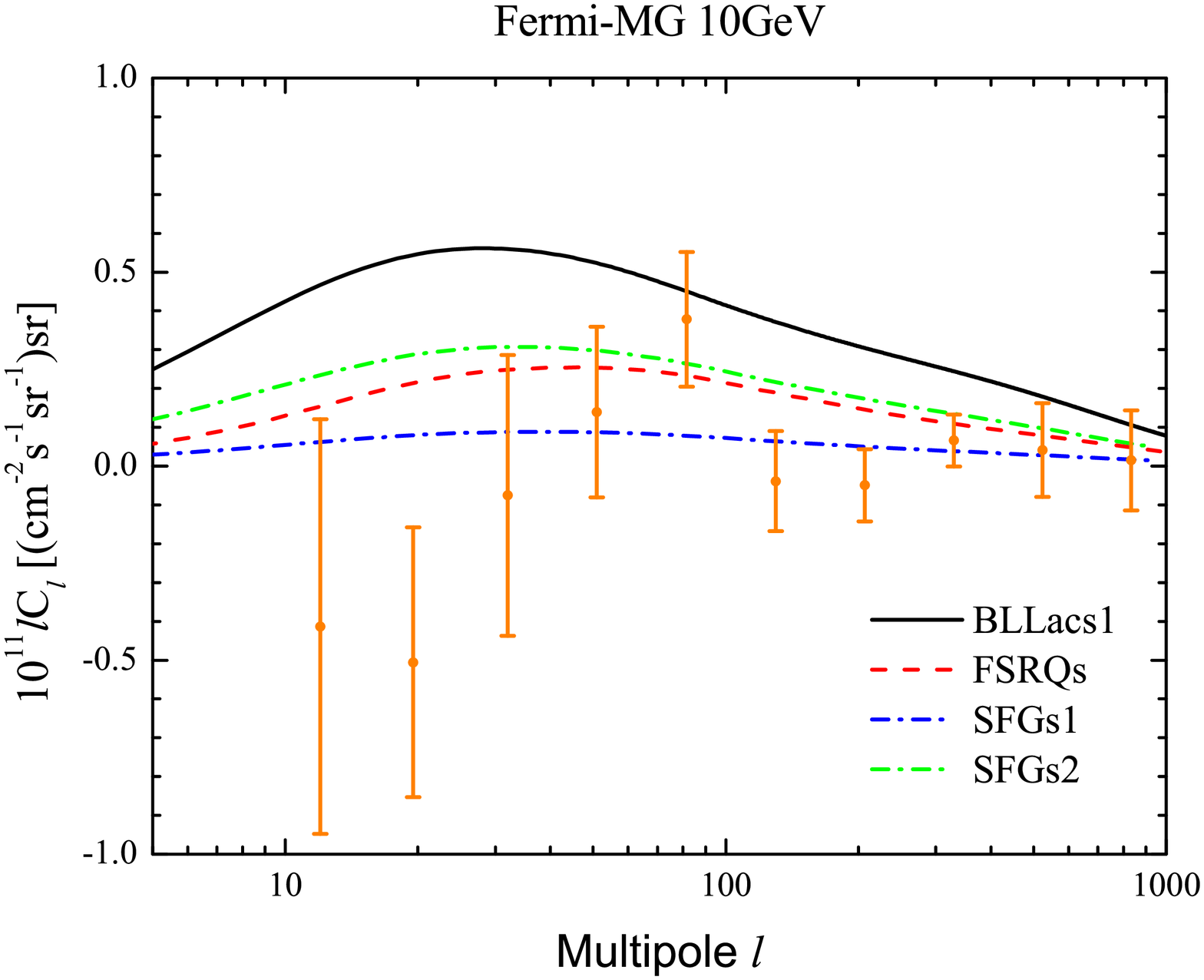, angle=0, width=0.33 \textwidth}
\centering \epsfig{file=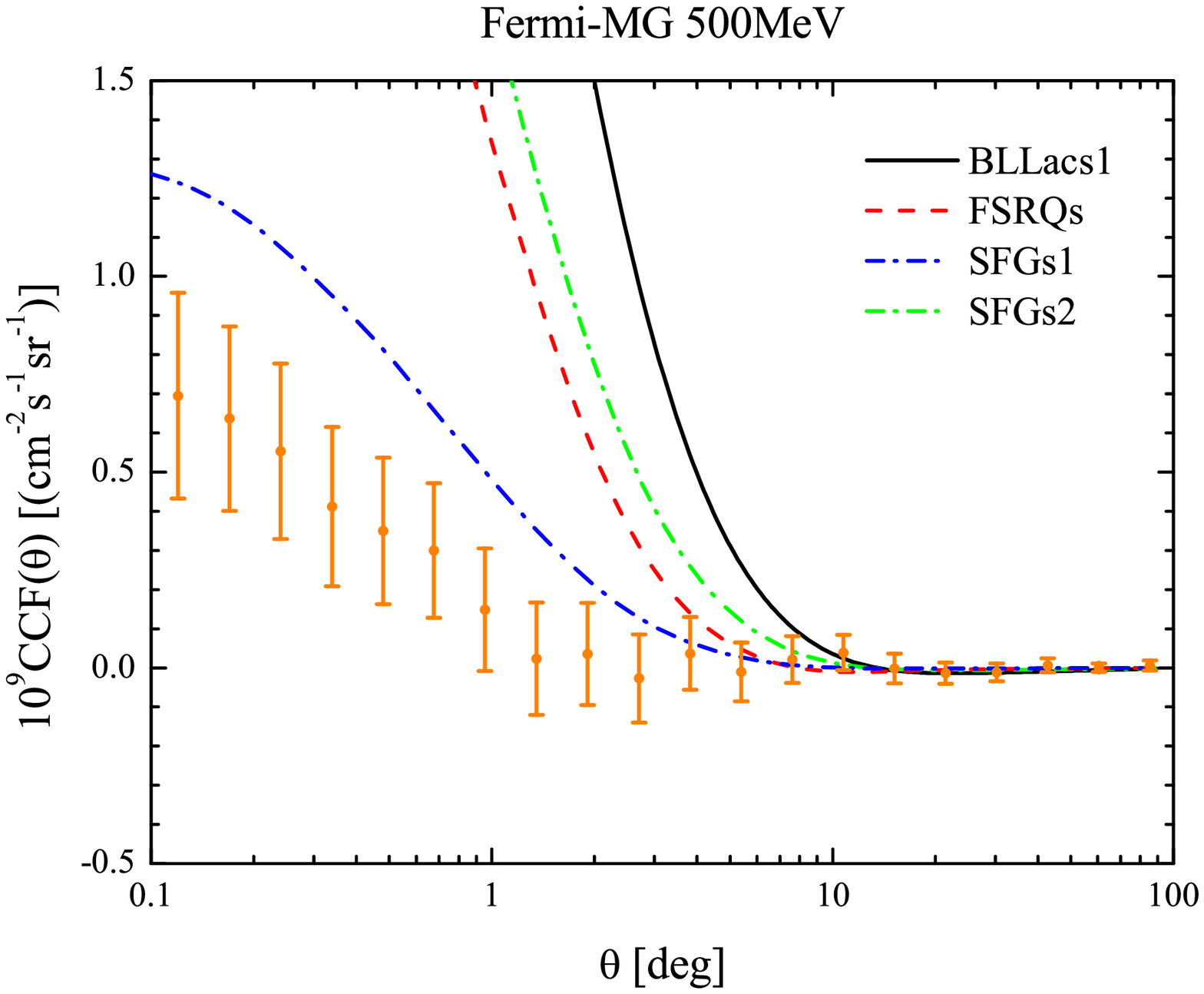, angle=0, width=0.33 \textwidth}
\centering \epsfig{file=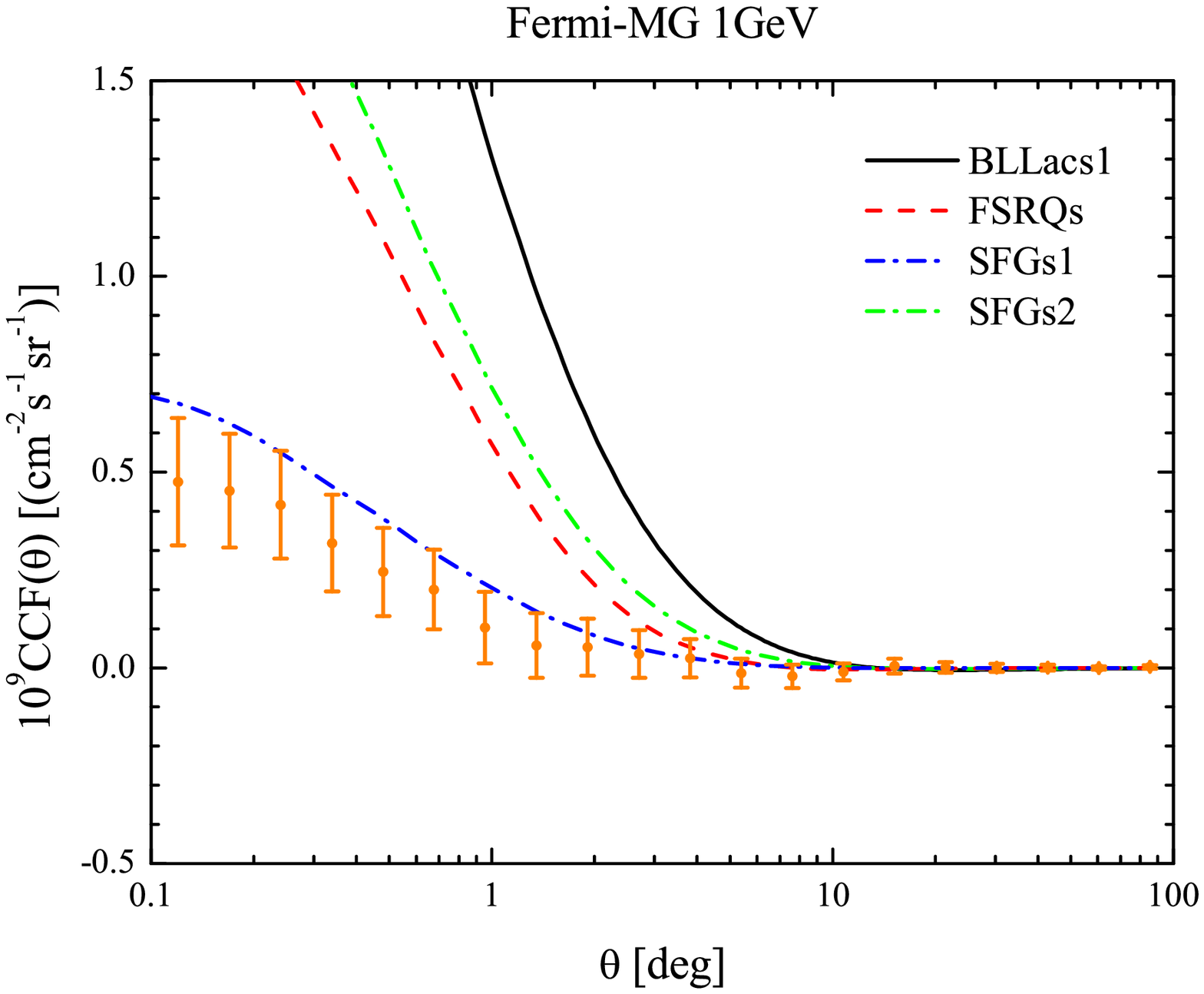, angle=0, width=0.33 \textwidth}
\centering \epsfig{file=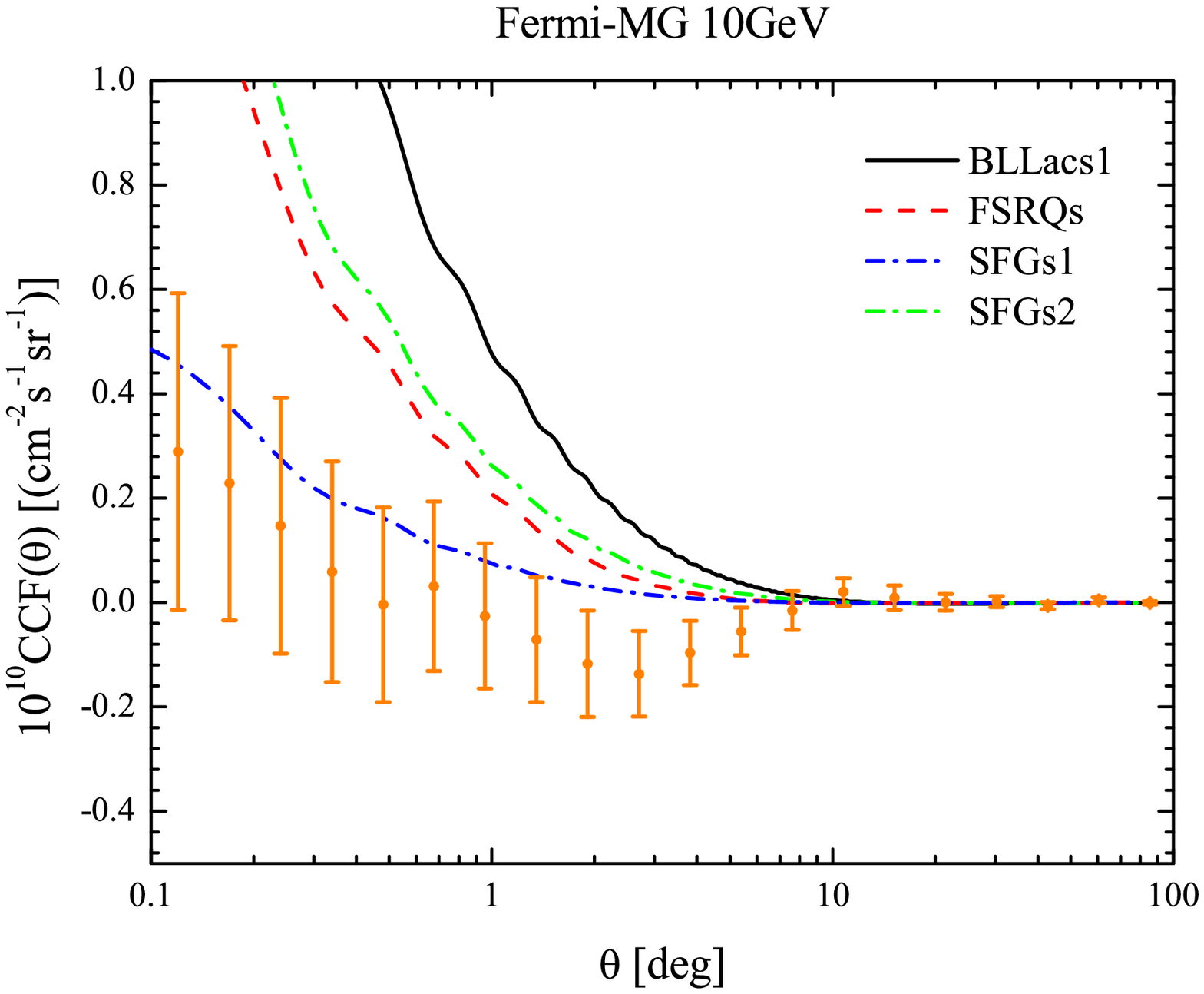, angle=0, width=0.33 \textwidth}
\caption{Analogous to fig.~\ref{fig:qso_ccf_fermi} using SDSS DR8 main galaxy sample}
\label{fig:mg_ccf_fermi}
\end{figure*}

\subsection{Cross-correlation with NVSS galaxies}
\label{sec:ccfnvss}

Fig.~\ref{fig:nvss_ccf_fermi} shows the results obtained by cross-correlating {\it Fermi} maps with the NVSS galaxy catalog.
In this case we detect a CCF signal with a strong significance of about $\sim8.0\, \sigma$
both for $E>500$ MeV and $E>1$ GeV
on small ($\theta<1^{\circ}$) angular scales.
In fact, we detect a strong signal also in the highest energy bin ($\sim5.0\, \sigma$), though only at $\theta<0.2^{\circ}$.
The fact that the peak in the CCF narrows with increasing energy is quite informative
and  indicates that the signal is intrinsically  confined  to very small $\theta$ and
it extends to larger $\theta$ values only because of the spreading out effect
by the LAT PSF, especially at low energies.
The width of the peak,  $\sim 1.5^{\circ}$,  $\sim 1.0^{\circ}$  and $\sim 0.2^{\circ}$
for the CCF at $E>0.5, 1, 10$ GeV is, indeed, also compatible with the
width of the LAT PSF at these energies.
The CAPS  gives similar significances and provides  additional information on the characteristics and possible origin
of the signal.
Differently from the CAPS with other catalogs, in fact, the CAPS with NVSS
is characterized by a strong signal at very high multipoles (up to $l\sim1000$)
confirming that the signal comes mostly from small scales.

\begin{table*}
\begin{center}
\caption{Significance of the CCFs cross-correlations for each energy bin and catalog calculated using the SFGs1 model with free normalization.
For each case, the best fit $\chi^2_{\rm bf}$, the significance $\sigma$ and the test statistics TS values are reported.
Each fit has 9 degrees of freedom (10 bins - 1 free parameter). For the NVSS case a further model, PSF, is tested.  \label{tab:sigmasCCF}}
\footnotesize{
\begin{tabular}{l|ccc|ccc|ccc|ccc|ccc|ccc}
\hline\hline
CCF &   \multicolumn{3}{|c|}{2MASS} & \multicolumn{3}{c|}{SDSS-MG}  & \multicolumn{3}{c|}{SDSS-LRG}  & \multicolumn{3}{c|}{SDSS-QSO}  & \multicolumn{3}{c|}{NVSS (LSS)}  & \multicolumn{3}{c}{NVSS (PSF)}  \\
\cline{2-19}
  &  $\chi^2_{\rm bf}$ & $\sigma$ & TS    & $\chi^2_{\rm bf}$ & $\sigma$ & TS  & $\chi^2_{\rm bf}$ & $\sigma$ & TS &  $\chi^2_{\rm bf}$ & $\sigma$ & TS  &  $\chi^2_{\rm bf}$ & $\sigma$ & TS   &  $\chi^2_{\rm bf}$ & $\sigma$ & TS\\
\hline
{\footnotesize $E >$ 500 MeV}  & 6.2  & 3.6  & 12.9 & 2.6 & 2.7 & 7.4 & 4.5 &  0.3 & 0.1 & 9.0 & 4.5 & 21 & 30.2 & 8.0 & 64.9 & 3.6 & 9.9 & 97.3  \\
{\footnotesize $E>$ 1 GeV}       & 10.6 & 4.4 & 19.4 & 2.1 & 3.0 & 9.3  & 4.6 & 0.4 & 0.2 & 3.5 & 2.3 & 5.1 & 45.1 & 8.6 & 73.6 & 4.9 & 10.3 & 106.4    \\
{\footnotesize $E>$ 10 GeV}     & 2.0   & 2.1 & 4.5   & 6.2 & 0.7 & 0.5  & 2.6 &  0.2 & 0.1 & 4.8 & 1.6 & 2.6 & 40.4 & 5.1 & 25.6 & 5.8 & 7.7 & 59.4   \\
\hline
\end{tabular}}
\end{center}
\end{table*}

\begin{table*}
\begin{center}
\caption{Same as Table~\ref{tab:sigmasCCF} but using CAPS.  \label{tab:sigmasCAPS}}
\footnotesize{
\begin{tabular}{l|ccc|ccc|ccc|ccc|ccc|ccc}
\hline\hline
CAPS &   \multicolumn{3}{|c|}{2MASS}   & \multicolumn{3}{c|}{SDSS-MG}   & \multicolumn{3}{c|}{SDSS-LRG}  & \multicolumn{3}{c|}{SDSS-QSO}   & \multicolumn{3}{c|}{NVSS (LSS)}   & \multicolumn{3}{c}{NVSS (PSF)}  \\
\cline{2-19}
  &  $\chi^2_{\rm bf}$ & $\sigma$ & TS    & $\chi^2_{\rm bf}$ & $\sigma$ & TS  & $\chi^2_{\rm bf}$ & $\sigma$ & TS  &  $\chi^2_{\rm bf}$ & $\sigma$ & TS  &  $\chi^2_{\rm bf}$ & $\sigma$ & TS   &  $\chi^2_{\rm bf}$ & $\sigma$ & TS\\
\hline
{\footnotesize $E >$ 500 MeV}  & 8.3  & 3.4 & 11.5 & 4.5 & 3.5 & 12.1 & 3.5 &  0.0 & 0.0 & 9.7 & 5.3 & 28.6 & 30.1 & 8.3 & 71.3 & 7.3 & 9.6 & 92.3  \\
{\footnotesize $E>$ 1 GeV}       & 3.7  & 3.6 & 12.8 & 3.9 & 3.3 & 11.2  & 5.4 & 0.4 & 0.2 & 7.6 & 3.3 & 10.9 & 23.1 & 8.4 & 70.7 & 5.3 & 9.1 &  82.8    \\
{\footnotesize $E>$ 10 GeV}     & 5.1  & 1.6 & 2.7   & 8.4 & 0.7 & 0.6   & 4.4 &  0.7 & 0.5 & 4.6 & 2.7 & 7.3   & 21.0 & 3.4 & 11.8& 9.3 & 4.8 & 23.2   \\
\hline
\end{tabular}}
\end{center}
\end{table*}

All models provide a good match to the data at large ($\theta>1^{\circ}$) angular scales. At smaller separations, however, the
observed signal overshoots model predictions, especially in the SFGs1 case,
as confirmed by the high $\chi^2_{\rm bf}$ in Tables~\ref{tab:sigmasCCF} and \ref{tab:sigmasCAPS}.
This excess signal  correlation on small scales does not seem to be related to the large-scale clustering of astrophysical sources. 
Instead, this correlation seems to be better described by a   {\it 1-halo}-like term.
Indeed, as  seen in section \ref{sec:1halodata}, the NVSS case is the only one where a {\it 1-halo}-like term is strongly detected. 
Using a {\it 1-halo}-like term as an alternative model to calculate the significance of the signal
improves the quality of the fit to both CAPS and CCF, as confirmed by the decrease in the $\chi^2_{\rm bf}$ values
and the corresponding increase in statistical significance to $\sim10.0\, \sigma$.

It is unclear if this small-scale signal is due to a pure {\it 1-halo} term or to the possibility that
a significant fraction of NVSS sources  might also be $\gamma$-ray emitters.
Indeed, the fact that NVSS sources are known to be good candidates for $\gamma$-ray emission
 and that this catalog  is routinely searched to identify the counterparts of $\gamma$-ray sources \citep{2012ApJS..199...31N,3FGL}
 is an argument in favor of the second possibility.

A further explanation, possibly not entirely independent of the previous ones, is the presence of  duplicate objects in the NVSS catalog.
It is well known that a  large fraction of close pairs are in fact single objects with a prominent radio jet wrongly classified
as a separate, companion object \citep{overzier03}. The net result is an excess of pairs at small angular separations
which is responsible for an unphysical large auto-correlation signal at small angles \citep{overzier03},
and could thus induce a corresponding cross-correlation excess.

For all the above reasons, we adopt a conservative approach and consider
 the NVSS cross-correlation at angles $\theta<1^{\circ}$ and multipoles $l>100$ as
 arising from physical processes that are not associated to the large-scale structures
 and will ignore it in the $\chi^2$ analysis performed in the next section.

\subsection{Cross-correlation with SDSS DR8 LRGs galaxies}
\label{sec:ccflrg}

The results of the cross-correlation analysis between the SDSS DR8 LRGs and the {\it Fermi}-LAT maps are shown in Fig.~\ref{fig:lrg_ccf_fermi}.
In this case we do not detect any correlation signal. In fact the CCF drops below zero at very small angular separations,
although the significance of this feature is very weak.
A possible reason for this surprising behaviour is the aggressive procedure used to remove possible systematics
from the raw LRGs data \citep{2012ApJ...761...14H} which might remove the genuine correlation signal together with the spurious one.
At larger ($\theta>0.2^{\circ}$) separations the correlation signal is consistent with zero.
This is in agreement with the model predictions for SGFs1 and, to a lesser extent, to  BLLacs1 and SFGs2. This is not surprising since the
$dI(>E)/dz$ of these sources barely overlap with the narrow $dN/dz$ distribution of the LRGs.
On the contrary, the FSRQ model predicts a significant cross-correlation, which is at variance with the
data.

\subsection{Cross-correlation with SDSS-DR8 main galaxy sample}
\label{sec:ccfsdssgal}

In Fig.~\ref{fig:mg_ccf_fermi} we plot the estimated CAPS and CCF between the {\it Fermi}-LAT maps and the SDSS-DR8 galaxies
in the main sample. We observe a correlation signal at small angles
at about $\gtrsim 3 \, \sigma$ level  for both the $E>500$ MeV and the $E>1$ GeV cases, similarly to the 2MASS case.
The observed CCF is marginally consistent, for $E>1$ GeV,  with theoretical predictions if the sources of the IGRB are SFGs
in model SFGs1.
In all AGN-based models the predicted cross-correlation signal is much higher than the observed one.
This is similar to the 2MASS case except that the $dN/dz$ of the DR8 galaxies peaks at significantly higher redshift than 2MASS galaxies.
We conclude that in this case SFGs  provide a significant contribution to the IGRB not only locally but also
at $z\sim 0.3$ and that their contribution is more important than that of
BL Lacs and FSRQs.
In the case of the SFGs2 model, instead,  SFGs are predicted to have a small contribution,
similar to the one of blazars.

\section{$\chi^2$ analysis}
\label{sec:chi2}

To quantify the qualitative conclusions drawn from the inspection of
the correlation analysis performed in the previous section we now perform
a  $\chi^2$ comparison between model predictions discussed in Section~\ref{sec:theory}
and the CCF and CAPS estimates presented in Section~\ref{sec:results}.
The aim is to estimate the free parameters of the models, i.e., to quantify the relative
contribution of different types of potential sources to the IGRB and to assess the goodness
of the fit, from which we can infer  which is the most likely mix of source candidates
responsible for the observed IGRB.
Here we present only the results of the CCF analysis since those obtained with the CAPS are fully consistent with those shown below.

\begin{table*}
\begin{center}
\caption{Minimum $\chi^2$ for the one-, two-, and three source models
and best fit values for the free parameters corresponding to
the fraction of the IGRB contributed by SFGs, BL Lacs and  FSRQs.  \label{tab:tabchi2}}
\begin{tabular}{|l|cccc|cccc|}
\hline\hline
 &   \multicolumn{4}{|c|}{BLLacs1}   & \multicolumn{4}{c|}{BLLacs2} \\
\hline
 &   $\chi^2$ & $f_{\rm SFGs}$ & $f_{\rm BLLs}$       &  $f_{\rm FSRQs}$ &    $\chi^2$ & $f_{\rm SFGs}$ & $f_{\rm BLLs}$ &  $f_{\rm FSRQs}$  \\
\hline
SFGs1 & 35.3  & 0.72 &        &         & 35.3  & 0.72 &       &       \\
BLLacs &  44.3 &       &  0.08 &        & 43.1   &       & 0.18 &     \\
FSRQs & 48.8  &        &        & 0.24 & 48.8   &        &        & 0.24\\
\hline
SFG1s +  BLLacs  &  35.3  & 0.72 & 0.0 &         &   35.3    & 0.72 & 0.0 &       \\
FSRQs +  SFGs1  &  35.3  &  0.72 &       & 0.0    &    35.3 & 0.72 &         & 0.0   \\
FSRQs + BLLacs  &  42.0  &          & 0.06 & 0.10 &    43.1 &        &  0.18 & 0.0 \\
\hline
FSRQs + BLLacs +  SFGs1   &   35.3  & 0.72 & 0.0 & 0.0  &  35.3  & 0.72 & 0.0 & 0.0  \\
\hline
FSRQs + BLLacs +  SFGs2   &   41.7  & 0.14 & 0.0 & 0.12 &  41.7  & 0.14 & 0.0 & 0.06  \\
\hline
\end{tabular}
\end{center}
\end{table*}

For each CCF estimated by comparing a galaxy catalog  and a {\it Fermi}-LAT
map above a given energy threshold we compute the following $\chi^2$ statistics:
\begin{equation}
   \chi^2 = \sum_{i\, j}   (d_i -m_i(\alpha) )  C^{-1}_{\theta_i\theta_j}  (d_j -m_j(\alpha) ) \, ,
\label{eq:chi2}
\end{equation}
where $C_{\theta_i\theta_j}$ is the covariance matrix computed using PolSpice that
quantifies the covariance among  different angular bins $\theta_i$,
 $d_i$ represents the data, i.e., the
CCF measured at the angular bin $i$, and
$m_i(\alpha)$ is the model prediction  which depends from a set of parameters $\alpha$.
We note that it is important to use the full covariance matrix since the
different bins are significantly correlated, a feature which is
typical of CCF measurements.
Instead, the covariance matrix of the CAPS is to a better approximation diagonal
(although some sizable correlations are nonetheless present, in particular for low and high multipoles),
at the price, however, of making the interpretation less intuitive
since it lacks the immediate identification
of the scale(s) responsible for the correlation, which is instead present for the
analysis in real space with the CCFs.
Thanks to the fact that we are considering $\gamma$-ray flux maps rather than fluctuation maps
we can express the model CCF as a sum of different contributions corresponding to the CCF
of different source types:
$m_i=f_{\rm sfg}\, c_{\rm sfg} (\theta_i) + f_{\rm bllac}\, c_{\rm bllac} (\theta_i)+ f_{\rm fsrq}\, c_{\rm fsrq} (\theta_i)$,
where $c_{\alpha} (\theta_i)$ is the model CCF for a given type of source when it represents 100 \% of the IGRB and
$f_\alpha$ is a free parameter that quantifies the actual IGRB fraction contributed by the source.
Note that, in our analysis, we do not require that $\sum f_{\alpha}=1$. Instead we verify that this condition is verified {\it a posteriori}.

In Eq.~\ref{eq:chi2} the sum extends over 10 angular bins logarithmically spaced between $\theta= 0.1^\circ$ and $100^{\circ}$.
We use logarithmic bins  to emphasize small scales where the signal-to-noise is higher while the choice
of 10 bins is dictated by the  compromise between  the need to robustly invert  the covariance matrix
(an operation that becomes unstable when too many correlated bins are considered) and that of maximizing the
available information. The number of bins used in the $\chi^2$ analysis (10) is smaller than that
used in the CCF plots shown in the previous Section (20), in which the aim was to illustrate the qualitative agreement
between models and data.

The total $\chi^2$ accounts for the contributions from the CCF of all catalogs.
The only exception
is the CCF of NVSS sources for which we have ignored separations $\theta<1^{\circ}$ since, as discussed in the
previous section, the signal in that range is likely not related to large-scale structure clustering.
In particular, the total $\chi^2$ for the $E>$500 MeV band is the quantity that we use to perform the bulk of the analysis detailed below.
However, we have also considered  the cases in which the $\chi^2$ only accounts
for the CCF of a subset of catalogs and/or of different energy bands.
Investigating the contribution to  $\chi^2$ by different catalogs is important to
illustrate the tomographic nature of our analysis. On the contrary, considering different energy cuts
turns out to be not very revealing.
In principle, breaking out the  total $\chi^2$ by energy bands
provides additional information to identify the contribution to the IGRB by
different sources.
However, the
constraints derived from the 2 higher energy bands
are weak and the $\chi^2$ discriminating power is dominated by the $E>500$ MeV band.
The practical outcome is that
the results obtained by considering photons with $E>1$ GeV or $E>10$ GeV are fully consistent with those
obtained with the $E>500$ MeV energy band.

\begin{figure*}
\raggedright  \epsfig{file=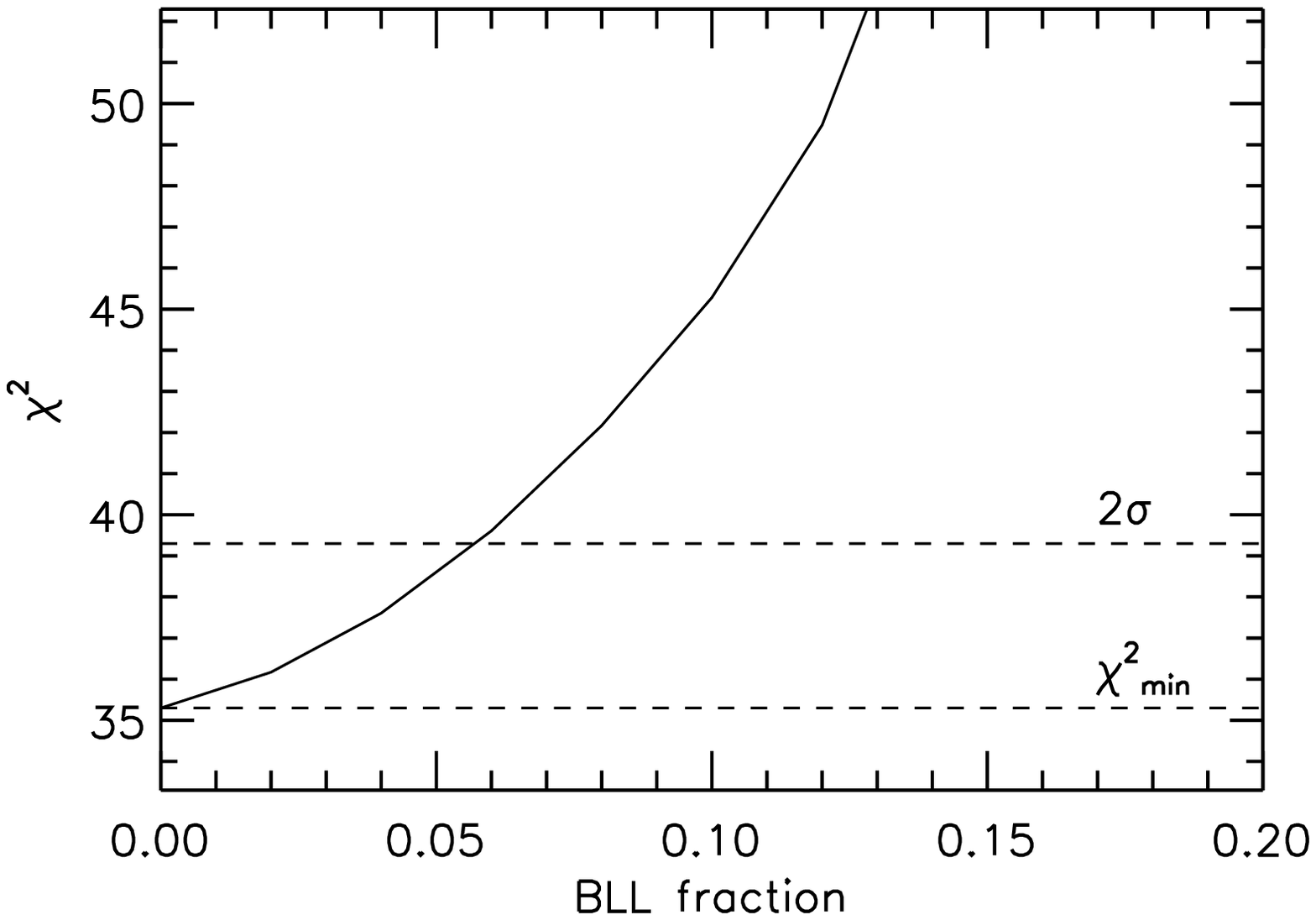, angle=0, width=0.33 \textwidth}
\hspace{5.88cm} \epsfig{file=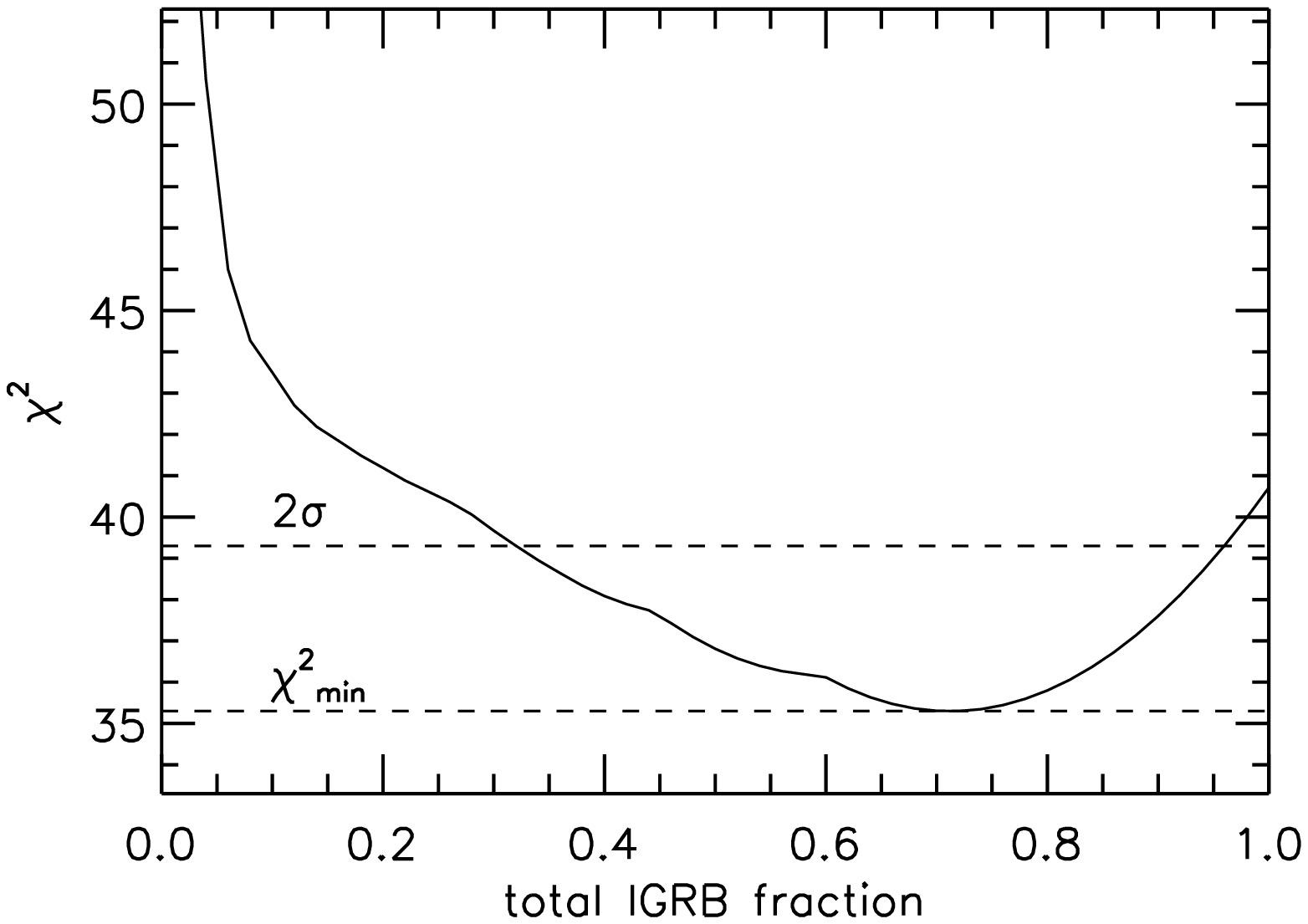, angle=0, width=0.33 \textwidth} \\
\raggedright   \epsfig{file=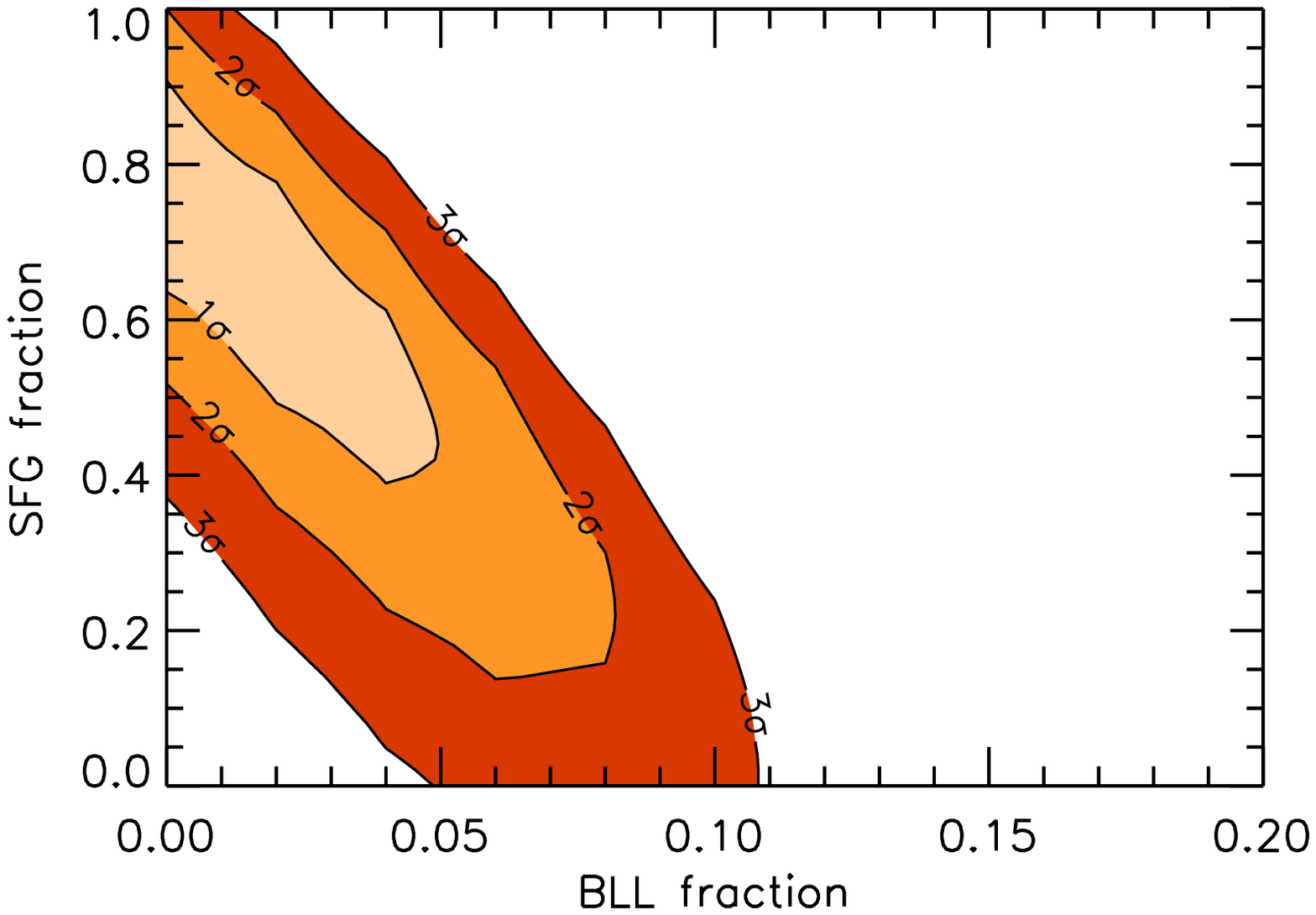, angle=0, width=0.33 \textwidth}
\raggedright   \epsfig{file=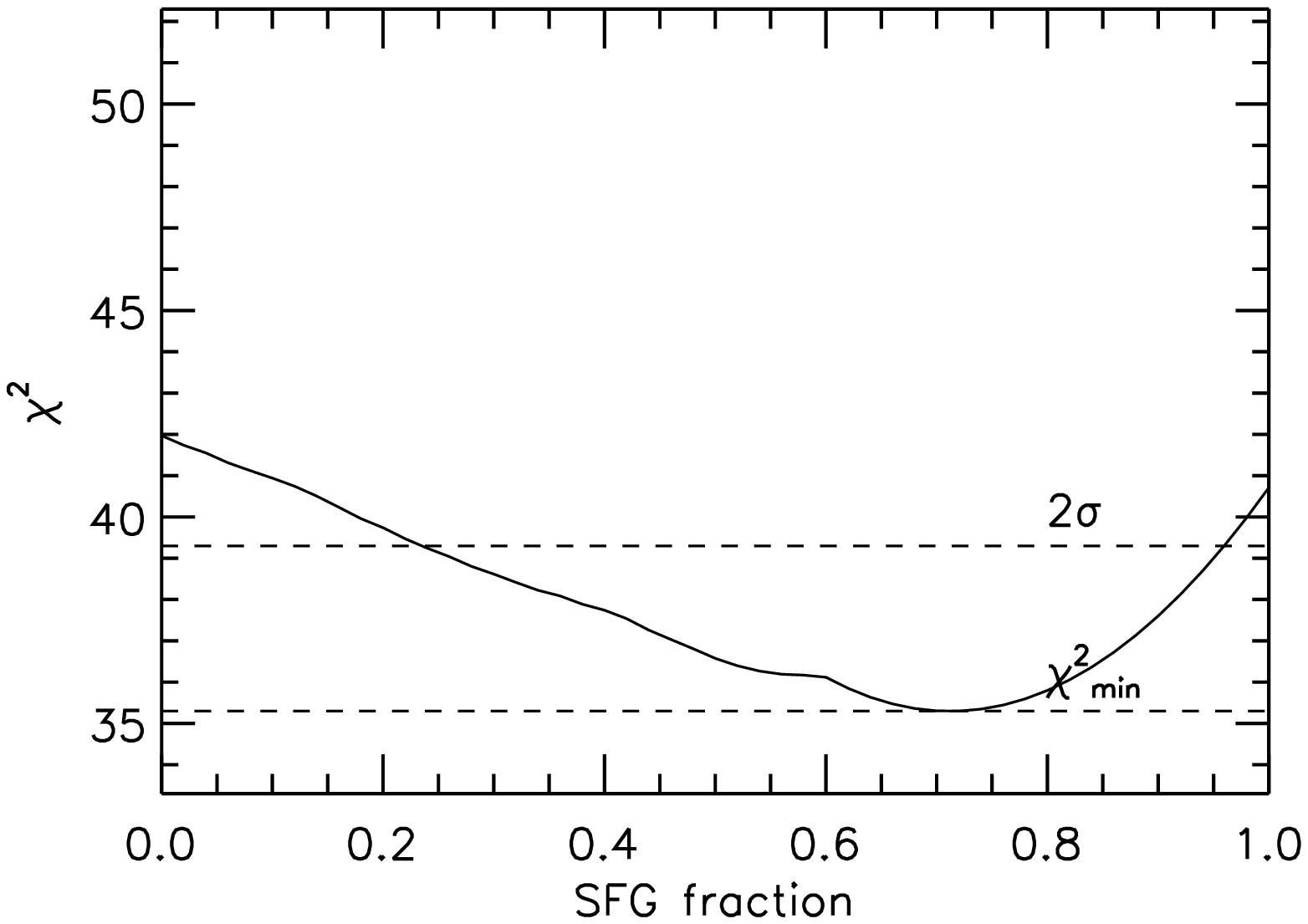, angle=0, width=0.33 \textwidth}  \\
%
\centering \epsfig{file=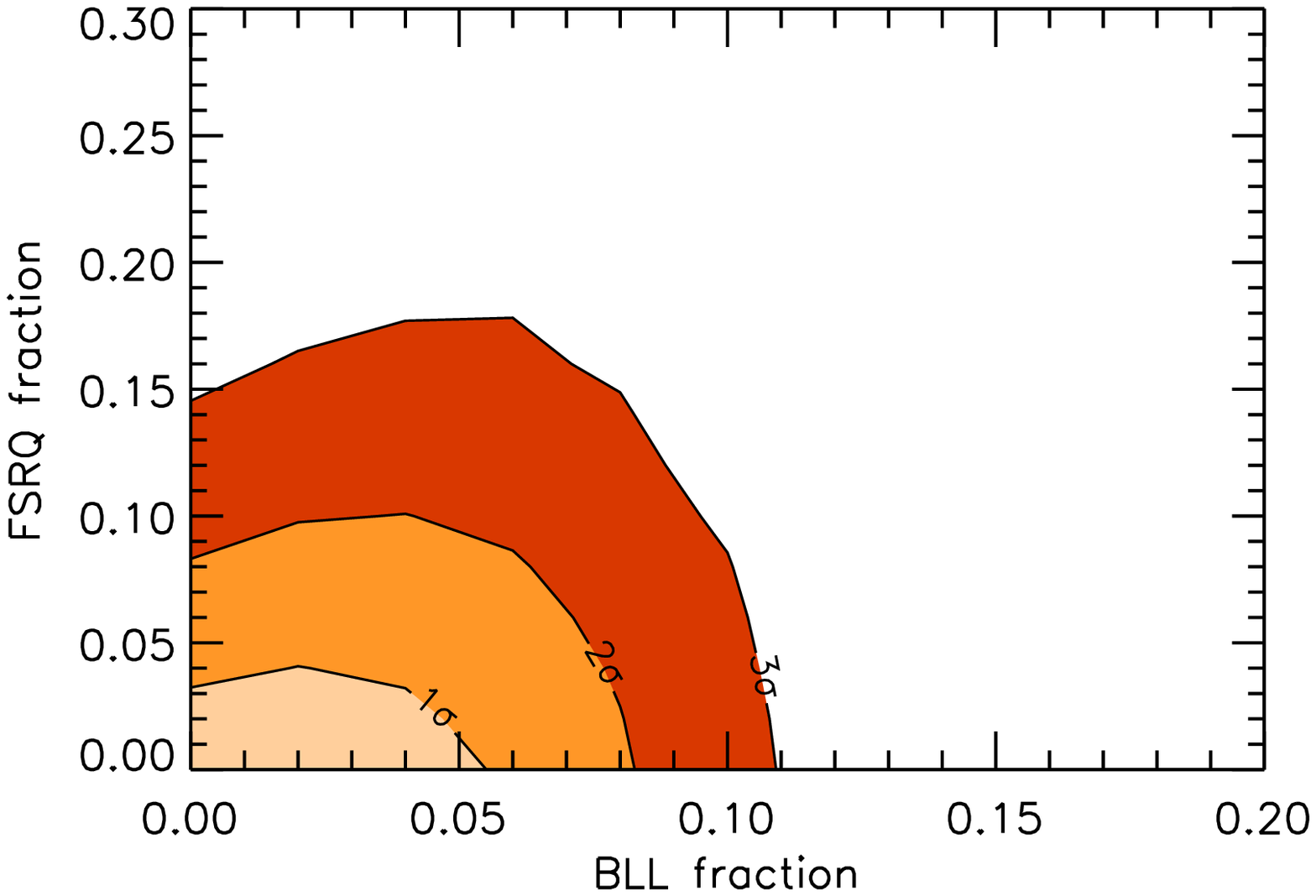, angle=0, width=0.33 \textwidth}
\centering \epsfig{file=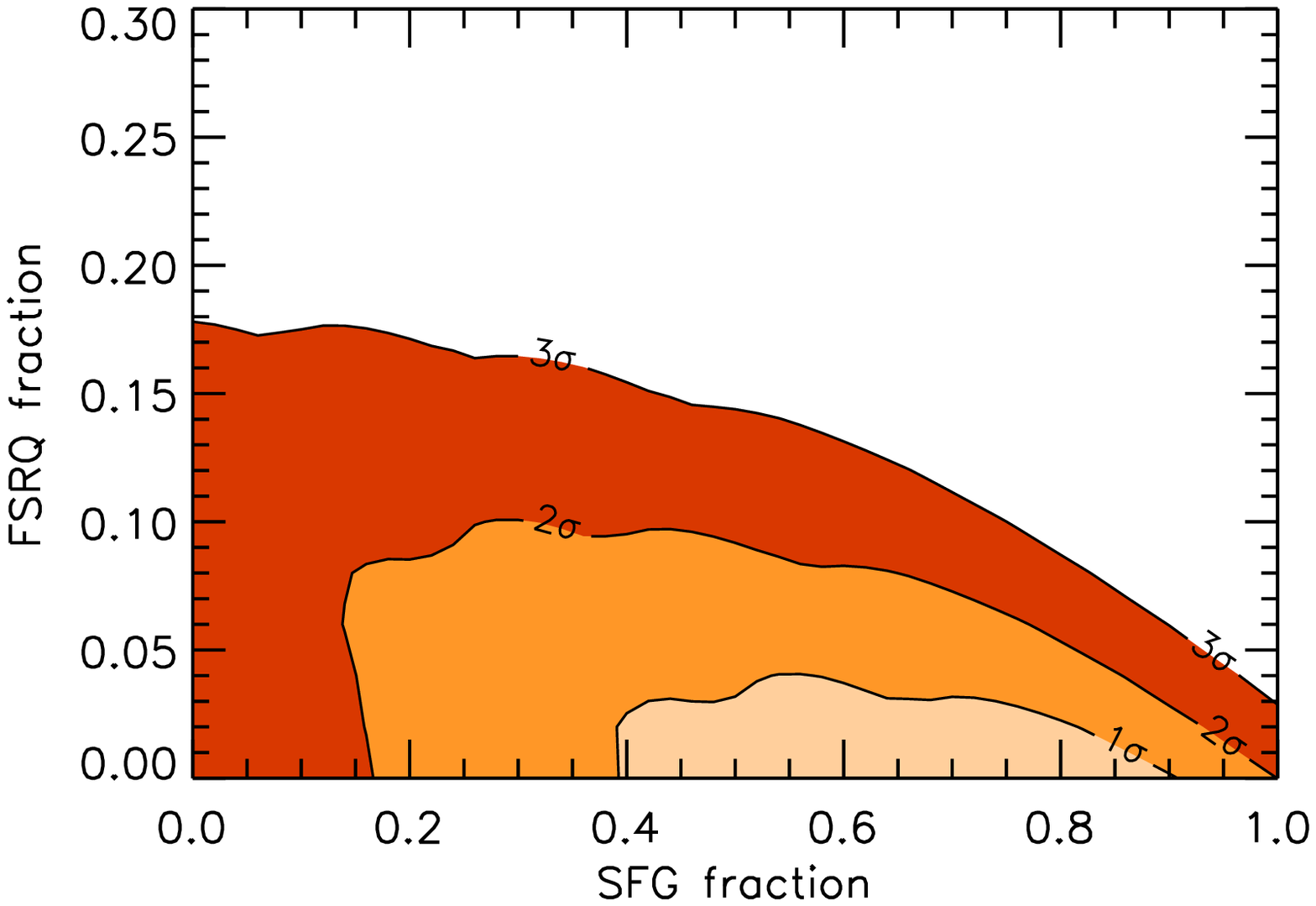, angle=0, width=0.33 \textwidth}
\centering \epsfig{file=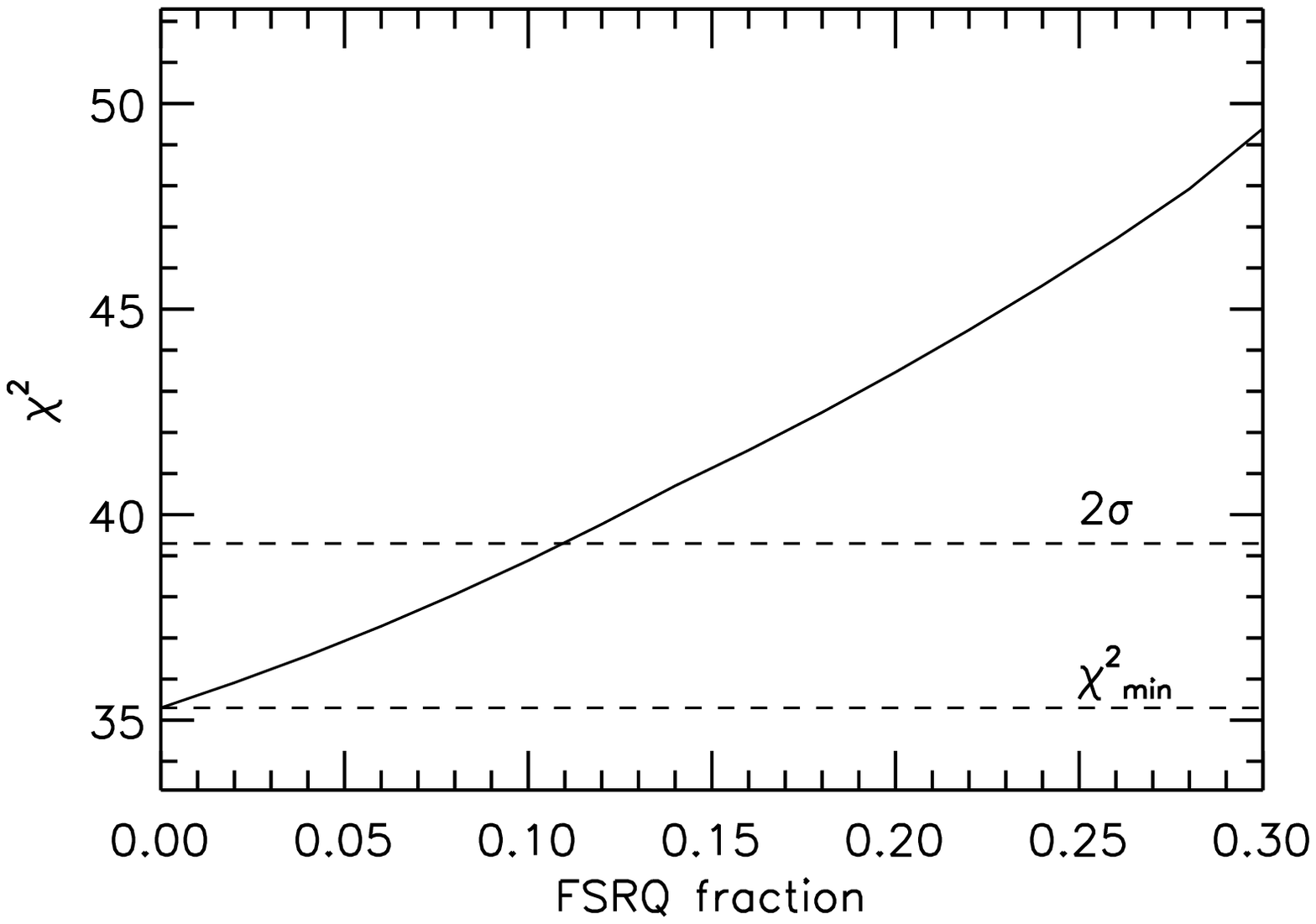, angle=0, width=0.33 \textwidth}

\caption{Plot matrix showing the 1-dimensional profile likelihoods for each component
and contours of the 2-dimensional profile likelihoods
for the three-component fit  (BL Lacs, FSRQs and SFGs) to all the experimental CCFs (i.e., all catalogs and all energy ranges).
The plots refer to the model BLLacs1 and SFGs1. The plot in the upper-right corner shows the profile likelihood for
the total IGRB fraction.
\label{fig:chi2s_CCF_bll1_all_energy}}
\end{figure*}

\begin{figure*}
\raggedright  \epsfig{file=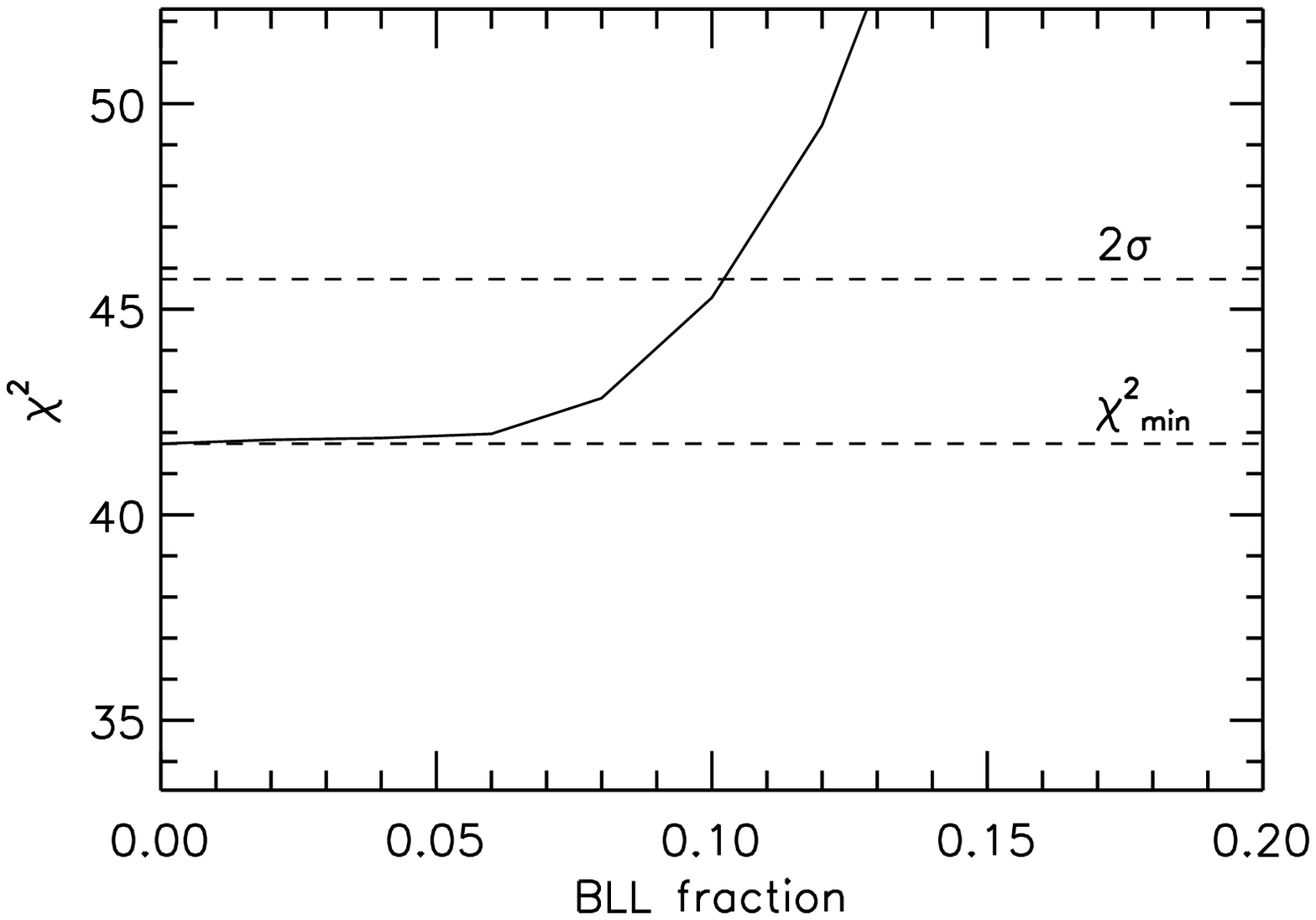, angle=0, width=0.33 \textwidth}
\hspace{5.88cm} \epsfig{file=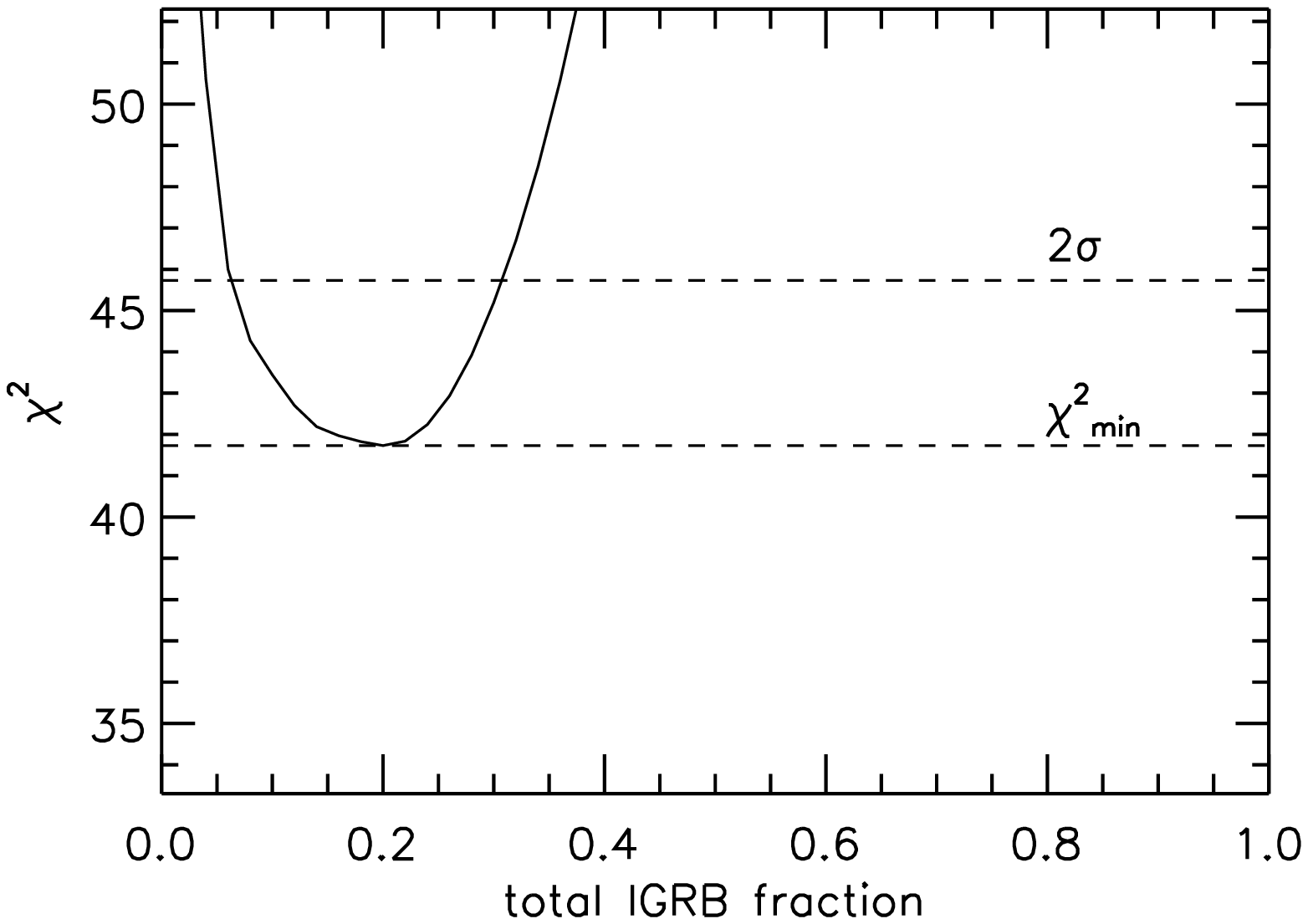, angle=0, width=0.33 \textwidth} \\
\raggedright   \epsfig{file=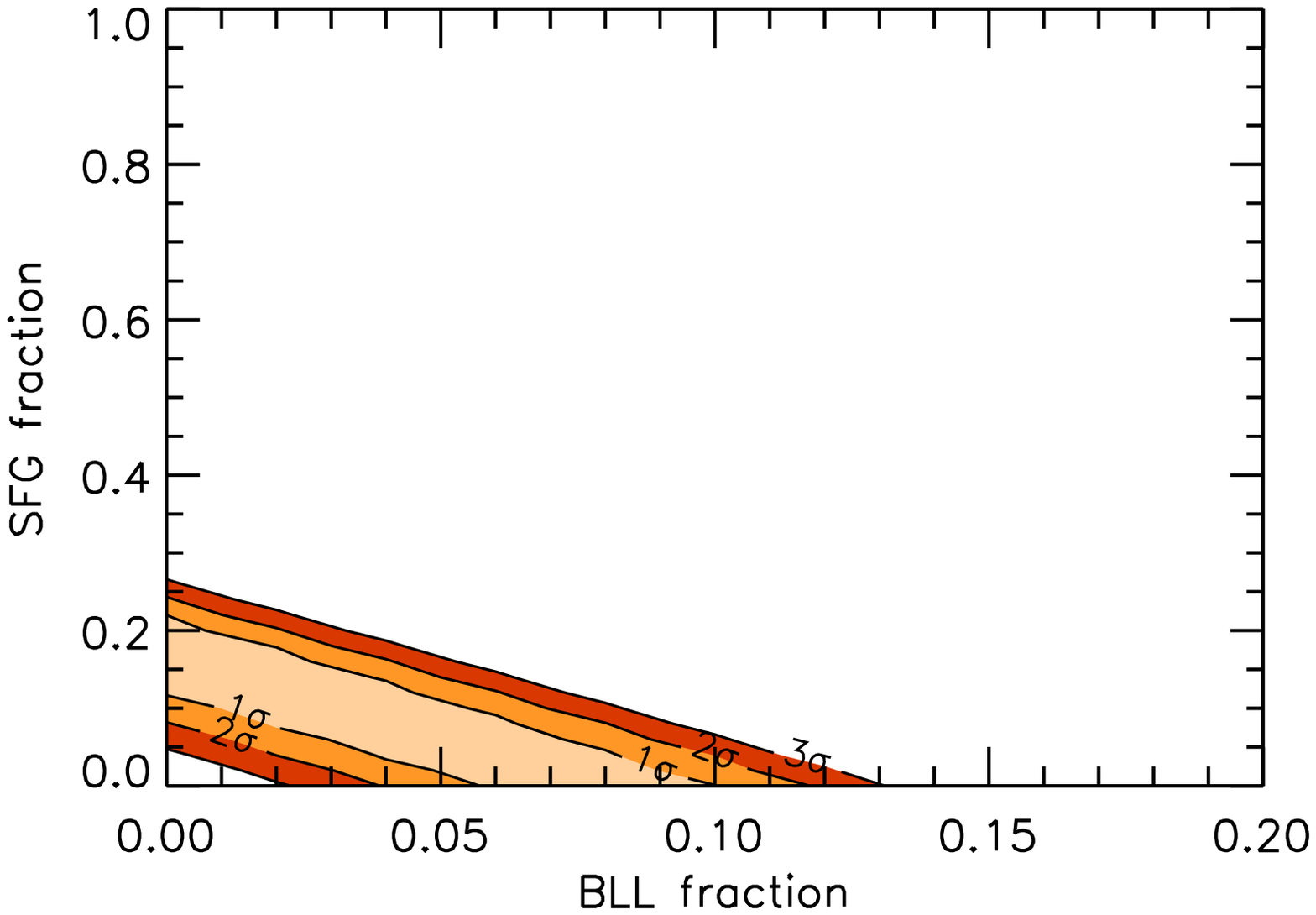, angle=0, width=0.33 \textwidth}
\raggedright   \epsfig{file=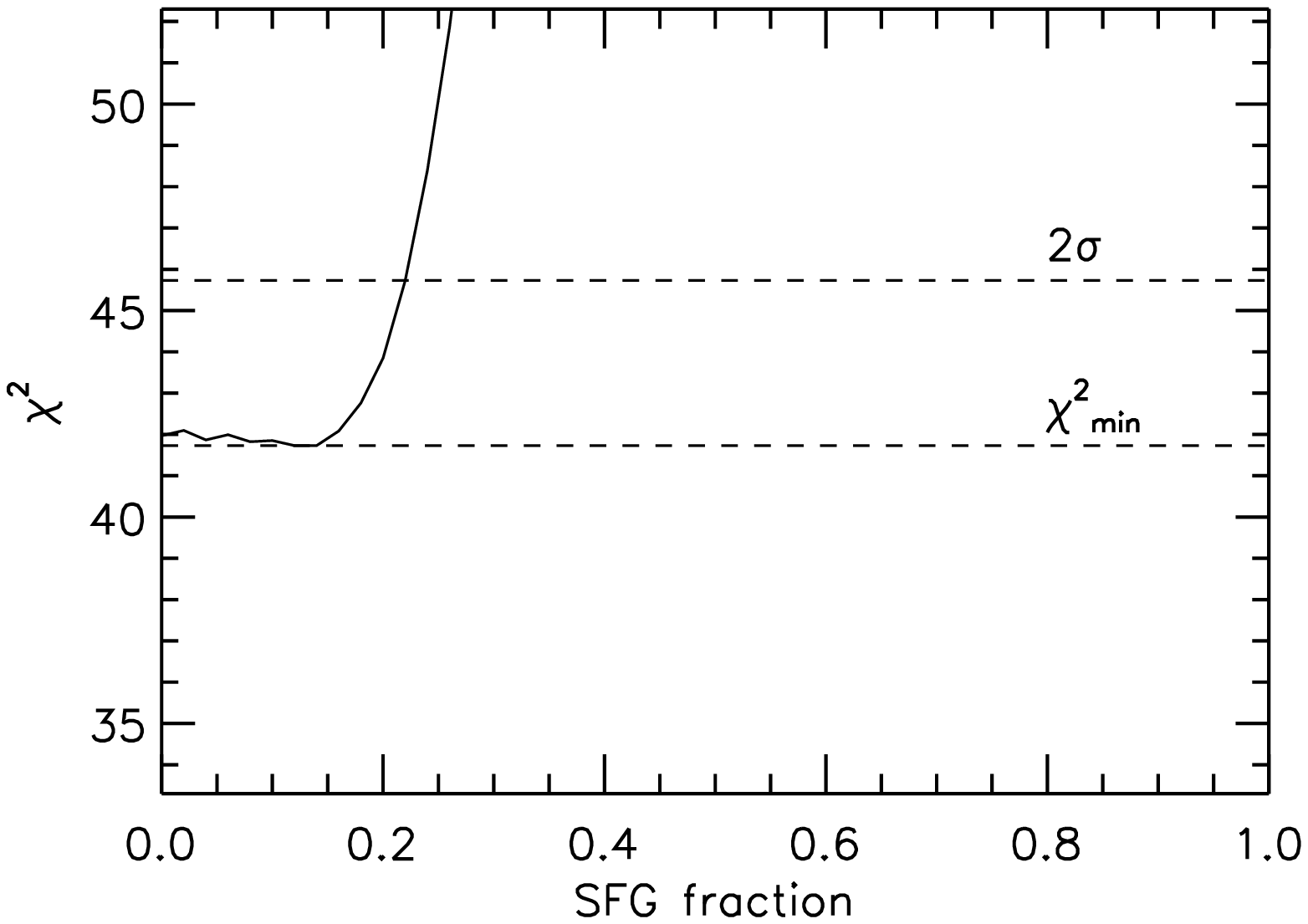, angle=0, width=0.33 \textwidth}  \\
%
\centering \epsfig{file=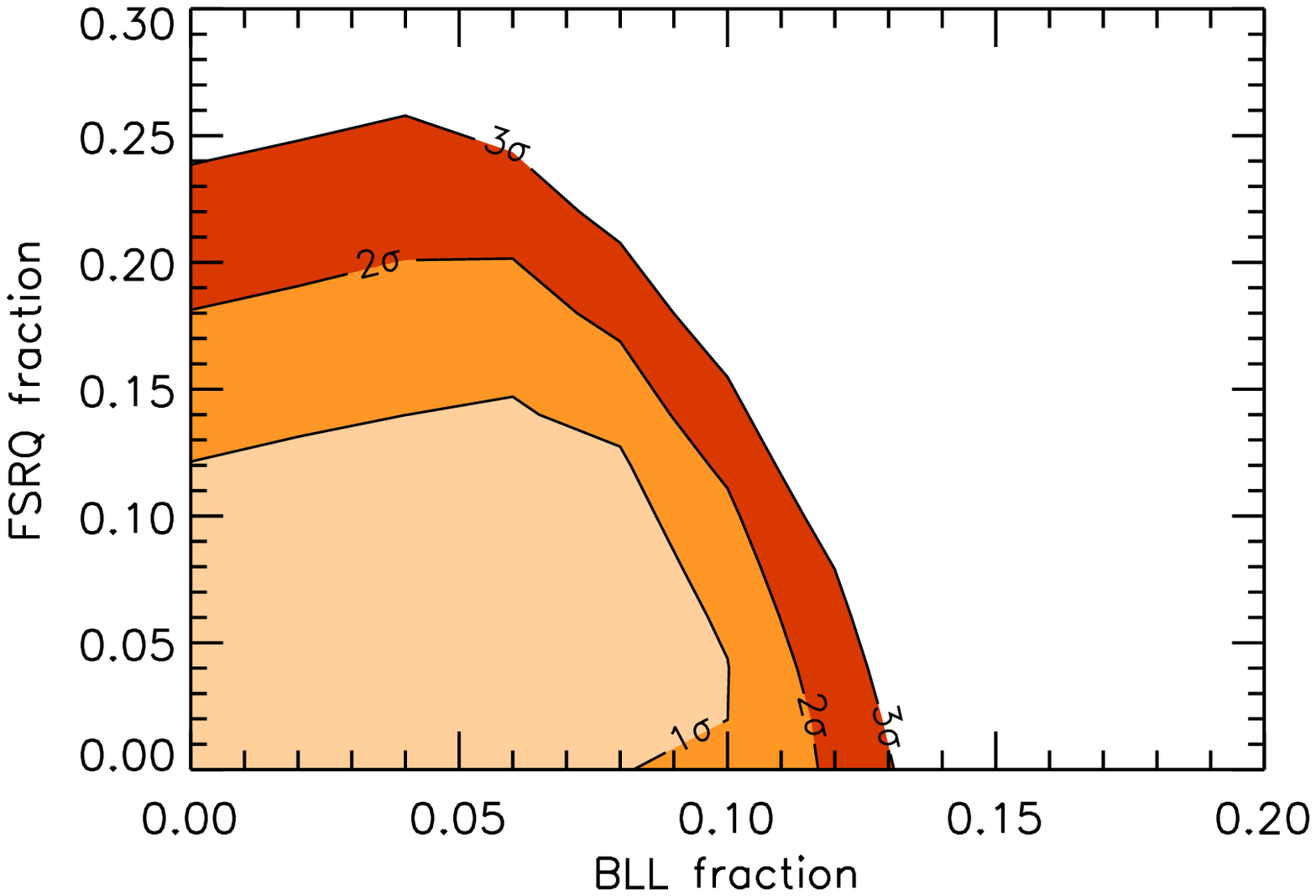, angle=0, width=0.33 \textwidth}
\centering \epsfig{file=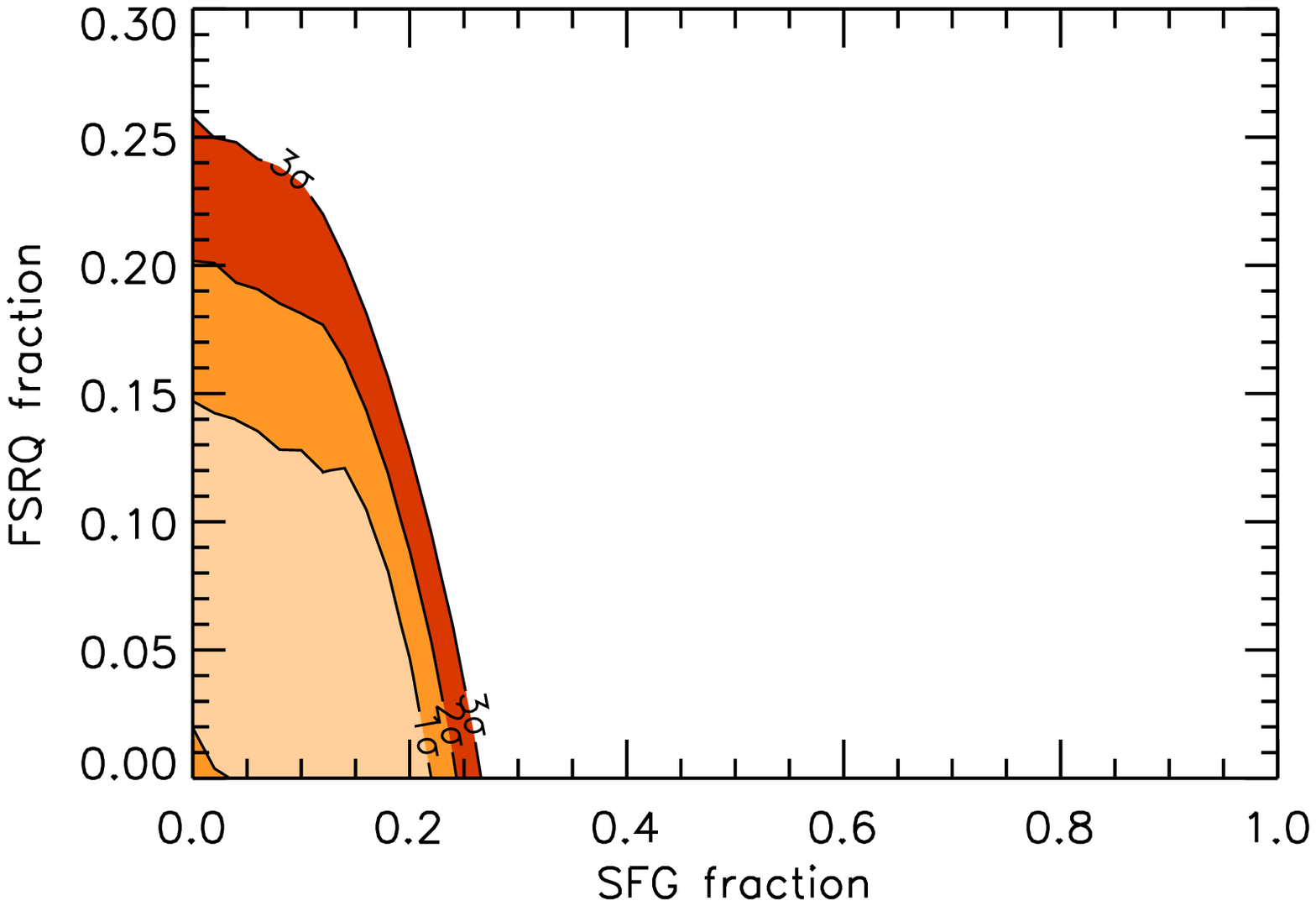, angle=0, width=0.33 \textwidth}
\centering \epsfig{file=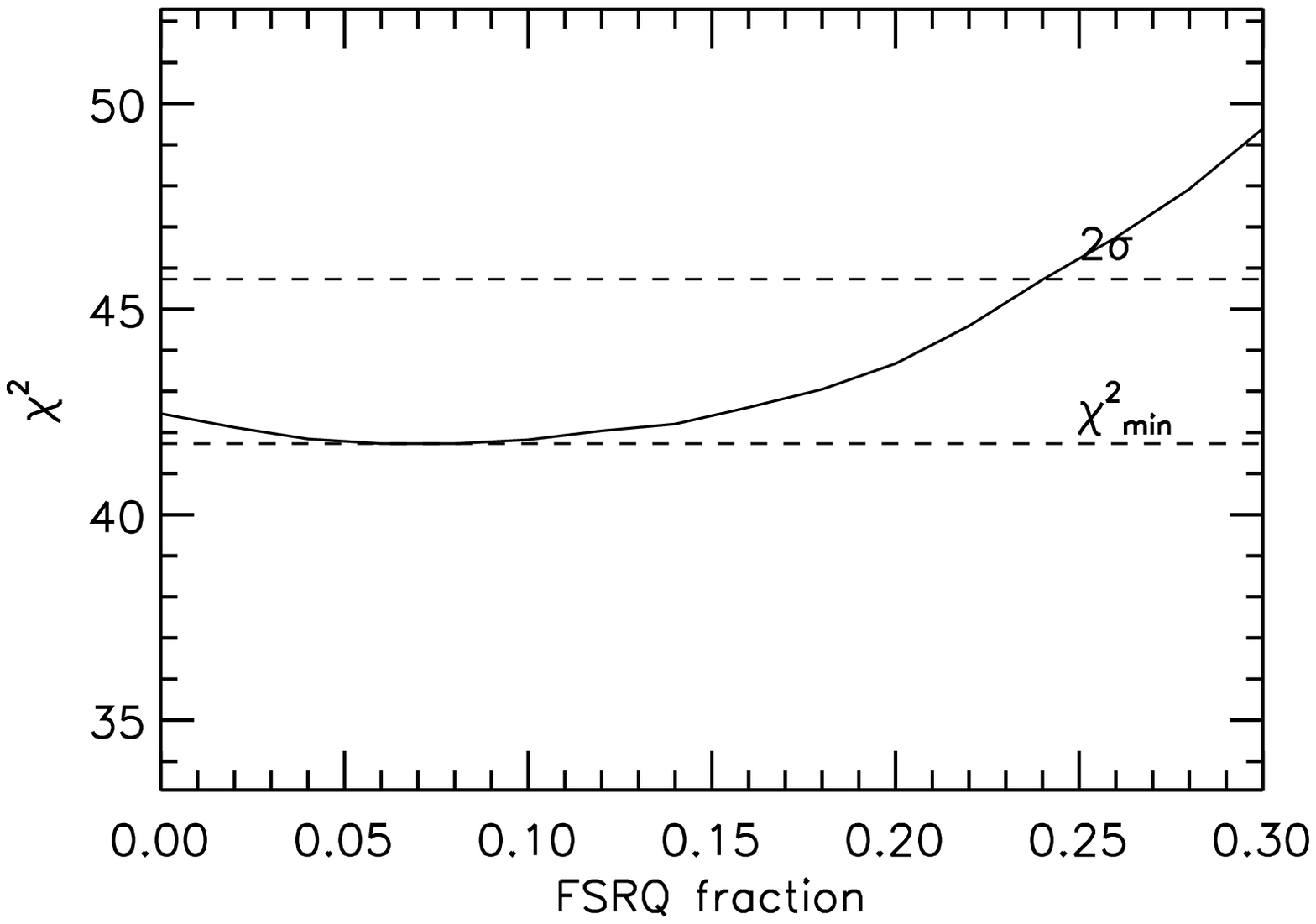, angle=0, width=0.33 \textwidth}

\caption{Same as Fig.~\ref{fig:chi2s_CCF_bll1_all_energy}
but for models BLLacs1 and SFGs2.
\label{fig:chi2s_CCF_bll2_all_energy}}
\end{figure*}

\begin{table*}
\begin{center}
\caption{Contribution to the best fit $\chi^2$ from the single catalog CCFs with the $E>$500 MeV $\gamma$-ray map,
for the two models FSRQs + BLLacs1 + SFGs1  and FSRQs + BLLacs1 + SFGs2 .
The numbers in parenthesis are the number of $\theta$ bins used to calculate the $\chi^2$.  \label{tab:chi2catalogs}}
\footnotesize{
\begin{tabular}{|l|c|c|c|c|c|}
\hline\hline
&   {\footnotesize 2MASS (10)} &  NVSS (6 )& SDSS-MG (10) & SDSS-LRG (10) & SDSS-QSO (10) \\
\hline
{\footnotesize FSRQs + BLLacs1 + SFGs1}  & 6.4 & 1.5 & 3.6 & 7.7 & 16.1 \\
\hline
{\footnotesize FSRQs + BLLacs1 + SFGs2} & 6.2 & 1.5 & 3.1 & 6.6  & 24.3 \\
\hline
\end{tabular}}
\end{center}
\end{table*}

We have performed our  $\chi^2$ analysis in three steps in which we increase the complexity
of the IGRB model: {\it i)}  the one-source scenario, in which we assume that only one
type of source, FSRQs, BL Lacs, or SFGs, contributes to the IGRB,  {\it ii)} the two-source
scenario, in which we allow for two possible contributors to the IGRB and {\it iii)} the three-source
scenario in which FSRQs, BL Lacs, and SFGs can contribute to the diffuse background.
As we already noted in all three cases the overall normalization is free, i.e., we do not impose that
the overall contribution should sum up to, or not exceed the observed IGRB. Instead we have checked that, after minimizing
the $\chi^2$, this condition is  satisfied in all cases explored.

The results of the $\chi^2$ minimization are summarized in Table \ref{tab:tabchi2} for the
one-source (upper part), two-source (middle section) and three-source (bottom) scenarios.
In the Table we list the minimum $\chi^2$ value together with the three best fit parameters, in parentheses,
 i.e., the IGRB   fraction contributed by SFGs, BL Lacs and FSRQs, respectively.
 In the one- and two-source scenarios the values of the sources not considered in the fit are set equal to zero,
 and the related space in the table is left blank for clarity.
 The two columns refer to the two different BL Lac models that we have considered (BLLacs1 and BLLacs2).
Note that we quote total $\chi^2$ values rather reduced ones, since it is not straightforward
to calculate the number of degrees of freedom involved.
This quantity, in fact, is not simply equal to the number of bins over which the $\chi^2$
is calculated due to the presence of correlation among
different catalogs, since
their redshift distributions and angular coverages  overlap significantly.
Instead, to assess the goodness of fit, we quote, for the case of the three-parameter models, the  best-fit $\chi^2$ values
for the cross-correlation between each single catalog and the {\it Fermi}
$E>$500 MeV map.
These $\chi^2$ values are presented in Table  \ref{tab:chi2catalogs},
which can be compared with the number of degrees of freedom,
given approximately by the number of bins used minus the number of fit parameters.
The results indicate that the fit to each catalog for the three-parameters models
is adequate except for a tension with the QSO CCF in the SFGs1
model that is even more prominent in the SFGs2 model.
The tension among  the models results in the under-prediction of the observed
correlation.

The main results of the $\chi^2$ analysis are:
\begin{itemize}
\item All models including a contribution from SFGs provide a
significantly better fit than those in which the IGRB is contributed to by AGN only.
 \item Model SFGs1 performs better than SFGs2. The main issue with the  SFGs2 model is
that it provides a poor fit to the CCF of the SDSS-QSOs,
as indicated in Table \ref{tab:chi2catalogs}, while for all other datasets the
two SFG models fit the data equally well, although with different overall normalizations of the SFG signal.
\item In all models explored the IGRB contribution from AGN is subdominant.
When SFGs are included among the IGRB sources, the AGN
contribution is consistent with zero.
The consistency with zero simply reflects the limited accuracy of our analysis which does not
account for the fact that, based on the observed number count distribution of the resolved $\gamma$-ray sources,
some contribution from AGN is to be expected \citep{AjelloFSRQs,AjelloBLLs}.
\item BLLacs1 and BLLacs2 models have similar $\chi^2$ values
although the normalization of the BLLacs2 component is approximately
a factor of 2 higher than the BLLacs1 model.
\end{itemize}

The uncertainties in the estimates of the parameters can be appreciated from the sets of panels
shown in Figs.~\ref{fig:chi2s_CCF_bll1_all_energy} (SFGs1 model) and \ref{fig:chi2s_CCF_bll2_all_energy}
(SFGs2 model).
We do not show results for the BLLacs2 case, since they are very similar to those of the BLLacs1
case, when one rescales the BL Lac component by a factor of $\sim$2.
Among the plots, those with the 1-dimensional $\chi^2$
represent the contribution to the IGRB from a specific type of source
that we obtain \emph{profiling} \citep{Rolke:2004mj}
over the other contributors. For example in the case of SFGs, this is the function
obtained after minimizing the $\chi^2(f_{\rm sfg},f_{\rm bllac},f_{\rm fsrq})$
with respect to $f_{\rm bllac}$ and $f_{\rm fsrq}$.
The plots show the $\chi^2$ together with the 2 $\sigma$ significance level.
The plot in the upper-right corner also shows  the derived quantity  $f_{\rm tot}$,
 i.e., the total IGRB fraction, $f_{\rm tot} = f_{\rm sfg} + f_{\rm bllac} + f_{\rm fsrq}$.
The 2D contours refer to the function obtained by profiling over
only one parameter.
In this case the contours are drawn in correspondence to the 1, 2 and 3 $\sigma$ confidence levels.
The constraints on the SFGs are rather broad: they show that, within  2 $\sigma$,
the contribution to the IGRB of these sources varies between  25-95 \%.
The constraints are tighter in the case of BLLacs2 model (50-95 \%) not shown
in the plots.
A scenario with no SFG contribution is rejected
with high statistical significance.
The contribution  of  AGNs is consistent with zero but, within 2 $\sigma$, can be as large as
 5-7\%  and 7-8\% for the FSRQs and BL Lacs, respectively.

For the SGFs2 model, which provides a worse fit than  SFGs1,
we only obtain upper limits: the SFG contributes $\le 20$ \%
to the IGRB and the contributions of BLLacs1 and FSRQs are limited to  10\% and 20\%, respectively.
Note that in this case the contribution from  FSRQs is larger than in the SFGs1 model.
Another difference between model SFGs1 and SFGs2 is the total contribution to the IGRB.
The three-source model SFGs1+ BLLacs1 + FSRQs model is able to account for
a large fraction of the IGRB, about 70-80\%,
while  the model SFGs2+ BLLacs1 + FSRQs model can only be responsible for
$\sim$20-30\% of the IGRB.
Table  \ref{tab:chi2catalogs} emphasizes the main issue with  the model
SFG2+ BLLacs1 + FSRQs, which is the poor fit to the QSO CCF.
Indeed, visual inspection of the CCF confirms that this model cannot
explain the amplitude of the measured cross-correlation signal.
Instead, model SFGs1+ BLLacs1 + FSRQs provides a better fit
to the data apart still a small residual underestimate of the QSO
correlation signal.

\begin{figure}
\centering
\epsfig{file=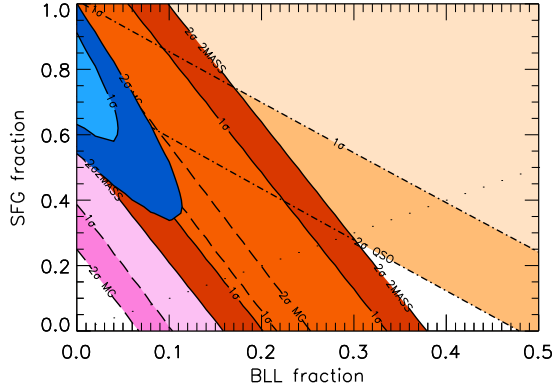, angle=0, width=0.45 \textwidth}
\caption{$\chi^2$  contours for the 2 component model SFG+BLLacs2 with respect to the 3 subsets of CCFs
with the SDSS QSOs (pale brown, dot-dashed), the 2MASS (red, solid) and the SDSS MG (pink, dashed) catalogs,
and the combined dataset (blue, solid).
The complementary tomographic information from the different catalogs helps to break the degeneracies present when using a single catalog.
\label{fig:tomography}}
\end{figure}

Finally, another interesting feature of the 2D $\chi^2$ contours is the non-negligible correlation among
the contributing fractions. The fact that these contours are not completely degenerate, however,
is a non-trivial result that we have obtained by cross-correlating the {\it Fermi} maps with different
catalogs of objects spanning different redshift ranges or, in other words, to the tomographic
nature of our $\chi^2$ analysis.
To illustrate this point, let us consider the simple two-source model  BLLacs2+SFGs1
instead of the 3-source one,  BLLacs2+SFGs1+FSRQs. The advantage is that
in this case the 2D function $\chi^2(f_{\rm sfg},f_{\rm bllac})$ encodes all information which, instead,
is partially lost when one profiles the  three-source model.
We show in Fig.~\ref{fig:tomography} the 1 and 2 $\sigma$ contours
of the SFG+BL Lac contributions  superimposed to
the 1 and 2 $\sigma$ contours obtained when only one type of catalog is considered.
Here we show the  $\chi^2$  values obtained by cross correlating the {\it Fermi} maps with
SDSS QSOs, SDSS galaxies and 2MASS galaxies.
Individual constraints are fully degenerate, as one can only constrain the ratio of the
two contributions. In particular, constraints from objects at low redshifts, like
2MASS and SDSS galaxies, would only narrow the width of the uncertainty strip, without removing
the degeneracy. It is only when we consider SDSS QSOs which  convey information on
clustering at high redshifts that we are able to remove part of the degeneracy.
Note that the combined constraints are consistent with those obtained from the analyses of the
individual catalogs.

\section{Discussion and Conclusions}
\label{sec:discussion}

In this work we have cross-correlated the {\it Fermi}-LAT sky maps obtained in 60 months
of observations with the angular positions of several types of extragalactic objects
at different redshifts. The aim is to constrain the origin of the IGRB
 under the hypothesis that it is constituted by unresolved
astrophysical sources that can be traced by but do not necessarily coincide with the
objects in the catalogs. The benefit of performing a cross-correlation analysis of the IGRB
rather than considering its mean amplitude or its auto-correlation properties is that
the cross-correlation is less prone to systematic effects
that may arise from the inaccurate cleaning of the $\gamma$-ray maps, such as
imperfect subtractions of the diffuse Galactic foreground, contributions from charged particles
and so on.

For this purpose we rely on two complementary statistical tools: the angular two-point correlation
function and its Legendre transform: the angular power spectrum. The results of our cross-correlation
analysis were compared with theoretical predictions in which one assumes that the IGRB is constituted,
in full or in part, by any of these potential candidates: SFGs,  BL Lacs and FSRQs.

The main results of our analysis are:

\begin{itemize}

\item We observe a significant ($> 3$-$4 \, \sigma$)
signal in the angular cross-correlation function of 2MASS galaxies, NVSS galaxies, and QSO with the IGRB on scales smaller
than $1^{\circ}$.
A weaker signal, $\sim 3 \, \sigma$, is also observed for SDSS main sample galaxies.
While in the case of 2MASS the cross-correlation signal is observed in all energy bands and seems to be
genuinely related to the underlying clustering properties of matter, in the case of NVSS
we interpret the CCF signature as not originating from the large-scale structures.
The NVSS signature is likely attributed, at least in part,
to undetected $\gamma$-ray sources
that have counterparts in the NVSS catalog and to spurious close pairs in the catalog that are, in fact, a single object.
The fact that a  cross-correlation signal on small scales is also observed when we consider SDSS galaxies and
SDSS QSOs  is a very interesting result as it suggests that the CCF signal does not solely originate
at redshift $z\lesssim 0.1$ , where 2MASS and SDSS galaxies are found, but is also contributed by high redshift ($z>1$) clustering
that is traced by the QSOs in the SDSS catalog.

\item The fact that we now observe a signal in several cross-correlation analyses is beyond the original expectations
of \cite{xia11} who performed an analysis similar to the one presented here using {\it Fermi}-LAT two-year maps.
Their forecast, however, was based on the assumptions that errors were dominated by Poisson noise from discrete
photon counts. Our positive result relies on several improvements that have enabled us to
efficiently remove potential sources of systematic and random errors. In particular:  {\it i)} in the map cleaning procedure we
have used three different models for the Galactic diffuse foreground that update and improve the one used in
the original analysis,  {\it ii)} we were able to excise a larger number of individually resolved sources from the $\gamma$-ray maps
thanks to the  most recent LAT source catalogs,  {\it iii)} the addition of another set of discrete sources, the SDSS Main Galaxy
catalog, in our cross-correlation analyses and  {\it iv)} a better characterization of the  PSF of the LAT.
The latter improvement is probably the most crucial since it not only allows us a better comparison between model and data, but also
allows to push our analysis down to 500 MeV, significantly increasing the photon statistics and reducing the amplitude of
statistical errors, and to smaller angular scales than \cite{xia11}, below $1^{\circ}$, where
the signal is the most prominent.

\item  We have verified with a series of dedicated tests that the
 results of our cross-correlation analysis are robust to {\it i)} the cleaning procedure of the {\it Fermi} maps, {\it ii)} the
subtraction of the Galactic Diffuse Foreground,  {\it iii)} the removal of the resolved $\gamma$-ray sources, {\it iv)}
 the choice of the mask,  {\it v)}  the $\gamma$-ray conversion layer in the LAT,  {\it vi)} the statistical estimator used to measure the angular cross-correlation function and {\it vii)}
the method adopted to assess the uncertainties in the CCF and CAPS and their covariance.
In addition, we have verified that our characterisation and treatment of the PSF of the telescope is good and does not introduce
any significant systematic error in the comparison between models and data.

\item The comparison between measured cross-correlation signals and model predictions indicates that the best-fit to the data
is obtained when SFGs are the main, if not the only, contributors
to the IGRB (possibly degenerate with MAGN, see below) and AGN provide a minor, possibly
negligible, contribution.
We have explored different combinations of sources and different models for the $\gamma$-ray contribution from
BL Lacs and SFGs. Models that include SFGs outperform those that consider AGN only.
And among the SFG models explored, the one proposed by   \cite{2010arXiv1003.3647F} that includes
the effect of gas quenching and its redshift dependence provides a better fit to the data than the one proposed
by \cite{tamborra014} which, instead, ignores this effect.

 Our $\chi^2$ analysis makes these statements more quantitative
and shows that, for the model that provides the best-fit,
SFGs contribute to $72^{+23}_{-37}$ \% (but see our discussion of MAGN below) of the IGRB (2 $\sigma$ confidence interval) whereas
BL Lac and FSRQs provide similar contributions ranging from 0 to 8\% each.
In none of our best fit models does the contribution to the IGRB total up to 100 \%. This is an interesting result
that keeps open the possibility that other types of sources could contribute
to the $\gamma$-ray background. In the framework of the cross-correlation analysis one
 can only speculate on the nature of these sources. Among the
different options, the possibility that they consist of astrophysical sources at high redshifts, that would not
be detected by our cross-correlation analysis, or that  they originate from the annihilation or decay of
dark matter particles, are especially intriguing and will be investigated in future analyses.

\item Model predictions depend on a number of parameters, including
the bias relation of the mass tracer. Current models of galaxy evolution do not
provide reliable predictions for the bias
of BL Lacs, FSRQs and SFGs  which are only weakly constrained by observations.
For this reason we run a series of robustness tests in which we have considered
the alternative bias models described in Section~\ref{sec:model}.

For BL Lacs and FSRQs we have considered the case of constant bias $b_{FSRQ}=b_{BLLac}=1.04$
as well as that of a $z-$dependent bias matching that of a $10^{13} \, {\rm M}_{\odot}$ dark matter halo.
 The first scenario predicts a larger cross-correlation signal
for low-redshift objects (i.e., 2MASS and SDSS galaxies) than in the reference case.
It is a $\sim 20$\% effect that
improves the match between FSRQ model and data. At higher redshift the cross-correlation
slightly decreases. However the effect is very small and doesn't affect the outcome of the
model vs. data  comparison.
In the second scenario the bias is systematically larger than the reference one at all redshifts,
significantly increasing the amplitude of the predicted cross-correlation. The net result is that, in this
rather extreme case, our conclusion that SFGs contribute to the bulk of the IGRB still holds.
The major change is that the  predicted contributions from BL Lacs and FSRQs are unlikely to differ from zero.

For the SFGs we have considered an alternative model in which the bias is set equal to that of
 a $10^{12}\, {\rm M}_{\odot}$ dark matter halo. The bias of this object is larger than the reference value
 $b_{SFG}=1$ at all redshifts. As a result the  amplitude of the cross-correlation signal is expected to increase.
 However,
the fractional increase is very small ($<$10\%), and,
due to
the large error bars of observed cross-correlation data points, this change does not significantly affect our
main conclusion that the IGRB is mainly produced by the SFGs.

\item Our results seem to be consistent, within the uncertainties, with the outcome of different, independent
analyses.
\cite{AjelloBLLs} have been able to estimate the contribution of unresolved BL Lacs to the IGRB from their $\gamma$-ray
luminosity function measured from LAT data, and found that they do not account for more than
10-15 \% of the IGRB signal, consistent with our results. In similar analyses focused on the
FSRQs \cite{dermer07}, \cite{inoue09}, \cite{inoue10}, and \cite{AjelloFSRQs} have found that these objects provide a similar contribution
($\sim 10$ \%) to the IGRB, again in agreement with our results, possibly increasing to $\sim 20$ \%
when one accounts for objects with misaligned jets.

As for SFGs, the estimate of \cite{Achermann12} of a 4-23\% contribution to the IGRB
is consistent with our estimate of 20-95\%, which, although favors  a higher value,
has a  large uncertainty.
A larger contribution from SFGs has also been recently suggested by \cite{tamborra014}
which might be due to accounting for
SFGs at $z>2$ in the IR LF that have been
recently observed by by Herschel PEP/HerMES \citep{gruppioni013}.
It is also worth noticing that, within the 2 $\sigma$ error bar, our results are also consistent with those of
\cite{2010arXiv1003.3647F} that, based on the extrapolation of the $\gamma$-ray production in the MW, find that
SFG contribute to $\sim 50$~\% of the IGRB (with rather large uncertainties).

\item
In our analysis we have ignored MAGNs, even if they are likely to contribute the IGRB and its fluctuations, and
restricted our modeling to SFG and blazars.
The reason for this is that the expected  cross-correlation signal from these sources is robust to uncertainties in their bias parameters, whereas
in the MAGN case model predictions are much more sensitive to both their contribution to IGRB at high redshift and their (large) bias.
Indeed, we find that, within the current uncertainties, their contribution to CCF and CAPS is degenerate with that of SFGs.
One should keep this in mind when interpreting the results of our cross-correlation analysis. It might underestimate the
expected cross-correlation signal at high redshift and, consequently, overestimate the SFG contribution whereas, in fact, part of the
observed cross-correlation may be due to MAGNs.
Possible ways to isolate the contribution of MAGNs are more stringent observational or theoretical constraints on their bias and
cross-correlation analyses with catalogs of high redshift objects.

\end{itemize}

The results of our work indicate possible directions for future research.
Our analysis, which is
mostly sensitive to sources at $z<2$ suggests that the combined emission from SFGs, BL Lacs and  FSRQs
within this redshift does not completely account for the whole diffuse  IGRB signal.  Extending our cross-correlation analysis to
higher redshifts, using deeper catalogs of extragalactic sources can provide
further information to clarify this scenario.

While we observe a significant cross-correlation signal, the amplitude of the errors is still too large to efficiently discriminate
among alternative IGRB models. We have learned that {\it Fermi} IGRB maps
improve in both accuracy and precision with time, not only because of
the better photon statistics but also thanks to the revised Galactic Diffuse model, better characterization
of the LAT PSF and to the identification and subtraction of an increasing number of point sources.
 We therefore expect that
errors will be further reduced with the next {\it Fermi} data releases.
Major improvements are also expected from multiwavelength catalogs,
 since the next few years will see the advent of next-generation galaxy redshift catalogs
like eBOSS\footnote{ http://www.sdss3.org/future/}, DESI \citep{schlegel11} and Euclid \citep{euclidRB}
extending over a large faction of the sky and containing
tens of millions to billions of objects with spectroscopic or photometric redshifts.
With these future surveys, we not only expect to reduce the uncertainties in the cross-correlation analysis but also to
be able to fully exploit their tomographic potential which we have only started exploring in this work.

\section*{acknowledgments}
JX is supported by the National Youth Thousand Talents Program, the National Science Foundation of China under Grant No. 11422323 and the Strategic Priority Research Program ``The Emergence of Cosmological Structures'' of the Chinese Academy of Sciences, Grant No. XDB09000000. EB and MV are supported by INFN-PD51 INDARK.
MV is supported by  FP7 ERC starting grant "cosmoIGM" and by PRIN INAF "A complete view on the first 2 billion years of galaxy formation"  and PRIN MIUR.
EB  acknowledges the financial support provided by
MIUR PRIN 2011 'The dark Universe and the cosmic evolution of baryons: from current surveys to Euclid'
and  Agenzia Spaziale Italiana for financial support from the agreement ASI/INAF/I/023/12/0.
The work of AC is supported by the research grant {\sl Theoretical Astroparticle Physics} number 2012CPPYP7 under the program PRIN 2012  funded by the Ministero dell'Istruzione, Universit\`a e della Ricerca (MIUR), by the research grant {\sl TAsP (Theoretical Astroparticle Physics)}
funded by the Istituto Nazionale di Fisica Nucleare (INFN).
We wish to thank Keith Bechtol, Seth Digel, Luca Latronico Philippe Bruel, Nicola Omodei,
Eric Charles, Luca Baldini and Massimo Persic for a careful reading of the manuscript and for providing
useful comments.

The \textit{Fermi} LAT Collaboration acknowledges generous ongoing support
from a number of agencies and institutes that have supported both the
development and the operation of the LAT as well as scientific data analysis.
These include the National Aeronautics and Space Administration and the
Department of Energy in the United States, the Commissariat \`a l'Energie Atomique
and the Centre National de la Recherche Scientifique / Institut National de Physique
Nucl\'eaire et de Physique des Particules in France, the Agenzia Spaziale Italiana
and the Istituto Nazionale di Fisica Nucleare in Italy, the Ministry of Education,
Culture, Sports, Science and Technology (MEXT), High Energy Accelerator Research
Organization (KEK) and Japan Aerospace Exploration Agency (JAXA) in Japan, and
the K.~A.~Wallenberg Foundation, the Swedish Research Council and the
Swedish National Space Board in Sweden.

Additional support for science analysis during the operations phase is
gratefully acknowledged from the Istituto Nazionale di Astrofisica in
Italy and the Centre National d'\'Etudes Spatiales in France.

Some of the results in this paper have been derived using the HEALPix \citep{2005ApJ...622..759G}
package.

\bibliographystyle{apj}
\bibliography{version5.11}

\end{document}